\documentclass[10pt]{article}
\usepackage{amsfonts,amsmath,color,graphicx,stmaryrd,mathabx,amssymb,mathdots,esint,tikz}

\DeclareMathAlphabet{\mathpzc}{OT1}{pzc}{m}{it}

\makeatletter
\@addtoreset{equation}{section}
\makeatother
\newtheorem{theorem}{Theorem}[section]
\newtheorem{conjecture}[theorem]{Conjecture}
\newtheorem{proposition}[theorem]{Proposition}
\newtheorem{lemma}[theorem]{Lemma}
\newtheorem{corollary}[theorem]{Corollary}
\newtheorem{porism}[theorem]{Porism}
\newtheorem{example}{Example}[section]
\newtheorem{definition}[example]{Definition}

\newtheorem{remark}[example]{Remark}
\newtheorem{hypothesis}[example]{Hypothesis}
\def\br{\begin{remark}\rm\small}
\def\er{\end{remark}}
\def\bt{\begin{theorem}}
\def\et{\end{theorem}}
\def\bcj{\begin{conjecture}}
\def\ecj{\end{conjecture}}
\def\bd{\begin{definition}}
\def\ed{\end{definition}}
\def\bp{\begin{proposition}}
\def\ep{\end{proposition}}
\def\bl{\begin{lemma}}
\def\el{\end{lemma}}
\def\bc{\begin{corollary}}
\def\ec{\end{corollary}}
\def\bh{\begin{hypothesis}}
\def\eh{\end{hypothesis}}
\def\beaq{\begin{eqnarray}}
\def\eeaq{\end{eqnarray}}

\newcommand{\proof}[1]{{\noindent \bf proof:}\par
{#1} $\square$}

\newcommand{\ii}{{\mathrm{i}}}
\newcommand{\sheet}[2]{\stackrel{{#2}}{#1}}
\newcommand{\om}{\omega}
\newcommand{\ee}[1]{{{\rm e}^{#1}}}
\newcommand{\Tr}{\mathrm{Tr}\,}
\newcommand{\beq}{\begin{equation}}
\newcommand{\eeq}{\end{equation}}
\newcommand{\bea}{\begin{eqnarray}}
\newcommand{\eea}{\end{eqnarray}}

\newcommand{\dd}{\mathrm{d}}

\newcommand{\Res}{\mathop{\,\rm Res\,}}

\newcommand{\Ms}{\mathpzc{M}}
\newcommand{\Hs}{\mathpzc{H}}
\newcommand{\Ls}{\mathpzc{L}}

\definecolor{rouge}{rgb}{0.84,0.18,0.07}
\definecolor{bleu}{rgb}{0.22,0.41,0.74}
\definecolor{vertf}{rgb}{0.08,0.46,0.07}
\usepackage[pdftex]{hyperref}
\hypersetup{colorlinks,urlcolor=vertf,citecolor=rouge,linkcolor=bleu,filecolor=black}

\begin{document}

\sloppy

\pagestyle{empty}
\addtolength{\baselineskip}{0.20\baselineskip}
\begin{center}

\vspace{26pt}

{\large \textbf{Abstract loop equations, topological recursion and applications}}

\vspace{26pt}

\textsl{Ga\"etan Borot}\footnote{D\'epartement de Math\'ematiques, Universit\'e de Gen\`eve. \href{mailto:gaetan.borot@unige.ch}{\textsf{gaetan.borot@unige.ch}}}, \textsl{Bertrand Eynard}\footnote{Institut de Physique Th\'eorique, CEA Saclay. Centre de Recherche Math\'ematiques, Montr\'eal, QC, Canada. \href{mailto:bertrand.eynard@cea.fr}{\textsf{bertrand.eynard@cea.fr}}}, \textsl{Nicolas Orantin}\footnote{Department of Mathematics, Instituto Superior T\'ecnico, Lisboa. \href{mailto:nicolas.orantin@math.ist.utl.pt}{\textsf{nicolas.orantin@math.ist.utl.pt}}}
\end{center}

\vspace{20pt}

\begin{center}
\textbf{Abstract}
\end{center}

\noindent \textsf{We formulate a notion of "abstract loop equations", and show that their solution is provided by a topological recursion under some assumptions, in particular the result takes a universal form. The Schwinger-Dyson equation of the one and two hermitian matrix models, and of the $O(n)$ model appear as special cases. We study applications to repulsive particles systems, and explain how our notion of loop equations are related to Virasoro constraints. Then, as a special case, we study in detail applications to enumeration problems in a general class of non-intersecting loop models on the random lattice of all topologies, to $\mathrm{SU}(N)$ Chern-Simons invariants of torus knots in the large $N$ expansion. We also mention an application to Liouville theory on surfaces of positive genus.}

%





\vspace{26pt}
\pagestyle{plain}
\setcounter{page}{1}


\section{Introduction}

The topological recursion \cite{EOFg,EORev} is a universal structure, formulated axiomatically in terms of algebraic geometry on a curve. To the data of a complex curve $\mathcal{C}$, a meromorphic function $x$ on $\mathcal{C}$, a meromorphic $1$-form $\omega_1^0$, a meromorphic symmetric $2$-form $\omega_2^0$ on $\mathcal{C}^2$, it associates a sequence of meromorphic, symmetric $n$-forms $\omega_n^g(z_1,\ldots,z_n)$ on $\mathcal{C}^n$ (the correlators), and a sequence of numbers $F^g$ (the free energies), which are "symplectic invariants" of the initial data. They are in a certain sense the unique solution to a hierarchy of linear and quadratic loop equations, which are closely related to Virasoro constraints \cite{M91}.

It has been first identified as the underlying structure of the large $N$ expansion in the $1$-hermitian matrix models \cite{E1MM}, as a culmination of the moment method developed in \cite{ACM92,ACKM,ACKMe,Ake96,AMM04,AMM07,AMM072,AMMP09}. In this case, $y^2 = \prod_{i = 1}^{2n} (x - a_i)$ and $\mathcal{C}$ was a hyperelliptic curve and $\omega_1^0=y \dd x$. Then, it has been shown to hold in the same form in the $2$-matrix models \cite{EORes,CEO06} and in the chain of hermitian matrices \cite{EORev}. In this case, $x$ and $y$ can be arbitrary meromorphic functions on a compact Riemann surface $\mathcal{C}$, and $\omega_1^0=y\dd x$. Then it was observed that the topological recursion makes sense and some of its properties are preserved with weaker assumptions on the triple $(\mathcal{C},x,y)$ called \emph{spectral curve}. It has been found in applications to enumerative geometry of moduli spaces \cite{Ekappa,BEMS,EMS,Norbu,MuPen,Einter}, especially in Gromov-Witten theory on toric Calabi-Yau $3$-folds \cite{BKMP,EOBKMP} where the relevant spectral curves are such that $e^{x}$ and $e^{y}$ are meromorphic on a compact Riemann surface $\mathcal{C}$. In combinatorics, the topological recursion structure has also been shown to solve the problem of enumerating maps \cite{Ebook}, and more recently enumerating maps carrying certain statistical physics models like the Ising model \cite{EORes} or self-avoiding loops models \cite{BEOn}. In the latter case, the relevant spectral curve $\mathcal{C}$ is a torus but $y$ is a multivalued function on $\mathcal{C}$. A deformation of the topological recursion by a parameter $\hbar$ has also been defined. The case of $\hbar \ll 1$ was treated in \cite{CE06,C06}: it remains in the framework of algebraic geometry, and governs the large $N$ expansion of the beta ensembles for fixed $\beta \neq 0$ (see \cite{WZ06} for definition and references therein), with identification:
\beq
\hbar = \frac{1}{N}\Big(\sqrt{\frac{\beta}{2}} - \sqrt{\frac{2}{\beta}}\Big).
\eeq
The case of $\hbar \in O(1)$ rather lives in the realm of geometry of $D$-modules, is currently being developed \cite{EMq0,CEMq1,CEMq2,CEMP}, and has potential applications in refined topological string theories. 

In this article, we will show that the topological recursion also governs the large $N$ expansion in generalized matrix models, which have been called "repulsive particle systems" in the recent work \cite{Venker1}. Those are statistical mechanical models whose partition function can be written, if we have only one species of particles,
\beq
Z_N = \int  \prod_{1 \leq i < j \leq N} R_0(\lambda_i,\lambda_j)
\prod_{i = 1}^N \dd\lambda_i\,e^{-N\,V(\lambda_i)}
\eeq
considered as a convergent integral. We usually assume that $R(\lambda_i,\lambda_j)\,\propto\,(\lambda_i - \lambda_j)^{\beta}$ at short distances, for some $\beta > 0$. We recall that $\hbar = 0$ corresponds to $\beta = 2$. Such models are ubiquitous in theoretical physics and enumerative geometry, even for $\beta = 2$. They appear in statistical physics on the random lattice \cite{PZJ6v,SportiSF}, in the theory of random partitions \cite{Epart,Eplanepart}, in supersymmetric gauge theories \cite{HKK,DiVa,Nekra,Sulko}, in topological strings \cite{HiggN}, in Chern-Simons theory \cite{Marinounp,Law,MarinoCSM}, etc.

This article begins with shaping a notion of "abstract $\hbar = 0$ loop equations", and show that they are solved by the same topological recursion which was formulated in \cite{EOFg} (Section~\ref{S1}). The key result about abstract loop equations is Proposition~\ref{2222}, and we show that many of the properties of the usual topological recursion are preserved\footnote{For the skilled reader, we anticipate by saying that the symplectic invariance is not expected to hold, as we comment in Conclusion.}. The initial data for this recursion is a 1-form $\omega_1^0$ (which was $y\dd x$) and a symmetric 2-form $\omega_2^0$ (which was a fundamental $2$-form of the $2^{\mathrm{nd}}$ kind, also called "Bergman kernel", when $\mathcal{C}$ is a curve).  In this way, we retrieve all previous avatars, like the 1-hermitian matrix model, the 2-hermitian matrix model (see \S~\ref{2mm}), in one-cut or multi-cut regimes. Their solution by a topological recursion had been obtained case by case so far, but the reason for existence of a universal solution was still missing. This article, especially Section~\ref{S7}-\ref{S5}-\ref{2mm}, solves this puzzle by putting the Schwinger-Dyson equations of those matrix models under the same roof. We also find interesting new applications. We illustrate the theory on four such new examples, and find that the topological recursion governs:
\begin{itemize}
\item[$\bullet$] the $1/N$ expansion of systems of repulsive particles, when it exists (Section~\ref{S3}). For fixed $\beta > 0$ different from $2$, a generalization along the lines of \cite{CE06} is possible, but is left aside in order to keep this article focused. It is natural to include several species of particles, which have species-dependent pairwise interactions. We shall see that an assumption of strict convexity of the pairwise interaction plays a key role (Definition~\ref{stconv}) in the construction of the relevant spectral curve. The proof of existence of a full asymptotic expansion in $1/N$ in such models in the one-cut regime or the multi-cut regime with fixed filling fractions is the matter of another ongoing work \cite{BGK}, and is not the concern of this article, which takes it (unless mentioned) as an assumption. Our main results are formulated in Proposition~\ref{kiosa} and Corollary~\ref{4333} (resp. Proposition~\ref{mik} and Corollary~\ref{aporo} for the multi-species case).
\item[$\bullet$] the enumeration of maps with a loop configuration, in all topologies, with uniform boundary conditions (Section~\ref{S5}). This is the "formal integral" counterpart of systems of repulsive particles, and the introduction of several species of particles has also a natural combinatorial origin in the model, as introducing colors for domains separated by the loops. Contrary to the convergent case, the proof of our results here is complete and does not rely on extra assumptions. We also treat height models on maps of all topologies with boundaries of fixed heights in \S~\ref{SQADE}, where heights take values at vertices of a ADET or $\hat{\mathrm{A}}\hat{\mathrm{D}}\hat{\mathrm{E}}$ Dynkin diagram. We indeed observe that among heights models, those are special because they lead to strictly convex pairwise interactions (Lemma~\ref{uj}).
\item[$\bullet$] the large $N$ expansion of torus knot invariants computed in $\mathrm{U}(N)$ Chern-Simons theory (Section~\ref{STknot}), where we justify a proposal of \cite{BEMknots}. Here we also justify the existence of the $1/N$ expansion.
\item[$\bullet$] the large impulsion expansion of Liouville correlation functions on a surface of positive genus (Section~\ref{SkI}), for which we stay in this article at a formal level. In particular, we do not address important issues of convergence, choice of contours of integrations and characterization of the cuts of the spectral curve.
\end{itemize}
On the way, we explain in Section~\ref{S7} how abstract loop equations for repulsive particle systems can be identified with Virasoro constraints after a non-linear change of times. We also illustrate in \S~\ref{S455}-\ref{SkI} concerning repulsive particle systems on positive genus surfaces, that ad hoc definitions of the correlators can sometimes simplify the analysis. We present our conclusions in Section~\ref{conclu}. A table of notations is collected in Appendix~\ref{appB}.

\section{Abstract loop equations}
\label{S1}
\subsection{Notion of domain}
\label{ss1}
We first collect some notations and definitions. A closed arc (resp. an open arc) is a piecewise smooth embedding of $\mathbb{S}_1$ (resp. of $[0,1]$) to a Riemann surface. Let $U$ be an open subset of a Riemann surface and $p$ be a point. We denote in particular:
\begin{definition}
\label{dea1}
\begin{itemize}
\item[$\phantom{\bullet}$] 
\item[$\bullet$] $\Ms(U)$ (resp. $\Hs(U)$), the space of meromorphic (resp. holomorphic) $1$-forms on $U$;
\item[$\bullet$] $\Ms'(\{p\})$ (resp. $\Hs'(\{p\})$), the space of germs of meromorphic (resp. holomorphic) $1$-forms at $p$ ;
\item[$\bullet$] $\Ms'_{-}(\{p\})$, the quotient space $\Ms'(\{p\})/\Hs'(\{p\})$.
\end{itemize}
\end{definition}
If $\xi$ is a local coordinate around $p$ such that $\xi(p) = 0$, $\Ms'_-(\{p\})$ can be identified with the space of polynomials in $(\xi(p))^{-1}$.
\begin{definition}
\label{domainn}We call $U$ a \emph{domain} if $\partial U$ consists of a nonempty, finite disjoint union of smooth, closed arcs $(\gamma_j)$, and is equipped with an involutive, orientation reversing diffeomorphism $\iota$.
\end{definition}

\begin{figure}
\begin{center}
\includegraphics[width=\textwidth]{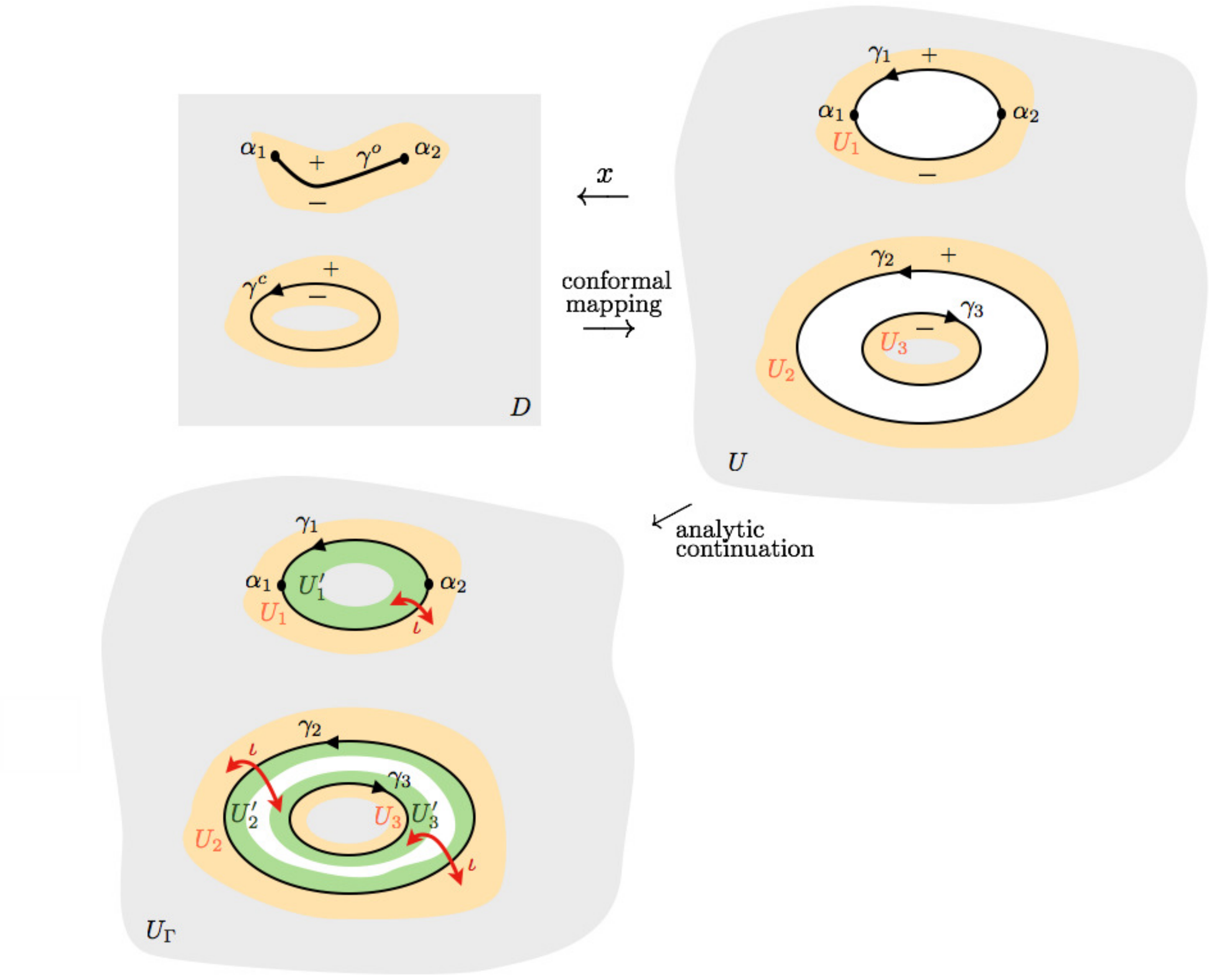}
\caption{\label{Fig1} Construction of a domain, and its continuation across $\Gamma$. $\gamma_1$ (resp. $\gamma_2 \dot{\cup} \gamma_3$) is the image of the cut $\gamma^o$ (resp. $\gamma^c$) after opening the cut by conformal mapping. $V_j$ is the neighborhood of $\gamma_j$ in $U_{\Gamma}$ obtained as the union of $U_j$ (in yellow), $U_j'$ (in green) and $\gamma_j$.}
\end{center}
\end{figure}

The main example of domains we have in mind can be constructed from an oriented Riemann surface $\Sigma$ which may have smooth boundaries, and a collection of open arcs $(\gamma_j^{o})_{1 \leq j \leq r_o}$ and closed arcs $(\gamma_{j}^{c})_{1 \leq j \leq r_c}$ on $\Sigma$ (see Fig.~\ref{Fig1}). We consider $D = \Sigma\setminus\big(\bigcup_{j = 1}^{r_o} \gamma_j^{o}\cup\bigcup_{j = 1}^{r_c} \gamma_{j}^c\big)$. The topological boundary of $D$ is the disjoint union of $r = r_0 + 2r_c$ connected components coming either from open arcs and closed arcs:
\begin{itemize}
\item[$\bullet$] Open arcs yield a component $\gamma_{j} = (\gamma_{j}^o)_1\amalg_{E_j} (\gamma_j^o)_2$, where $(\gamma_j^{o})_{a}$ for $a = 1,2$ are two copies of $\gamma_j^{o}$, and $E_j = \partial\gamma_j^o$ is the set of endpoints of $\gamma_j$. It is naturally equipped with an involution $\iota_j^{o}$ which sends a point of $(\gamma_j^o)_{1}$ to the same point on $(\gamma_j^o)_2$;
\item[$\bullet$] Closed arcs yield two components $(\gamma_j^{c})_1$ and $(\gamma_j^{c})_2$ corresponding to the exterior and the interior of $\gamma_j^{c}$. Their disjoint union is naturally equipped with an involution $\iota_{j}^{c}$, which sends a point on $(\gamma_j^o)_{1}$ to the same point on $(\gamma_j^{c})_2$.
\end{itemize}
We denote $\overline{D} = D\cup\partial D$ the topological completion of $D$. Then, one can always find a conformal mapping from $D$ to some $U$ with everywhere smooth boundary, which extends to a homeomorphism from $\overline{D}$ to $\overline{U}$, but behaves as a squareroot near $E_j \hookrightarrow \overline{D}$. Thus, $U$ is a domain, with an orientation reversing involution $\iota$ defined globally on $\partial U$. By similar uniformization arguments, we could also allow $\Sigma$ to have only piecewise smooth boundaries.

Let $U$ be a domain, and $\Gamma = \bigcup_{j = 1}^{r} \gamma_j$ be a subunion of connected components of $\partial U$ (it will be $\partial U$ itself unless precised). Let $U_j \subseteq U$ be an annular open neighborhood of $\gamma_j$, and $U'_j$ be another copy of $U_j$. We may glue $U_j'$ to $U$ along $\gamma_j$ to define a Riemann surface $U_{\Gamma}$, and identify $U$ and $\gamma_j$ to their image in $U_{\Gamma}$ by inclusion. $V_j = U_j \amalg_{\gamma_j} U_j'$ is an open neighborhood of $\gamma_j$ in $U_{\Gamma}$. The orientation reversing involution $\iota$ initially defined on $\Gamma$ can be extended uniquely to a holomorphic involution on $V = \coprod_{j = 1}^r V_j$. Later, we need to introduce smooth arcs $\gamma_j^{\mathrm{ext}} \subseteq U_j$ which are homotopic in $V_j$ to $\gamma_j$ (by definition, they do not intersect $\gamma_j$), and their image $\gamma_j^{\mathrm{int}} = \iota(\gamma_j^{\mathrm{ext}})$. We also introduce $\Gamma^{\mathrm{ext}} = \bigcup_{j = 1}^r \gamma_j^{\mathrm{ext}}$.

\subsection{Spaces of continuable functions}

\begin{definition}
\label{def22} Let $\Hs_{\Gamma}(U)$ be the space of \emph{continuable} $1$-forms across the boundary components $\Gamma$. It is defined as the space of holomorphic $1$-forms on $U$, which can be extended as meromorphic $1$-forms on $U_{\Gamma}$ defined for small enough neighborhoods $U_j$:
\beq
\Hs_{\Gamma}(U) = \Hs(U)\cap\Big(\mathop{\underrightarrow{\lim}}_{U\amalg\Gamma \hookrightarrow U_{\Gamma}} \Ms(U_{\Gamma})\Big).
\eeq
\end{definition}
We make an abuse of notations and identify a $1$-form $f \in \Hs_{\Gamma}(U)$ with the unique meromorphic $1$-form on some $U_{\Gamma}$ which coincides with $f$ on $U \subseteq U_{\Gamma}$. The involution $\iota$ acting on $V$ can be used to define linear operators $\Delta,\mathcal{S}\,:\,\Ms(V) \rightarrow \Ms(V)$ by the formulas:
\beq
\Delta f (z) = f(z) - (\iota^*f)(z),\qquad \mathcal{S} f(z) = f(z) + (\iota^* f)(z).
\eeq
By abuse of notations, we shall write later $f(\iota(z)) \equiv \iota^* f(z)$. Conversely, we have:
\beq
\forall f \in \Hs_{\Gamma}(U),\quad \forall z \in V \, , \qquad f(z) = \frac{\mathcal{S} f(z) + \Delta f(z)}{2},\qquad f(\iota(z)) = \frac{\mathcal{S} f(z) - \Delta f(z)}{2}.
\eeq
Since $\iota$ is defined on $V =  \coprod {V_j}$, one may consider it as a collection of local involutions $\iota_j = \iota|_{V_j}$ by restricting the involution to $V_j$ for $j \in \ldbrack 1,r \rdbrack$. In the same way, one can view $\Delta$ and $\mathcal{S}$ as a collection of local operators $\Delta_j$ and $\mathcal{S}_j$, defined on $\Ms(V_j)$. We also point out a polarization formula which becomes useful later: for any $1$-forms $f,g \in \Ms(V)$, we have:
\beq
\label{pola} f(z)g(\iota(z)) + f(\iota(z))g(z) = \frac{1}{2}\big(\mathcal{S}f(z)\,\mathcal{S}g(z) - \Delta f(z)\,\Delta g(z)\big).
\eeq
We say that $f \in \Hs(U)$ extends continuously to $\Gamma \subseteq \partial U$ when, for any coordinate $x$ on $\Sigma$ locally defined on an open set $O \subseteq U$ such that $\overline{O}$ intersects $U$, $f/\dd x$ initially defined and holomorphic on $O\cap U$ extends to a continuous function on $O \cap \overline{U}$.
\begin{definition}
\label{dea2}By Schwarz reflection principle, the subspace of $\Hs(\coprod_{j = 1}^r U_j)$ consisting of $1$-forms which extend continuous, $\iota$-invariant functions on $\Gamma$, can be identified with the subspace of $\iota$-invariant holomorphic $1$-forms in $V$. We denote it $\Hs_{\Gamma}^{\mathrm{inv}}(V)$.
\end{definition}
We want to consider classes of continuable $1$-forms for which a Cauchy residue formula holds. This leads us to:
\begin{definition}
\label{Cau}A \emph{local Cauchy kernel} $G(z,z_0)$ is a meromorphic $1$-form in $z_0 \in U_{\Gamma}$ and a meromorphic function in $z \in V$, with only singularity in its first variable a simple pole at $z = z_0$, such that locally:
\beq
G(z,z_0) \sim \frac{dz_0}{z_0-z} + {\rm analytical}
\eeq
\end{definition}
\begin{definition} \label{reapo}\label{normdef}A subspace $\mathcal{H}$ of $\Ms(U_{\Gamma})$ is normalized if $\mathcal{H}\cap\Hs(U_{\Gamma}) = \{0\}$.
\end{definition}
\begin{definition}
\label{repa}A subspace $\mathcal{H}$ of $\Hs_{\Gamma}(U)$ is \emph{representable by residues} if there exists a local Cauchy kernel $G(z,z_0)$ such that, for any $f$ in $\mathcal{H}$, the function $\tilde{f}$ in $\Ms(U_{\Gamma})$ defined by:
\beq
\label{resp}\tilde{f}(z_0) = \sum_{p \in V} \Res_{z \rightarrow p} G(z,z_0)\,f(z).
\eeq
has the same poles as $f$, i.e. $(f - \tilde{f}) \in \Hs(U_{\Gamma})$. 
\end{definition}
By definition, a $1$-form belonging to $\mathcal{H}$ can only have a finite number of poles in $V$ chosen small enough, so the sum in \eqref{resp} is finite.
\begin{definition}
\label{dea3}We define the subspace $\Ls_{\Gamma}(U) \subseteq \Hs_{\Gamma}(U)$ consisting of $1$-forms $f$ such that
$\mathcal{S} f \in \Hs_{\Gamma}^{\mathrm{inv}}(V)$.
\end{definition}
The key property of a $1$-form $f \in \Ls_{\Gamma}(U)$ is that $\mathcal{S} f$ is holomorphic in a neighborhood of $\Gamma$ in $U_{\Gamma}$. Thus, its behavior at poles is determined by that of $\Delta f(z)$.
\begin{lemma}
\label{mais}Assume we have a subspace $\mathcal{H} \subseteq \Ls_{\Gamma}(U)$ representable by residues. Then:
\beq
\label{reqi}\forall f \in \mathcal{H},\quad \forall z_0 \in U_{\Gamma},\qquad \tilde{f}(z_0) = \sum_{p \in V} \Res_{z \rightarrow p} \frac{\Delta^{z} G(z,z_0)}{4}\,\Delta f(z),
\eeq
where the superscript $z$ indicates the variable on which the operator $\Delta$ acts. 
\end{lemma}
\textbf{Proof.} Let $f \in \mathcal{H}$. We remark that $p \in V$ is a pole of $f$ iff $\iota(p)$ is a pole of $f$, and we compute from \eqref{resp}:
\bea
\tilde{f}(z_0) & = & \frac{1}{2} \sum_{p \in V} \Res_{z \rightarrow p} \mathcal{S}^{z}[G(z,z_0)\,f(z)] \nonumber \\
& = & \frac{1}{4} \sum_{p \in V} \Res_{z \rightarrow p} [\mathcal{S}^{z} G](z,z_0)\,\mathcal{S} f(z) + [\Delta^{z} G](z,z_0)\,\Delta f(z) \nonumber \\
& = & \sum_{p \in V} \Res_{z \rightarrow p} \frac{[\Delta^{z} G](z,z_0)}{4}\,\Delta f(z).
\eea
Indeed, since $f \in \Ls_{\Gamma}(U)$ is holomorphic at the poles, the last term does not contribute to the residue. \hfill $\Box$

\subsection{Properties of representable subspaces}
\label{Htil}
We denote $\tilde{\mathcal{H}}$ the subspace of $\Ms(U_{\Gamma})$ spanned by $\tilde{f}$ when $f$ runs  in $\mathcal{H}$.

\begin{lemma}\label{l21} If $\mathcal{H}$ is representable by residues, so are $\tilde{\mathcal{H}}$ and $\mathcal{H} + \tilde{\mathcal{H}}$ with same local Cauchy kernels. Besides, for any $f \in \mathcal{H}$, $\tilde{\tilde{f}} = \tilde{f}$. Hence, $\tilde{\mathcal{H}}$ is normalized and $\mathcal{H} + \tilde{\mathcal{H}}$ is normalized iff $\mathcal{H}$ is normalized.
\end{lemma}
\textbf{Proof.} For any $f \in \mathcal{H}$, since $(\tilde{f} - f)$ is holomorphic in a neighborhood of $\Gamma$:
\beq
\tilde{f}(z_0) = \sum_{p \in V} \Res_{z \rightarrow p} G(z,z_0)\,f(z) = \sum_{p \in V} \Res_{z \rightarrow p} G(z,z_0)\,\tilde{f}(z).
\eeq
Hence $\tilde{\mathcal{H}}$ is representable by residues with local Cauchy kernel $G$, and the right-hand side coincides with $\tilde{\tilde{f}}$ by definition. \hfill $\Box$

\vspace{0.2cm}

\noindent Given a local Cauchy kernel, we consider the linear map:
\beq
\label{eq29} \begin{array}{rcl} \bigoplus_{p \in V} \Ms'_{-}(\{p\}) & \longrightarrow & \Ms(U_{\Gamma}) \\
(g_{p})_{p} & \longmapsto & \tilde{g}(z_0) = \sum_{p \in V} \Res_{z \rightarrow p} G(z,z_0)\,g_{p}(z), \end{array}
\eeq
and denote $\mathcal{H}_{G}$ its image. The $1$-forms on the right-hand side behave like $g_{p}(z_0)$ when $z_0 \rightarrow p$, $p$ being a point in $\Gamma$.
\begin{lemma}
$\mathcal{H}_{G}$ is representable by residues with local Cauchy kernel $G(z,z_0)$, normalized, and is maximal within such subspaces.
\end{lemma}
\textbf{Proof.} 
We need to compute:
\beq
\tilde{\tilde{g}}(z_0) = \sum_{p \in V} \Res_{z \rightarrow p} G(z,z_0)\Big(\sum_{q \in V} \Res_{\xi \rightarrow q} G(\xi,z)\,g_{q}(\xi)\Big).
\eeq
Notice that the poles are isolated, so the sum over $p$ is finite. Taking into account the pole of $G(\xi,z)$ at $z = \xi$, we may exchange the residues in $z$ and $\xi$:
\bea
\tilde{\tilde{g}}(z_0) & = & \sum_{(p,q) \in V^2} \Res_{\xi \rightarrow q} \big(\Res_{z \rightarrow p} + \delta_{p,q}\Res_{z \rightarrow \xi}\big)G(z,z_0)G(\xi,z)\,g_{q}(\xi). \nonumber
\eea
The first term does not contribute since $G(z,z_0)G(\xi,z)$ is holomorphic when $z \rightarrow p$, while the second term gives:
\beq
\tilde{\tilde{g}}(z_0) = \sum_{p \in V} \Res_{\xi \rightarrow p} G(\xi,z_0)\,g_{p}(\xi),
\eeq
which coincides with $\tilde{g}(z_0)$. \hfill $\Box$

\vspace{0.2cm}

\noindent For large enough normalized representable subspaces, we do not have the choice of a Cauchy kernel.
\begin{lemma}
If $G_1$ and $G_2$ are two local Cauchy kernels for $\mathcal{H}$ containing a sequence of function $(f_{k,j})_{k \geq 1,j \in \ldbrack 1,r \rdbrack}$ with a pole of order $k$ at a given point $p_j \in V_j$, and $\mathcal{H}$ is normalized, then $G_1 \equiv G_2$.
\end{lemma} 
\textbf{Proof.} The assumption implies that, for any $f \in \mathcal{H}$ and any $j$, $\Res_{z \rightarrow p_j} (G_{1}(z,z_0) - G_{2}(z,z_0))f(z) = 0$. By specializing to $f = f_{j,k}$, we find that the Taylor expansion of $(G_{1}(z,z_0) - G_{2}(z,z_0))$ at $z = z_j$ vanishes identically. By the principle of isolated zeroes, we must have $G_{1}(z,z_0) = G_{2}(z,z_0)$ for any $z \in V_j$, thus any $z \in V$. \hfill $\Box$

\subsection{Residues as contour integrals in the physical sheet}

Equivalently, we may rewrite \eqref{reqi} as a contour integral in $U_j$ only:
\begin{lemma}
\label{cont}
Assume $\mathcal{H} \subseteq \Ms(U_{\Gamma})$ is representable by residues. For a given $z_0 \in U_{\Gamma}$, we choose arcs $\gamma_j^{\mathrm{ext}} \subseteq U_j$ as in \S~\ref{ss1}, oriented like $\gamma_j$, so that $z_0$ or $\iota(z_0)$ does not lie between $\gamma_j^{\mathrm{ext}}$ and $\gamma_j$. Remind the notation $\Gamma^{\mathrm{ext}} = \bigcup_{j = 1}^r \gamma_j^{\mathrm{ext}}$.
\beq
\forall f \in \mathcal{H},\quad \forall z_0 \in U_{\Gamma},\qquad \tilde{f}(z_0) = \frac{1}{2{\rm i}\pi}\oint_{\Gamma^{\mathrm{ext}}} [\Delta^{z} G](z,z_0)\,f(z) + G(\iota(z),z_0)\,\mathcal{S} f(z).
\eeq
\end{lemma}
\textbf{Proof.} The orientation of $\gamma_j$ can be carried by homotopy to an orientation of $\gamma_j^{\mathrm{ext}}$. Then, the orientation of $\gamma_j^{\mathrm{int}} = \iota_j(\gamma_j^{\mathrm{ext}})$ is opposite to the orientation carried by homotopy from $\iota(\gamma_j)$. These arcs allows to represent the residues. Setting $\Gamma^{\mathrm{int}} = \iota(\Gamma^{\mathrm{ext}})$, we have:
\bea
\tilde{f}(z_0) & = & \frac{1}{2{\rm i}\pi}\Big(\oint_{\Gamma^{\mathrm{ext}}} + \oint_{\Gamma^{\mathrm{int}}}\Big) G(z,z_0)\,f(z) \nonumber \\
& = & \frac{1}{2{\rm i}\pi} \oint_{\Gamma^{\mathrm{ext}}} \mathcal{S}^{z}[G(z,z_0)\,f(z)] \nonumber \\
& = & \frac{1}{2{\rm i}\pi} \oint_{\Gamma^{\mathrm{ext}}} [\Delta^{z} G](z,z_0)\,f(z) + G(\iota(z),z_0)\,\mathcal{S} f(z),
\eea
the last line being a mere rewriting of the previous one.
\hfill $\Box$

\vspace{0.2cm}

\noindent A similar computation shows:
\begin{lemma}
Assume $\mathcal{H} \subseteq \Ls_{\Gamma}(U)$ representable by residues. Then, for any $z_0 \in U_{\Gamma}$:
\beq
\forall f \in \mathcal{H},\qquad \tilde{f}(z_0) = \frac{1}{2{\rm i}\pi}\oint_{\Gamma^{\mathrm{ext}}} [\Delta^{z} G](z,z_0)\Big(f(z) - \frac{\mathcal{S} f(z)}{2}\Big).
\eeq
\hfill $\Box$
\end{lemma}

\subsection{Loop equations and topological recursion}
\label{htil2}

\begin{definition}
\label{afix}We denote $\Gamma^{\mathrm{fix}}$ the set of fixed points of $\Gamma$ under $\iota$. Elements of $\Gamma^{\mathrm{fix}}$ are called \emph{ramification points}.
\end{definition}
Notice that in the Example given in \S~\ref{ss1}, ramification points only arise from ends $\alpha \in E_j$ of open arcs.
\begin{definition}
\label{offc}We say that a $1$-form $f \in \Ls(U)$ is \emph{off-critical} if the zeroes of $\Delta f(z)$ in $V$ only occur at ramification points, and their order is exactly $2$.
\end{definition}
If $\mathcal{H}$ is a vector space of $1$-forms, we denote $\mathcal{H}_n$ the space of symmetric $n$-forms $f(z_1,\ldots,z_n)$, such that $f(\cdot,z_2,\ldots,z_n) \in \mathcal{H}$ for any $z_2,\ldots,z_n$ away from poles of $f$.
We consider in this paragraph a family $\omega_{\bullet}^{\bullet} = (\omega_n^{g})_{n,g}$ of meromorphic, symmetric $n$-forms ($n \geq 1$) on $U^n$, indexed by an integer $g \geq 0$. In other words, $\omega_n^{g} \in \Ms_n(U_{\Gamma})$ with our notations.
\begin{definition}
\label{stable}We say that a couple $(n,g)$ is \emph{stable} if $2g - 2 + n > 0$, i.e. $(n,g) \neq (1,0),(2,0)$.
\end{definition}
The main topic of this article is to study families $\omega_{\bullet}^{\bullet}$ which satisfy certain constraints, that we will call "loop equations".
\begin{definition}
\label{def10}We say that $\omega_{\bullet}^{\bullet}$ satisfies \emph{linear loop equations} if:
\begin{itemize}
\item[(i)] $\omega_1^0 \in \Ls_{\Gamma}(U)$ is an off-critical $1$-form.
\item[(ii)] $G(z,z_0) = -\int^{z} \omega_2^0(z_0,\cdot)$ defines a local Cauchy kernel. 
\item[(iii)] For any stable $(n,g)$, we have $\omega_n^{g} \in \Hs_{\Gamma}(U)$, and for any $z_2,\ldots,z_n$ which are not poles of $\omega_n^g$, \mbox{$\mathcal{S}\omega_n^{g}(\cdot,z_2,\ldots,z_n) \in \Hs^{\mathrm{inv}}_{\Gamma}(V)$}.
\end{itemize}
We say that those loop equations are solvable when (iii) is replaced by:
\begin{itemize}
\item[(iv)] For any stable $(n,g)$, $\omega_n^{g} \in (\mathcal{H}_{G})_n$, i.e. $\omega_n^g(\cdot,z_2,\ldots,z_n)$ belongs to the maximal normalized subspace of $\Ls_{\Gamma}(U)$ which is representable by residues for the local Cauchy kernel of (ii).
\end{itemize}
\end{definition}
\begin{definition}
\label{def20}We say that $\omega_{\bullet}^{\bullet}$ satisfies \emph{quadratic loop equations} if, for any stable $(n,g)$, 
\beq
\label{qua}\mathcal{Q}_{n}^{g}(z;z_I) = \omega_{n + 1}^{g - 1}(z,\iota(z),z_I) + \sum_{J \subseteq I,\,\, 0 \leq h \leq g} \omega_{|J| + 1}^{h}(z,z_J)\,\omega_{n - |J|}^{g - h}(\iota(z),z_{I\setminus J})
\eeq
is a quadratic differential in $z \in V$ with double zeroes at ramification points.
\end{definition}
Assuming linear loop equations, there are equivalent ways to write the quadratic loop equation. For instance, using the polarization formula given in \eqref{pola}, we can recast \eqref{qua} as:
\beq
\label{216}\frac{1}{2}\,\Delta\omega_1^0(z)\,\Delta^{z}\omega_n^{g}(z,z_I) = \mathcal{E}_n^{g}(z,\iota(z);z_I) + \tilde{\mathcal{Q}}_{n}^{g}(z;z_I),
\eeq
where we have introduced:
\beq
\label{edef}\mathcal{E}_n^{g}(z,z';z_I) = \omega_{n + 1}^{g - 1}(z,z',z_I) + \sum_{\substack{J \subseteq I,\,\,0 \leq h \leq g \\ (J,h) \neq (\emptyset,0),(I,g)}} \omega_{|J| + 1}^{h}(z,z_J)\,\omega_{n - |J|}^{g - h}(z',z_{I\setminus J}),
\eeq
and $\tilde{\mathcal{Q}}_{n}^{g}(z;z_I) = \frac{1}{2}\,\mathcal{S}\omega_1^0(z)\,\mathcal{S}\omega_n^{g}(z,z_I) - \mathcal{Q}_{n}^{g}(z;z_I)$. Since $\mathcal{S}\omega_n^g(\cdot,z_I) \in \Hs_{\Gamma}^{\mathrm{inv}}(V)$, it must have at least a simple zero at ramification points. Therefore, $\tilde{Q}_{n}^{g}(z;z_I)$ has double zeroes at ramification points iif $\mathcal{Q}_{n}^g(z;z_I)$ does.

Here is the central result of the theory, whose applicability will be illustrated in the remaining of the article:
\begin{proposition}
\label{2222}If $\omega_{\bullet}^{\bullet}$ satisfies solvable linear and quadratic loop equations, then for any stable $(n,g)$, the poles of $\omega_n^{g}(\cdot,z_2,\ldots,z_n)$ in $U_{\Gamma}$ occur only at ramification points, and we have the \emph{topological recursion formula}:
\beq
\label{topore}\omega_n^{g}(z,z_I) = \sum_{\alpha \in \Gamma^{\mathrm{fix}}} \Res_{z \rightarrow \alpha} K(z_0,z)\,\mathcal{E}_n^{g}(z,\iota(z);z_I),
\eeq
where $\mathcal{E}_n^{g}$ was defined in \eqref{edef} and we introduced the \emph{recursion kernel}:
\beq
\label{reck}K(z_0,z) = \frac{-\frac{1}{2}\int_{\iota(z)}^{z} \omega_2^0(z_0,\cdot)}{\omega_1^0(z) - \omega_1^0(\iota(z))}.
\eeq
\end{proposition}
We observe that any ramification point $\alpha$ actually belongs to some contour $\gamma_j$ such that $\iota(\gamma_j) = \gamma_j$: if $z$ is near $\alpha$, so is $\iota(z)$ and therefore the path of integration from $\iota(z)$ to $z$ remains in a neighborhood of $\alpha$, and thus $\int_{\iota(z)}^{z}$ is well-defined. The key point to use this proposition in practice is to show that the linear loop equations are solvable, i.e. show that $\omega_n^g$ can be represented by a residue formula for a certain Cauchy kernel\footnote{Here, we described a situation where the Cauchy kernel does not depend on $n$ and $g$, but this assumption might be relaxed if needed in some applications.}. The purpose of Section~\ref{S3} is to provide a non-trivial class of examples where solvable linear and quadratic loop equation arise.

\vspace{0.2cm}

\noindent \textbf{Proof.} Let us fix a family of spectator variable $z_I = (z_2,\ldots,z_n) \in U^{n - 1}$. Since they are chosen away from $\Gamma$, we can always assume that $z_i \notin V$ for any $i \in \ldbrack 2,n \rdbrack$. Firstly, we notice that, from linear loop equations, $\omega_n^g(z,z_I)$ has the same poles with respect to $z \in V$ as $\Delta^{z}\omega_n^{g}(z,z_I)$, and our definition of solvability implies that these are the only possible poles of $\omega_n^g(z,z_I)$ for $z \in U_{\Gamma}$. Besides, property $(ii)$ in Definition~\ref{def10} imposes that the only singularity of $\omega_2^0(z_0,z) = -\dd_{z} G(z_0,z)$ in the range $(z_0,z) \in U_{\Gamma}\times V$ is a double pole without residues at $z_0 = z$. We prove the statement about the location of the poles of $\omega_n^g$ by recursion of $\chi_{n}^{g} = 2g - 2 + n > 0$. Indeed, the right-hand side of the decomposition \eqref{216} of $\omega_n^{g}$ involves $\omega_{n'}^{g'}$ with $\chi_{n'}^{g'} < \chi_{n}^{g}$, and the unknown but regular $\tilde{\mathcal{Q}}_n^{g}$. At level $\chi = 1$, we have to consider $(n,g) = (3,0)$ or $(1,1)$. As regards $\omega_3^0$, we have a decomposition:
\beq
\Delta^{z}\omega_3^0(z,z_2,z_3) = \frac{2}{\Delta \omega_1^0(z)}\Big(\omega_2^0(z,z_2)\omega_2^0(\iota(z),z_3) + \omega_2^0(\iota(z),z_2)\omega_2^0(z,z_3) + \tilde{\mathcal{Q}}_{3}^0(z;z_2,z_3)\Big).
\eeq
The numerator of the right-hand side is regular for $z \in V$, and the denominator may create poles at zeroes of $\Delta \omega_1^0(z)$. Since $\omega_1^0$ is assumed off-critical, they can only occur at ramification points. Thus $\omega_3^0$ has poles at ramification points only. As regards $\omega_1^{1}$, we have a decomposition:
\beq
\Delta^{z} \omega_1^1(z) = \frac{2}{\Delta \omega_1^0(z)}\Big(\omega_2^0(z,\iota(z)) + \tilde{\mathcal{Q}}_{1}^1(z)\Big).
\eeq
$\omega_2^0(z,\iota(z))$ has a pole of order $2$ at ramification points, and since $\omega_1^0$ is off-critical and $\mathcal{Q}_1^1(z)$ regular, we deduce that $\omega_1^1(z)$ has poles at ramification points only. If we assume the property true for $\omega_{n'}^{g'}$ such that $\chi_{n',g'} < \chi$, one shows by the same arguments that any $\omega_n^{g}(z,z_I)$ with $\chi_{n}^{g} = \chi$ has poles at ramification points only. Therefore, we know that $\omega_n^{g}$ has poles only at ramification points. Since $\omega_n^g$ satisfy solvable linear loop equations, we can use Lemma~\ref{mais}, and find:
\beq
\omega_n^g(z_0,z_I) = \sum_{\alpha \in \Gamma^{\mathrm{fix}}} \Res_{z \rightarrow \alpha} \frac{-\frac{1}{2}\int^{z}_{\iota(z)} \omega_2^0(z_0,\cdot)}{\omega_1^0(z) - \omega_1^0(\iota(z))}\Big(\mathcal{E}_n^g(z,\iota(z);z_I) + \tilde{\mathcal{Q}}_{n}^g(z;z_I)\Big).
\eeq
The quadratic loop equations provide exactly the condition under which $\frac{\tilde{\mathcal{Q}}_{n}^g(z;z_I)}{\omega_1^0(z) - \omega_1^0(\iota(z))}$ is regular at ramification points. Therefore, this term does not contribute to the residue, and we find \eqref{topore}. \hfill $\Box$

There is a converse to Proposition~\ref{2222}:
\begin{proposition}
If $\omega_1^0$ and $\omega_2^0$ satisfy $(i)$-$(ii)$ of Definition~\ref{def10}, the topological recursion formula \eqref{topore} defines $\omega_n^g$ for stable $n,g$, which are elements of $\Ms_n(U_{\Gamma})$, satisfying solvable linear loop equations (i.e. $(iv)$ of Definition~\ref{def10}) and quadratic loop equations (Definition~\ref{def20}).
\end{proposition}
\textbf{Proof.} We first mention that the formula \eqref{topore} has a diagrammatic representation \cite{CEO06,EOFg}, i.e. $\omega_n^g$ can be written as a sum over skeleton graphs of a Riemann surface of genus $g$ with $n$ punctures. It is quite useful to prove elementary properties of the topological recursion. For instance, repeating the diagrammatic proof of \cite{EORev}, one can show directly that, despite the special role played by $z_0$ outwardly, Eqn.~\ref{topore} does produce a symmetric $n$-form. Then, $\mathcal{S}\omega_n^g(\cdot,z_I) \in \Hs_{\Gamma}^{\mathrm{inv}}(V)$ follows from \eqref{topore} and the fact that $\mathcal{S}G(z,\cdot) \in \Hs_{\Gamma}^{\mathrm{inv}}(V)$, and solvability follow from Lemmate~\ref{mais}-\ref{l21}. To establish the quadratic loop equations, we write thanks to Lemma~\ref{mais}:
\beq
\omega_n^g(z,z_I) = \sum_{\alpha \in \Gamma^{\mathrm{fix}}} \Res_{z \rightarrow \alpha} \frac{\Delta^{z} G(z,z_0)}{4}\,\Delta^{z}\omega_n^g(z,z_I).
\eeq
And, comparing to the topological recursion formula \eqref{topore}, we find:
\beq
\sum_{\alpha \in \Gamma^{\mathrm{fix}}} \Res_{z \rightarrow \alpha} \frac{\frac{1}{2}\Delta^z G(z,z_0)}{\Delta \omega_1^0(z)}\Big(\frac{1}{2}\Delta \omega_1^0(z)\omega_n^g(z,z_I) - \mathcal{E}_n^{g}(z,\iota(z);z_I)\Big) = 0,
\eeq
where $\mathcal{E}_n^g$ defined in \eqref{edef}, and considering the limit when $z_0$ approaches one of the ramification points, this equation implies that $\tilde{Q}_n^g$ defined in \eqref{216} has a zero at ramification points. Since it is invariant under $\iota$, this zero has even order. Since we already have proved linear loop equations, this implies quadratic loop equations in their equivalent form noticed after Definition~\ref{def20}. \hfill $\Box$

We now come to properties of the topological recursion formula which were already identified in \cite{EOFg}: behavior under variations of the initial data in \S~\ref{spsa}, and singular limits in \S~\ref{sings}. We also give definition in a minimal framework of numbers $\omega_0^g$ (the free energies, in \S~\ref{wdv}), so that the formulas for the variation of the initial data continue to hold.

\subsection{Variations of initial data}
\label{spsa}
Let $\Omega^*$ be a cycle in $U_{\Gamma}$, which lies outside a compact neighborhood of the ramification points. Assume we are given an initial data consisting of $\Gamma$, $\omega_1^0 \in \Ls_{\Gamma}(U)$ and $\omega_2^0 \in \Ms_2(U_{\Gamma})$ so that $G(z,z_0) = -\int^{z}\omega_2^0$ is a local Cauchy kernel. We call this data a \emph{spectral curve}. Then, we can define $\omega_n^g$ by the topological recursion formula \eqref{topore}.

In this paragraph, we discuss the effect of an infinitesimal variation of the spectral curve. More specifically:
\begin{definition}
\label{diq}Let $\Omega^*$ be a path in $U_{\Gamma}$ which lies outside a compact neighborhood of the ramification points, and $\Lambda_{\Omega}$ a germ of holomorphic function on $\Omega^*$. We consider a variation of the form:
\bea
\label{cqyq}\delta_{\Omega} \omega_1^0(z) & = & \Omega(z) = \int_{\Omega^*} \Lambda_{\Omega}(\cdot)\,\omega_2^0(\cdot,z,), \\
\label{cqyq2}\delta_{\Omega} \omega_2^0(z_1,z_2) & = & \int_{\Omega^*} \Lambda_{\Omega}(\cdot)\,\omega_3^0(\cdot,z_1,z_2).
\eea
We call $\delta_{\Omega}$ a \emph{WDVV-compatible variation}.
\end{definition}
The reason for this denomination will appear in \eqref{ejj} below. $\delta_{\Omega}$ is a derivation on the space of functionals of $\omega_1^0$ and $\omega_2^0$. Then, we can deduce:
\begin{theorem}
For any stable $n,g$:
\beq
\label{vqrsqa}\delta_{\Omega}\omega_n^g(z_1,\ldots,z_n)  = \int_{\Omega^*} \Lambda_{\Omega}(\cdot)\,\omega_{n + 1}^{g}(\cdot,z_1,\ldots,z_n).
\eeq
\end{theorem}
This result has a nice diagrammatic interpretation, and the proof is identical to that in  \cite{EOFg}.
 
\subsection{Definition of free energies}
\label{wdv}
We define the \emph{stable free energies}:
\begin{definition}
For any $g \geq 2$, we define the number:
\beq
\omega_0^{g} = F^{g} = \frac{1}{2 - 2g} \sum_{\alpha \in \Gamma^{\mathrm{fix}}}  \Res_{z \rightarrow \alpha} \omega_1^g(z)\Big(\int^{z}\omega_1^0\Big),
\eeq
where $o$ is an arbitrary base point and $\Gamma^{pole}$ is the set of poles of $\om_1^g$.
\end{definition}
By integrating theorem~\ref{vqrsqa}, a straightforward computation shows:
\begin{corollary}
\label{corof}Eqn.~\eqref{vqrsqa} holds also for $n = 0$ and any $g \geq 2$. \hfill $\Box$
\end{corollary}
If we have two $1$-forms $\Omega$ and $\Omega'$ defining variations $\delta_{\Omega}$ and $\delta_{\Omega'}$, the fact that stable $\omega_n^g$ do not have poles in $U_{\Gamma}$ except at branchpoints imply that $\delta_{\Omega}\delta_{\Omega'} = \delta_{\Omega'}\delta_{\Omega}$. Therefore, if we have a smooth family of spectral curves depending on parameters $(t_i)_i$ around some initial value $(t_i^0)_i$, so that $\partial_{t_i}$ can be realized as $\delta_{\Omega_i}$ satisfying \eqref{cqyq}-\eqref{cqyq2}, we may define \emph{unstable free energies} as follows:
\begin{definition}
We define $\omega_0^1 \equiv F^1$ as the function of $(t_i)_i$ modulo a constant, at least locally in the neighborhood of $(t_i^0)_i$ such that:
\beq
\partial_{t_i} F^1 = \int_{z \in \Omega^*_i} \Lambda_{\Omega}(z)\,\omega_1^1(z) .
\eeq
\end{definition}
This definition makes sense because the derivative of the right-hand side with respect to $t_j$ is symmetric by exchange of $i$ and $j$, due to the fact that $\omega_2^1$ is a symmetric $2$-form which is regular across $\Omega^*_i \times \Omega_j^*$ since those paths remain away from the ramification points where $\omega_2^1$ has its poles. Similarly:
\begin{definition}
We define $\omega_0^0 = F^0$ as the function of $(t_i)_i$ modulo a quadratic form, at least locally in the neighborhood of $(t_i^0)_i$ such that:
\beq
\partial_{t_i}\partial_{t_j}\partial_{t_k} F^0 = \int_{z_1 \in \Omega^*_i}\Lambda_{\Omega}(z_1)\int_{z_2 \in \Omega^*_j}\Lambda_{\Omega}(z_2)\int_{z_3 \in \Omega_{k}^*} \Lambda_{\Omega_k}(z_3)\,\omega_3^0(z_1,z_2,z_3).
\eeq
\end{definition}
Therefore, we conclude that \eqref{vqrsqa} holds for any $n,g \geq 0$ at least with $\Omega$ is equal to some $\Omega_i$.
We can compute from the residue formula:
\bea
\label{ejj} \partial_{t_i}\partial_{t_j}\partial_{t_k} F^0 & = & \sum_{\alpha \in \Gamma^{\mathrm{fix}}} \frac{\Omega_i(\underline{\alpha})\,\Omega_j(\underline{\alpha})\,\Omega_k(\underline{\alpha})}{2\rho_{\alpha}},
\eea
where
\beq
\Omega_i(\underline{\alpha}) = \frac{\Omega_i}{\dd \xi}(\alpha),\qquad \omega_1^0(z) - \omega_1^0(\iota(z)) \mathop{\sim}_{z \rightarrow \alpha} 2\rho_{\alpha}\,\xi(z)\dd\xi(z),
\eeq
and $\xi$ is a local coordinate near $\alpha$ so that $\xi(\alpha) = 0$ and $\xi(\iota(z)) = -\xi(z)$. This representation of third derivatives of $F^0$ as a sum of cubic terms is closely related to WDVV equations \cite{Dub}. Remark that, according to those definitions, \eqref{vqrsqa} holds for any $n,g \geq 0$ such that $(n,g) \neq (0,0)$ and $(1,0)$.

\subsection{Singular limits}
\label{sings}
A family of spectral curve parametrized by $t \in ]0,t_0]$ is said singular at $t = 0$ if ramification points collide at $t = 0$, or a singularity of $\omega_1^0$ collides with at least one ramification point at $t = 0$. The topological recursion formula usually diverges when $t \rightarrow 0$, but we can control precisely how it diverges in terms of the blow-up of the singularity. This blow-up curve contains only the \emph{singular} ramification points, i.e. those where a singularity arise in the limit $t \rightarrow 0$. In the topological recursion formula \eqref{topore}, since the computation of residues in the topological recursion is a local operation, we find that the contribution of the residues at non singular ramification points remains finite, and for the computation of the residues at singular ramification points, we may replace $\omega_1^0$ and $\omega_2^0$ by their blow-up. Therefore, we find:

\begin{proposition}
\label{singp} Assume $(\omega_1^0)_{t}(z_{t}) \sim t^{\alpha}\,(\omega_1^0)^*(\zeta)$ and $(\omega_2^0)_{t}(z_{1,t},z_{2,t}) \sim (\omega_2^0)^*(\zeta_1,\zeta_2)$ when $t \rightarrow 0$, and $z_t,z_{i,t}$ denote family of points approaching the point $\zeta,\zeta_i$ in the blow-up curve. Then, for any stable $(n,g)$:
\beq
(\omega_n^g)_{t}(z_{1,t},\ldots,z_{n,t}) = t^{\alpha(2 - 2g - n)}\,(\omega_n^g)^*(\zeta_1,\ldots,\zeta_n) + o(t^{\alpha(2 - 2g - n)}).
\eeq
\hfill $\Box$
\end{proposition}

\subsection{Spectral curves with automorphisms}
\label{autmor}

Eventually, we explain for spectral curves with symmetry, how the symmetry carries to the $\omega_n^g$. This remark has not appeared yet in the literature, and is noteworthy in recent applications of the topological recursion to knot theory.

Let $U$ be a domain and $\Gamma$ be a subset of the connected components of $\partial U$. Let $\mathcal{H}$ be a subspace of $\Hs_{\Gamma}(U)$ representable by residues, and $G$ its local Cauchy kernel. In this paragraph, we imagine that we have a finite degree, holomorphic map $\sigma\,:\,U \rightarrow U$ such that $\sigma_{|\partial U}$ commutes with the involution $\iota$ on $\Gamma$. If we assume furthermore that $\Gamma$ avoids fixed points of $\sigma$, the quotient $\pi\,:\,U \rightarrow U/\sigma$ is smooth in the vicinity of the image of $\Gamma$. For any $f \in \Hs(U)$, we define:
\beq
f^{\sigma}(x) = \sum_{y \in \sigma^{-1}\{x\}} f(y).
\eeq
$\sigma$ can be extended to a holomorphic (resp. anti-holomorphic) finite degree map, hence we have a covering $\pi\,:\,U_{\Gamma} \rightarrow U_{\Gamma}/\sigma$. If $f \in \Hs_{\Gamma}(U)$, then $f^{\sigma} \in \Hs_{\pi(\Gamma)}(U/\sigma)$, and the subspace $\mathcal{H}^{\sigma}$ is representable by residues with local Cauchy kernel:
\beq
G^{\sigma}(z_0,z) = \sum_{\zeta_0 \in \sigma^{-1}\{z_0\}} G(z_0,z).
\eeq
It is easy to see that:
\begin{proposition}
$\omega_{\bullet}^{\bullet}$ satisfies linear loop equations (resp. solvable linear loop equations, and quadratic loop equations) if and only if $(\omega_{\bullet}^{\bullet})^{\sigma}$ satisfies linear loop equations (resp. solvable linear loop equations, and quadratic loop equations).
\end{proposition}
In other words, the topological recursion commutes with the quotient operation, provided the quotient is smooth near the ramification points.

\section{Repulsive particles systems}
\label{S3}

Here we consider an important class of applications of the previous formalism.

\subsection{The model}
\label{S31}
Let $\Gamma_0$ be a union of arcs and open arcs in $\widehat{\mathbb{C}}$. We consider a $N$-point process in $\Gamma_0$ with joint distribution of the form:
\bea
\label{mes2}\dd\varpi(\lambda_1,\ldots,\lambda_N) & = &  \prod_{1 \leq i < j \leq N} |\lambda_i - \lambda_j|^{\beta}\,\prod_{1 \leq i,j \leq N} \big(R(\lambda_i,\lambda_j)\big)^{\rho/2}\,\prod_{i = 1}^N e^{-N\,\mathcal{V}(\lambda_i)}\dd\lambda_i,  \\
Z_N & = & \int_{(\Gamma_0)^N} \dd\varpi(\lambda_1,\ldots,\lambda_N).
\eea
When $\beta = 2$, it can be realized as the eigenvalue distribution of a random normal matrix $M$ with spectrum included in $\Gamma_0$:
\beq
\label{ee33}\dd\varpi(M) = \dd M\,e^{-N \Tr \mathcal{V}(M) + \frac{\rho}{2}\,\Tr \ln R(M\otimes\mathbf{1}_N,\mathbf{1}_N\otimes M)},
\eeq
where $\mathbf{1}_{N}$ is the identity matrix of size $N \times N$. Although we borrow a probabilistic language, $\dd\varpi$ can be a signed or complex measure, and even a formal measure in this definition. By \emph{formal measure}, we mean that:
\beq
\mathcal{V}(x) = \frac{1}{t}\Big(\frac{x^2}{2} - \mathcal{U}(x)\Big),\qquad \mathcal{U}(x) = \sum_k t_k\,\mathcal{U}_k(x),
\eeq
where $t_k$ are formal variables, and $Z_N$ or any expectation value with respect to $\dd\varpi$ is considered as a generating series in the formal variables $t_k$. We have factored the distribution \eqref{mes2}, because we will later assume that $R(x,y)$ does not vanish for $(x,y) \in \Gamma_0^2$. It is thus characterized by a repulsion at short distance between two particles $i$ and $j$, proportional to $|\lambda_i - \lambda_j|^{\beta}$. We call $\mathcal{V}$ the potential, $R$ the two-point interaction, and $\beta$ the \emph{Dyson index}. We may assume without restriction that $R(x,y) = R(y,x)$. It was convenient to introduce a redundant, free parameter $\rho$ in the model. 

In the context of formal integrals, we shall review in Section~\ref{S5} that this model describes the statistical physics of self-avoiding loops on random lattices, i.e the general $O(-\rho)$-loop model on random maps \cite{GK}. For $\rho = -1$, it contains for instance the Ising model on faces of random triangulations \cite{Eformal}. In the context of convergent integrals, such a model has also appeared in the context of quantum entanglement \cite{BNBures}, and in relation with dynamics of fluid interfaces \cite{BDJ12}.

We denote $M = \mathrm{diag}(\lambda_1,\ldots,\lambda_N)$, and
\beq
\langle f(M) \rangle = \frac{1}{Z_N}\int_{(\Gamma_0)^N} \dd\varpi(\lambda_1,\ldots,\lambda_N)\,f(\lambda_1,\ldots,\lambda_N).
\eeq
We are interested in computing the partition function $Z_N$, and the connected \emph{correlators}:
\beq
\label{conera}W_n(x_1,\ldots,x_n) = \Big\langle \prod_{i = 1}^n \mathrm{Tr}\,\frac{1}{x_i - M} \Big\rangle_c,
\eeq
where $c$ stands for "cumulant". Equivalently,
\beq
\overline{W}_n(x_1,\ldots,x_n) \equiv \Big\langle \prod_{i = 1}^n \mathrm{Tr}\,\frac{1}{x_i - M} \Big\rangle = \sum_{J_1 \dot{\cup} \cdots \dot{\cup} J_r = \ldbrack 1,n \rdbrack} \prod_{i  = 1}^r W_{|J_i|}\big((x_{j_i})_{j_i \in J_i}\big).
\eeq
In the following, we assume that $\mathcal{V}$, $R$ and $\Gamma_0$ are such that $Z_N$ exists and $Z_N\neq 0$. Then, $W_n(x_1,\ldots,x_n)$ defines a holomorphic function in the domain $(\mathbb{C}\setminus\Gamma_0)^n$, and a priori, $W_n(x_1,\ldots,x_n)$ has a discontinuity when one of the $x_i$'s crosses $\Gamma_0$.

\subsection{Some results of potential theory}
\label{Popo}
\subsubsection*{Preliminaries}

In this paragraph, we focus on convergent integrals and assume $\Gamma_0 \subseteq \mathbb{R}$, and non-negative distribution $\varpi(\lambda_1,\ldots,\lambda_N)$. We use potential theory to prove useful technical results.

We denote $\mathcal{P}_1(\Gamma_0)$ (resp. $\mathcal{P}_{0}(\Gamma_0)$) the convex set of probability measures (resp. signed measures of total mass $0$) on $\Gamma_0$. Those sets are equipped with weak-* topology, which means that:
\beq
\lim_{n \to \infty} \mu_n = \mu_{\infty}\qquad \Longleftrightarrow \qquad\forall f \in \mathcal{C}_{b}^{0}(\Gamma_0),\quad \lim_{n \rightarrow \infty} \Big(\int_{\Gamma_0} \dd\mu_n(x)\,f(x)\Big) = \int_{\Gamma_0} \dd\mu_{\infty}(x)\,f(x),
\eeq
where $\mathcal{C}_b^0(\Gamma_0)$ denotes the space of bounded continuous functions on $\Gamma_0$. We introduce the functional $\mathcal{E}$ on $\mathcal{P}_{1}(\Gamma_0)$:
\beq
\mathcal{E}[\mu] = -\iint_{\Gamma_0^2} \dd\mu(x)\dd\mu(y) \ln R_0(x,y) + \int_{\Gamma_0} \dd\mu(x)\mathcal{V}(x),
\eeq
where $R_0(x,y) = |x - y|^{\beta/2}\,\big(R(x,y)\big)^{\rho/2}$. Since $\ln R_0$ and $\mathcal{V}$ can have singularities, there might exist probability measures for which it is infinite or left undefined. Let $\Gamma_0^{o} = \{\left. x \in \Gamma_0\right| \mathcal{V}(x) < +\infty\}$, and let $a$ be an endpoint of $\Gamma_0^{o}$. 
\begin{definition}
\label{sconf}We say that $(\mathcal{V},R_0)$ defines a \emph{strongly confining interaction} at a point $a$ if there exist $M\,:\,\Gamma_0^{o} \rightarrow \mathbb{R}_+^*$ such that:
\beq
\label{confay} \ln R_0(x,y) \leq M(x) + M(y),\qquad \liminf_{x \rightarrow a}\big(\mathcal{V}(x) -  2M(x)\big) = +\infty.
\eeq
We say that $(\mathcal{V},R_0)$ defines a strongly confining interaction if this is true for any endpoint $a$ of $\Gamma_0^{o}$.
\end{definition}

\begin{definition}
\label{stconv}We say that $R_0$ defines a \emph{strictly convex interaction} if, for any signed measure $\nu$ such that $\nu(\Gamma_0) = 0$,
\beq
\label{hold}\iint_{\Gamma_0^2} \dd\nu(x)\dd\nu(y)\,\ln R_0(x,y) \leq 0,
\eeq
with equality iff $\nu = 0$.
\end{definition}
This implies in particular that $\mathcal{P}^{\mathcal{E}}_{1}(\Gamma_0) = \{\mu \in \mathcal{P}_1(\Gamma_0),\quad \mathcal{E}[\mu] < +\infty\}$ is a convex set, on which $\mathcal{E}$ is strictly convex. Besides,
\begin{lemma}
\label{leoa}
If $R_0$ defines a strictly convex interaction, then for any \emph{complex} measure $\nu$ such that $\nu(\Gamma_0) = 0$,
\beq
\label{holdon} \iint_{\Gamma_0^2} \dd\nu(x)\big(\dd\nu(y)\big)^*\ln R_0(x,y) \leq 0.
\eeq
\end{lemma}
\textbf{Proof.} Since $\ln R_0(x,y) = \ln R_0(y,x)$ is real-valued by assumption, the left-hand side is real-valued. Therefore:
\beq
\iint_{\Gamma_0^2} \dd\nu(x)\big(\dd\nu(y)\big)^*\ln R_0(x,y) = \iint_{\Gamma_0^2} \big[\mathrm{Re}\,\dd\nu(x)\,\mathrm{Re}\,\dd\nu(y) + \mathrm{Im}\,\dd\nu(x)\,\mathrm{Im}\,\dd\nu(y)\big]\ln R_0(x,y).
\eeq
Since $\mathrm{Re}\,\dd\nu$ and $\mathrm{Im}\,\dd\nu$ are signed measure with mass $0$, \eqref{hold} applies to each term. \hfill $\Box$ 

When $(\mathcal{V},R_0)$ defines a strongly confining interaction, we may adopt a slightly weaker definition of strictly convex interaction, by restricting oneself to $\nu$ with support included in a compact of $\Gamma_0^{o}$. Examples of strictly convex interactions are given in Appendix~\ref{appA}. They include $R_0(x,y) = |x - y|^{\beta}$, its trigonometric and elliptic analog.

\subsubsection*{Equilibrium measures}

Our goal is to establish that, with help of the functional $\mathcal{E}$ for well-chosen potentials $\mathcal{V}$ and some extra assumptions, one can define subspaces of $\Hs_{\Gamma}(U)$ which are representable by residues and normalized. They will play an important role in the analysis of the $1/N$ expansion of Schwinger-Dyson equations.

We consider the following set of assumptions:
\begin{hypothesis}
\label{auq}\begin{itemize}
\item[$\phantom{aaa}$] \phantom{,,,}
\item[$(i)$]  $(\mathcal{V},R_0)$ is strongly confining ;
\item[$(ii)$] $R_0$ is a strictly convex interaction ;
\item [$(iii)$] $\mathcal{V}\,:\,\Gamma_0^{o} \rightarrow \mathbb{R}$ is real-analytic ;
\item[$(iv)$] $\ln R\,:\,(\Gamma_0^{o})^2 \rightarrow \mathbb{R}$ is real-analytic.
\end{itemize}
\end{hypothesis}
\begin{proposition}
\label{poeq} If Hypothesis~\ref{auq} holds, then $\mathcal{E}$ admits a unique minimizer $\mu_{\mathrm{eq}} \in \mathcal{M}_{1}(\Gamma_0)$. It is characterized by the existence of a constant $C$ such that:
\beq
\label{saqq}2\int_{\Gamma_0} \dd\mu_{\mathrm{eq}}(y)\ln R_0(x,y) \leq \mathcal{V}(x) + C,
\eeq
with equality $\mu_{\mathrm{eq}}$-almost everywhere. The support $\Gamma$ of $\mu_{\mathrm{eq}}$  is included in a compact of $\Gamma_0^{o}$, consists of the disjoint union of segments $(\gamma_j)_{1 \leq j \leq r}$, and has continuous density in $\mathring{\Gamma}$. Besides, if $\tilde{\Gamma}_0 = \bigcup_{j = 1}^{r} \tilde{\gamma_j}$ is a disjoint union of segments so that $\tilde{\gamma}_j$ is a neighborhood of $\gamma_j$ in $\Gamma_0$, $\mu_{\mathrm{eq}}$ is also the unique minimizer of $\mathcal{E}$ on $\mathcal{M}_1(\tilde{\Gamma}_0)$.
\end{proposition}
\begin{proposition}
\label{sajuq}
If Hypothesis~\ref{auq} holds, the random empirical measure $\frac{1}{N}\sum_{i = 1}^N \delta_{\lambda_i}$ whose distribution is induced by \eqref{mes2} converges almost surely and in expectation to $\mu_{\mathrm{eq}}$, and $\lim_{N \rightarrow \infty} \frac{1}{N^2} \ln Z_N = -\mathcal{E}[\mu_{\mathrm{eq}}]$.
\end{proposition}
\textbf{Proofs.} These are classical results of potential theory in the case $R(x,y) \equiv 1$ \cite{SaffTotik,Defcours,BookAG}, that we actually do not state with optimal assumptions. The proof can easily be generalized to a strictly convex interaction $R_0$, since this assumption guarantees the uniqueness of a minimizer of $\mathcal{E}$. See e.g. \cite{Venker1} for some details when $R(x,y) \neq 1$. \hfill $\Box$

The assumption $(i)$ on strong confinement was chosen to simplify the presentation. The same conclusion holds if it is weakened so as to keep the support compact. The case of non-compact supports is interesting but beyond the scope of this article, see e.g. \cite{Hard} for potential-theoretic results for $\mathbb{R} \equiv 1$. The assumption $(ii)$ on strictly convex interactions is a convenient framework under which existence and unicity of the equilibrium measure is guaranteed, but might be relaxed if the latter can be established by other means. The assumptions $(iii)$ and $(iv)$ about analyticity of $\mathcal{V}$ and $\ln R$ are only used to ensure that $\Gamma$ is a finite union of segments, as can be observed on \eqref{sqisol} below. Within the present Section~\ref{Popo}, we may replace them by requiring directly that $\Gamma$ is a finite union of segments.

\subsubsection*{Stieltjes transform and analytical continuation}

A complex measure (a fortiori a probability measure) $\mu$ on $\Gamma_0$ can be characterized by its Stieltjes transform:
\beq
\omega(x) = \Big(\int_{\Gamma_0} \frac{\dd\mu(y)}{x - y}\Big)\dd x.
\eeq
$\omega(x)$ is a holomorphic $1$-form in $\widehat{\mathbb{C}}\setminus\mathrm{supp}\,\mu$, which behaves as $\omega(x) \sim \frac{\dd x}{x}$ when $x \rightarrow \infty$ away from $\mathrm{supp}\,\mu$. Equivalently, for any $x \in \Gamma_0$, we have in the sense of distributions:
\beq
2{\rm i}\pi\,\dd\mu(x) = \omega(x - {\rm i}0) - \omega(x + {\rm i}0).
\eeq
In particular, $\omega(x)$ is discontinuous at any interior point of $\mathrm{supp}\,\mu$.

The singular integral equation satisfied by the equilibrium measure \eqref{saqq} can be rewritten in terms of its Stieltjes transform $\omega_{\mathrm{eq}}$:
\beq
\label{funcq}\forall x \in \mathring{\Gamma},\qquad \omega_{\mathrm{eq}}(x + {\rm i}0) + \omega_{\mathrm{eq}}(x - {\rm i}0) + \frac{2\rho}{\beta}\,\frac{1}{2{\rm i}\pi}\oint_{\Gamma} \dd_x\ln R(x,y)\omega_{\mathrm{eq}}(y) = \dd \mathcal{V}(x).
\eeq
Given that $\ln R(x,y)$ is holomorphic in a neighborhood of $\Gamma^2$, the last term in the left-hand side is a holomorphic $1$-form in $x$ in a neighborhood of $\Gamma$.

$\mathbb{C}\setminus\Gamma$ (resp. $\widehat{\mathbb{C}}\setminus\Gamma$) defines a domain $U$ (resp. $\widehat{U}$) in the sense of \S~\ref{ss1}, and we now use extensively the notations of Section~\ref{S1}. The coordinate $x$ defines a function $x\,:\,\widehat{U}_{\Gamma} \rightarrow \widehat{\mathbb{C}}$ which is $\iota$-invariant. In general, we identify a function (or a $1$-form $f$) defined on $\widehat{\mathbb{C}}\setminus\Gamma$ and its pullback $x^*f$, which is a function (or a $1$-form) defined on $U \subseteq U_{\Gamma}$. The assumption $(iii)$ (resp. $(iv)$) allows to define $\mathcal{V}$ (resp. $\ln R$) as a $\iota$-invariant holomorphic function on $V$ (resp. in $V^2$). Therefore, the functional equation \eqref{funcq} shows that $\omega_{\mathrm{eq}} \in \Hs_{\Gamma}(U)$ (see Definition~\ref{def22}) and defines a $1$-form $\omega_{\mathrm{eq}} \in \Ms(U_{\Gamma})$, which satisfies:
\beq
\forall z \in V,\qquad \omega_{\mathrm{eq}}(z) + \omega_{\mathrm{eq}}(\iota(z)) + \frac{2\rho}{\beta}\,\mathcal{O}\omega_{\mathrm{eq}}(z) = \dd \mathcal{V}(z),
\eeq
where\footnote{For \label{totor} the definition to make sense, we restricted ourselves to a subspace $\Ms^*(V)$ of $\Ms(V)$ consisting of $1$-forms which do not have poles on $\Gamma^{\mathrm{ext}}$. Since $\Gamma^{\mathrm{ext}}$ is a floating contour, this means that we require poles in $\coprod_{j = 1}^r V_j$ only arise in the complement of a neighborhood of $\Gamma$. This technicality is not important, except in the proof of  Proposition~\ref{Berg} where it will be pointed out, so the reader may also consider that $\Ms^* \approx \Ms$.} we have introduced an operator $\mathcal{O}\,:\,\Ms^*(V) \rightarrow \Hs_{\Gamma}^{\mathrm{inv}}(V)$ by:
\beq
\label{OOdef}\mathcal{O}f(z) = \frac{1}{2{\rm i}\pi}\oint_{\Gamma} \dd_{z} \ln R(z,\zeta)\,f(\zeta).
\eeq
As a consequence of Proposition~\ref{sajuq}:
\begin{corollary}
\beq
\lim_{N \rightarrow \infty} \frac{W_1(x)}{N} = \frac{\omega_{\mathrm{eq}}(x)}{\dd x}
\eeq
and the convergence is uniform for $x$ in any compact of $\mathbb{C}\setminus\Gamma_0$. \hfill $\Box$
\end{corollary} 
For generic $\mathcal{V}$, the $1$-form $\omega_1^0 = \omega_{\mathrm{eq}} \in \Ls_{\Gamma}(U)$ will be off-critical, in the sense of Definition~\ref{offc}.

\subsubsection*{Regularity and fixed filling fractions}

The \emph{filling fractions} are the partial masses of $\mu_{\mathrm{eq}}$ on the connected components of the support: $\epsilon_j^* = \mu_{\mathrm{eq}}(\gamma_j)$. Notice that $\sum_{j = 1}^r \epsilon_j^* = 1$. We would like to study variations of $\mu_{\mathrm{eq}}$ with respect to the potential, and when $r \geq 2$, with respect to filling fractions as well. If we vary the potential, the filling fractions will change. We also prefer to disentangle those variations.

Let $\tilde{\gamma}_j$ be a neighborhood of $\gamma_j$ in $\Gamma$, and set $\tilde{\Gamma} = \coprod_{j = 1}^{r} \tilde{\gamma}_j$. Let $h\,:\,\tilde{\Gamma} \rightarrow \mathbb{R}$ be a real-analytic function on $\tilde{\Gamma}$. Let $\sigma$ be the simplex $\{\epsilon \in (\mathbb{R}_+^\times)^r,\quad \sum_{j = 1}^r \epsilon_j = 1\}$. If $\epsilon \in \sigma$, we denote $\mathcal{P}_{\epsilon}(\tilde{\Gamma})$ the set of probability measures $\mu$ on $\tilde{\Gamma}$ so that $\mu[\tilde{\gamma}_j] = \epsilon_j$ for any $j \in \ldbrack 1,r \rdbrack$. Restricted to this convex set, the strictly convex functional $\mathcal{E}[\mu]$ has a unique minimizer, that we denote momentarily $\mu_{\mathrm{eq}}[V,\epsilon]$. We believe that, for generic $\mathcal{V}$ and $\epsilon$ generic, the equilibrium measure is $\mathcal{C}^1$ with respect to potential and filling fractions. Since we do not have an explicit formula to describe the equilibrium measure, a proof would involve more functional analysis, so we only present this proposal as a conjecture:

\begin{conjecture}
\label{theconj} If Hypothesis~\ref{auq} holds, for $\mathcal{V}$ and $\epsilon$ generic, there exists a linear map $\mu'_{\mathrm{eq}}[V,\epsilon]$, defined over triples $(h,\delta,f)$ consisting of an admissible function $h\,:\,\tilde{\Gamma} \rightarrow \mathbb{R}$, a vector $\delta \in \mathbb{R}^r$ so that $\sum_{j = 1}^r \delta_j$, and a bounded continuous function $f\,:\,\tilde{\Gamma} \rightarrow \mathbb{R}$, so that:
\beq
\mu'_{\mathrm{eq}}[V,\epsilon]\cdot(h,\delta,f) = \lim_{t \rightarrow 0} \frac{1}{t}\int_{\tilde{\Gamma}} f(x)\,\dd\mu_{\mathrm{eq}}[V + th,\epsilon + t\delta](x)
\eeq
\end{conjecture}
By linearity, $\mu'_{\mathrm{eq}}[V,\epsilon]$ can be extended to complex-valued $h$ whose real and imaginary part are admissible, and complex valued $f$. The difficult point in Conjecture~\ref{theconj} is to justify differentiability of $\mu_{\mathrm{eq}}[V + th,\epsilon + t\delta]$ and regularity of the support $\Gamma[V,\epsilon]$ when $t$ is small enough. Then, we can differentiate the relations:
\bea
\forall x \in \mathring{\Gamma}[V,\epsilon] & \quad & \int 2\partial_{x}\ln R_0(x,y)\dd\mu_{\mathrm{eq}}[V,\epsilon](y) = V'(x) \\
\forall j \in \ldbrack 1,r \rdbrack & \quad & \int_{\tilde{\Gamma}_j} \dd\mu_{\mathrm{eq}}[V,\epsilon](y) = \epsilon_j 
\eea
to find the functional equation for $\mu'_{\mathrm{eq}}$:
\bea
\forall x \in \mathring{\Gamma}[V,\epsilon] & \quad & \mu_{\mathrm{eq}}'[V,\epsilon]\cdot(2\,\partial_{x}\ln R_0(x,\bullet),\delta,h) = h'(x) \nonumber \\
\label{fuco}\forall j \in \ldbrack 1,r \rdbrack & \quad & \mu_{\mathrm{eq}}'[V,\epsilon]\cdot(\mathbf{1}_{\gamma_{j}},\delta,h) = \delta_{j}
\eea
We also give two useful results relating $\mu'_{\mathrm{eq}}$ to the second derivative of the energy functional:
\begin{lemma}
\label{pqsy} If Conjecture~\ref{theconj} holds, we have, for any admissible $f$:
\beq
\mu'_{\mathrm{eq}}[V,\epsilon]\cdot(h,0,f) = \mu'_{\mathrm{eq}}[V,\epsilon]\cdot(f,0,h) = \frac{\partial^2}{\partial t\partial s}\Big|_{\substack{t = 0 \\ s = 0}} \mathcal{E}\big[\mu_{\mathrm{eq}}[V + th + sf,\epsilon]\big]
\eeq
\end{lemma}
\textbf{Proof.} Let us define:
\beq
F(t,s) = \mathcal{E}[\mu_{\mathrm{eq}}[V + th + sf,\epsilon]\big]
\eeq
We have:
\bea
F(t,s) & = & - \iint_{\Gamma_0^2} \dd\mu_{\mathrm{eq}}[V + th + sf,\epsilon](x)\dd\mu_{\mathrm{eq}}[V + th + sf,\epsilon](y)\,\ln R_0(x,y) \nonumber \\
& & + \int_{\Gamma_0} \dd\mu_{\mathrm{eq}}[V + th + sf,\epsilon](x)\,(V + th + sf)(x) 
\eea
and if Conjecture~\ref{theconj} holds, we can differentiate for $(t,s)$ small enough:
\bea
\partial_t F(t,s) = & = & \mu_{\mathrm{eq}}'[V + th + sf,\epsilon]\Big(h,0,V - 2\int_{\Gamma_0} \dd\mu_{\mathrm{eq}}[V + th + sf,\epsilon](y)\ln|\bullet - y|\Big) \nonumber \\
& & + \int_{\Gamma_0} \dd\mu_{\mathrm{eq}}[V + th + sf,\epsilon](x)\,h(x) \nonumber \\
& = & \int_{\Gamma_0} \dd\mu_{\mathrm{eq}}[V + th + sf,\epsilon](x)\,h(x) 
\eea
because of the characterization of $\mu_{\mathrm{eq}}$. Similarly, we can compute $\partial_{s} F(t,s)$. The answer is $\mathcal{C}^1$, hence $F$ is $\mathcal{C}^2$ near $(0,0)$. In particular, we find:
\beq
\partial_{s = 0}\partial_{t = 0} F(t,s) = \mu'_{\mathrm{eq}}[V](f,0,h) = \partial_{t = 0}\partial_{s = 0} F(t,s) = \mu'_{\mathrm{eq}}[V](h,0,f).
\eeq
\hfill $\Box$.

\subsubsection*{$1^{\mathrm{st}}$ kind differentials}
\label{1stk}

When $r \geq 2$, we introduce the basis of \emph{$1$st kind differentials} as the Stieltjes transform of variations of $\mu_{\mathrm{eq}}$ with respect to filling fractions.
\begin{definition}
If we let $(\mathrm{e}_{1},\ldots,\mathrm{e}_{r})$ the canonical basis of $\mathbb{R}^r$, we define the holomorphic $1$-forms $(h_i)_{1 \leq i \leq r - 1}$ by:
\beq
h_{i}(x) = \mu_{\mathrm{eq}}'[V,\epsilon]\cdot(0,\mathrm{e}_{i} - \mathrm{e}_{r},w_{x})\,\dd x,\qquad w_{x}(\xi) = \frac{1}{x - \xi}
\eeq
\end{definition}
It is uniquely characterized by the functional relation:
\beq
\forall x \in \mathring{\Gamma},\qquad \int 2 h_{i}(y)\,\partial_{x} \ln R_0(x,y) = 0
\eeq
The functional equation deduced from \eqref{fuco} allows to upgrade $h_i$ to a holomorphic $1$-form on $\widehat{U}_{\Gamma}$, such that:
\beq
\forall z \in V,\qquad h_i(z) + h_i(\iota(z)) + \frac{2\rho}{\beta}\,\mathcal{O}h_i(z) = 0,
\eeq
and:
\beq
\forall (i,j) \in \ldbrack 1,r - 1 \rdbrack \times \ldbrack 1,r \rdbrack,\qquad \frac{1}{2{\rm i}\pi}\oint_{\gamma_j^{\mathrm{ext}}} h_i = \delta_{j,i} - \delta_{j,r} .
\eeq
Notice that the cycle $\sum_{j = 1}^{r - 1}\gamma_j^{\mathrm{ext}}$ is homologically equivalent in $\widehat{U}_{\Gamma}$ to $-\gamma_{r}^{\mathrm{ext}}$.

\subsubsection*{Fundamental $2$-form of the $2^{\mathrm{nd}}$ kind and local Cauchy kernel}

\begin{definition}
\label{defBerg}
A \emph{fundamental $2$-form of the $2^{\mathrm{nd}}$ kind} is a meromorphic $2$-form in $(z_0,z) \in \widehat{U}_{\Gamma}$, denoted $B(z_0,z)$, such that:
\begin{itemize}
\item[$\bullet$] $B(z_0,z) = B(z,z_0)$.
\item[$\bullet$] The only singularity of $B(z_0,z)$ is a double pole at $z = z_0$ with leading coefficient $1$ and without residue.
\item[$\bullet$] It satisfies the functional equation, for any $z_0 \in U$ and $z \in V$:
\beq
\label{338}B(z_0,z) + B(z_0,\iota(z)) + \frac{2\rho}{\beta}\,\mathcal{O}^{z}B(z_0,z) = \frac{\dd x(z_0)\,\dd x(z)}{(x(z_0) - x(z))^2}.
\eeq
\end{itemize}
We say it is \emph{normalized} on $(\gamma_j)_j$ if, for any $j \in \ldbrack 1, r \rdbrack$, $\oint_{\gamma_j^{\mathrm{ext}}} B(z_0,\cdot) = 0$.
\end{definition}
\begin{lemma}
\label{Corpo}
$G(z_0,z) = -\int^{z} B(z_0,\cdot)$ is a local Cauchy kernel, which satisfies, for all $z \in U$ and $z_0 \in V$:
\beq
\mathcal{S}^{z_0}G(z_0,z) + \frac{2\rho}{\beta}\,\mathcal{O}^{z_0}G(z_0,z) = \frac{\dd x(z_0)}{x(z_0) - x(z)} + \mathrm{constant}.
\eeq
\end{lemma}
\textbf{Proof.} It follows from the description of the singularities of $B(z_0,z)$, and the functional equation for $B(z_0,z) = B(z,z_0)$ with respect to the variable $z_0$. \hfill $\Box$

\begin{proposition}
\label{Berg} If Hypothesis~\ref{auq} and Conjecture~\ref{theconj} hold, there exists a fundamental $2$-form of the $2^{\mathrm{nd}}$ kind (it will be proved to be unique in Corollary~\ref{coa2}).
\end{proposition}
 
\noindent \textbf{Proof}. Let $x_0 \in \mathbb{C}\setminus\Gamma$. Thanks to Conjecture~\ref{theconj}, we can compute the variation of the equilibrium measure with respect to the function $w_{x_0} = \frac{1}{x - x_0}$, and then its Stieltjes transform, i.e. we define, for $x \in \mathbb{C}\setminus\Gamma$:
\beq
\label{stis} \tilde{B}(x_0,x) = \mu'_{\mathrm{eq}}[V,\epsilon]\cdot(w_{x_0},0,w_x)\,\dd x_0\dd x
\eeq
By Lemma~\eqref{pqsy}, we deduce that $\tilde{B}(x_0,x) = \tilde{B}(x,x_0)$. By construction in \eqref{stis}, $\tilde{B}(x_0,\cdot)$ is a holomorphic $1$-form on $\mathbb{C}\setminus\Gamma$, and even on $\widehat{\mathbb{C}}\setminus\Gamma$ since $\mu_{\mathrm{eq},g_{x_0}}'$ and $\mu_{\mathrm{eq},h_{x_0}}'$ have zero total mass. This implies by symmetry that $\tilde{B}$ is a holomorphic $2$-form in $(\widehat{\mathbb{C}}\setminus\Gamma)^2$. The characterization \eqref{338} becomes, in terms of Stieltjes transform: for any $x_0 \in \mathbb{C}\setminus\Gamma$ and $x \in \mathring{\Gamma}$,
\beq
\tilde{B}(x_0,x + {\rm i}0) + \tilde{B}(x_0,x - {\rm i}0) + \frac{2\rho}{\beta}\,\mathcal{O}^{x}\tilde{B}(x_0,x) = -\frac{\dd x_0\,\dd x}{(x - x_0)^2}.
\eeq
By previous arguments, it can be upgraded to a meromorphic $1$-form $\tilde{B}(z_0,z)$ for $(z_0,z) \in \widehat{U}\times \widehat{U}_{\Gamma}$, which satisfies, for any $(z_0,z) \in U\times V$:
\beq
\label{lavoila}\mathcal{S}^{z}\tilde{B}(z_0,z) + \frac{2\rho}{\beta}\,\mathcal{O}^{z}\tilde{B}(z_0,z) = -\frac{\dd x(z)\,\dd x(z_0)}{(x(z_0) - x(z))^2}.
\eeq
where $\mathcal{O}^{z}$ denotes the operator defined in \eqref{OOdef}, acting on the variable $z$.

The candidate for the fundamental $2$-form of the $2^{\mathrm{nd}}$ kind is:
\beq
\label{juqi}B(z_0,z) = \tilde{B}(z_0,z) + \frac{\dd x(z_0)\,\dd x(z)}{(x(z_0) - x(z))^2}.
\eeq
Since the last term added in \eqref{juqi} is holomorphic in a neighborhood of $\Gamma$ and $\iota$-invariant, it is annihilated by $\mathcal{O}^{z}$ and we deduce from \eqref{lavoila}, for any $(z_0,z) \in U\times V$:
\beq
\label{jki0}\mathcal{S}^{z}B(z_0,z) + \frac{2\rho}{\beta}\,\mathcal{O}^{z}B(z_0,z) = \frac{\dd x(z_0)\,\dd x(z)}{(x(z_0) - x(z))^2}.
\eeq
By construction from \eqref{fuco}, the periods when $z$ goes around $\gamma_j^{\mathrm{ext}}$ must vanish. It remains to show that $B(z_0,z)$ can be extended to $\widehat{U}_{\Gamma}^2$ and to describe its singularities. If $z_0 \in U_j$, we may move the contour $\Gamma^{\mathrm{ext}}$ in $\mathcal{O}^{z}$ so as to surround $z_0$. Taking into account the double pole of $B(z_0,z)$ at $z = z_0$, coming from the last term added in \eqref{juqi}, we find:
\beq
\mathcal{O}^{z}B(z_0,z) = -\dd_{z}\dd_{z_0} \ln R(z,z_0) + \oint_{\Gamma^{\mathrm{ext}}\cup\{z_0\}} \!\!\!\!\!\!\!\!\!\!\!\!\!\dd_z\ln R(z,\zeta)B(z_0,\zeta).
\eeq
Thus, we can analytically continue $B(z_0,z)$ to $z_0 \in U_j'$ with the formula:
\beq
\label{jki}\mathcal{S}^{z}B(z_0,z) + \frac{2\rho}{\beta}\,\mathcal{O}^{z}B(z_0,z) = \frac{\dd x(z_0)\,\dd x(z)}{(x(z_0) - x(z))^2} + \dd_{z_0}\dd_{z}\ln R(z_0,z).
\eeq
In this way, we have defined $B(z,z_0)$ as a meromorphic $2$-form in $(z,z_0) \in \widehat{U}_{\Gamma}^2$. Since $B(z,z_0) = B(z_0,z)$ for $(z,z_0) \in \widehat{U}^2$, the symmetry must hold for $(z,z_0) \in \widehat{U}_{\Gamma}^2$. We already know that the only singularity of $B(z_0,z)$ when $z \in \widehat{U}$ is a double pole at $z = z_0$ with leading coefficient $1$ and without residue. Remind that in $\widehat{U}_{\Gamma}$, $V_j$ can be described as the gluing along $\gamma_j$ of $U_j$ and $U_j'$ (see Fig.~\ref{Fig1}). Let $z_0 \in \widehat{U}$, and consider $z \in U_j'$. Using \eqref{jki0}, we find:
\beq
B(z_0,z) = -B(z_0,\iota(z)) - \mathcal{O}^{z} B(z_0,z) + \frac{\dd x(z_0)\,\dd x(z)}{(x(z_0) - x(z))^2}.
\eeq
Since $\mathcal{O}^{z} B(z_0,z)$ is regular when $z \in V_j$, the only other singularity of $B(z_0,z)$ could be a double pole at $z = \iota_j(z_0)$, but it does not occur since the first term in the right-hand side has leading coefficient $-1$ at $z = \iota_j(z_0)$, while the last term has leading coefficient $1$ and both have no residue. The last case to study is $z_0 \in U_{j_0}'$ and $z \in U_{j}'$, for which we can use \eqref{jki} to write:
\beq
B(z_0,z) = -B(z_0,\iota(z)) - \mathcal{O}^{z}B(z_0,z) + \dd_{z_0}\dd_{z}\ln R(z_0,z) + \frac{\dd x(z_0)\,\dd x(z)}{(x(z_0) - x(z))^2}.
\eeq
Since $\iota(z) \in U$ and $z_0 \in U_{j_0}'$, we deduce by using symmetry and the property we just proved that the first term in the right-hand side is regular. The only singularity of the right-hand side comes from the last term, and is a double pole at $z = z_0$ with leading coefficient $1$ and no residue, so the proof is complete.
\hfill $\Box$

\subsection{Representation by residues}
\label{repres}
The inhomogeneous linear equations of the form:
\beq
\label{TTZ}\forall z \in V,\qquad \mathcal{S} f(z) + \frac{2\rho}{\beta}\,\mathcal{O} f(z) = T(z),
\eeq
where $T \in \Hs_{\Gamma}^{\mathrm{inv}}$, plays a key role in our construction. This equation was closely related to a saddle point condition for the functional $\mathcal{E}$. An easy particular solution of \eqref{TTZ} is $T(z)/2$, and $\breve{f}(z) = f(z) - T(z)/2$ now solves the homogeneous linear equation, i.e. with vanishing right-hand side. Therefore, we would like to describe the subspace $\mathcal{H}$ of $\Hs_{\Gamma}(\widehat{U})$ consisting of $1$-forms $f$ satisfying:
\beq
\label{homoe}\forall z \in V,\qquad \mathcal{S} f(z) + \frac{2\rho}{\beta}\,\mathcal{O} f(z) = 0.
\eeq
\begin{proposition}
\label{410} If Hypothesis~\ref{auq} holds, $\mathcal{H}$ is representable by residues, with local Cauchy kernel $G(z_0,z)$ defined in Lemma~\ref{Corpo}.
\end{proposition}
\textbf{Proof.} For any $f \in \mathcal{H}$, consider the $1$-form:
\beq
\tilde{f}(z_0) = \sum_{\alpha \in \Gamma^{\mathrm{fix}}} \Res_{z \rightarrow \alpha} \Big(-\int^{z} B(z_0,z)\Big)f(z) = \sum_{\alpha \in \Gamma^{\mathrm{fix}}} \Res_{z \rightarrow \alpha} \Big(-\int^{z} \breve{B}(z_0,z)\Big)f(z).
\eeq
Since $\mathcal{S} f(z)$ is regular at the ramification points, we could replace $B(z_0,z)$ by:
\beq
\breve{B}(z_0,z) = B(z_0,z) - \frac{1}{2}\,\frac{\dd x(z_0)\,\dd x(z)}{(x(z_0) - x(z))^2}
\eeq
without affecting the residues. By construction, $\breve{B}(z_0,z)$ satisfies the homogeneous linear equation with respect to its variable $z$. Hence, $\tilde{f} \in \mathcal{H}$, and since $G(z_0,z)$ is a local Cauchy kernel, $\tilde{f} - f \in \Hs(U_{\Gamma})$. \hfill $\Box$

\vspace{0.2cm}

\noindent When the support $\Gamma$ consists of $r \geq 2$ segments, we cannot hope $\mathcal{H}$ to be normalized. Indeed, we have $1$-forms of the $1^{\mathrm{st}}$ kind, which are non-zero holomorphic elements of $\mathcal{H}$. However, we claim:
\begin{lemma}
\label{ksq}
Assume $R_0$ is a strictly convex interaction. Let $f \in \Hs(\widehat{U}_{\Gamma})\cap\mathcal{H}$ such that, for any $j \in \ldbrack 1,r \rdbrack$, $\oint_{\gamma_j} f = 0$. Then $f \equiv 0$.
\end{lemma}
This leads us to introduce:
\beq
\mathcal{H}_0 = \Big\{f \in \mathcal{H},\qquad \forall j \in \ldbrack 1,r \rdbrack,\quad \oint_{\gamma_j^{\mathrm{ext}}} f = 0 \Big\}.
\eeq
\begin{corollary}
\label{coa}
If Hypothesis~\ref{auq} holds, $\mathcal{H}_0$ is normalized, and $\mathcal{H} = \mathrm{span}(h_1,\ldots,h_{r - 1}) \oplus \mathcal{H}_0$.
\end{corollary}
Eventually, we may give an alternative characterization of the Cauchy kernel adapted to the subspace $\mathcal{H}$.
\begin{corollary}
\label{coa2} If Hypothesis~\ref{auq} holds, there is a unique fundamental $2$-form of the $2^{\mathrm{nd}}$ kind normalized on $(\gamma_j)_j$ in the sense of Definition~\ref{defBerg}.
\end{corollary}

\noindent \textbf{Proof of Lemma~\ref{ksq}.} Let $f \in \Hs(\widehat{U}_{\Gamma})\cap\mathcal{H}$. It can be represented as the Stieltjes transform of a complex measure supported on $\Gamma$, namely $\dd\nu_{f}(x) = \frac{1}{2{\rm i}\pi}\big(f(x - {\rm i}0) - f(x + {\rm i}0)\big)$ if we identify $f$ to an element of $\Hs(\widehat{\mathbb{C}}\setminus\Gamma)$. Integrating \eqref{homoe} with respect to $x$ and rewriting in terms of $\nu$, we obtain
\beq
\forall x \in \Gamma,\qquad \beta\int_{\Gamma} \dd\nu_f(\xi)\big(2\ln|x - \xi| + \frac{2\rho}{\beta}\,\ln R(x,\xi)\big) = C_j
\eeq
for some constant $C_j$. Let us integrate this relation against the complex conjugate of $\dd\nu_{f}(x)$ over $\gamma_j$, and sum over $j$. We find:
\beq
\iint_{\Gamma^2} (\dd\nu_f(x))^*\dd\nu_f(\xi)\big(2\ln|x - \xi| + \frac{2\rho}{\beta}\,\ln R(x,\xi)\big) = \sum_{j = 1}^r C_j \Big(\int_{\gamma_j} \dd\nu_f(x)\Big)^*,
\eeq
and notice that $\int_{\gamma_j} \dd\nu_f(x) = \frac{1}{2{\rm i}\pi}\oint_{\gamma_j^{\mathrm{ext}}} f$. Hence, assuming that $f$ has vanishing periods around $\gamma_j$ imply that $\nu_{f}(\gamma_j) = 0$, and a fortiori $\nu_{f}(\Gamma_0) = 0$. By strict convexity (see Lemma~\ref{leoa}), we deduce that $\nu_f \equiv 0$, hence $f \in \Hs_{\Gamma}^{\mathrm{inv}}$. Thus, $\mathcal{O} f(z) \equiv 0$ and \eqref{homoe} implies $f \equiv 0$. \hfill $\Box$

\vspace{0.2cm}

\noindent \textbf{Proof of Corollary \ref{coa}.} Notice that the cycle $\sum_{j = 1}^r \gamma_j^{\mathrm{ext}}$ of $\widehat{U}$ is homologically equivalent to the trivial cycle (we may contract it through the $\infty$ point). Hence, the $1^{\mathrm{st}}$ kind differentials are linearly dependant: $\sum_{j = 1}^r h_j = 0$, but $(r - 1)$ of them are independent. For any $\omega \in \mathcal{H}$, if we denote $\epsilon_j = \frac{1}{2{\rm i}\pi}\oint_{\gamma_j} \omega$, we have $(\omega - \sum_{j = 1}^r \epsilon_j h_j) \in \mathcal{H}_0$. \hfill $\Box$

\vspace{0.2cm}

\noindent \textbf{Proof of Corollary \ref{coa2}.} If $B_1$ and $B_2$ are two such $2$-forms, $B_1(z,z_0) - B_2(z,z_0)$ satisfies the homogeneous equation \eqref{homoe} with respect to $z$, has vanishing periods around the $\gamma_j$, and is holomorphic in $\widehat{U}$. According to Lemma~\ref{ksq}, we must have $B_1 \equiv B_2$. \hfill $\Box$

\subsection{Schwinger-Dyson equations}
\label{SDSQ}
Schwinger-Dyson equations can be derived by integration by parts, or change of variables in the integrals. They are exact for any finite $N$ and do not depend on the contour $\Gamma_0$. To simplify the exposition, we assume that there is no boundary terms. It happens for instance when $\Gamma_0$ is a union of arcs, and the interactions are strongly confining at the endpoints of $\Gamma_0$ in the sense of Definition~\ref{sconf}. It would not be difficult to include effects of boundaries in the equations below, and our conclusion would hold the same (in the case $R \equiv 1$, see for instance \cite{C06}).

\begin{lemma}
\label{SD} For any $x,x_2,\ldots,x_n$ in $\mathbb{C}\setminus\Gamma_0$:
\bea
\label{aaua} -\Big(1 - \frac{2}{\beta}\Big)\Big\langle \Tr \frac{1}{(x - M)^2} \prod_{i \in I} \Tr \frac{1}{x_i - M}\Big\rangle_c + \Big\langle \Big(\Tr\frac{1}{x - M}\Big)^2\,\prod_{i \in I} \Tr \frac{1}{x_i - M}\Big\rangle_c & & \\
+ \sum_{J \subseteq I} \Big\langle \Tr \frac{1}{x - M}\prod_{j \in J}\Tr\frac{1}{x_j - M}\Big\rangle_c\Big\langle\Tr\frac{1}{x - M}\prod_{j' \in I \setminus J} \Tr\frac{1}{x_{j'} - M}\Big\rangle_c  & & \nonumber  \\ 
- \frac{2}{\beta}\Big\langle \Tr\frac{N\,\mathcal{V}'(M)}{x - M} \prod_{i \in I} \Tr\frac{1}{x_i - M}\Big\rangle_c + \frac{2}{\beta}\sum_{i \in I} \Big\langle \Tr\,\frac{1}{(x - M)(x_i - M)^2} \prod_{j \in I\setminus\{i\}} \Tr\,\frac{1}{x_j - M}\Big\rangle_c  & & \nonumber  \\
+ \frac{2\rho}{\beta}\Big\langle \Big(\Tr \,\frac{(\partial_1\ln R)(M\otimes\mathbf{1}_N,\mathbf{1}_N\otimes M)}{x - M}\Big)\prod_{i \in I} \Tr\frac{1}{x_i - M}\Big\rangle_c & = & 0.  \nonumber 
\eea
\end{lemma}

\noindent \textbf{Sketch of proof.} Eqn.~\ref{aaua} for $n = 1$ is obtained by performing the infinitesimal change of variable $\lambda_i \rightarrow \lambda_i + \frac{\varepsilon}{x - \lambda_i} + O(\epsilon^2)$ in the integral $Z_N$ (which is invariant since we assumed the absence of boundary terms). Eqn.~\ref{aaua} for $n \geq 2$ is then deduced by writing the equation for $n = 1$ but for a new potential $\mathcal{V}(\lambda) + \sum_{i \in I} \frac{\varepsilon_i}{x_i - \lambda}$ for $i \in I$, and collecting the terms of order $\prod_{i \in I} \varepsilon_i$. All these steps can be justified both for formal integrals, and case by case for convergent integrals. For instance, in the case of convergent integrals over $\Gamma_0 = \mathbb{R}$, we may use, for any smooth function $h\,:\,\mathbb{R} \rightarrow \mathbb{R}$ going to $0$ at $\pm\infty$ and with bounded derivative, the change of variable $\lambda_i \rightarrow \lambda_i + \varepsilon\,h(\lambda_i)$, which is well defined for $\varepsilon$ small enough. And then, we specialize to $h(\lambda) = \mathrm{Re}\,\frac{1}{x - \lambda}$ and $\mathrm{Im}\,\frac{1}{x - \lambda}$ for a given $x \in \Gamma_0$. \hfill $\Box$

\vspace{0.2cm}

\noindent When we assume $\ln R$ analytic in a neighborhood of $\Gamma_0^2$, we can rewrite Eqn.~\ref{aaua} completely in terms of the correlators, with contour integrals around $\Gamma_0$.
\bea
\label{SDeq}\Big(1 - \frac{2}{\beta}\Big) \partial_x W_{n}(x,x_I) + W_{n + 1}(x,x,x_I) + \sum_{J} W_{|J| + 1}(x,x_J)\,W_{n - |J|}(x,x_{I\setminus J}) & & \\
- \frac{2}{\beta}\oint_{\Gamma_0} \frac{\dd\xi}{2{\rm i}\pi}\,\frac{N\,\mathcal{V}'(\xi)\,W_n(\xi,x_I)}{x - \xi} + \frac{2}{\beta}\sum_{i \in I} \oint_{\Gamma_0}\frac{\dd\xi}{2{\rm i}\pi}\,\frac{W_{n - 1}(\xi,x_{I\setminus\{i\}})}{(x - \xi)(x_i - \xi)^2} & & \nonumber \\
+ \frac{2\rho}{\beta}\oint_{\Gamma_0^2} \frac{\dd\xi\,\dd\eta}{(2{\rm i}\pi)^2}\,\frac{(\partial_{\xi} \ln R)(\xi,\eta)}{x - \xi}\Big(W_{n + 1}(\xi,\eta,x_I) + \sum_{J \subseteq I} W_{|J| + 1}(\xi,x_J)W_{n - |J|}(\eta,x_{I\setminus J})\Big) & = & 0. \nonumber
\eea
We call $n$ the \emph{rank} of the equation.

\subsection{Topological expansion of the correlators and loop equations}
\label{topor}
\begin{definition}
\label{top1}The correlators have a large $N$ expansion of \emph{topological type} if, for any $n \geq 1$, 
\beq
\label{exp}W_n(x_1,\ldots,x_n) = \sum_{g \geq 0} N^{2 - 2g - n}\,W_n^h(x_1,\ldots,x_n),
\eeq
where $W_n^g(x_1,\ldots,x_n)\dd x_1\cdots\dd x_n$ is an element of $\Hs(\widehat{U})$ independent of $N$, and the meaning of the right-hand side is either a formal series, or an asymptotic series with uniform convergence for $x_1,\ldots,x_n$ in compact subsets of $\widehat{U} = \widehat{\mathbb{C}}\setminus\Gamma$.
\end{definition}

The goal of this article is not to discuss general conditions which guarantee the existence of such an expansion. For formal integrals, $W_n$ is by construction defined as a formal power series, with a $1/N$ behavior of the form \eqref{exp}, see Section~\ref{S5}. For convergent integrals, with Hypotheses~\ref{auq} and the assumption that $\mathcal{V}$ is off-critical, \eqref{exp} would have to be justified. It is expected to hold in two cases:
\begin{itemize}
\item[$\bullet$] when the support $\Gamma$ is connected ($r = 1$).
\item[$\bullet$] when $\Gamma$ consists of $r$ segments, but in a model with fixed filling fractions.
\end{itemize}
It is clear from the Schwinger-Dyson equations that \eqref{exp} is possible only for $\beta = 2$, otherwise one would find all powers of $1/N$ in the expansion. In the multi-cut case, we do not expect a $1/N$ expansion, but rather \eqref{exp} where the coefficients $W_n^h$ are bounded but featuring fast modulations with $N$, and the heuristic argument of \cite{Ecv} for hermitian matrix models can easily be adapted to describe precisely those coefficients for general systems of repulsive particles.

We would like in the present section to forget about Hypotheses~\ref{auq}, and we shall rather be working with:
\begin{hypothesis}
\label{poiu} 
\begin{itemize}
\item[\phantom{$(o)$}]
\item[$(i)$] $\beta = 2$ ;
\item[$(ii)$] $\Gamma = \bigcup_{j = 1}^r \gamma_j$ and $\gamma_j$ are disjoint bounded intervals of $\mathbb{R}$ ;
\item[$(iii)$] $\mathcal{V}$ is analytic in a neighborhood of $\Gamma$ ;
\item[$(iv)$] $\ln R$ is real-analytic in a neighborhood of $\Gamma^2$ ;
\item[$(v)$] The correlators have a large $N$ expansion of topological type ;
\item[$(vi)$] $W_1^0$ is discontinuous at any interior point of $\Gamma$.
\item[$(vii)$] $\omega_1^0(x) = W_1^0(x)\dd x$ is an off-critical $1$-form.
\end{itemize}
\end{hypothesis}
We may wish to add a stronger condition at some point, so we introduce as well:
\begin{hypothesis}
\label{bus}
$(i)-(vii)$ of Hypothesis~\ref{poiu}, and
\begin{itemize}
\item[$(viii)$] $R_0(x,y) = |x - y|\big(R(x,y)\big)^{\rho/2}$ is a strictly convex interaction.
\end{itemize}
\end{hypothesis}
$(i)$ is mandatory if we want to restrict ourselves to expansions of topological type, and not general $1/N$ expansions. $(ii)-(iv)$ is implied by $(i),(ii)$ and $(iv)$ of Hypothesis~\ref{auq}. $(vi-vii)$ amounts to saying that the density of $\mu_{\mathrm{eq}}$ remains positive on the interior of $\Gamma$ and behaves as a squareroot at the edges, and is satisfied for generic potentials. $(vii)$ in Hypothesis~\ref{bus} is $(iii)$ of Hypothesis~\ref{auq}: it is a convenient framework to analyze the question of uniqueness of solutions of \eqref{homoe}, i.e. to prove that $\mathcal{H}_0$ is normalized. It can be relaxed if one can show normalization by other means.  It is useful even in the context of formal integrals, but it is a technical assumption that one would like to relax in some applications, for instance, in the $O(-\rho)$-model (i.e. $R(x,y) = (x + y)$) with $|\rho| > 2$ \cite{EKOn2}. Eventually, $(v)$ includes the assumption that the leading order of $W_2$ (denoted $W_2^0$) exists, and given that, we do not need to assume Conjecture~\ref{theconj}.

\begin{proposition}
\label{kiosa}
Let us assume Hypothesis~\ref{poiu}, and define $\omega_n^{g} \in \Hs_n(U)$ by the formulas:
\beq
\label{defsq}\omega_n^{g}(z_1,\ldots,z_n) = W_n^{g}(x(z_1),\ldots,x(z_n))\dd x(z_1)\cdots\dd x(z_n) + \delta_{n,2}\delta_{g,0}\,\frac{\dd x(z_1)\,\dd x(z_2)}{(x(z_1) - x(z_2))^2}.
\eeq
Then, $\omega_{\bullet}^{\bullet}$ satisfies linear and quadratic loop equations. More precisely, they satisfy, for any $n,g$, any $z_I = (z_2,\ldots,z_n) \in U^{n - 1}$,
\beq
\label{sjq} \forall z \in V,\qquad \mathcal{S}^{z}\omega_n^{g}(z,z_I) + \rho\,\mathcal{O}^{z}\omega_n^g(z,z_I) = \delta_{g,0}\Big(\delta_{n,1}\dd \mathcal{V}(z) + \delta_{n,2}\frac{\dd x(z)\,\dd x(z_2)}{(x(z) - x(z_2))^2}\Big).
\eeq
\end{proposition}
This proposition is proved below in \S~\ref{rsqp}. In other words, $\omega_n^g \in \mathcal{H}_n$, where $\mathcal{H}$ is the subspace of $\Ms(U_{\Gamma})$ consisting of $1$-forms $f$ satisfying:
\beq
\forall z \in V,\qquad \mathcal{S} f(z) + \rho\,\mathcal{O} f(z) = 0,
\eeq
and $\mathcal{H}_n$ is its $n$-variable analog. We insist on the following intermediate result:\begin{porism}
\label{019}$\omega_2^0(z_0,z)$ is a fundamental $2$-form of the $2^{\mathrm{nd}}$ kind, $G(z_0,z) = -\int^{z} \omega_2^0(z_0,z)$ is a local Cauchy kernel, and $\mathcal{H}$ is representable by residues. \hfill $\Box$
\end{porism}
Those two results hold without assumptions about unicity of solutions of \eqref{homoe}, and we prove them in Section~\ref{rsqp} below. Then, it shows that the topological recursion formula holds in all models of the form \eqref{mes2}:
\begin{corollary}
\label{4333}Let us assume Hypothesis~\ref{bus} and for any stable $n,g$ and any $j \in \ldbrack 1,r \rdbrack$, $\oint_{\gamma_j} \omega_n^{g}(\cdot,z_I) = 0$. Then, stable $\omega_n^g$ can be computed by the topological recursion:
\beq
\omega_n^{g}(z_0,z_I) = \sum_{\alpha \in \Gamma^{\mathrm{fix}}} \Res_{z \rightarrow \alpha} K(z_0,z)\Big(\omega_{n + 1}^{g - 1}(z,\iota(z),z_I) + \sum_{\substack{J \subseteq I,\,\,0 \leq h \leq g \\ (J,h) \neq (\emptyset,0),(I,g)}} \omega_{|J| + 1}^{h}(z,z_J)\,\omega_{n - |J|}^{g - h}(\iota(z),z_{I\setminus J})\Big), \nonumber
\eeq
where the recursion kernel is:
\beq
K(z_0,z) = \frac{-\frac{1}{2}\int_{\iota(z)}^z \omega_2^0(z_0,\cdot)}{\omega_1^0(z) - \omega_1^0(\iota(z))}.
\eeq
\end{corollary}
\textbf{Proof.} Since $R_0$ is a strictly convex interaction, we deduce as in the proof of Corollary~\ref{coa} that the subspace:
\beq
\mathcal{H}_0 = \big\{f \in \mathcal{H}\quad \forall j \in \ldbrack 1,r \rdbrack,\qquad \oint_{\gamma_j^{\mathrm{ext}}} f = 0\big\}
\eeq
is normalized. If the stable $\omega_n^g$ have vanishing periods around $\gamma_j^{\mathrm{ext}}$, they belong to $(\mathcal{H}_0)_n$, which means that the linear loop equations are solvable. Thus, we can apply Proposition~\ref{2222}. \hfill $\Box$

In Corollary~\ref{019}, if the stable $\omega_{n}^{g}(z_0,z_I)$ had non-vanishing periods when $z_0$ goes around $\gamma_j^{\mathrm{ext}}$, it could be computed by the residue formula (yielding an element of $\mathcal{H}_0)$ shifted by a linear combination (with coefficients depending on $z_I$) of $1^{\mathrm{st}}$ kind differentials introduced in \S~\ref{Popo}, so as to achieve the correct periods in $z_0$. Within the Hypothesis~\ref{bus}, it is thus clear, by recursion, that the knowledge of all periods of $\omega_n^{g}$ allows to determine it uniquely and explicitly.

\subsection{Topological expansion of the partition function}

One can also have access to derivatives of the partition function $Z_N$ with respect to any parameters of the potential. The partition function itself may have a prefactor which does not depend on perturbations of the potential $\mathcal{V} \rightarrow \mathcal{V} + t\varphi$ where $\varphi$ is real-analytic on $\Gamma$ and $t$ is small enough, but depends on $N$, so that $\ln Z_N \equiv W_0$ does not necessary has an expansion of topological type.
\begin{proposition}
With the assumptions of Corollary~\ref{4333},
\beq
W_0=F = C_N + \sum_{g \geq 0} N^{2 - 2g}\,F^g
\eeq
where, for any $g \geq 2$, 
\beq
F^g =  \frac{1}{2 - 2g} \sum_{\alpha \in \Gamma^{\mathrm{fix}}}  \Res_{z \rightarrow \alpha} \omega_1^g(z)\Big(\int^{z}\omega_1^0\Big),
\eeq
which does not depend on the choice of primitive for $\omega_1^0$, and $C_N$ does not depend on real-analytic perturbations of $\mathcal{V}$.
\end{proposition}
We refer to \S~\ref{wdv} for the discussion about $F^0$ and $F^1$. We do not address here the computation of the constant $C_N$ which depends on the applications. For applications to topological string theories and for having some symmetry properties, the computation of this constant has been fixed explicitly for example in \cite{BS11,Bouchetal}.

\noindent\textbf{Proof.} One has to check that the derivative of both sides match, with respect to the parameter $t$ shifting the potential to $\mathcal{V}_t = \mathcal{V} + t\varphi$, where $\varphi$ is a real-analytic function on $\Gamma_0$. If the model with potential $\mathcal{V}$ satisfies the assumptions of Corollary~\ref{4333}, so does the model with potential $\mathcal{V}_t$ for $t$ small enough. We know from first principles:
\bea
\label{wu0} \partial_t F & = & -\oint_{\Gamma_0} \frac{\dd\xi}{2{\rm i}\pi}\,\varphi(\xi)\,W_1(\xi), \\
\label{wu1} \partial_t W_1(x) & = & -\oint_{\Gamma_0} \frac{\dd\xi}{2{\rm i}\pi}\,\varphi(\xi)\,W_2(\xi,x), \\
\label{wu2} \partial_t W_2(x_1,x_2) & = & -\oint_{\Gamma_0} \frac{\dd\xi}{2{\rm i}\pi}\,\varphi(\xi)\,W_3(\xi,x_1,x_2).
\eea
If we plug the topological expansion for $W_1$ in \eqref{wu0}, we find that $\partial_t F$ has an expansion of topological type:
\beq
\label{Fexp}\partial_t F = \sum_{g \geq 0} N^{2 - 2g}\Big(-\oint_{\Gamma_0} \frac{\dd\xi}{2{\rm i}\pi}\,W_1^g(\xi)\Big).
\eeq
Besides, if we consider only the leading term of \eqref{wu1}-\eqref{wu2} when $N$ is large, we deduce:
\bea
\partial_t \omega_1^0(z) & = & -\frac{1}{2{\rm i}\pi}\oint_{\Gamma^{\mathrm{ext}}} \varphi(\xi)\,\omega_2^0(\xi,z), \\
\partial_t \omega_2^0(z_1,z_2) & = & -\frac{1}{2{\rm i}\pi}\oint_{\Gamma^{\mathrm{ext}}} \varphi(\xi)\,\omega_3^0(\xi,z_1,z_2).
\eea
Hence, $\partial_t$ is a WDVV-compatible variation in the sense of Definition~\ref{diq}. Then, Corollary~\ref{corof} tells us, for any $g \geq 2$:
\beq
\label{juy}\partial_t F^{g} = \oint_{\Gamma^{\mathrm{ext}}} \frac{\dd\xi}{2{\rm i}\pi}\,\varphi(\xi)\,\omega_1^g(\xi).
\eeq
We thus identify, for any $g \geq 2$, the $N^{2 - 2g}$ term in \eqref{Fexp} with \eqref{juy}. \hfill $\Box$

\subsection{Proof of Proposition~\ref{kiosa}}
\label{rsqp}
We first work with functions on $\mathbb{C}\setminus\Gamma$ rather than with $1$-forms on the domain $U$. For any holomorphic function $f$ defined in a neighborhood of $\Gamma$ in $\mathbb{C}\setminus\Gamma$, we define the function:
\beq
\label{odef}\mathcal{O}f(x) = \frac{1}{2{\rm i}\pi}\oint_{\Gamma} (\partial_x \ln R(x,\xi))f(\xi)\dd\xi,
\eeq
which is holomorphic in a neighborhood of $\Gamma$, and:
\beq
\Delta f(x) = f(x + {\rm i}0) - f(x - {\rm i}0),\qquad \mathcal{S} f(x) = f(x + {\rm i}0) + f(x - {\rm i}0),
\eeq
which are defined for $x \in \Gamma$. We are going to prove, by recursion on $\chi_{n}^{g} = 2g - 2 + n \geq -1$:

\begin{lemma}
\label{lsa}For any $x \in \mathring{\Gamma}$, any $x_I = (x_2,\ldots,x_n) \in (\mathbb{C}\setminus\Gamma)^{n - 1}$,
\beq
\label{sqqq}\mathcal{S}^{x}W_n^{g}(x,x_I) + \rho\,\mathcal{O}^{x}W_n^{g}(x,x_I) = \delta_{g,0}\Big(\delta_{n,1} \mathcal{V}'(x) + \frac{\delta_{n,2}}{(x - x_2)^2}\Big),
\eeq
where the superscript ${}^x$ stresses that $\mathcal{S}$ and $\mathcal{O}$ acts on the variable $x$.
\end{lemma}
\textbf{Proof of the lemma.} At level $\chi = -1$, we just have $(n,g) = (1,0)$. The rank $1$ Schwinger-Dyson equation to leading order in $N$ gives:
\beq
\big(W_1^0(x)\big)^2 + \frac{1}{2{\rm i}\pi}\oint_{\Gamma} \frac{\dd\xi}{x - \xi}\big(\rho\,\mathcal{O}W_1^0(\xi) - \mathcal{V}'(\xi)\big)W_1^0(\xi) = 0.
\eeq
If we take the discontinuity of this equation at $x \in \mathring{\Gamma}$ (i.e. specialize to $x \pm {\rm i}0$ and substract), we find:
\beq
\Delta W_1^0(x) \big(\mathcal{S}W_1^0(x) + \rho\,\mathcal{O}W_1^0(x) - \mathcal{V}'(x)\big) = 0.
\eeq
Thanks to $(vi)$, we arrive to:
\beq
\label{kjb}\forall x \in \mathring{\Gamma},\qquad \mathcal{S}W_1^0(x) + \rho\,\mathcal{O}W_1^0(x) = \mathcal{V}'(x).
\eeq
At level $\chi = 0$, we have $(n,g) = (2,0)$. The rank $2$ Schwinger-Dyson equation to leading order in $N$ gives:
\bea
2W_2^0(x,x_2)W_1^0(x) + \frac{1}{2{\rm i}\pi}\oint_{\Gamma} \frac{\dd\xi}{x - \xi}\,\frac{W_1^0(\xi)}{(\xi - x_2)^2} & & \\
+ \frac{1}{2{\rm i}\pi}\oint_{\Gamma} \frac{\dd \xi}{x - \xi}\Big[\big(\rho\,\mathcal{O}W_1^0(\xi) - \mathcal{V}'(\xi)\big)W_2^0(\xi,x_2) + \rho\,W_1^0(\xi)\,\mathcal{O}^{\xi}W_2^0(\xi,x)\Big] & = & 0. \nonumber 
\eea
We compute its discontinuity at $x \in \mathring{\Gamma}$:
\bea
\Delta^{x} W_2^0(x,x_2)\Big[\mathcal{S}W_1^0(x) + \rho\,\mathcal{O}W_1^0(x) - \mathcal{V}'(x)\Big] & & \\
+ \Delta W_1^0(x)\Big[\mathcal{S}^{x}W_2(x,x_2) + \rho\,\mathcal{O}^xW_2^0(x,x_2) + \frac{1}{(x - x_2)^2}\Big] & = & 0.
\eea
The first line vanishes since we already showed \eqref{kjb}, and thanks to $(vi)$, we find:
\beq
\label{pqu}\mathcal{S}^{x}W_2^0(x,x_2) + \rho\,\mathcal{O}^{x}W_2^0(x,x_2) = -\frac{1}{(x - x_2)^2}.
\eeq
Then, let $\chi \geq 1$, let us assume that \eqref{sqqq} holds for all $n',g'$ such that $\chi_{n'}^{g'} > \chi$, and let $n,g$ such that $\chi_{n}^{g} = \chi$. We collect the term of order $N^{2g - 1 + n}$ in the rank $n$ Schwinger-Dyson equation:
\bea
W_{n + 1}^{g - 1}(x,x,x_I) + \sum_{J \subseteq I,\,\,0 \leq h \leq g} W_{|J| + 1}^{h}(x,x_J)W_{n - |J|}^{g - h}(x,x_{I\setminus J}) & & \\ 
+ \frac{\rho}{2{\rm i}\pi}\oint_{\Gamma} \frac{\dd \xi}{x - \xi}\Big(\mathcal{O}^{\xi,2}W_{n + 1}^{g - 1}(\xi,\xi,x_I) + \sum_{J \subseteq I,\,\,0 \leq h \leq g} W_{|J| + 1}^{h}(\xi,x_J)\,\mathcal{O}^{\xi}W_{n - |J|}^{g - h}(\xi,x_{I\setminus J})\Big) & & \nonumber \\
+ \frac{1}{2{\rm i}\pi}\oint_{\Gamma} \frac{\dd\xi}{x - \xi}\Big(-\mathcal{V}'(\xi)W_n^g(\xi,x_I) + \sum_{i \in I} \frac{W_{n - 1}^{g}(\xi,x_{I\setminus\{i\}})}{(\xi - x_i)^2}\Big) & = & 0, \nonumber
\eea
where $\mathcal{O}^{x,i}$ means that $\mathcal{O}$ acts on the $i^{\mathrm{th}}$ variable only of the function to its right. We organize the computation of the discontinuity of this equation as follows:
\label{las}\bea
\Delta^{x,1}\big[\mathcal{S}^{x,2}W_{n + 1}^{g - 1}(x,x,x_I) + \rho\,\mathcal{O}^{x,2}W_{n + 1}^{g - 1}(x,x,x_I)\big] && \\
+ \sum_{J \subseteq I,\,\,0 \leq h \leq g}^{''} \Delta^{x} W_{|J| + 1}^{h}(x,x_J)\big[\mathcal{S}^{x}W_{n - |J|}^{g - h}(x,x_{I\setminus J}) + \rho\,\mathcal{O}^xW_{n - |J|}^{g - h}(x,x_{I\setminus J})\big] & & \nonumber \\
+ \sum_{i \in I} \Delta^{x} W_{n - 1}^{g}(x,x_{I\setminus\{i\}})\Big[\mathcal{S}^{x}W_2^0(x,x_i) + \rho\,\mathcal{O}^{x}W_2^0(x,x_i) + \frac{1}{(x - x_2)^2}\Big] & & \nonumber \\
+ \Delta^{x}W_n^g(x,x_I)\big[\mathcal{S}W_1^0(x) + \rho\,\mathcal{O}W_1^0(x) - \mathcal{V}'(x)\big] & & \nonumber \\
+ \Delta W_1^0(x)\big[\mathcal{S}^xW_n^g(x,x_I) + \rho\,\mathcal{O}^{x}W_n^g(x,x_I)\big] & = & 0, \nonumber 
\eea
In the third line, $\sum^{''}$ means that we excluded the temrs $(J,h) = (\emptyset,0), (I,g)$ and $(I\setminus\{i\},g)$ from the sum. According to the recursion hypothesis, the brackets in the first four lines vanish, with a word of caution for the first line when $(n,g) = (1,1)$. In this case, we may rewrite:
\bea
& & \Delta^{x,1}\big[\mathcal{S}^{x,2}W_{2}^0(x,x) + \rho\,\mathcal{O}^{x,2}W_2^0(x,x)\big] \\
& = & \lim_{y \rightarrow x} \Delta^{y}\big[\mathcal{S}^{x}W_2^0(y,x) + \rho\,\mathcal{O}^{x}W_2^0(y,x)\big] \nonumber \\
& = & \lim_{y \rightarrow x} \Delta^{y}\Big[\mathcal{S}^x W_2^0(y,x) + \rho\,\mathcal{O}^x W_2^0(y,x) + \frac{1}{(x - y)^2}\Big],
\eea
which indeed vanishes since we already proved \eqref{pqu}. Only the last line of \eqref{las} remains, and thanks to $(vi)$:
\beq
\forall x \in \mathring{\Gamma},\qquad \mathcal{S}^{x}W_n^{g}(x,x_I) + \rho\,\mathcal{O}^{x}W_n^{g}(x,x_I) = 0.
\eeq
By induction, this proves the lemma for any $n,g$. \hfill $\Box$

\vspace{0.2cm}

\noindent Now that we have Lemma~\ref{lsa}, we come back to the proof of Proposition~\ref{kiosa}. Those functional relations imply that, for any $z_I = (z_2,\ldots,z_n) \in \widehat{U}^{n - 1}$, $\omega_n^{g}(\cdot,z_I)$ defined by \eqref{defsq} are continuable $n$-forms across $\Gamma$ (see Definition~\ref{def22}), which satisfy, for any $z \in V$,
\beq
\label{nju}\mathcal{S}^{z}\omega_n^{g}(z,z_I) + \mathcal{O}^{z}\omega_n^{g}(z,z_I) = \delta_{g,0}\Big(\delta_{n,1}\,\dd \mathcal{V}(z) + \delta_{n,2}\,\frac{\dd x(z)\,\dd x(z_2)}{(x(z) - x(z_2))^2}\Big).
\eeq
In particular, for $(n,g) = (2,0)$, notice the change of sign in the right hand side due to the shift between $W_2^0$ and $\omega_2^0$.

Let us have a look at the Schwinger-Dyson equation for $W_1^0(x)$, that we rewrite:
\beq
\label{SDQ}\big(W_1^0(x)\big)^2 - \mathcal{V}'(x)W_1^0(x) + \frac{1}{2{\rm i}\pi}\oint_{\Gamma} \frac{\dd\xi}{x - \xi}\big(\mathcal{V}'(x) - \mathcal{V}'(\xi) + \rho\,\mathcal{O}W_1^0(\xi)\big) = 0.
\eeq
Although the last term is unknown, we know that it is holomorphic in a neighborhood of $\Gamma$. Thus, considering \eqref{SDQ} as a quadratic equation for $W_1^0(x)$, we may solve it\footnote{We choose the sign of the square root such that $W_1^0(x)$ has the right behavior as $x \to \infty$.}:
\beq
\label{sqisol}W_1^0(x) = \frac{\mathcal{V}'(x)}{2} - \sqrt{\frac{(\mathcal{V}'(x))^2}{4} - \frac{1}{2{\rm i}\pi}\oint_{\Gamma} \frac{\dd \xi}{x - \xi}\Big(\mathcal{V}'(x) - \mathcal{V}'(\xi) + \rho\,\mathcal{O}W_1^0(\xi)\Big)}.
\eeq
Let $a$ be a ramification point in $U_{\Gamma}$, i.e. such that $x(a) \in \Gamma$. We infer from \eqref{sqisol} that $W_1^0(x)$ is finite when $x$ approaches $x(a)$, hence $\omega_1^0 \in \Ls_{\Gamma}(U)$. Besides, $W_1^0(x) - W_1^0(x(a)) \in O(\sqrt{x - x(a)})$. We recall that $\sqrt{x - a}$ is a local coordinate in $U_{\gamma}$ near $a$, thus $\Delta\omega_1^0(z)$ has at least double zero at ramification points. And, since $\Delta W_1^0(x)$ does not vanish in the interior of $\Gamma$, this ensures that $\omega_1^0$ is off-critical. Besides, the computation given in the proof of Proposition~\ref{Berg} relies on this functional relation \eqref{nju} for $\omega_2^0$ only (replace $B(z_0,z)$ there by our present $\omega_2^0(z_0,z)$), and shows that $G(z_0,z) = -\int^{z} \omega_2^0(z_0,z)$ is a local Cauchy kernel. Last but not least, since $\mathcal{O}^{z}\omega_n^{g}(z,z_I)$ is holomorphic for $z \in V$, we deduce from \eqref{nju} that $\mathcal{S}\omega_n^{g}(\cdot,z_I) \in \Hs_{\Gamma}^{\mathrm{inv}}$. Therefore, we established linear loop equations.

Now, we are going to recast the Schwinger-Dyson equations so as to obtain quadratic loop equations. As before, we first work in $\mathbb{C}\setminus\Gamma$ with the functions $W_n^g$. Since $W_n^{g}(\xi,x_I) \in O(1/\xi)$ when $\xi \rightarrow \infty$, we may represent:
\bea
& & \mathcal{O}^{x,2}W_{n + 1}^{g - 1}(x,x,x_I) + \sum_{J \subseteq I,\,\,0 \leq h \leq g} W_{|J| + 1}^{h}(x,x_J)\mathcal{O}^{x}W_{n - |J|}^{g - h}(x,x_{I\setminus J}) \\
& = & \frac{1}{2{\rm i}\pi}\oint_{\Gamma} \frac{\dd \xi}{x - \xi}\Big(\mathcal{O}^{x,2}W_{n + 1}^{g - 1}(\xi,x,x_I) + \sum_{J \subseteq I,\,\, 0 \leq h \leq g} W_{|J| + 1}^{h}(\xi,x_J)\mathcal{O}^{x,2}W_{n - |J|}^{g - h}(x,x_{I\setminus J})\Big). \nonumber 
\eea
Thus, Eqn~\ref{SDeq} can be decomposed:
\bea
\label{jui}W_{n + 1}^{g - 1}(x,x,x_I) + \sum_{J \subseteq I,\,\,0 \leq h \leq g} W_{|J| + 1}^{h}(x,x_J)W_{n - |J|}^{g - h}(x,x_{I\setminus J}) & & \\
+ \rho\Big(\mathcal{O}^{x,2}W_{n + 1}^{g - 1}(x,x,x_I) + \sum_{J \subseteq I,\,\,0 \leq h \leq g} W_{|J| + 1}^{h}(x,x_J)\mathcal{O}^{x}W_{n - |J|}^{g - h}(x,x_{I\setminus J})\Big) & & \nonumber \\
- \mathcal{V}'(x)W_n^{g}(x) + \sum_{i \in I} \frac{W_{n - 1}^{g}(x,x_{I\setminus\{i\}})}{(x - x_i)^2} + P_{n}^g(x;x_I) & = & 0, \nonumber
\eea
where:
\bea
\label{3899} P_n^{g}(x;x_I) & = & \frac{1}{2{\rm i}\pi}\oint_{\Gamma} \frac{\mathcal{V}'(x) - \mathcal{V}'(\xi)}{x - \xi}\,W_n^{g}(\xi)\dd\xi - \sum_{i \in I} \frac{\dd}{\dd x_i}\Big(\frac{W_{n - 1}^{g}(x,x_{I\setminus\{i\}})}{x - x_i}\Big) \\
& & + \frac{\rho}{2{\rm i}\pi}\oint_{\Gamma} \frac{\dd \xi}{x - \xi}\Big(\mathcal{O}^{x,2}W_{n + 1}^{g - 1}(\xi,x,x_I) + \sum_{J \subseteq I,\,\,0 \leq h \leq g} W_{|J| + 1}^{h}(\xi,x_J)\mathcal{O}^{x}W_{n - |J|}^{g - h}(x,x_{I\setminus J}) \nonumber \\
& & - \mathcal{O}^{\xi,2}W_{n + 1}^{g - 1}(\xi,\xi,x_I) - \sum_{J \subseteq I,\,\,0 \leq h \leq g} W_{|J| + 1}^{h}(\xi,x_J)\mathcal{O}^{\xi}W_{n - |J|}^{g - h}(\xi,x_{I\setminus\{i\}})\Big). \nonumber
\eea
The relevance of this decomposition comes from the observation that $P_n^{g}(x;x_I)$ is a holomorphic function of $x$ in a neighborhood of $\Gamma$. Now, let us multiply by $(\dd x)^2\prod_{i \in I}\dd x_i$ and translate this equation in the realm of differential forms in the domain $U$. In particular, we can define a quadratic differential $\mathcal{P}_{n}^{g}(z;z_I)$ for $z \in V$, such that $\mathcal{P}_n^{g}(z,z_I) = P_n^{g}(x(z);x(z_I))(\dd x)^2\prod_{i \in I} \dd x(z_i)$ when $z \in U$. It has double zeroes at ramification points, coming from the zeroes of $(\dd x)^2$. One also has to take into account the shift between $W_2^0$ and $\omega_2^0$ (see \eqref{defsq}). When $(n,g) \neq (1,1)$, we find that \eqref{jui} becomes, for $z \in \coprod_{j = 1}^r U_j$ and $z_I \in U^{n - 1}$:
\bea
\label{qui}\omega_{n + 1}^{g - 1}(z,z,z_I) + \sum_{J \subseteq I,\,\,0 \leq h \leq g} \omega_{|J| + 1}^{h}(z,z_J)\omega_{n - |J|}^{g - h}(z,z_{I\setminus J}) - 2 \sum_{i \in I} \omega_{n - 1}^{g}(z,z_{I\setminus\{i\}})\,\frac{\dd x(z)\,\dd x(z_2)}{(x(z) - x(z_2))^2} & & \\
+ \rho\Big(\mathcal{O}^{z,2}\omega_{n + 1}^{g - 1}(z,z,z_I) + \sum_{J \subseteq I,\,\,0 \leq h \leq g} \omega_{|J| + 1}^{h}(z,z_J)\mathcal{O}^{z}\omega_{n - |J|}^{g - h}(z,z_{I\setminus J})\Big) & & \nonumber \\
- \dd \mathcal{V}(z)\omega_{n}^{g}(z,z_I) + \sum_{i \in I} \frac{\dd x(z)\,\dd x(z_i)}{(x(z) - x(z_i))^2}\,\omega_{n - 1}^{g}(z_{I\setminus\{i\}})+ \tilde{\mathcal{P}}_{n}^{g}(z;z_I) & = & 0, \nonumber
\eea
where we included all the terms having double zeroes at ramification points in $\tilde{\mathcal{P}}_n^{g}(z;z_I)$, namely:
\beq
\tilde{\mathcal{P}}_{n}^g(z;z_I) = \mathcal{P}_{n}^{g}(z;z_I) + \delta_{g,0}\Big(\delta_{n,2}\,\frac{\dd x(z)\,\dd x(z_2)}{(x(z) - x(z_2))^2} + \delta_{n,3}\delta_{g,0}\,\frac{2\,(\dd x(z))^2\,\dd x(z_2)\,\dd x(z_3)}{(x(z) - x(z_2))^2\,(x(z) - x(z_3))^2}\Big).
\eeq
The operator $\mathcal{O}$ was defined in \eqref{OOdef} in the framework of $1$-forms, and are just the translation of $\mathcal{O}$ defined in \eqref{odef} in the framework of functions. Notice that the third term in the first line of \eqref{qui} combines with the middle term of the third line, and just amounts to change the sign of the latter. We now use the linear loop equation in the form \eqref{nju} to replace in \eqref{qui} the quantities involving $\mathcal{O}^{z}$ by quantities involving $\iota(z)$ only, for $z \in \coprod_{j = 1}^r U_j$. We find:
\beq
-\omega_{n + 1}^{g - 1}(z,\iota(z),z_I) - \sum_{J \subseteq I,\,\,0 \leq h \leq g} \omega_{|J| + 1}^{h}(z,z_J)\omega_{n - |J|}^{g - h}(\iota(z),z_{I\setminus\{i\}}) + \tilde{\mathcal{P}}_{n}^{g}(z;z_I) = 0.
\eeq
Hence, the quadratic differential $\mathcal{Q}_{n,j}(z;z_I)$ defined in \eqref{qua} coincides with $\tilde{\mathcal{P}}_{n,j}^{g}(z;z_I)$, so has double zeroes at ramification points. For $(n,g) = (1,1)$, we must be careful because of the double pole in $\omega_2^0$ at coinciding points. We start with \eqref{jui}:
\bea
W_2^0(x,x) + 2W_1^1(x)W_1^0(x) - \mathcal{V}'(x)W_1^1(x) & & \\
+ \rho\Big(\mathcal{O}^{x,2}W_2^0(x,x) + \mathcal{O}^{x}W_1^0(x)\,W_1^1(x) + W_1^0(x)\,\mathcal{O}^{x}W_1^1(x)\Big) + P_1^1(x) & = & 0. \nonumber 
\eea
For $z \in U_j$, we first compute:
\bea
W_{2}^0(x(z),x(z)) & = & \lim_{z' \rightarrow z} W_2^0(x(z'),x(z)) = \lim_{z' \rightarrow z} \frac{1}{\dd x(z)\dd x(z')}\Big(\omega_2^0(z',z) - \frac{\dd x(z')\,\dd x(z)}{(x(z') - x(z))^2}\Big) \nonumber \\
& = & -\lim_{z' \rightarrow z} \frac{\omega_2^0(z',\iota(z)) + \rho\,\mathcal{O}^{z}\omega_2^0(z',z)}{\dd x(z)\,\dd x(z')}\nonumber \\
& = & -\frac{\omega_2^0(z,\iota(z)) + \rho\,\mathcal{O}^{z,2}\omega_2^0(z,z)}{(\dd x(z))^2},
\eea
and we use this expression to replace $W_2^0$. We obtain:
\beq
-\omega_2^0(z,\iota(z)) - \omega_1^0(z)\omega_1^1(\iota(z)) - \omega_1^0(\iota(z))\omega_1^1(z) + \tilde{\mathcal{P}}_{1}^{1}(z) = 0,
\eeq
hence $\mathcal{Q}_{1}^{1}(z) = \tilde{\mathcal{P}}_{1}^{1}(z)$ again, and it has double zeroes at ramification points. Therefore, we have obtained the quadratic loop equations.

\subsection{The model with several species of particles}

Our results can be extended to repulsive systems of particles of $s \geq 2$ different species. We consider $N_k$ particles of type $k$ with $k \in \ldbrack 1,s \rdbrack$, and denote $\lambda_{i,k}$ their position on some arc $\Gamma_{0,k}$, and $N = \sum_{k = 1}^s N_k$ the total number of particles. We consider the model:
\bea\label{defmeasureseveral}
\!\!\!\!\!\!\! & & \dd\varpi(\lambda) \\
\!\!\!\!\!\!\! & = & \int \Big[\prod_{1 \leq k,l \leq s} \prod_{i = 1}^{N_k}\prod_{j = 1}^{N_l} \big(R_{k,l}(\lambda_{i,k},\lambda_{j,l})\big)^{\rho_{k,l}/2}\Big] \prod_{k = 1}^s \Big[\prod_{1 \leq i < j \leq N_k} |\lambda_{i,k} - \lambda_{j,k}|^{\beta_k}\,\prod_{i = 1}^{N_k} e^{-N\,\mathcal{V}_k(\lambda_{i,k})}\dd\lambda_{i,k}\Big], \nonumber
\eea
where the integral runs over $\prod_{k = 1}^s (\Gamma_{0,k})^{N_k}$. In the decomposition of the measure, it is understood that the two-point interactions $R_{k,l}$ will be regular on $\Gamma_{0,k}\times\Gamma_{0,l}$ for any $k,l \in \ldbrack 1,s \rdbrack$. We may assume without restriction that $R_{k,l}(x,y) = R_{l,k}(y,x)$ and $\rho_{k,l} = \rho_{l,k}$. We denote again $Z_N$ the partition function of $\varpi$.

In the context of formal integrals, we shall see in Section~\ref{S5} that they describe the statistical physics of self-avoiding loops on random lattices, where the symmetry between the inner and the outer domains delimited by the loops is broken. It contains the $\rho^2$-Potts model on general random maps as a special case \cite{BBG3}.

Let us denote:
\beq
M_k = \mathrm{diag}(\lambda_{i,k})_{1 \leq i \leq N_k},\qquad M = \mathrm{diag}(M_1,\ldots,M_s).
\eeq
We now want to consider observables distinguishing the type of particles. For any $k_1,\ldots,k_n \in \ldbrack 1,s \rdbrack$, we define the \emph{refined correlators}:
\beq
\label{deq}W_{n}(\sheet{x_1}{k_1},\ldots,\sheet{x_n}{k_n}) = \Big\langle \prod_{i = 1}^{n} \Tr\,\frac{1}{x_i - M_{k_i}}\Big\rangle_c.
\eeq
For each variable $x_i$, we use the notation $\sheet{x_i}{k_i}$ to indicate to which type of particles it is coupled, but it should not hide the fact that $W_n(\sheet{x_1}{k_1},\ldots,\sheet{x_n}{k_n})$ is a different function of $x_1,\ldots,x_n$ for each $k_1,\ldots,k_n$. If we want to sum over all type of particles, we rather write:
\beq
\label{rceo}W_n(x_1,\sheet{x_2}{k_2},\ldots,\sheet{x_n}{k_n}) = \sum_{k_1 = 1}^s W_{n}(\sheet{x_1}{k_1},\sheet{x_2}{k_2},\ldots,\sheet{x_n}{k_n}) = \Big\langle \Tr \frac{1}{x_1 - M} \prod_{i = 2}^{n} \Tr\frac{1}{x_i - M_{k_i}}\Big\rangle_c.
\eeq
If $(x_i)_{i \in I}$ is a set of variables and $(k_i)_{i \in I}$ a sequence in $\ldbrack 1,s \rdbrack$, we also write collectively $W_n(\sheet{x_I}{k_I})$. This function is holomorphic in the domain $\prod_{i = 1}^n (\mathbb{C}\setminus\Gamma_{0,k_i})$, and, a priori, has a discontinuity when one of the $x_i$'s crosses $\Gamma_{0,k_i}$.

In the original model \eqref{mes2}, we considered the cases where the support $\Gamma$ of the equilibrium measure $\mu_{\mathrm{eq}}$, was the union of $r$ disjoint segments $\gamma_k$. Such a regime can be elegantly described as a model with $r$ species of particles, where:
\begin{itemize}
\item[$\bullet$] $\beta_k \equiv \beta$, $\mathcal{V}_k \equiv \mathcal{V}$, $R_{k,l} \equiv R$ and $\rho_{k,l} \equiv \rho$ do not depend on $k,l \in \ldbrack 1,r \rdbrack$.
\item[$\bullet$] $N_k = \lfloor N\mu_{\mathrm{eq}}(\gamma_{k}) \rfloor$, and $\Gamma_{0,k}$ is a small neighborhood\footnote{The partition function of \eqref{mes2} on $(\Gamma_0)^N$ differs from that on $\bigcup_{k = 1}^r (\Gamma_{0,k})^{N_k}$ by exponentially small corrections when $N$ is large, which are irrelevant from the point of view of $1/N$ expansions.} of $\gamma_k$ in $\Gamma_0$.
\end{itemize}
The refined correlators $W_n(\sheet{x}{k},\sheet{x_I}{a_I})$ are then obtained by projecting $W_n(x,\sheet{x_I}{a_I})$ on the space of holomorphic functions having a discontinuity on $\Gamma_{0,k}$, i.e.:
\beq
W_n(\sheet{x}{k},\sheet{x_I}{a_I}) = \frac{1}{2{\rm i}\pi}\oint_{\Gamma_{0,k}} \frac{\dd\xi\,W_{n}(\xi,\sheet{x_I}{a_I})}{x - \xi}.
\eeq

\subsection{Generalization of the method}
\label{38}
The analysis of the model with $r$ species is very similar to the case $r = 1$. The only difference is that we have to deal with vectors \mbox{$[W_{n}(\sheet{x}{k},\sheet{x_I}{k_I})]_{1 \leq k \leq s}$}, and the operator $\mathcal{O}$ becomes a matrix of operators $(\mathcal{O}_{k,l})_{k,l}$. Hence, we will not reproduce all the details, and rather summarize the main steps leading to the results of Section~\ref{juko} below.

\subsubsection*{Results from potential theory and analytic continuation}

Let $\epsilon_k = N_k/N$ be fixed and positive. We are led to introduce the following functional on $\prod_{k = 1}^s \mathcal{M}_{\epsilon_k}(\Gamma_{0,k})$:
\bea
\mathcal{E}[\mu] & = &  - \frac{1}{2}\sum_{k = 1}^s \beta_k \iint_{\Gamma_{0,k}^2} \dd\mu_k(x)\dd\mu_k(y)\,\ln |x - y|  - \frac{1}{2} \sum_{k,l = 1}^s \rho_{k,l} \iint_{\Gamma_{0,k}\times\Gamma_{0,l}} \!\!\!\!\!\!\!\!\!\!\!\!\!\dd\mu_k(x)\dd\mu_l(y)\,\ln R_{k,l}(x,y)\nonumber \\
& & + \sum_{k = 1}^s \int_{\Gamma_{0,k}} \dd\mu_k(x)\,\mathcal{V}_k(x).
\eea
Critical points of this functional are characterized by the following equations: for any $k \in \ldbrack 1,s \rdbrack$, for $x \in \Gamma_{0,k}$ $\mu_k$-almost everywhere:
\beq
\label{sad}\beta_k \int_{\Gamma_{0,k}} \dd\mu_k(\xi)\,\ln|x - \xi| + \sum_{l = 1}^s \rho_{k,l}\int_{\Gamma_{0,l}} \dd\mu_l(\xi)\,\ln R_{k,l}(x,\xi) = \mathcal{V}_k(x) + C_k
\eeq
for some constant $C_k$.

\begin{definition}
\label{stricq}We say the interactions defined by the data $(R_{k,l},\rho_{k,l},\beta_k)_{k,l}$, are strictly convex if, for any vector of complex measures $\nu = (\nu_1,\ldots,\nu_s)$ such that $\nu_k(\Gamma_{0,k}) = 0$ for any $k \in \ldbrack 1,s \rdbrack$, we have:
\beq
\label{hodlin}\Big(\beta_k \iint_{\Gamma_{0,k}^2} \dd\nu_k(x)(\dd\nu_k(y))\,\ln|x - y| + \sum_{l = 1}^s \rho_{k,l}\iint_{\Gamma_{0,k}\times\Gamma_{0,l}} \!\!\!\!\!\!\!\!\!\!\!\!\!\dd\nu_k(x)(\dd\nu_l(y))\ln R_{k,l}(x,y)\Big) \leq 0,
\eeq
with equality iff $\nu \equiv 0$.
\end{definition}
As in Lemma~\ref{leoa}, it is equivalent to use complex measures instead of signed measures in this definition, provided the measure on $y$ in \eqref{hodlin} is replaced by its complex conjugate. In particular, this implies that $\mathcal{E}$ is strictly convex. The natural set of assumptions is now:
\begin{hypothesis}
\label{qwe}\begin{itemize}
\item[$\phantom{(o)}$]
\item[$(i)$] $\mathcal{V}_k\,:\,\Gamma_{0,k} \rightarrow \mathbb{R}$ are real-analytic.
\item[$(ii)$] $\ln R_{k,l}\,:\,\Gamma_{0,k}\times\Gamma_{0,l} \rightarrow \mathbb{R}$ are real-analytic.
\item[$(iii)$] The interactions are strictly convex.
\item[$(iv)$] If some $\Gamma_{0,k}$ is unbounded, the potentials are strongly confining.
\end{itemize}
\end{hypothesis}
If Hypothesis~\ref{qwe} holds, the existence and uniqueness of a vector of equilibrium measures is guaranteed, and $\Gamma_k = \mathrm{supp}\,\mu_k$ will be a union of $r_k$ segments: $\Gamma_k = \bigcup_{j = 1}^{r_k} \gamma_{k,j}$. We can define domains $U_k$ (resp. $\widehat{U}_k$) which maps bijectively to $\mathbb{C}\setminus\Gamma_{k}$ (resp. $\widehat{\mathbb{C}}\setminus\Gamma_k$), and include them in Riemann surfaces $(U_{k})_{\Gamma_k}$ as in~\ref{ss1}. Let $V_{k,j}$ be annular neighborhoods of $\gamma_{k,j}$ in $(U_k)_{\Gamma_k}$, let $V_k = \bigcup_{j = 1}^{r_k} V_{k,j}$ and $\iota_k$ its holomorphic involution. The potentials $\mathcal{V}_k(x)$ can be promoted to a sequence $\mathcal{V}_k(z)$ of $\iota_k$-invariant holomorphic functions of $z \in V_k$. Similarly, $\ln R_{k,l}(x,y)$ defines a sequence $\ln R_{k,l}(z,w)$ of holomorphic functions of $(z,w) \in V_k\times V_k$, which are $\iota_{k}$-invariant in $z$ and $\iota_{l}$-invariant in $w$. We are led to define the operators\footnote{See footnote page \pageref{totor} for the meaning of $\Ms^*$.} $\mathcal{O}_{k,l}\,:\,\Ms^*(V_l) \rightarrow \Hs_{\Gamma_k}^{\mathrm{inv}}$ by
\beq
\label{iksq}\mathcal{O}_{k,l} f(z) = \frac{1}{2{\rm i}\pi}\oint_{\Gamma_k^{\mathrm{ext}}} \dd_z \ln R_{k,l}(z,\xi)\,f(\xi).
\eeq
The Stieltjes transform of the equilibrium measures:
\beq
\omega_1^0(\sheet{z}{k}) = \Big(\int_{\Gamma_k} \frac{\dd\mu_{\mathrm{eq},k}(\xi)}{x - \xi}\Big)\dd x(z)
\eeq
are initially defined as holomorphic $1$-forms in $U_k$. From the saddle point equation \eqref{sad}, we deduce that they can be continued across $\Gamma_k$ (i.e. $\omega_1^0(\sheet{\cdot}{k}) \in \Hs_{\Gamma_k}(U_k)$), and they satisfy the functional equation:
\beq
\forall z \in V_{k},\qquad \Delta_{k}\omega_1^0(\sheet{z}{k}) + \sum_{l = 1}^s \frac{2\rho_{k,l}}{\beta_k}\,\mathcal{O}_{k,l} \omega_1^0(\sheet{z}{l}) = \frac{2}{\beta_k}\,\dd \mathcal{V}_k(z).
\eeq
Besides, assuming an analog of Conjecture~\ref{theconj} and repeating the steps of \S~\ref{Popo}, we can show the existence of a sequence of a fundamental $2$-form of the $2^{\mathrm{nd}}$ kind, denoted $B(\sheet{\cdot}{k_0},\sheet{\cdot}{k}) \in \Ms(\widehat{U}_{\Gamma_{k_0}}\times\widehat{U}_{\Gamma_k})$, such that:
\begin{itemize}
\item[$\bullet$] $B(\sheet{z_0}{k_0},\sheet{z}{k}) = B(\sheet{z}{k},\sheet{z_0}{k_0})$.
\item[$\bullet$] for any $z \in V_{k}$ and $z_0 \in U_{k_0}$:
\beq
\mathcal{S}_{k}^{z}B(\sheet{z_0}{k_0},\sheet{z}{k}) + \sum_{l = 1}^s \rho_{k,l}\,\mathcal{O}_{k,l}^{z} B(\sheet{z_0}{k_0},\sheet{z}{l}) = \delta_{k,k_0}\,\frac{\dd x(z_0)\,\dd x(z)}{(x(z_0) - x(z))^2}.
\eeq
\item[$\bullet$] for any $j \in \ldbrack 1,r_k \rdbrack$, $\oint_{\gamma_{k,j}^{\mathrm{ext}}} B(\sheet{z_0}{k_0},\sheet{\cdot}{k}) = 0$.
\end{itemize}
Similarly, we can also construct $1^{\mathrm{st}}$ kind differentials, i.e. a sequence $h_{m,j}(\sheet{\cdot}{k}) \in \Hs((U_k)_{\Gamma_k})$, indexed by $k,m \in \ldbrack 1,s \rdbrack$ and $j \in \ldbrack 1,r_m \rdbrack$, such that:
\begin{itemize}
\item[$\bullet$] for any $z \in V_{k}$:
\beq
\mathcal{S}_{k} h_{\bullet}(\sheet{z}{k}) + \sum_{l = 1}^s \frac{2\rho_{k,l}}{\beta_k}\,\mathcal{O}_{k,l} h_{\bullet}(\sheet{z}{l}) = 0.
\eeq
\item[$\bullet$] for any $j' \in \ldbrack 1,r_k \rdbrack$, $\frac{1}{2{\rm i}\pi}\oint_{\gamma_{k,j'}^{\mathrm{ext}}} h_{m,j}(\sheet{\cdot}{k}) = \delta_{k,m}\delta_{j,j'}$.
\end{itemize}

\subsubsection*{Representation by residues}

We define $\mathcal{H}$, the subspace of $\bigoplus_{k = 1}^s \Hs_{\Gamma_k}(\widehat{U}_k)$ consisting of vectors of $1$-forms $f$ which satisfy, for any $k \in \ldbrack 1,s \rdbrack$ and $z \in V_{k}$,
\beq
\mathcal{S}_{k} f(\sheet{z}{k}) + \sum_{l = 1}^s \frac{2\rho_{k,l}}{\beta_k}\,\mathcal{O}_{k,l} f(\sheet{z}{l}) = 0,
\eeq
and its subspace of forms with vanishing periods:
\beq
\mathcal{H}_0 = \Big\{f \in \mathcal{H},\qquad \forall k \in \ldbrack 1,s \rdbrack,\quad \forall j \in \ldbrack 1,r_k \rdbrack,\quad \oint_{\gamma_{k,j}^{\mathrm{ext}}} f(\sheet{\cdot}{k}) = 0 \Big\}.
\eeq
We also define:
\beq
G(\sheet{z_0}{k_0},\sheet{z}{k}) = -\int^{z} B(\sheet{z_0}{k_0},\sheet{\cdot}{k}).
\eeq
Notice that $\bigoplus_{k = 1}^s \Hs_{\Gamma_k}(\widehat{U}_k) \simeq \Hs_{\Gamma_k}\big(\coprod_{k = 1}^s \widehat{U}_k\big)$, where we insist that the right hand side involves the disjoint union of $\widehat{U}_k$, so that we are still in the framework on \S~\ref{ss1}. The proofs in \S~\ref{repres} can be adapted to show:

\begin{proposition}
If Hypothesis~\ref{qwe} holds, $\mathcal{H}$ is representable by residues, with local Cauchy kernel $G(\sheet{z_0}{k_0},\sheet{z}{k})$. Besides:
\beq
\mathcal{H} = \mathcal{H}_0 \oplus\Big(\bigoplus_{k = 1}^s \mathop{\mathrm{span}}_{\substack{1 \leq m \leq s \\ 1 \leq j \leq r_m}} h_{m,j}(\sheet{\cdot}{k})\Big),
\eeq
and $\mathcal{H}_0$ is normalized. \hfill $\Box$
\end{proposition}

\subsubsection*{Schwinger-Dyson equations}

Schwinger-Dyson equations for the refined correlators in the multi-species case can be derived as in \S~\ref{SDSQ}. When we assume $\ln R_{k,l}(x,y)$ analytic for $(x,y)$ in a neighborhood of $\Gamma_{0,k}\times\Gamma_{0,l}$, the rank $n$ Schwinger-Dyson equation without boundary terms reads, for any $k$ and $k_I = (k_2,\ldots,k_n) \in \ldbrack 1,s \rdbrack$, for any $x \in \mathbb{C}\setminus\Gamma_{0,k}$ and $x_I = (x_2,\ldots,x_n) \in \prod_{i \in I} (\mathbb{C}\setminus\Gamma_{0,k_i})$,
\bea
\Big(1 - \frac{2}{\beta_k}\Big)\partial_x W_{n}(\sheet{x}{k},\sheet{x_I}{k_I}) + W_{n + 1}(\sheet{x}{k},\sheet{x}{k},\sheet{x_I}{k_I}) + \sum_{J \subseteq I} W_{|J| + 1}(\sheet{x}{k},\sheet{x_J}{k_J})W_{n - |J|}(\sheet{x}{k},\sheet{x_{I\setminus J}}{k_{I\setminus J}}) & & \nonumber \\
- \frac{2}{\beta_k} \oint_{\Gamma_{0,k}} \frac{\dd\xi}{2{\rm i}\pi}\,\frac{N\,\mathcal{V}_k'(\xi)\,W_n(\sheet{\xi}{k},\sheet{x_I}{k_I})}{x - \xi} + \frac{2}{\beta_k} \sum_{i \in I} \oint_{\Gamma_{0,k}} \frac{\dd\xi}{2{\rm i}\pi}\,\frac{W_{n - 1}(\sheet{\xi}{k},\sheet{x_{I\setminus\{i\}}}{k_{I\setminus\{i\}}})}{(x - \xi)(x_i - \xi)^2} & &  \nonumber \\
+ \sum_{l = 1}^s \frac{2\rho_{k,l}}{\beta_k} \oiint_{\Gamma_{0,k}\times\Gamma_{0,l}}\!\! \frac{\dd\xi\,\dd\eta}{(2{\rm i}\pi)^2}\,\frac{\partial_{\xi} \ln R_{k,l}(\xi,\eta)}{x - \xi}\Big(W_{n + 1}(\sheet{\xi}{k},\sheet{\eta}{l},\sheet{x_I}{k_I}) + \sum_{J \subseteq I} W_{|J| + 1}(\sheet{\xi}{k},\sheet{x_J}{k_J})W_{n - |J|}(\sheet{\eta}{l},\sheet{x_{I\setminus J}}{k_{I\setminus J}})\Big) & \!\!\!\!\!\!\!= \!\!\!\!\!\!\! & 0. \nonumber \\
\label{317} & &
\eea
The coupling between different species of particles only occur in the last line, through the off-diagonal terms of the matrix $\rho = (\rho_{k,l})_{k,l}$.

\subsection{Results}
\label{juko}

We introduce:
\begin{hypothesis}
\label{ppp}
\begin{itemize}
\item[$\phantom{(o)}$] 
\item[$(i)$] $\beta_k \equiv 2$.
\item[$(ii)$] $\Gamma_k = \bigcup_{j = 1}^{r_k} \gamma_{k,j}$ is a disjoint union of bounded intervals $\gamma_{k,j}$.
\item[$(iii)$] $\mathcal{V}_k$ is analytic in a neighborhood of $\Gamma_k$.
\item[$(iv)$] $\ln R_{k,l}$ is analytic in a neighborhood of $\Gamma_{k}\times\Gamma_{l}$.
\item[$(v)$] The refined correlators of \eqref{rceo} have a large $N$ expansion of topological type.
\item[$(vi)$] $W_1^0(\sheet{\cdot}{k})$ is discontinuous at any interior point of $\Gamma_k$.
\item[$(vii)$] $\omega_1^0(\sheet{x}{k}) = W_1^0(\sheet{x}{k})\dd x$ are off-critical $1$-forms.
\end{itemize}
\end{hypothesis}

\begin{proposition}
\label{mik}Let us assume Hypothesis~\ref{ppp}, and define $\omega_n^g \in \bigoplus_{k_1,\ldots,k_n = 1}^s \Hs(\prod_{i = 1}^n U_{k_i})$ by the formulas:
\beq
\omega_n^g(\sheet{z_1}{k_1},\ldots,\sheet{z_n}{k_n}) = W_n^{g}(\sheet{x(z_1)}{k_1},\ldots,\sheet{x(z_n)}{k_n})\prod_{i = 1}^n \dd x(z_i) + \delta_{n,2}\delta_{g,0}\delta_{k_1,k_2}\,\frac{\dd x(z_1)\,\dd x(z_2)}{(x(z_1) - x(z_2))^2}.
\eeq
Then, $\omega_{\bullet}^{\bullet}$ satisfies linear and quadratic loop equations. More precisely, for any $n,g$, any $k,k_I = (k_2,\ldots,k_n)$, any $z_I = (z_2,\ldots,z_n) \in \prod_{i = 2}^n U_{k_i}$, and for any  $z \in V_{k}$, we have:
\beq
\label{poux}\mathcal{S}^z_{k}\omega_n^g(\sheet{z}{k},\sheet{z_I}{k_I}) + \sum_{l = 1}^s \rho_{k,l}\,\mathcal{O}_{k,l}\omega_n^g(\sheet{z}{k},\sheet{z_I}{k_I}) = \delta_{g,0}\Big(\delta_{n,1}\,\dd \mathcal{V}_k(z) + \delta_{n,2}\delta_{k,k_2}\,\frac{\dd x(z)\,\dd x(z_2)}{(x(z) - x(z_2))^2}\Big).
\eeq
\hfill $\Box$
\end{proposition}

\begin{hypothesis}
\label{4331}\begin{itemize}
\item[$(vii)$] The interactions are strictly convex, in the sense of Definition~\ref{stricq}.
\end{itemize}
\end{hypothesis}
Notice that $(ii)$-$(iv)$ of Hypothesis~\ref{ppp} and $(vii)$ of Hypothesis~\ref{4331} are implied by Hypothesis~\ref{qwe}.

\begin{corollary}
\label{aporo}Let us assume Hypothesis~\ref{4331} and for any stable $n,g$, any $k,k_I$ and any $j \in \ldbrack 1,r_k \rdbrack$, $\oint_{\gamma_{k,j}^{\mathrm{ext}}} \omega_n^g(\sheet{\cdot}{k},\sheet{z_I}{k_I}) = 0$. Then, $\omega_{n}^{g}$ can be computed by the topological recursion:
\bea
\omega_n^g(\sheet{z_0}{k_0},\sheet{z_I}{k_I}) & = & \sum_{\alpha \in \Gamma_{k_0}^{\mathrm{fix}}} \Res_{z \rightarrow \alpha} K_{k_0}(z_0,z) \\
& & \Big(\omega_{n + 1}^{g - 1}(\sheet{z}{k_0},\iota_{k_0}(\sheet{z}{k_0}),\sheet{z_I}{k_I}) 
+ \sum_{J \subseteq I,\,\,0 \leq h \leq g} \omega_{|J| + 1}^{h}(\sheet{z}{k_0},\sheet{z_J}{k_J})\omega_{n - |J|}^{g - h}(\iota_{k_0}(\sheet{z}{k_0}),\sheet{z_{I\setminus J}}{k_{I\setminus J}})\Big), \nonumber
\eea
where the recursion kernel is given by:
\beq
K_{k_0}(z_0,z) = \frac{-\frac{1}{2}\int_{\iota_{k_0}(z)}^{z} \omega_2^0(\sheet{z_0}{k_0},\sheet{\cdot}{k_0})}{\omega_1^0(\sheet{z}{k_0}) - \omega_1^0(\iota_{k_0}(\sheet{z}{k_0}))}.
\eeq
\hfill $\Box$
\end{corollary}
We remark that the recursion kernel only involves $\omega_2^0(\sheet{z_0}{k_0},\sheet{z}{k})$ for $k = k_0$. After summing over $k$, the topological recursion takes the usual form \eqref{topore}.

\section{Virasoro constraints, graphs and loop equations}
\label{S7}
In this section, we rewrite the Schwinger-Dyson equations of the repulsive particle with $s$ species in terms of the partition function, with index $\beta_k \equiv 2$ for any $k \in \ldbrack 1,s \rdbrack$. We show that they can be obtained by a canonical transformation mixing $s$ independent copies of a set of Virasoro constraints, making the connection e.g. with \cite{kostovhirota} or the work of \cite{AMM072}. This decomposition can be seen as a realization of Givental formula expressing the value of the potential of a Frobenius manifold as the result of the action of a canonical transformation on a product of Kontsevich integrals \cite{G01}. From this point of view, the coefficients of the potentials ${\cal V}$ are interpreted as flat coordinates on the Frobenius manifold, the interaction between the eigenvalues corresponding to a motion in this manifold. The topological recursion therefore gives the $1/N$ expansion of those deformed Virasoro constraints, provided the one and two-points functions are known to leading order in $N$.

In this section, we consider all quantities as formal series in coefficients of the potentials and the two-point interaction. The method to define properly such \emph{formal matrix models} has been reviewed in detail in \cite{Eformal}.

\subsection{Virasoro contraints and the one hermitian matrix model}

Considering a family of formal parameters $\mathbf{t} = (t_k)_{k \geq 0}$, we say that a formal series $f_N(\mathbf{t})$ satisfies \emph{Virasoro constraints} if:
\beq
\forall m \geq -1,\qquad L_m[\mathbf{t},N]\cdot f(\mathbf{t}) = 0,
\eeq
where for $m=-1,0,1$:
\beq
L_{-1}[\mathbf{t},N]= t_1\,\frac{\partial}{\partial t_{0}} + \sum_{j \geq 2} (j -1)t_j\,\frac{\partial}{\partial t_{j-1}},
\eeq
\beq
L_0[\mathbf{t},N] = \frac{1}{N^2} \,\frac{\partial^2}{\partial t_0^2} + \sum_{j \geq 1} j\,t_j\,\frac{\partial}{\partial t_{ j}}.
\eeq
\beq
L_1[\mathbf{t},N] = \frac{2}{N^2} \,\frac{\partial}{\partial t_0}\,\frac{\partial}{\partial t_1} + \sum_{j \geq 1} (j + 1)t_j\,\frac{\partial}{\partial t_{j+1}}.
\eeq
and for $m\geq 2$
\beq
L_m[\mathbf{t},N] = \frac{2m}{N^2} \,\frac{\partial}{\partial t_0}\,\frac{\partial}{\partial t_{m}}+ \frac{1}{N^2} \sum_{j = 1}^{m-1} j(m - j)\,\frac{\partial}{\partial t_j}\frac{\partial}{\partial t_{m - j}} + \sum_{j \geq 1} (j + m)t_j\,\frac{\partial}{\partial t_{m + j}}.
\eeq
The name "Virasoro" comes from the commutation relations satisfied by those formal differential operators:
\beq
\big[L_m[\mathbf{t},N],L_{m'}[\mathbf{t},N]\big] = (m - m')L_{m + m'}[\mathbf{t},N].
\eeq
For further convenience, when there is no confusion, if $f(\mathbf{t})$ is a formal series in $\mathbf{t}$, we denote:
\beq
\forall \mathbf{j} \in \mathbb{N}^n,\qquad f_{\mathbf{j}}(\mathbf{t}) = \frac{\partial^n f(\mathbf{t})}{\partial t_{j_1}\cdots\partial t_{j_n}}.
\eeq
It is well-known \cite{AJM90,GMM91,MM90,M91} (see Lemma \ref{lemmaVir1MM} below) that the partition function of a one hermitian matrix model satisfies Virasoro constraints. More precisely, consider the \emph{local partition function}:
\beq
\label{locap}Z[\mathcal{V},N] = \int_{\Gamma_0^{N}} \prod_{1 \leq i < j \leq N} (\lambda_i - \lambda_j)^2\,\prod_{i = 1}^{N} e^{-N\mathcal{V}(\lambda_i)}\dd\lambda_{i},
\eeq
with a potential of the form:
\beq
\mathcal{V}(x) = \mathcal{V}^{(0)}(x) -t_0 - \sum_{j \geq 1} \frac{t_j}{j}\,(x - \Lambda)^j.
\eeq
$\Gamma_0$ is a contour ending at $\infty$ in the complex plane, $\mathcal{V}^{(0)}$ is analytic near $\Lambda \in \mathbb{C}$ with a Taylor expansion of the form:
\beq
\mathcal{V}^{(0)}(x) = -t_0^{(0)}-\sum_{j \geq 1} \frac{t_{j}^{(0)}}{j}\,(x - \Lambda)^j,
\eeq
and $(\Gamma_0,\mathcal{V}^{(0)})$ is chosen such that $Z[\mathcal{V}^{(0)},N]$ is a convergent integral. Here, $\mathbf{t} = (t_j)_{j \geq 1}$ is a set of formal parameters. If $x_i\notin \Gamma_0$, is far enough from $\Gamma_0$, writing that
\beq
\Tr \frac{1}{x_i-M} = \Tr \frac{1}{x_i-\Lambda-(M-\Lambda)}= \sum_{j \geq 0} \frac{\Tr(M-\Lambda)^j}{(x_i-\Lambda)^{j+1}}, 
\eeq
we observe that the $n$-point correlators of \eqref{conera} near $x_i = \Lambda \in \Gamma_0$ are generating series of $n$-th order derivatives of the partition function:
\beq
\label{1psqo}W_n(x_1,\ldots,x_n) = \frac{\delta_{n,1}}{(x_1 - \Lambda)} + \sum_{j_1,\ldots,j_n \geq 1} \prod_{i = 1}^{n} \frac{j_i/N}{(x_i - \Lambda)^{j_i + 1}}\,(\ln Z)_{\mathbf{j}}[\mathcal{V},N].
\eeq

We then have:
\begin{lemma}\label{lemmaVir1MM}
The 1-Matrix model satisfies the Virasoro constraints:
\beq
\forall m \geq -1,\qquad L_m[\mathbf{t}^{(0)} + \mathbf{t},N]\cdot Z[\mathcal{V},N] = 0,
\eeq
\end{lemma}

\proof{The Virasoro constraints are a mere rewriting of the rank $1$ Schwinger-Dyson equation (specialization of Lemma~\ref{SD} for $R_0 \equiv 1$ and $n = 1$) for the correlators, as a set of differential equations satisfied by the partition function.
}

\subsection{Several species of particles and sum over graphs}

One can generalize the preceding section by considering the model with $s$ species of particles defined in \eqref{defmeasureseveral}, with $\beta_k \equiv 2$:
\beq
\dd\varpi(\lambda) \propto \int\Big[\prod_{1 \leq k,l \leq s} \prod_{i = 1}^{N_k}\prod_{j = 1}^{N_l} \big(R_{k,l}(\lambda_{i,k},\lambda_{j,l})\big)^{\rho_{k,l}/2}\Big] \prod_{k = 1}^s \Big[\prod_{1 \leq i < j \leq N_k} (\lambda_{i,k} - \lambda_{j,k})^{2}\,\prod_{i = 1}^{N_k} e^{-N\,\mathcal{V}_k(\lambda_{i,k})}\dd\lambda_{i,k}\Big],
\eeq
where the range of integration is $\prod_{k = 1}^s \Gamma_{0,k}^{N}$ with $\Gamma_{0,k}$ some contour in the complex plane. We denote $\mathbf{R} = (R_{k,l})_{k,l}$, $\mathbf{\mathcal{V}} = (\mathcal{V}_{k})_k$ and $\mathbf{N} = ((N_k)_{k},N)$, and $Z[\mathbf{R},\mathbf{\mathcal{V}},\mathbf{N}]$ the partition function of such a model. Without loss of generality we assume that $R_{k,l}(x,y)=R_{l,k}(y,x)$ and $\rho_{l,k}=\rho_{k,l}$.

As before, we take a point $\Lambda_k \in \Gamma_{k,0}$ around which performing Taylor expansions, and we take as potential:
\beq
\mathcal{V}_k(x) = \mathcal{V}_k^{(0)}(x) - t_{k,0}- \sum_{j \geq 1} \frac{t_{k,j}}{j}\,(x - \Lambda_k)^j,
\eeq
where $(\mathcal{V}_k^{(0)},\Gamma_{k,0})_k$ is chosen so that $Z[\mathbf{R},\mathcal{V}^{(0)},\mathbf{N}]$ is a convergent integral. We also Taylor expand the two-point interaction as follows:
\beq
\label{defrea} {\rho_{k,l}}\,\ln R_{k,l}(x,y) = \sum_{i,j \geq 0} \frac{\mathcal{R}_{k,l;i,j}}{ij}\,(x - \Lambda_k)^i(y - \Lambda_l)^{j},
\qquad \quad \mathcal{R}_{k,l;i,j} = \mathcal{R}_{l,k;j,i}
\eeq
with the convention that $\frac{1}{ij} \equiv 1$ if $i = 0$ or $j = 0$. Again, the refined correlators (defined in \eqref{deq}) are related to derivatives of the partition function:
\beq
\label{1psqoq}W_n(\sheet{x_1}{k_1},\ldots,\sheet{x_n}{k_n}) = \frac{\delta_{n,1}\,N_{k_1}}{(x_1 - \Lambda_{k_1})} + \sum_{\substack{(j_1,k_1),\ldots,(j_n,k_n) \\ j_1,\ldots,j_n \geq 1}} \Big[\prod_{i = 1}^n \frac{j_i/N}{(x_i - \Lambda_{k_i})^{j_i + 1}}\Big] (\ln Z)_{\mathbf{j}}[\mathbf{R},\mathcal{V},\mathbf{N}].
\eeq
The Taylor expansion \eqref{defrea}, implies that one can build the partition function out of local partition functions \eqref{locap}, by "mixing" them with some differential operator \cite{DBOSS12,Einter,Kostov:2010fk}
\beq
Z[\mathbf{R},\mathbf{\mathcal{V}},\mathbf{N}] = \exp\Big(\frac{1}{2N^2}\sum_{1\leq k,l \leq s} \sum_{i,j \geq 0} \mathcal{R}_{k,l;i,j}\,\frac{\partial^2}{\partial t_{k,i}\partial t_{l,j}}\Big)\prod_{k = 1}^s Z\Big[\frac{N}{N_k}\,\mathcal{V}_k,N_k\Big].
\eeq
The action of this quadratic differential operator can be written as usual with Wick's theorem as a sum over graphs with vertices weighted by derivatives of the local free energies $F[\mathcal{V},N] = \ln Z[\mathcal{V},\mathbf{N}]$, and edges weighted by the two-point interaction:
\bea
Z[\mathbf{R},\mathbf{{\cal V}},\mathbf{N}] 
&=& \sum_{\substack{{\cal G},\,\,\textrm{s-colored} \\ \mathrm{graph}}}  \frac{N^{-2\#\,E(\mathcal{G})}}{\#{\rm Aut}({\cal G})} \, \prod_{v \in V({\cal G})} F_{\mathbf{j}(\mathbf{e}(v))}\Big[\frac{N}{N_{c(v)}}\,\mathbf{\mathcal{V}}_{c(v)},N_{c(v)}\Big] \cr
&& \prod_{e \in E({\cal G})} {\cal R}_{c(v_0(e)),c(v_1(e));j(v_0(e),e),j(v_1(e),e)}.
\eea
The \emph{$s$-colored graphs} over which the sum ranges are described as follows:
\begin{itemize}
\item[$\bullet$] $\mathcal{G}$ is an oriented graph (maybe disconnected).
 We denote $V(\mathcal{G})$ (resp. $E(\mathcal{G})$), its set of vertices (resp. edges).  
\item[$\bullet$] Vertices $v$ are decorated by a color in $\ldbrack 1,s \rdbrack$, denoted $c(v)$.
\item Half-edges $(v,e)$, where $v$ is a vertex and $e$ an edge adjacent to $v$, are decorated by a positive integer, denoted $j(v,e)\geq 0$. If $e$ is an edge, we denote $v_0(e)$ and $v_1(e)$ the 2 adjacent vertices. 
\item[$\bullet$] If $v$ is a vertex, we denote $\mathbf{e}(v)$ the set of edges adjacent to $v$, and $\mathbf{j}(\mathbf{e}(v))$ the set of $j(v,e)$ for $e \in \mathbf{e}(v)$.
\end{itemize}
One can go further and make explicit the dependance of the partition function in terms of the times by introducing the set of $s$-colored graphs with leaves. We call \emph{leave} an open half-edge, i.e. a pair consisting in a marked univalent vertex and an edge which connects it to a regular vertex $v \in E(\mathcal{G})$. We denote $L(\mathcal{G})$ the set of leaves of such a graph.
Then, we have:
\bea
Z[\mathbf{R},\mathbf{\mathcal{V}},\mathbf{N}] &=& \sum_{\substack{{\cal G}\,\,s\textrm{-colored}\,\,\mathrm{graph} \\ \mathrm{with}\,\,\mathrm{leaves}}}\,\,\,\, \frac{N^{-2|E({\cal G})|}}{|{\rm Aut}({\cal G})|} \,\prod_{v \in V({\cal G})} F_{\mathbf{j}({\bf e}(v))}\Big[\frac{N}{N_{c(v)}}\,\mathcal{V}^{(0)}_{c(v)},N_{c(v)}\Big]\, \prod_{\ell \in L({\cal G})} t_{c(\ell),j(\ell)} \nonumber \\
&& \qquad\qquad\qquad \times \prod_{e \in E({\cal G})} {\cal R}_{c(v_0(e)),c(v_1(e));j(v_0(e),e),j(v_1(e),e)},
\eea
where the vertices are weighted by
\beq
{\cal F}_{\bf j}[(N/N_c)\mathcal{V}^{(0)}_c,N_c]= \frac{\partial^n \ln Z[(N/N_c)\mathcal{V}_c,N_c]}{\partial t_{j_1}\cdots \partial t_{j_n}} \Big|_{\mathbf{t} = 0}.
\eeq

\begin{figure}
\begin{center}\includegraphics[width=10cm]{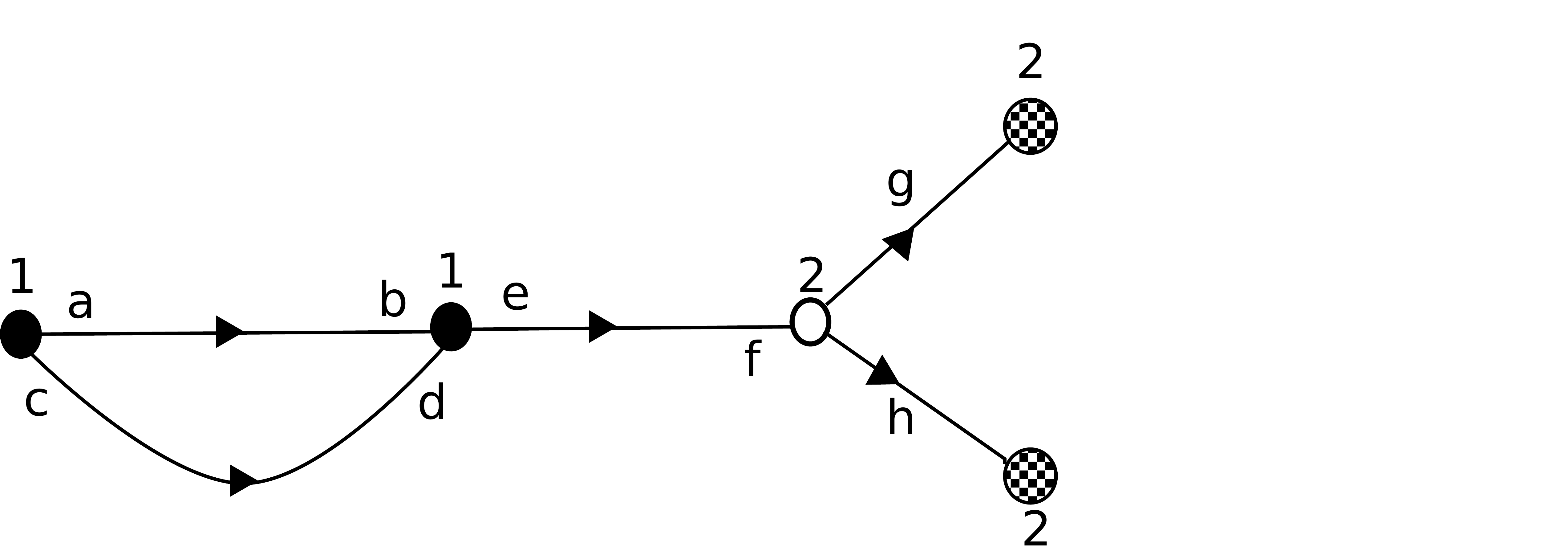}\end{center}
\caption{\label{Figgraph} Example of $2$-colored graph with two leaves. We indicate the two colors as black and white, and the leaves as shaded. This graph contributes to the global partition function with a weight $F_{a,c}[(N/N_1){\cal V}_1^{(0)},N_1] {\cal R}_{1,1;a,b} {\cal R}_{1,1;c,d} F_{b,d,e}[(N/N_1){\cal V}_1^{(0)},N_1]\,{\cal R}_{1,2;e,f}\,F_{f,g,h}[(N/N_2){\cal V}_2^{(0)},N_2]\,t_{2,g}t_{2,h}$.}
\end{figure}

\subsection{Virasoro constraints and loop equation}

In this section, we show how the Virasoro constraints for the local partition functions imply $s$ independent sets of Virasoro constraints for $Z[\mathbf{R},\mathcal{V},\mathbf{N}]$. If one comes back to Schwinger-Dyson equations in terms of correlators, it explains how the Schwinger-Dyson equations of the repulsive particle model with $s$ species can be deduced from the Schwinger-Dyson equations of the one-hermitian matrix model.

Observe that, if $f(t,\tilde t)$ is a function of 2 variables, one has
\bea
e^{\frac{{\cal R}+\tilde {\cal R}}{2}\,\frac{\partial}{\partial t}\,\frac{\partial}{\partial \tilde t}}\,f(t,\tilde{t})
&=& \sum_{k=0}^\infty \,\frac{({\cal R}+\tilde {\cal R})^k}{2^k\,\,k!}\,\,\frac{\partial^k}{\partial t^k}\,\,\frac{\partial^k}{\partial \tilde t^k} f(t,\tilde t) \nonumber \\
&=& \sum_{k=0}^\infty \left( \sum_{j=0}^k \frac{{\cal R}^j\,\tilde{\cal R}^{k-j}\,\,k!}{j!\,(k-j)!}\right) \,\frac{1}{2^k\,\,k!}\,\,\frac{\partial^k}{\partial t^k}\,\,\frac{\partial^k}{\partial \tilde t^k} f(t,\tilde t) \nonumber \\
&=& \sum_{i=0}^\infty\sum_{j=0}^\infty   \frac{(i+j)!}{i!\,j!}\,\frac{{\cal R}^{i}\,\tilde{\cal R}^j}{2^{i+j}\,\,(i+j)!}\,\,\frac{\partial^{i+j}}{\partial t^{i+j}}\,\,\frac{\partial^{i+j}}{\partial \tilde t^{i+j}} f(t,\tilde t) \nonumber \\
&=& \sum_{i=0}^\infty\sum_{j=0}^\infty  \frac{({\cal R}/2)^i\,\,(\partial/\partial \tilde t)^i}{i!}\,\frac{(\tilde{\cal R}/2)^j\,(\partial/\partial t)^j}{j!}\,\,\frac{\partial^{i}}{\partial t^{i}}\,\,\frac{\partial^{j}}{\partial \tilde t^{j}} f(t,\tilde t) \nonumber \\
& = & f(t+\frac{{\cal R}}{2}\,\frac{\partial}{\partial \tilde t},\tilde t+\frac{\tilde{\cal R}}{2}\,\frac{\partial}{\partial t})
\eea
where the last line is a convenient notation. This shows that the full partition function $Z[\mathbf{R},\mathcal{V},\mathbf{N}]$ is obtained from the one particle one by formal substitution:
\beq
\label{reusub}Z[\mathbf{R},\mathcal{V},\mathbf{N}] = \prod_{k = 1}^s Z\Big[\frac{N}{N_k}\,\tilde{\mathcal{V}}_k[\mathbf{R},\mathcal{V},\mathbf{N}],N\Big],
\eeq
where the potential $\tilde{\mathcal{V}}_k[\mathbf{R},\mathcal{V},N]$ is obtained by shifting the coefficients of the Taylor expansion of $\mathcal{V}_k$ by a differential operator:
\bea
\label{shifr}t_{k,i} \rightarrow \tilde{t}_{k,i}[\mathbf{R},\mathcal{V},N] & = & t_{k,i} + \sum_{l=1}^s \frac{1}{N^2}\Big(\sum_{j \geq 0} \frac{1}{2}{\cal R}_{k,l;i,j} \frac{\partial}{\partial t_{l,j}}\Big).
\eea
By convention, those differential operators are pushed to the left of the product of local partition functions.
We observe that this shift of times \eqref{shifr} is closely related to the Taylor expansion of the operators $\mathcal{O}_{k,l}$ introduced in \eqref{iksq} and which appeared in the loop equations studied in Section~\ref{S3}. Indeed, using the formal expansion of the $1$-point correlator \eqref{1psqoq}, recalling that $\Lambda_k \in \Gamma_{k,0}$ and the definition of the coefficients $\mathcal{R}_{k,l;i,j}$ in \eqref{defrea}:
\bea
& & \sum_{i \geq 0} \big(\tilde{t}_{k,i}[\mathbf{R},\mathcal{V},\mathbf{N}] - t_{k,i}\big)(x - \Lambda_k)^{i - 1} Z[\mathbf{R},\mathcal{V},\mathbf{N}] \nonumber \\
& = &  -\frac{1}{2N} \sum_{l = 1}^s \sum_{i,j \geq 0} \frac{\mathcal{R}_{k,l;i,j}}{j}(x - \Lambda_k)^{i - 1}\big(\Res_{\xi \rightarrow \infty} (\xi - \Lambda_l)^{j}\,W_1(\sheet{\xi}{l})\dd\xi\big) Z[\mathbf{R},\mathcal{V},\mathbf{N}] \nonumber \\
& = & -\frac{1}{2N} \sum_{l = 1}^s \oint_{\Gamma_{l,0}} \frac{\dd\xi}{2{\rm i}\pi}\Big(\sum_{i,j \geq 0} \frac{\,\mathcal{R}_{k,l;i,j}}{j} (x - \Lambda_k)^{i - 1}(\xi - \Lambda_l)^{j}\Big)\,W_1(\sheet{\xi}{l})\dd \xi\,Z[\mathbf{R},\mathcal{V},\mathbf{N}] \nonumber \\
& = & -\frac{1}{2N} \sum_{l = 1}^s \rho_{k,l} \oint_{\Gamma_{l,0}} \frac{\dd\xi}{2{\rm i}\pi}\,\partial_{x} \ln R_{k,l}(x,\xi)\,W_1(\sheet{\xi}{k})\,Z[\mathbf{R},\mathcal{V},\mathbf{N}] \nonumber \\
& = & -\frac{1}{2N} \sum_{l = 1}^s \rho_{k,l} (\mathcal{O}_{k,l} W_1)(\sheet{x}{k})\,Z[\mathbf{R},\mathcal{V},\mathbf{N}].
\eea
By substitution in the Virasoro constraints satisfied by the local partition functions, we obtain:
\bl
The partition function in the repulsive particle model with $s$ species satisfies:
\beq
\forall k \in \ldbrack 1,s \rdbrack,\quad \forall m \geq -1,\qquad L_{k,m}[\mathbf{R},\mathbf{t},\mathbf{N}]\cdot
 Z[\mathbf{R},{\cal V},\mathbf{N}] = 0,
\eeq
where we have set:
\beq
L_{k,m}[\mathbf{R},\mathbf{t},\mathbf{N}] = L_{k,m}\Big[\frac{N}{N_k}\,\mathbf{t}_k,N_k\Big]
+ \sum_{l=1}^s \sum_{i,j \geq 0} (m+j)\frac{{\cal R}_{k,l;j,i}}{N^2}\,\frac{\partial^2}{\partial t_{k,i}\partial t_{l,m+j}}.
\eeq
These operators satisfy the commutation relation: 
\beq
\forall k,k' \in \ldbrack 1,s \rdbrack,\quad \forall m,m' \geq -1,\qquad \big[L_{k,m}[\mathbf{R},\mathbf{t},\mathbf{N}],L_{k',m'}[\mathbf{R},\mathbf{t},\mathbf{N}]\big]  = \delta_{k,k'} (m-m') L_{k,m + m'}[\mathbf{R},\mathbf{t},\mathbf{N}].
\eeq
\el
\noindent\textbf{Proof.} The fact that the operators $L_{k,m}$ annihilate the full partition function is a direct consequence of the substitution relation \eqref{reusub}. To establish the commutation relations, we start with:
\beq
\big[ L_m[\mathbf{t}_k,N_k], L_{m'}[\mathbf{t}_{k'},N_{k'}]\big] = \delta_{k,k'}\,(m-m')\,L_{m+m'}[\mathbf{t}_k,N_k],
\eeq
and:
\bea
& & \Big[\sum_{i,j \geq 0} (m+j)  \frac{{\cal R}_{k,l;j,i}}{N^2} \frac{\partial^2}{\partial t_{k,i}\partial t_{l,m+j}},  L_{m'}[\mathbf{t}_{k'},N_{k'}]\Big] \nonumber \\
& = & \delta_{k,k'} \sum_{i,j \geq 0} (m'+m+j)(m+j) \frac{{\cal R}_{k,l;j,i}}{N^2}\,\frac{\partial^2}{\partial t_{k,i}\partial t_{k',m'+m+j}} \nonumber \\
& & + \delta_{l,k'}\sum_{i,j \geq 0} (m+j) (m'+i) \frac{{\cal R}_{k,l;j,i}}{N^2}\,\frac{\partial^2}{\partial t_{k,m'+i}\partial t_{l,m+j}}.
\eea
Besides, we remark:
\beq
\Big[\sum_{i,j \geq 0} (m+j) {\cal R}_{k,l;j,i} \frac{\partial^2}{\partial t_{k,i}\partial t_{l,m+j}},\sum_{i',j' \geq 0} (m'+j') {\cal R}_{k',l';j',i'} \frac{\partial^2}{\partial t_{k',i'}\partial t_{l',m'+j'}} \Big] = 0.
\eeq
Therefore:
\bea
& & \big[ L_{k,m}[\mathbf{R},\mathbf{t},\mathbf{N}],L_{k',m'}[\mathbf{R},\mathbf{t},\mathbf{N}]\big] \nonumber \\
& = &  \big[L_{m}[(N/N_k)\mathbf{t}_k,N_k],L_{m'}[(N/N_{k'})\mathbf{t}_{k'},N_{k'}]\big] \nonumber \\
& & + \Big[L_{m}[(N/N_k)\mathbf{t}_k,N_k],\sum_{l = 1}^s \sum_{i,j \geq 0} (m' + j)\,\frac{\mathcal{R}_{k',l;j,i}}{N^2}\,\frac{\partial^2}{\partial t_{k',i} \partial t_{l,m' + i}}\Big] \nonumber \\
& & + \Big[\sum_{l = 1}^{s} \sum_{i,j \geq 0} (m + j)\,\frac{\mathcal{R}_{k,l;j,i}}{N^2}\,\frac{\partial^2}{\partial t_{k,i}\partial t_{l,m + j}},L_{m'}[(N/N_{k'})\mathbf{t}_{k'},N_{k'}]\Big] \nonumber \\
& = & \delta_{k,k'}(m - m')L_{m + m'}[\mathbf{t}_k,N_k] \nonumber \\
& & - \delta_{k,k'}\sum_{l = 1}^s \sum_{i,j \geq 0} (m' + m + j)(m' + j)\,\frac{\mathcal{R}_{k,l;j,i}}{N^2}\,\frac{\partial^2}{\partial t_{l,i}\partial t_{k,m + m' + j}} \nonumber \\
& & - \sum_{i,j \geq 0} (m' + j)(m + i)\,\frac{\mathcal{R}_{k',k;j,i}}{N^2}\,\frac{\partial^2}{\partial t_{k,m + i}\partial t_{k',m' + j}} \nonumber \\
& & + \delta_{k,k'} \sum_{l = 1}^s \sum_{i,j \geq 0} (m' + m + j)(m + j)\,\frac{\mathcal{R}_{k,l;j,i}}{N^2},\frac{\partial^2}{\partial t_{l,i}\partial t_{k,m' + m + j}} \nonumber \\
& & + \sum_{i,j \geq 0} (m + j)(m' + i)\,\frac{\mathcal{R}_{k,k';j,i}}{N^2}\,\frac{\partial^2}{\partial t_{k',m' + i}\partial t_{k,m + j}} \nonumber \\
& = & \delta_{k,k'}(m - m')L_{m + m'}^{(k)}[\mathbf{R},\mathbf{t},\mathbf{N}].
\eea
\phantom{sq}\hfill $\Box$

It is noteworthy to give the combinatorial interpretation of the Virasoro constraints. The Virasoro operator $L_{k,m}[\mathbf{R},\mathcal{V},\mathbf{N}]$ acts on a $s$-colored graph $\mathcal{G}$ as follows:
\begin{itemize}
\item[$\bullet$] The linear operator $(m+j) t_{k,j}\,\frac{\partial}{\partial t_{k,m+j}}$ replaces a leaf of degree $m$ and color $k$ by a leaf of degree $m - j$ and of same color.
\item[$\bullet$] The bilinear operator $\frac{1}{N}  j(m-j)  \frac{\partial}{\partial t_{k,j}} \frac{\partial}{\partial t_{k,m-j}}$ replaces a couple of leaves of same color $k$ and respective degree $j$ and $m-j$ by two vertices linked by an edge of weight 1, oriented from the first to the second.
\item[$\bullet$] The bilinear operator $(m+j)i {\cal R}_{k,l;j,i} \frac{\partial^2}{\partial t_{l,i} \, \partial t_{k,m+j}}$ replaces a couple of leaves of respective colors $k$ and $l$ and respective degree $m+j$ and $i$ by two vertices linked by an edge $e'$ of weight ${\cal R}_{k,l;j,i}$, oriented from the first to the second, and such that $j_0(e') = j$ and $j_1(e') = i$. The new vertices keep the color of the leaf they came from.
\end{itemize}
The Virasoro constraints can then be seen as a consequence of a bijection between sets of $s$-colored graphs.
This interpretation can be mapped to Tutte's equations for generating series of colored maps with tubes introduced in Section~\ref{61} below, by making the following correspondence for a given $s$-colored graph $\mathcal{G}$:
\begin{itemize}
\item[$\bullet$] Each $n$-valent vertex $v \in V(\mathcal{G})$ of color $k$ is mapped to a sum of maps of color $k$ with $n$ boundaries, whose respective lenghts are given by the indices $j_i(e)$ of the incident edges (i.e. $i = 1$ if the edge is pointing towards $v$, $i = 0$ else).
\item[$\bullet$] Each edge weighted by $\mathcal{R}_{k,l;i,j}$ is mapped to an annular face, the two boundaries of which have respective colors $k$ and $l$, and respective lengths $i$ and $j$.
\end{itemize}

\section{Enumeration of maps with decorations}
\label{S5}

In this section, we apply Section~\ref{S3} to the study of the enumerative problems emerging from the combinatorial interpretation of some formal matrix models. In a first step (\S~\ref{611}), we introduce a combinatorial model enumerating colored maps with tubes, and we explain in \S~\ref{612} an equivalent formulation in terms of maps with self-avoiding loops. We then show in \S~\ref{62} that the repulsive particle model (which is a kind of matrix model) with arbitrary two-point interaction generates such maps. We derive in \S~\ref{63} combinatorial
relations between generating series by the technique of substitution developed in \cite{BBG1,BBG2} for planar maps, and by Tutte's decomposition for higher genus maps. Those relations are actually equivalent to the loop equations satisfied the repulsive particle model. This allows us to compute the generating series of colored maps of all topologies with tubes by the topological recursion.
We complete this result by describing in \S~\ref{64} a technique to compute explicitly the coefficients of the generating series of such maps, since closed form for the generating series themselves is in general out of reach. Eventually, we describe in detail in \S~\ref{SQADE} the special case of maps carrying an ADE height model, which fits in our general formalism.

\subsection{Combinatorial models}

\label{61}

\subsubsection{Colored maps with tubes}
\label{611}
A \emph{map} (see e.g. \cite{ChapuyThese}) is an equivalence class modulo oriented homeomorphisms of proper embeddings of a finite graph in an oriented surface, such that:
\begin{itemize}
\item[$\bullet$] erasing the image of the graph in the surface yields a finite union of connected components (called \emph{faces}), which are homeomorphic to a disk.
\item[$\bullet$] each connected component of the surface has a non-empty intersection with the image of the graph.
\end{itemize}
We call length of a boundary of a face the number of edges forming this boundary, and the degree of a face is the length of its unique boundary. If $g,n \geq 0$ and $\ell = (\ell_1,\ldots,\ell_n)$ is a vector of positive integers, we define $\mathsf{M}_n^{g}(\ell)$, the set of maps drawn on a genus $g$ surface, with $n$ marked faces whose boundary have a marked edge, and respective lengths $\ell_i$. 
By convention, $\mathsf{M}_{1}^0(0)$ has a single element, which is the embedding of the graph with one vertex, and that is the only case where we encounter $0$ boundary lengths.
 
We now introduce the notion of $s$-colored maps with tubes, by replacing the first point by:
\begin{itemize}
\item[$\bullet$] faces are either homeomorphic to a disk (those have $1$ boundary, and are called \emph{simple faces}), or to an annulus (those have $2$ boundaries, and are called \emph{annular faces}).
\item[$\bullet$] the connected components of the graph carry a color, which is an integer between $1$ and $s$.
\end{itemize}
Notice that if the two boundaries of an annular face belong to the same connected component of the graph, they must carry the same color. We agree that simple faces receive the color of their boundary, while annular faces receive the couple of color of their two boundaries. 

Let $g,n \geq 0$, and $\mathbf{k} \in \ldbrack 1,s \rdbrack^n$ a vector of colors, $\mathbf{\ell} \in (\mathbb{N}^*)^n$ a vector of lengths. We define $s\mathsf{CMT}_{n}^g(\mathbf{k};\mathbf{\ell})$ as the set of $s$-colored maps which are connected surface of genus $g$ with $n$ marked simple faces of respective colors $(k_i)_i$ and lengths $(\ell_i)_{i}$, each carrying a marked edge on its boundary.

\begin{figure}[h!]
\begin{center}\includegraphics[width=5cm]{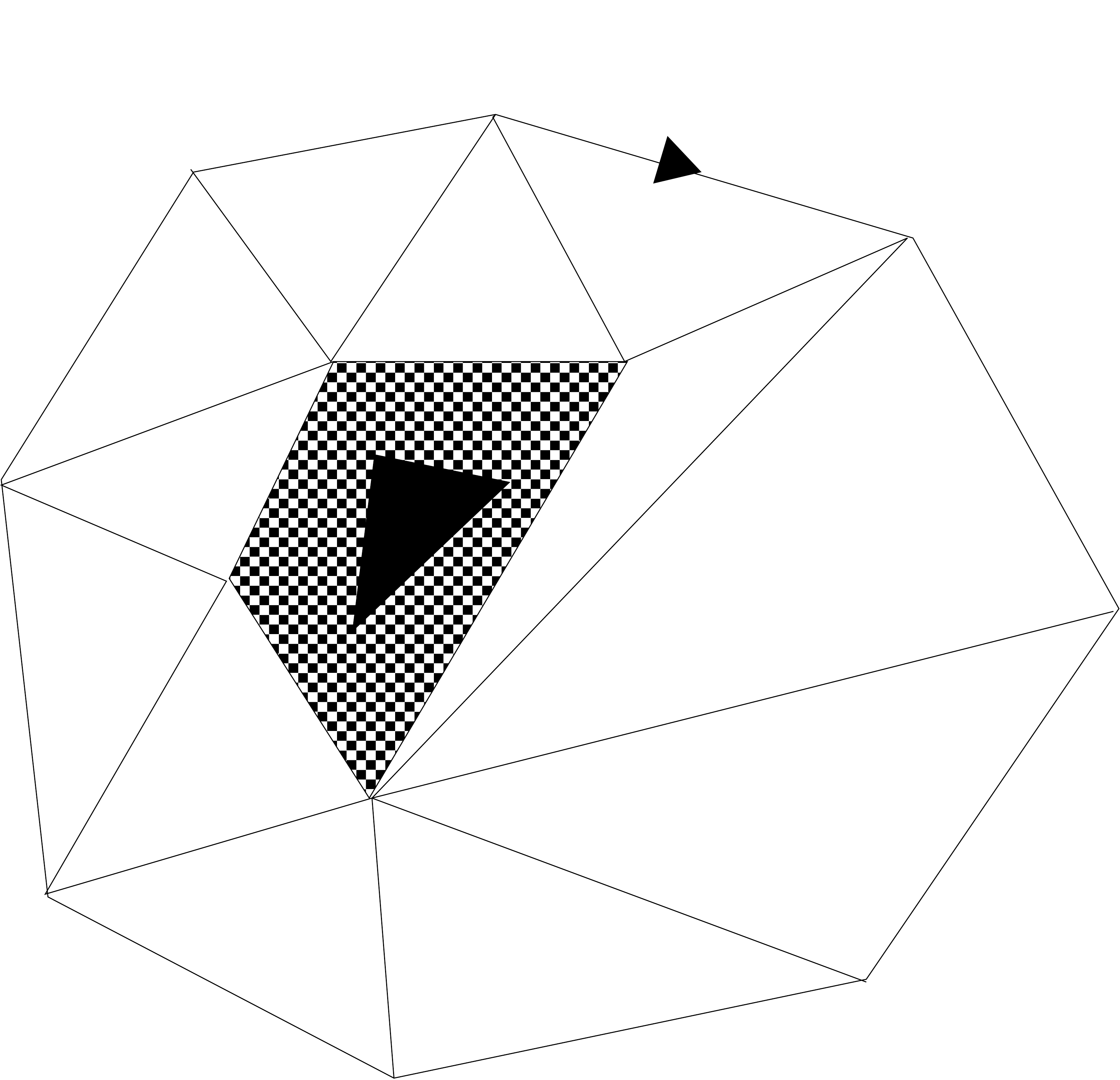}\end{center}
\caption{\label{mapex1} Example of genus $0$ map with two colors: one marked white face of degree 8 (octogon), 12 unmarked white simple faces of degree $3$ (triangles), one black simple face of degree $3$, and one annular face of degree $(4,3)$.}
\end{figure}

We want to consider a model where a map $\mathcal{M} \in s\mathsf{CMT}_{n}^g(\mathbf{k},\mathbf{l})$ receives the weight $w(\mathcal{M})$ obtained as a product of the following Boltzmann weights:
\begin{itemize}
\item[$\bullet$] Each vertex receives a local weight $u$.
\item[$\bullet$] Each vertex of color $k$ receives a local weight $u_{k}$.
\item[$\bullet$] Each unmarked simple face of color $k$ and degree $\ell$ receives a local weight $t_{k,\ell}$.
\item[$\bullet$] Each annular face of colors $(k_1,k_2)$ and degrees $(\ell_1,\ell_2)$ receives a local weight $\mathcal{R}_{k_1,k_2;\ell_1,\ell_2} = \mathcal{R}_{k_2,k_1;\ell_2,\ell_1}$.
\end{itemize}
We denote $|\mathrm{Aut}\,\mathcal{M}|$ the number of automorphisms of the map.

We fix once for all a sequence of real numbers $\Lambda = (\Lambda_1,\ldots,\Lambda_s)$. For $n = 0$ and any $g \geq 0$, we define the generating series:
\beq
F_g = \sum_{\mathcal{M} \in s\mathsf{CMT}_{0}^g} \frac{w(\mathcal{M})}{|\mathrm{Aut}\,\mathcal{M}|}.
\eeq
For any $n \geq 1$ and $g \geq 0$, we define a sequence of generating series indexed by $\mathbf{k} \in \ldbrack 1,s \rdbrack^n$:
\beq
\label{poiuq}W_{n}^g(\sheet{x_1}{k_1},\ldots,\sheet{x_n}{k_n}) = \frac{\delta_{n,1}\delta_{g,0}\,uu_{k_1}}{(x_1 - \Lambda_{k_1})} + \sum_{\ell_1,\ldots,\ell_n \geq 1} \left(\sum_{\mathcal{M} \in s\mathsf{CMT}_{n}^g(\mathbf{k};\mathbf{\ell})} \frac{w(\mathcal{M})}{|\mathrm{Aut}\,\mathcal{M}|}\right)\left[\prod_{i = 1}^n \frac{1}{(x_i - \Lambda_{k_i})^{\ell_i + 1}}\right].
\eeq
Eventually, we introduce formal series in a large parameter $N$ to collect all genera:
\beq
Z = e^{F},\qquad F = \sum_{g \geq 0} \Big(\frac{N}{u}\Big)^{2 - 2g}\,F^{g},\qquad W_n(\sheet{x_1}{k_1},\ldots,\sheet{x_n}{k_n}) = \sum_{g \geq 0} \Big(\frac{N}{u}\Big)^{2 - 2g - n}\,W_n^{g}(\sheet{x_1}{k_1},\ldots,\sheet{x_n}{k_n}).
\eeq
A standard counting argument using the Euler characteristics shows that, for any $v \geq 0$, for given $g,n$, there is only a finite number of maps in $s\mathsf{CMT}_{n}^g(\mathbf{k};\mathbf{\ell})$ with exactly $v$ vertices, so that the coefficient of $u^{v}$ is given by a finite sum \cite{Eformal,TheseGB}. Hence, $F_g$, $W_n^g$, $Z$, $F$ and $W_n$ are well-defined formal series in $u$.

It is convenient to introduce the following generating series of annular faces:
\beq
\label{diu} \big(R_{k_1,k_2}(x_1,x_2)\big)^{\rho_{k_1,k_2}/2} = \exp\Big(\sum_{\substack{i_1,i_2 \geq 0 \\ (i_1,i_2) \neq (0,0)}} \frac{\mathcal{R}_{k_1,k_2;i_1,i_2}}{i_1i_2} (x_1 - \Lambda_{k_1})^{i_1}(x_2 - \Lambda_{i_2})^{i_2}\Big).
\eeq
with the convention that $\frac{1}{i} = 1$ if $i = 0$.

\subsubsection{Maps carrying self-avoiding loop configuration}
\label{612}
Self-avoiding loop models play an important role in two dimensional statistical physics, because they allow to reach at the critical point a continuum of universality classes, parametrized by the fugacity given to a loop, and believed to be described by conformal field theories with central charge $c < 1$ \cite{PagesjaunesOn}. Their analog on maps (i.e. on a random two dimensional lattice) have also been studied \cite{KosStau}, and their relation at the critical point with the same model on the fixed lattice should be captured by the KPZ relations \cite{KPZ,Dav,DKLiou}.

Given a map $\mathcal{M}$, a \emph{loop configuration} is a collection of self- and mutually non intersecting cycles (also called \emph{loops}\footnote{This denomination has nothing to do with the usual name of "loop equations".}) drawn on the surface, avoiding the vertices and crossing edges at most once and transversally. The set of faces and edges crossed by a cycle are thus cyclically ordered, their union has the topology of an annulus. We call \emph{ring} this sequence of faces. Eventually, we define a \emph{$s$-colored map with a loop configuration} as a map with a loop configuration such that each connected component of the graph minus the edges crossed by a loop carries a color between $1$ and $s$. Faces which are not crossed by a loop thus receive the color of their boundary, and rings receive two colors (one for each boundary). When some faces are marked, we require that their boundary is not crossed by a loop\footnote{This means that we consider here only uniform boundary conditions for the maps. Maps where we allow open paths whose ends are located on boundaries of marked faces, can be decomposed upon removing the faces visited by these open paths, into a collection of maps with uniform boundary conditions. Their generating series, for a finite number of open paths realizing a given link pattern, can be computed by universal relations described in \cite[Chapitre 5]{TheseGB}}. The colors can be seen as colors of domains separated by loops, with the precision that if removing a loop did not disconnect the map, the colors of the domains on both sides of this loop should be the same.

As in \S~\ref{611}, we collect the sets of connected $s$-colored maps with a given topology and given length and colors for marked faces $s\mathsf{ML}_{n}^g(\mathbf{k};\mathbf{\ell})$.

We give to such a map a weight $w(\mathcal{M})$ obtained as the product of:
\begin{itemize}
\item[$\bullet$] a local weight $uu_k$ per vertex of color $k$.
\item[$\bullet$] a local weight $t_{k,\ell}$ per face of color $k$ which is not crossed by a loop.
\item[$\bullet$] a local weight $z_{k_1,k_2} = z_{k_2,k_1}$ per face of a ring of color $k_1,k_2$.
\item[$\bullet$] a local weight $g_{k_1,k_2;\ell_1,\ell_2} = g_{k_2,k_1;\ell_2,\ell_1}$ per face crossed by a loop consisting of a sequence of $\ell_1$ edges with color $k_1$, an edge crossed by the loop, another sequence of $\ell_2$ edges with color $k_2$, and another edge crossed by the loop, for some $\ell_1,\ell_2 \geq 0$.
\item[$\bullet$] a non-local weight $-\rho_{k,l} = -\rho_{l,k}$ per ring of color $(k,l)$.
\end{itemize}
And, we define as in \S~\ref{611} the generating series:
\beq
F^g = \sum_{\mathcal{M} \in s\mathsf{ML}_{0}^g} \frac{w(\mathcal{M})}{|\mathrm{Aut}\,\mathcal{M}|},
\eeq
and for any $n \geq 1$, the sequence of generating series indexed by a vector $\mathbf{k}$ of $n$ colors:
\beq
\label{juanito}W_n^{g}(\sheet{x_1}{k_1},\ldots,\sheet{x_n}{k_n}) = \frac{\delta_{n,1}\delta_{g,0}\,uu_k}{(x_1 - \Lambda_{k_1})} + \sum_{\ell_1,\ldots,\ell_n \geq 1}\Big(\sum_{\mathcal{M} \in s\mathsf{ML}_{n}^g(\mathbf{k},\mathbf{\ell})} \frac{w(\mathcal{M})}{|\mathrm{Aut}\,\mathcal{M}|}\Big)\Big[\prod_{i = 1}^n \frac{1}{(x_j - \Lambda_{k_j})^{\ell_j + 1}}\Big].
\eeq
Eventually, we introduce the generating series for all genera:
\beq
Z = e^{F},\qquad F = \sum_{g \geq 0} \Big(\frac{N}{u}\Big)^{2 - 2g}\,F^{g},\qquad W_n(\sheet{x_1}{k_1},\ldots,\sheet{x_n}{k_n}) = \sum_{g \geq 0} \Big(\frac{N}{u}\Big)^{2 - 2g - n}\,W_n^{g}(\sheet{x_1}{k_1},\ldots,\sheet{x_n}{k_n}).
\eeq
Again, a counting argument using Euler characteristics shows that these are well defined formal series in $u$ and $(z_{k,k'})_{k,k'}$.

This model is a particular case of the model of $s$-colored maps with tubes defined in \S~\ref{611}. Indeed, if $\mathcal{M}$ is an $s$-colored maps with a loop configuration, by erasing the edges crossed by the loops, we obtain an $s$-colored map with tubes. The weight given to annular faces via this bijection is encoded in the generating series (compare with \eqref{diu}):
\beq
\label{dsqi}R_{k_1,k_2}(x_1,x_2) = \frac{1}{z_{k_1,k_2}} + \sum_{\substack{\ell_1,\ell_2 \geq 0 \\ (\ell_1,\ell_2) \neq (0,0)}} g_{k_1,k_2;\ell_1,\ell_2} (x_1 - \Lambda_{k_1})^{\ell_1}(x_2 - \Lambda_{k_2})^{\ell_2}.
\eeq 
Conversely, general weights $g_{k_1,k_2;\ell_1,\ell_2}$ in the model of $s$-colored maps with a loop configuration allow to reproduce general weights $\mathcal{R}_{k_1,k_2;i,j}$ in the model of $s$-colored maps with tubes. Therefore, these two combinatorial models are actually equivalent, and both points of view are interesting: maps with tubes appear naturally in relation with Virasoro constraints, while maps with loop configurations appear naturally in the combinatorial interpretation of matrix models as we now review.

\subsection{Formal matrix model representations}
\label{62}

\subsubsection{Maps}

The relation between maps and hermitian matrix models was pioneered by Br\'ezin, Itzykson, Parisi and Zuber \cite{BIPZ}.
They have shown that the Feynman diagram expansion for the formal local partition function \eqref{locap}
\bea
Z & = & Z[\mathcal{V}_k,N_k] = \int_{\mathcal{H}_{N_k}} \dd M\,e^{-N_k\,\mathrm{Tr}\,\mathcal{V}_k(M)},\qquad \mathcal{V}_k(M) = \frac{1}{uu_k}\Big(\frac{(x - \Lambda_k)^2}{2} - \sum_{j \geq 3} \frac{t_j}{j}\,(M - \Lambda_k)^{j}\Big) \nonumber \\
\label{hjuq}& = & \sum_{n = 0}^{\infty} \frac{1}{n!} \Big(\frac{N}{u}\Big)^n \sum_{j_1,\ldots,j_n \geq 3} \int_{\mathcal{H}_{N_k}} \dd M \exp\Big(-\frac{N}{u}\,\frac{\Tr (M - \Lambda_k)^2}{2}\Big)\Big[\prod_{i = 1}^n \frac{t_{j_i}}{j_i}\,\Tr (M - \Lambda_k)^{j_i}\Big]
\eea
coincides with the generating series of maps:
\beq
\sum_{g \geq 0} \Big(\frac{N}{u}\Big)^{2 - 2g} \sum_{\mathcal{M} \in \mathsf{M}_{g,0}} \frac{w(\mathcal{M})}{|\mathrm{Aut}\,\mathcal{M}|}
\eeq
assuming $N_k = Nu_k$. The degree $j$ term in $\mathcal{V}_k$ generates faces of degree $j$ and color $k$, while the Gaussian matrix integral over $M_k$ is responsible for gluing faces.

\subsubsection{$s$-colored maps with a loop configuration}
\label{scolorl}
The same techniques have been used by Gaudin, Kostov and Mehta \cite{KosM,GK} to represent the generating series of maps with a loop configuration as a formal matrix model with several hermitian matrices, and it is straightforward to adapt them to $s$-colored maps with a loop configuration. Let us assume momentarily that $\rho_{k,l}$ are negative integers. Then, the generating series $Z$ (resp. \mbox{$W_n(\sheet{x_1}{k_1},\ldots,\sheet{x_n}{k_n})$}) introduced in \S~\ref{612} coincides with the partition function (resp. the correlators of the matrices $M_1,\ldots,M_s$) in the formal matrix model defined by the measure:
\bea
\dd\varpi & \propto & \prod_{k = 1}^s \dd M_k \exp\Big(-N\,\Tr\,\mathcal{V}_k(M_k)\Big)  \\
& & \times \prod_{1 \leq k \leq l \leq s} \prod_{a = 1}^{-\rho_{k,l}} \dd A_{k,l}^{(a)} \exp\left\{N\left(-\frac{\Tr (A_{k,l}^{(a)})^2}{2z_{k,l}} + \sum_{i,j \geq 0} \frac{g_{k,l;i,j}}{2}\,\Tr(M_{k}^{i} A_{k,l}^{(a)} M_{l}^j (A_{k,l}^{(a)})^{\dagger})\right)\right\}. \nonumber
\eea
Here, $M_k$ are hermitian matrices of size $N_k \times N_k$, $A_{k,k}^{(a)}$ are hermitian matrices of size $N_k \times N_k$ and $A_{k,l}$ are complex rectangular matrices of size $N_k \times N_l$, and $N_k = Nu_k$. The coupling between the matrices $M_k$ and $A_{k,l}^{(a)}$ generates faces with two distinguished edges, while the Gaussian matrix integral over $A_{k,l}^{(a)}$ is responsible for gluing such faces along the distinguished edges. Thus, if we draw a path which crosses the distinguished faces of such faces, it will form a loop. Since the matrices $A_{k,l}^{(a)}$ come with a Gaussian weight and the correlators we are interested in only involve the $M_k$'s, we can integrate them out. We find that it induces the measure on $M_k$'s:
\bea
\label{pouq}\dd\varpi & \propto & \prod_{k = 1}^s \dd M_k \exp\Big(- N\,\Tr\,\mathcal{V}_k(M_k)\Big) \nonumber \\
& & \times  \prod_{1 \leq k,l \leq s} \exp\Big[\frac{\rho_{k,l}}{2}\,\Tr\ln\big(R_{k,l}(M_k\otimes\mathbf{1}_{N_l},\mathbf{1}_{N_k}\otimes M_l)\big)\Big],
\eea
with $R_{k,l}$ given in \eqref{dsqi}. In this form, the measure makes sense for $\rho_{k,l}$ not restricted to be a negative integer, and it coincides with the formal model of repulsive particles introduced in \eqref{defmeasureseveral} for $\beta_{k} = 2$. We thus have obtained the combinatorial interpretation of this model. It is not difficult to extend this interpretation to values of $\beta_k$ by including non orientable maps (i.e. maps where some edges are M\"{o}bius strips) as in \cite{CEMq1}, but we shall not pursue this direction.

\subsection{Loop equations}
\label{63}

\subsubsection{Review for usual maps}
\label{maps2} In this paragraph, we shall reserve the notation
\beq
\label{jue}\widehat{W}_n^{g}[\mathcal{V}_k](\sheet{x_1}{k},\ldots,\sheet{x_n}{k}),\qquad \mathcal{V}_k(x) = \frac{1}{uu_k}\Big(-\sum_{j \geq 1} \frac{t_{k,j}}{j}\,(x - \Lambda_k)^j\Big)
\eeq
to the generating series of maps (in the usual sense) of given topology, i.e. $W_n^g(\sheet{x_1}{k},\ldots,\sheet{x_n}{k})$ defined in \eqref{juanito} where $s\mathsf{ML}$ is replaced by $\mathsf{M}$. $k$ denotes a color between $1$ and $s$, but does not play an important role for the moment, it intervenes in \eqref{jue} only through $\Lambda_k$. Maps with one marked face can be constructed recursively by removing the face adjacent to the marked edge on the marked face. For planar maps ($g = 0$), this results into the celebrated Tutte's equation \cite{TutteQ}:
\beq
\label{maseq}\big(\widehat{W}_1^0(\sheet{x}{k})\big)^2 - \oint \frac{\dd\xi}{2{\rm i}\pi}\,\frac{u\,\mathcal{V}'_k(\xi)\,\widehat{W}_1^0(\sheet{\xi}{k})}{x - \xi} = 0,
\eeq
where the contour integral is sufficiently far away from $\Lambda_k$. The same procedure can be worked out in any topology, and the result is
 \cite{Eformal,Ebook}:
\beq
\label{wyqu}\widehat{W}_2(\sheet{x}{k},\sheet{x}{k}) + \big(\widehat{W}_1(\sheet{x}{k})\big)^2- N_k \oint \frac{\dd\xi}{2{\rm i}\pi}\,\frac{\mathcal{V}'_k(x)\,\widehat{W}_1(\sheet{x}{k})}{x - \xi} = 0,
\eeq
and for any $n \geq 2$:
\bea
\label{qywu} \widehat{W}_{n + 1}(\sheet{x}{k},\sheet{x}{k},\sheet{x_I}{k}) + \sum_{J\subseteq I} \widehat{W}_{|J| + 1}(\sheet{x}{k},\sheet{x_{J}}{k})\widehat{W}_{n - |J|}(\sheet{x}{k},\sheet{x_{I\setminus J}}{k}) - N_k \oint \frac{\dd\xi}{2{\rm i}\pi}\,\frac{\mathcal{V}_k'(x)\,\widehat{W}_{n}(\sheet{\xi}{k},\sheet{x_I}{k})}{x - \xi} & & \\
 + \sum_{i \in I} \oint \frac{\dd\xi}{2{\rm i}\pi}\,\frac{\widehat{W}_{n - 1}(\sheet{\xi}{k},\sheet{x_{I\setminus\{i\}}}{k})}{(x - \xi)(x_i - \xi)^2} & = & 0, \nonumber 
\eea
where $\mathcal{V}_k$ is given in \eqref{hjuq}. As a matter of fact, \eqref{qywu} can be obtained from \ref{wyqu} by marking faces. Both equations can also be derived from the matrix model representation \eqref{hjuq} as Schwinger-Dyson equations.

Similarly, the Schwinger-Dyson equations of the formal repulsive particle model \eqref{317} provide functional relations between generating series of $s$-colored maps with a loop configuration:
\bea
\label{qywu2} W_{n + 1}(\sheet{x}{k},\sheet{x}{k},\sheet{x_I}{k_I}) + \sum_{J\subseteq I} W_{|J| + 1}(\sheet{x}{k},\sheet{x_{J}}{k})W_{n - |J|}(\sheet{x}{k},\sheet{x_{I\setminus J}}{k_{I\setminus J}}) - N \oint \frac{\dd\xi}{2{\rm i}\pi}\,\frac{\mathcal{V}_k'(x)\,W_{n}(\sheet{\xi}{k},\sheet{x_I}{k_I})}{x - \xi} & & \\
 + \sum_{l = 1}^s \rho_{k,l}\,\oiint \frac{\dd\xi\,\dd\eta}{(2{\rm i}\pi)^2}\,\frac{\partial_{\xi} \ln R_{k,l}(\xi,\eta)}{x - \xi}\Big(W_{n + 1}(\sheet{\xi}{k},\sheet{\eta}{l},\sheet{x_I}{k_I}) + \sum_{J \subseteq I} W_{|J| + 1}(\sheet{\xi}{k},\sheet{x_J}{k_J})W_{n - |J|}(\sheet{\eta}{l},\sheet{x_{I\setminus J}}{k_{I\setminus J}})\Big) & & \\
 + \sum_{i \in I} \oint \frac{\dd\xi}{2{\rm i}\pi}\,\frac{\widehat{W}_{n - 1}(\sheet{\xi}{k},\sheet{x_{I\setminus\{i\}}}{k_{I\setminus\{i\}}})}{(x - \xi)(x_i - \xi)^2} & = & 0. \nonumber 
\eea
 We now explain how they can be given a purely combinatorial proof. We reserve in this paragraph the notation $W_n$ for generating series of $s\mathsf{ML}$ (as opposed to $\mathcal{W}_n$ for $\mathsf{M}$). 
 
\subsubsection{Planar maps with a loop configuration by substitution}

The nested loop approach developed in \cite{BBG1,BBG2} allows to decompose $s$-colored maps with a loop configuration, into usual maps. It is enough to consider the case of $n = 1$ marked face, and we first focus on planar maps, i.e. $g = 0$. We summarize the argument of \cite{BBG1}. Given $\mathcal{M} \in s\mathsf{ML}_{1}^0$, let us  remove the faces crossed by the outermost loops from the point of view of the marked face: "outermost" here means the loops which can be reached by a continuous path on the surface, starting on the marked face without crossing any other loop. Since $\mathcal{M}$ is planar, every loop is separating, therefore the ring where it is drawn has an outer and an inner boundary. We mark an edge on such an inner boundary by an arbitrary but well-defined procedure\footnote{For instance, we can take the edge $e$ which is the closest in $\mathcal{M}$, for geodesic distance on the graph, to the marked edge $e_0$ on the marked face of $\mathcal{M}$ ; if two edges $e$ and $e'$ lie at same distance along two geodesics starting at $e_0$, we can choose the one reached by the geodesic which turns left at the first point when the two geodesics become distinct.}. The outer connected component, which contains the marked face, is a map $\mathcal{M}' \in \mathsf{M}_{1}^0$ called \emph{the gasket}, which has a definite color equal to that of the marked face. The unmarked faces of $\mathcal{M}'$ are either faces of $\mathcal{M}$, or large faces created by the removal. The other connected components are again $s$-colored maps with a loop configuration $\mathcal{M}'_i \in s\mathsf{ML}_1^0$. Conversely, the data of the rings removed, $\mathcal{M}'$ and $\mathcal{M}_i'$ allow to reconstruct the initial map $\mathcal{M}$. Therefore, we have a bijective decomposition of $\mathsf{ML}_{1}^0$.

At the level of generating series, this translates into:
\beq
\label{519}W_1^0(\sheet{x}{k}) = \widehat{W}_1^0[\tilde{\mathcal{V}}_k](\sheet{x}{k}),
\eeq
where the shifted potential is the generating series of weights for the faces of the gasket:
\beq
\label{shifeq}\tilde{\mathcal{V}}_k(x) = \mathcal{V}_k(x) - \sum_{l = 1}^s \rho_{k,l} \oint \ln R_{k,l}(x,\xi)\,W_1^0(\sheet{\xi}{l})\,\frac{\dd\xi}{2{\rm i}\pi}.
\eeq
The contour of integration is chosen sufficiently far away from $\Lambda_k$. $\rho_{k,l}\ln R_{k,l}(x,y)$ is the generating series of rings with inner boundary of color $k$ and outer boundary of color $l$. The last term of \eqref{shifeq} expresses the fact that large faces of the gasket are identified in the correspondence with the gluing of a ring with a $s$-colored map with a loop configuration. Therefore, \eqref{maseq} for $\widehat{W}_1^0$ implies the relation:
\beq
\label{disqaq}\big(W_1^0(\sheet{x}{k})\big)^2 + \sum_{l = 1}^s \rho_{k,l} \oint \frac{\dd\xi\,\dd\eta}{(2{\rm i}\pi)^2}\,\frac{\partial_{x} \ln R_{k,l}(\xi,\eta)}{x - \xi}\,W_1^0(\sheet{\xi}{k})\,W_1^0(\sheet{\eta}{l}) - uu_k\,\oint \frac{\dd\xi}{2{\rm i}\pi}\,\frac{\mathcal{V}'_k(\xi)\,W_1^0(\sheet{\xi}{k})}{x - \xi} = 0.
\eeq
This gives a combinatorial origin to the Schwinger-Dyson equations \eqref{317} for $(n,g) = (1,0)$. Going from $n = 1$ to arbitrary $n$ just amounts to mark $(n - 1)$ extra faces, and this process also has a clear combinatorial meaning.

The substitution procedure cannot be naively applied when $g \geq 1$, for two reasons. Firstly, outermost loops might be non-contractible, so the "outside" and the "inside" have no meaning, and removing them creates pairs of new boundaries for the gasket. Secondly, to retrieve the initial map after the removal, we may need to glue $s$-colored maps with a loop configuration having $n' \geq 2$ boundaries, into a set of $n$ large faces of the gasket. This implies that the weight of the gasket is not local anymore (distinct faces can be coupled in this way). Therefore, such configurations cannot be enumerated in general by a $\widehat{W}_{n'}^{g'}$ for a shifted potential.

\subsubsection{Tutte's method and higher genus}

Eqn.~\ref{disqaq} can be rederived directly by Tutte's method. For any $\mathcal{M} \in \mathsf{ML}_{1}^0(k;\ell)$ which is not reduced to a single vertex, if we remove the marked edge, three situations are possible. Either it was bordered on both sides by the marked face, in which case we obtain two maps $\mathcal{M}'_1 \in \mathsf{ML}_{1}^0(k;\ell_1)$ and $\mathcal{M}'_2 \in \mathsf{ML}_{1}^0(k;\ell_2)$, with $\ell_1 + \ell_2 = \ell - 2$. Or, it was bordered by a face of degree $j$ which is not crossed by a loop, in which case we obtain a map $\mathcal{M}' \in \mathsf{M}_{1}^0(k;\ell - 2 + j)$. Or, it was bordered by a face crossed by a loop, whose boundary is a sequence of $i$ edges with color $k_1$, followed by an edge crossed by the loop, followed by another sequence of $j$ edges with color $k_2$, and another edge crossed by the loop. We then obtain a map $\mathcal{M}''$ of genus $0$ with $1$ marked face of degree $\ell - 2 + (i + j + 2)$, where two edges on the boundary of the marked face are ends of an open path (the former loop), and separated by $j$. Such a map can be further decomposed into:
\begin{itemize}
\item[$\bullet$] the strip of color $(k,l)$ consisting of the faces crossed by the path, whose boundary of color $k$ has length $i'$ and boundary of color $l$ has length $j'$.
\item[$\bullet$] two maps $\mathcal{M}_1'' \in \mathsf{ML}_{1}^0(k;\ell - 2 + i + i')$ and $\mathcal{M}_2'' \in \mathsf{ML}_1^0(l;j + j')$. 
\end{itemize}
This decomposition is a bijection, and reminding that $R_{k,l}(x,y)$ is the generating series of faces crossed by a loop, it translates into the functional relation:
\bea
\label{hgy} xW_1^0(\sheet{x}{k}) - uu_k & = & \big(W_1^0(\sheet{x}{k})\big)^2 \nonumber \\
& & + \oint \frac{\dd \xi}{2{\rm i}\pi}\,\frac{\sum_{j \geq 3} t_{k,j}\xi^{j - 1}}{x - \xi}\,W_1^0(\sheet{\xi}{k}) \nonumber \\
& & + \sum_{l = 1}^s \rho_{k,l}\,\oiint \frac{\dd\xi\,\dd\eta}{(2{\rm i}\pi)^2}\,\frac{\partial_{\xi}\ln R_{k,l}(\xi,\eta)}{x - \xi}\,W_1^0(\sheet{\xi}{k})\,W_1^0(\sheet{\eta}{l}).
\eea
Reminding the definition of the potential:
\beq
\mathcal{V}_k'(x) = \frac{1}{uu_k}\Big(\frac{x^2}{2} - \sum_{j \geq 3} \frac{t_{k,j}}{j}\,\xi^{j - 1}\Big),
\eeq
we retrieve Eqn.~\ref{disqaq}.

The advantage of Tutte's method is that it can easily be adapted in genus $g \geq 1$. In the first situation (when the marked face borders the marked edge of both sides), we obtain either a connected map with one handle less, i.e. of genus $g - 1$, but two marked faces, or two connected components $\mathcal{M}_1'$ and $\mathcal{M}_2'$ with one marked face each and genera summing up to $g$. In the second situation, the topology is not changed. In the third situation, when the strip consisting of the faces crossed by the open path is cut out in $\mathcal{M}''$, we obtain either a connected map of genus $(g - 1)$ with $2$ marked faces, or two connected components $\mathcal{M}_1''$ and $\mathcal{M}_2''$ with one marked face each and whose genera must sum up to $g$. We thus find instead of \eqref{hgy}:
\bea
xW_1^g(\sheet{x}{k}) & = & W_2^{g - 1}(\sheet{x}{k},\sheet{x}{k}) + \sum_{h = 0}^g W_1^h(\sheet{x}{k})\,W_1^{g - h}(\sheet{x}{k}) \nonumber \\
& & + \oint \frac{\dd \xi}{2{\rm i}\pi}\,\frac{\sum_{j \geq 3} t_{k,j}\xi^{j - 1}}{x - \xi}\,W_1^g(\sheet{\xi}{k}) \nonumber \\
& & + \sum_{l = 1}^s \oiint \frac{\dd\xi\,\dd\eta}{(2{\rm i}\pi)^2}\,\frac{\partial_{\xi}\ln R_{k,l}(\xi,\eta)}{x - \xi}\Big(W_2^{g - 1}(\sheet{\xi}{k},\sheet{\eta}{l}) + \sum_{h = 0}^g W_1^h(\sheet{\xi}{k})\,W_1^{g - h}(\sheet{\eta}{l})\Big),
\eea
which is a genus expansion form of \eqref{qywu2} for $n = 1$. Once again, obtaining the equation for any $n \geq 2$ can be done by marking extra faces. This concludes our combinatorial proof of the loop equations.

\subsection{Solution by the topological recursion}

By construction, the generating series of $s$-colored maps with a loop configuration (or equivalently, with tubes) $W_n(\sheet{x_1}{k_1},\ldots,\sheet{x_n}{k_n})$ have a topological expansion in the sense of Definition~\eqref{top1}. In order to check the other points in Hypothesis~\ref{ppp}, we need to address the convergence property of $W_1^0(\sheet{x}{k})$. For this purpose, we can use its representation by substitution in $\widehat{W}_1^0(x)$, and the description of the cut locus of the latter, established in full generality in \cite[Section 6]{BBG2}. To state it, we need:
\begin{definition}
A family of nonnegative numbers $\mathbf{t} = (t_j)_j$ is \emph{admissible} if, for any $\ell \geq 1$, the generating series of planar maps with $1$ marked face of degree $\ell$, equipped with a marked point, and with local vertex weight $u$ and local face weights $t_j$, is finite.
\end{definition}
\begin{lemma}\cite{BBG2}\label{lcut} (One-cut lemma)
If the coefficients of the shifted potential \eqref{shifeq} are admissible, $W_1^0(\sheet{x}{k})$ has a non-zero radius of convergence around $x = \Lambda_k$. Besides, $W_1^0(\sheet{x}{k})$ defines a holomorphic function in $\mathbb{C}\setminus \Gamma_k$, where $\Gamma_k = [a_k,b_k]$ is a segment of the real axis containing $\Lambda_k$. And, $W_1^0(\sheet{x}{k})$ is discontinuous at every interior point of $\Gamma_k$, and $\tilde{\mathcal{V}}_k(x)$ is absolutely convergent at least in an open disk centered at $\Lambda_k$ and containing the interior of $\Gamma_k$. $a_k$ (resp. $b_k$) are strictly increasing functions of $u$, and all other parameters being fixed and $u > 0$, it is a power series in $(\rho_{k,l})_{k,l}$.
\end{lemma}
This shows that, when the weights $t_{k,l}$, $g_{k_1,k_2;\ell_1,\ell_2}$ and $-\rho_{k,l}$ are nonnegative and the generating series we want to compute are not $+\infty$, Hypothesis~\ref{ppp} is verified. We can apply Proposition~\ref{mik}, namely
\beq
\label{leab}\omega_n^{g}(\sheet{z_1}{k_1},\ldots,\sheet{z_n}{k_n}) = W_n^{g}(\sheet{x(z_1)}{k_1},\ldots,\sheet{x(z_n)}{k_n}) + \delta_{n,2}\delta_{g,0}\delta_{k_1,k_2}\,\frac{\dd x(z_1)\,\dd x(z_2)}{\big(x(z_1) - x(z_2)\big)^2}
\eeq
satisfies linear and quadratic loop equations in the sense of \S~\ref{def10}-\ref{def20}. Besides, there exists a critical value $u^* > 0$ so that, for any $u < u^*$, the disk of convergence of $\tilde{\mathcal{V}}_k$ contains $[a_k,b_k]$ itself, while $a_k$ or $b_k$ reach its boundary at $u = u^*$. It is also possible to consider negative weights or even complex weights, but the cut locus might be more complicated.
\begin{definition}
A family of complex numbers $\mathbf{t} = (t_j)_{j \geq 1}$ is \emph{sub-admissible} if $(|t_j|)_{j \geq 1}$ is admissible.
\end{definition}
\begin{lemma}\cite{BBG2} \label{lcut2}(One-cut lemma, weaker version)
If the coefficients of the shifted potential \eqref{shifeq} are sub-admissible, $W_1^0(\sheet{x}{k})$ has a non-zero radius of convergence around $x = \Lambda_k$. At least for $u$ small enough and real-valued weights, the conclusions of Lemma~\ref{lcut} still hold (except for monotonicity of $a_k$ and $b_k)$.
\end{lemma}
Then, we can deduce as in Section~\ref{S3} that $\omega_n^{g}$ satisfies \eqref{poux}. Let us introduce $\mathcal{H}$, the subspace of holomorphic $1$-forms in $\coprod_{k = 1}^s \widehat{\mathbb{C}}\setminus [a_k,b_k]$ satisfying:
\beq
\forall x \in ]a_k,b_k[,\qquad f(\sheet{x}{k} + {\rm i}0) + f(\sheet{x}{k} - {\rm i}0) + \sum_{l = 1}^s \rho_{k,l} \mathcal{O}_{k,l} f(\sheet{x}{l}) = 0,
\eeq
which are formal power series\footnote{It is an example where we can prove normalizability without establishing strict convexity of the two-point interaction. It would have been a condition on the weights concerning loops, i.e. $z_{k,l}$, $g_{k_1,k_2;\ell_1,\ell_2}$ and $\rho_{k,l}$, which is satisfied at least for real valued weights and $\rho_{k,l}$ small enough. However, we expect that the range of parameters for which the two-point interactions are strictly convex to determine the radius of convergence of those formal series.} in $(\rho_{k,l})_{k,l}$. 
\begin{lemma}
$\mathcal{H}$ is normalized in the sense of Definition~\ref{normdef}.
\end{lemma}
\textbf{Proof.} First, we claim that for any $u > 0$, $a_k$ and $b_k$ are formal power series in $\rho_{k,l}$ (it is an easy consequence of the substitution relation \eqref{519} and \S~\ref{comme} below). Let us denote $a_k(\mathbf{0})$ and $b_k(\mathbf{0})$ are the value of $a_k$ and $b_k$ at $\rho_{k,l} \equiv 0$. We introduce $\mathcal{H}(\mathbf{0})$, the space of holomorphic $1$-forms in $\coprod_{k = 1}^s \widehat{\mathbb{C}}\setminus[a_k,b_k]$ satisfying:
\beq
\forall x \in ]a_k(\mathbf{0}),b_k(\mathbf{0})[,\qquad f(x + {\rm i}0) + f(x - {\rm i}0) = 0.
\eeq
We claim that $\mathcal{H}(\mathbf{0})$ is representable by residues and normalized. Indeed, it is associated with the standard two-point interaction $R(x,y) = |x - y|^2$ which is strictly convex, so we can use Proposition~\ref{Berg} and Lemma~\ref{ksq} (see also \S~\ref{comme} below for explicit residue representations). Accordingly, the linear map which associates to $f \in \mathcal{H}$ its specialization at $\rho_{k,l} \equiv 0$ denote $f(\mathbf{0}) \in \mathcal{H}(\mathbf{0})$ is an isomorphism (see \S~\ref{comme} for the recursive determination of the coefficients of the power series from $f(\mathbf{0})$). Thus, $\mathcal{H}$ is also normalized. \hfill $\Box$

\vspace{0.2cm}

So, assuming further that $\ln R_{k,l}$ is analytic in a neighborhood of $[a_k,b_k]\times[a_l,b_l]$ (off-criticality assumption which amounts to Hypothesis~\ref{ppp}-$(iv)$), we can conclude with Proposition~\ref{mik}-Corollary~\ref{aporo} that $\omega_n^{g}$ for $n \geq 1$, $g \geq 0$ and $(n,g) \neq (1,0),(2,0)$ satisfy solvable linear and quadratic loop equations, and can be expressed solely from $\omega_1^0$ and $\omega_2^0$ by the topological recursion:
\bea
\label{toporeaq}\omega_n^g(\sheet{z_0}{k_0},\sheet{z_I}{k_I}) & = & \sum_{\alpha \in \Gamma_{k_0}^{\mathrm{fix}}} \Res_{z \rightarrow \alpha} K_{k_0}(z_0,z) \\
& & \Big(\omega_{n + 1}^{g - 1}(\sheet{z}{k_0},\iota_{k_0}(\sheet{z}{k_0}),\sheet{z_I}{k_I}) 
+ \sum_{J \subseteq I,\,\,0 \leq h \leq g} \omega_{|J| + 1}^{h}(\sheet{z}{k_0},\sheet{z_J}{k_J})\omega_{n - |J|}^{g - h}(\iota_{k_0}(\sheet{z}{k_0}),\sheet{z_{I\setminus J}}{k_{I\setminus J}})\Big), \nonumber
\eea
where the recursion kernel is given by:
\beq
K_{k_0}(z_0,z) = \frac{-\frac{1}{2}\int_{\iota_{k_0}(z)}^{z} \omega_2^0(\sheet{z_0}{k_0},\sheet{\cdot}{k_0})}{\omega_1^0(\sheet{z}{k_0}) - \omega_1^0(\iota_{k_0}(\sheet{z}{k_0}))}.
\eeq
Thus, we have reduced the problem of enumerating $s$-colored maps with a loop configuration in any topologies, to the problem of enumeration of disks ($\omega_1^0$) and cylinders ($\omega_2^0$). Following a previous remark, we observe that the recursion kernel only involves the generating series of cylinders whose marked faces have the same color.

\subsection{Comments on disk and cylinder generating series}
\label{comme}
Explicit formulas for $W_1^0$ and $W_2^0$ in the model of $s$-colored maps with a loop configuration is out of reach for general weights $g_{k,l;i,j}$. Indeed, one has to solve master loop equation\footnote{Computing the discontinuity of \eqref{disqaq}, we see that these equations holds as soon as the interior of the cut locus $\Gamma_k$ (which could be more complicated than a segment in general, see \cite[Eqn. 6.28]{BBG2}) is included in the open disk of convergence of $\mathcal{V}_k$ around $\Lambda_k$. In practice in loop models, one starts by assuming the position of the cut locus is known, then attempt to solve the equation, and eventually derive necessary condition of the weights for such an assumption to be possible.}, for any $k \in \ldbrack 1,s \rdbrack$:
\beq
\label{liners}\forall x \in \mathring{\Gamma}_k,\qquad W_1^0(\sheet{x}{k} + {\rm i}0) + W_1^0(\sheet{x}{k} - {\rm i}0) + \sum_{l = 1}^s \rho_{k,l} \oint_{\Gamma_k} \partial_x\ln R_{k,l}(x,\xi)\,W_1^0(\sheet{\xi}{l}) = \mathcal{V}'_k(x).
\eeq
Let us rewrite slightly differently this equation when faces crossed by loops have bounded degree. In this case, $R_{k,l}$ is a symmetric polynomial in $2$ variables, that we may factorize:
\beq
R_{k,l}(x,\xi) = S_{k,l}(x)\,\prod_{p = 1}^{d_l} (\xi - s_{k,l,p}(x))^{m_{p}}\quad \Longrightarrow\quad \partial_x \ln R_{k,l}(x,\xi) = \frac{S'_{k,l}(x)}{S_{k,l}(x)} - \sum_{p = 1}^{d_l} \frac{s_{k,l,p}'(x)}{\xi - s_{k,l,p}(x)}.
\eeq
The assumption $(iv)$ of Hypothesis~\ref{ppp} that all singularities of $\ln R_{k,l}$ lie away from $\Gamma_k \times \Gamma_l$ amounts to require that $s_{k,l,p}(\Gamma_k) \cap \Gamma_l = \emptyset$ for any $p \in \ldbrack 1,d_l \rdbrack$. Then, we can move the contour integral in \eqref{liners} to pick up residues at $s_j(x)$: for all $x \in \mathring{\Gamma}_k$,
\beq
\label{jueqq}W_1^0(\sheet{x}{k} + {\rm i}0) + W_1^0(\sheet{x}{k} - {\rm i}0) + \sum_{l = 1}^s \sum_{p = 1}^{d_l} \rho_{k,l}m_p\,s_{k,l,p}'(x)\,W_1^0(\sheet{s_{k,l,p}(x)}{l}) = \mathcal{V}'_k(x) + \sum_{l = 1}^s \sum_{p = 1}^{d_l} uu_l\,\rho_{k,l}\,\frac{S_{k,l}'(x)}{S_{k,l}(x)}.
\eeq
It is therefore very natural to work with the differential form $\omega_1^0(\sheet{x}{k}) = W_1^0(\sheet{x}{k})\dd x$ in order to absorb the function $s_{k,l,p}'(x)$: forall $x \in \mathring{\Gamma}_k$,
\beq
\label{linsa}\omega_1^0(\sheet{x}{k} + {\rm i}0) + \omega_1^0(\sheet{x}{k} - {\rm i}0) + \sum_{l = 1}^s \sum_{p = 1}^{d_l} \rho_{k,l}m_{p}\,\omega_1^0(\sheet{s_{k,l,p}(x)}{l}) = \dd\mathcal{V}'_k(x) + \sum_{l = 1}^s  uu_l\,\rho_{k,l}\dd\ln S_{k,l}(x).
\eeq
This equation comes with the condition that $\omega_1^0$ is holomorphic in $\mathbb{C}\setminus\Gamma_k$, with a simple pole of residue $-uu_k$ at $\infty$, and $W_1^0(\sheet{x}{k})$ remains bounded on $\Gamma_k$ (see \cite[Chapitre 6]{BBG2}).

Because of the third term in the left-hand side, this equation is highly non-local, and we cannot hope to solve it in full generality, even assuming that $\Gamma_k = [a_k,b_k]$ is a known segment. We notice that this non-local term only depends on the weights assigned to loops $g_{k,l;i,j}$, $z_{k,l}$ and $\rho_{k,l}$, while the weights for faces not crossed by a loop only appear in the right-hand side. For this reason, solving \eqref{linsa} is equally difficult (or easy) for all values of $t_{k,j}$. Similarly:
\beq
\check{\omega}_2^0(\sheet{x}{k},\sheet{x_2}{k_2}) = W_2^0(\sheet{x}{k},\sheet{x_2}{k_2})\,\dd x_1\dd x_2= \omega_2^0(\sheet{x}{k},\sheet{x_2}{k_2}) - \delta_{k,k_2}\,\frac{\dd x_1\,\dd x_2}{(x - x_2)^2}
\eeq
satisfies an equation of type \eqref{jueqq} with respect to $x$, with right hand side replaced by $$-\delta_{k,k_2}\dd x_1\dd x_2/(x - x_2)^2,$$ and extra conditions that it is holomorphic in $(\widehat{\mathbb{C}}\setminus\Gamma_k)\times(\widehat{\mathbb{C}}\setminus\Gamma_l)$, and it is holomorphic in a neighborhood of the image of $\Gamma_k$ in the Riemann surface $U_k$ as defined in Section~\ref{S1}. On the contrary, the nature of the solution will depend much on the "group" generated by the $s_{k,l,p}$, and thus is a non trivial way on the parameters $z_{k,l}$ and $g_{k,l;i,j}$.

Actually, it is enough to find $\omega_2^0$, since it gives access to a local Cauchy kernel (see Definition~\ref{Cau} and \S~\ref{38})
\beq
\label{taq}G(\sheet{x}{k},\sheet{x_2}{k_2}) = - \int^{x} \omega_2^0(\sheet{\cdot}{k},\sheet{x_2}{k_2}).
\eeq
We can use its properties to show that:
\beq
\label{tildeaq}\tilde{\omega}_1^0(\sheet{x}{k}) = \frac{1}{2}\dd\mathcal{V}_k(x) - \frac{1}{2}\,\frac{1}{2{\rm i}\pi}\oint_{\Gamma_k} G(\sheet{\cdot}{k},\sheet{x}{k})\,\dd\mathcal{V}_k(\xi)
\eeq
satisfies the same equation and extra conditions as $\omega_1^0$. If the solution of such constraints is unique (this is the case when the two-point interaction defined by $(R_{k,l})_{k,l}$ is strictly convex), we conclude that $\omega_1^0(\sheet{x}{k})$ is given by the expression \eqref{tildeaq}. For instance, when faces which are not crossed by loops have bounded degree, $\mathcal{V}_k$ is a polynomial, thus the contour integral can be moved to pick up a residue at $\infty$.

At present, the solution of such equations is known in very few cases assuming $\Gamma_k$ is known. When there is only one color ($s = 1$), and $R(x,\xi) = (1/z + g_{1,0}(x + \xi))$, this is the $O(-\rho)$ model where loops live only on triangles initially considered by Gaudin and Kostov \cite{GK}. Eqn. \eqref{linsa} with general right hand side has been solved in \cite{EKOn,EKOn2,BEOn}. These techniques were adapted in \cite{BBG1,BBG2} to solve the model where $R$ is an homography (which must be involutive by symmetry), which describes a model where loops cross only triangles, but an extra weight is introduced to take into account curvature of the loops. They were extended in a straightforward way \cite{BBG3} to $s = 2$ colors and $R_{1,1} = R_{2,2} = 0$, while $R_{1,2}$ is an homography, which describe a $O(-\rho)$ model where loops cross only triangles, separate domains of different colors, and also receive a weight taking into account curvature. The special case where all faces are crossed by loops (the fully-packed model) was shown to include the Potts model on general random maps, i.e. a model of maps with faces of arbitrary degree but all weighted $1$, whose vertices are equipped with $Q = \rho^2$ Potts variables. In all those cases, $\Gamma_k = [a_k,b_k]$ are determined by implicit equations in terms of the weights, which are sometimes amenable to an explicit solution.

\subsection{Computing explicitly the coefficients of the generating series}
\label{64}
Although computing $\omega_1^0(\sheet{x}{k})$ and $\omega_2^0(\sheet{x}{k},\sheet{x_2}{k_2})$ explicitly is in general out of reach, we recall that they were defined as formal series in some parameters. In combinatorics, one is interested in numbers of $s$-colored maps with a loop configuration, with a given number of vertices, a given number of faces of each type, a given number of loops, etc. We explain below that these numbers can be determined effectively for any topology. We illustrate the method by considering the generating series $\omega_n^{g}(\mathbf{m}\,;\,\sheet{x_I}{k_I})$ of $\mathcal{M} \in s\mathsf{ML}$ with a given number of loops $\mathbf{m} = (m_{k,l})_{k,l}$ separating domains of color $k$ and $l$. By construction:
\beq
\omega_n^g(\sheet{x_I}{k_I}) = \sum_{\mathbf{m} \in \mathbb{N}^{s^2}} \prod_{1 \leq k \leq l \leq s} (-\rho_{k,l})^{m_{k,l}}\,\omega_n^{g}(\mathbf{m}\,;\,\sheet{x_I}{k_I}).
\eeq
The $\mathbf{m} = \mathbf{0}$ term corresponds to generating series of maps without loops, i.e. it vanishes if two colors $k_i$ and $k_j$ are different, and in terms of \eqref{jue}:
\beq
\omega_n^g(\mathbf{0}\,;\,\sheet{x_1}{k},\ldots,\sheet{x_n}{k}) = \widehat{W}_n^{g}[\mathcal{V}_k](\sheet{x_1}{k},\ldots,\sheet{x_n}{k})\dd x_1\cdots\dd x_n+ \delta_{n,2}\delta_{g,0}\,\frac{\dd x_1\dd x_2}{(x_1 - x_2)^2}.
\eeq
If we want to compute $\omega_n^g(\mathbf{m},\sheet{x_I}{k_I})$ for a given $\mathbf{m}$, thanks to the topological recursion \eqref{toporeaq} and the remark made in \eqref{tildeaq}, we just need to compute $\omega_2^0(\mathbf{m}'\,;\,\sheet{x_1}{k_1},\sheet{x_2}{k_2})$ for $\mathbf{m} = (m'_{k,l})_{k,l}$ with $m_{k,l}' \leq m_{k,l}$. This can be done recursively.

For the initialization, it is a classical result \cite{dFZJ} that:
\bea
\omega_1^0(\mathbf{0}\,;\,\sheet{x}{k}) & = & \frac{1}{2}\,\mathcal{V}_k'(x) - \frac{1}{2}\oint \frac{\dd \xi}{2{\rm i}\pi}\,\frac{\mathcal{V}'_k(x) - \mathcal{V}'_k(\xi)}{x - \xi}\,\frac{\sqrt{(x - a_k(\mathbf{0}))(x - b_k(\mathbf{0}))}}{\sqrt{(\xi - a_k(\mathbf{0}))(\xi - b_k(\mathbf{0}))}}, \nonumber \\
\omega_2^0(\mathbf{0}\,;\,\sheet{x}{k},\sheet{x_2}{k_2}) & = & \delta_{k,k_2}\,\frac{\dd x\,\dd x_2}{2(x - x_2)^2}\,\frac{(x - \Lambda_{k})(x_2 - \Lambda_{k}) - (x + x_2 - 2\Lambda_k)\frac{a_k(\mathbf{0}) + b_k(\mathbf{0})}{2} + a_k(\mathbf{0})b_k(\mathbf{0})}{\sqrt{(x - a_k(\mathbf{0}))(x - b_k(\mathbf{0}))(x_2 - a_k(\mathbf{0}))(x_2 - b_k(\mathbf{0}))}}. \nonumber
\eea
$\mathbf{a}_k(\mathbf{0})$ and $b_{k}(\mathbf{0})$ are order $0$ terms in $a_k$ and $b_k$ considered as a power series in $(\rho_{k',l})_{k',l}$, and are determined by the equations \cite[Section 6]{BBG2}:
\bea
\frac{a_k(\mathbf{0}) + b_k(\mathbf{0})}{2} & = & \frac{1}{2{\rm i}\pi} \oint \frac{\dd\xi}{2{\rm i}\pi} \frac{\sum_{j \geq 1} t_{k,j}(\xi - \Lambda_k)^{j - 1}}{\sqrt{(\xi - a_k(\mathbf{0}))(\xi - b_k(\mathbf{0}))}}, \nonumber \\
\frac{a_k(\mathbf{0})^2 + b_k(\mathbf{0})^2 + 3a_k(\mathbf{0})b_k(\mathbf{0})}{8} & = & 2uu_k + \frac{1}{2{\rm i}\pi}\oint \frac{\xi \dd\xi}{2{\rm i}\pi}\,\frac{\sum_{j \geq 1} t_{k,j}(\xi - \Lambda_k)^{j - 1}}{\sqrt{(\xi - a_k(\mathbf{0}))(\xi - b_k(\mathbf{0}))}}.
\eea
where the contour integral surrounds $\Gamma_k(\mathbf{0}) = [a_k(\mathbf{0}),b_k(\mathbf{0})]$ and lies in the domain of analyticity of $\mathcal{V}_k(x)$. Those expressions can be obtained by solving directly \eqref{linsa} for $\rho_{k,l} \rightarrow 0$. If we introduce a uniformization variable $\zeta$ so that:
\beq
x = \frac{a_k(\mathbf{0}) + b_k(\mathbf{0})}{2} + \frac{b_k(\mathbf{0}) - a_k(\mathbf{0})}{4}\Big(\zeta + \frac{1}{\zeta}\Big).
\eeq
We actually have:
\beq
\omega_2^0(\mathbf{0}\,;\,\sheet{x(\zeta)}{k},\sheet{x(\zeta_2)}{k_2}) = \delta_{k,k_2}\,\frac{\dd\zeta_1\dd\zeta_2}{(\zeta_1 - \zeta_2)^2},
\eeq
and it is more convenient to use such a variable for further computations. Then, the generating series of cylinders with finite number of loops can be obtained by solving recursively:
\bea
\omega_{2}^0(\mathbf{m}\,;\,\sheet{x}{k},\sheet{x_2}{k_2})  & = & \delta_{k,k_2}\delta_{\mathbf{m},\mathbf{0}}\,\frac{\dd\zeta\,\dd\zeta_2}{(\zeta - \zeta_2)^2} \\
\nonumber &  & + \sum_{l_1,l_2 = 1}^s \!\!\!\!\!\sum_{\substack{0 \leq p_{k,l} \leq  m_{k,l} \\ \prod_{k,l} p_{k,l}(m_{k,l} - p_{k,l}) \neq 0}} \!\!\!\!\!\!\!\!\!\!\!\oint_{\Gamma_{l_1}(\mathbf{0}) \times \Gamma_{l_2}(\mathbf{0})} \!\!\!\!\!\!\!\!\!\omega_2^{0}(\mathbf{p}\,;\,\sheet{x}{k},\sheet{\xi_1}{l_1})\,\ln R_{l_1,l_2}(\xi_1,\xi_2)\,\omega_2^0(\mathbf{m} - \mathbf{p}\,;\,\sheet{\xi_2}{l_2},\sheet{x_2}{k_2}).
\eea
Similarly, \eqref{tildeaq} leads to:
\bea
\omega_1^0(\mathbf{m}\,;\,\sheet{x}{k}) & = & \omega_1^0(\mathbf{0}\,;\,\sheet{x}{k}) \\
\nonumber & & + \sum_{l_1,l_2 = 1}^s \!\!\!\!\!\sum_{\substack{0 \leq p_{k,l} \leq m_{k,l} \\ \prod_{k,l} p_{k,l}(m_{k,l} - p_{k,l}) \neq 0}} \!\!\!\!\!\!\!\!\!\!\!\oint_{\Gamma_{l_1}(\mathbf{0}) \times \Gamma_{l_2}(\mathbf{0})} \!\!\!\!\!\!\!\!\!\omega_2^{0}(\mathbf{p}\,;\,\sheet{x}{k},\sheet{\xi_1}{l_1})\,\ln R_{l_1,l_2}(\xi_1,\xi_2)\,\omega_1^0(\mathbf{m} - \mathbf{p}\,;\,\sheet{\xi_2}{l_2}).
\eea
Then, one can plug the series $\omega_1^0(\sheet{x}{k})$ and $\omega_2^0(\sheet{x}{k},\sheet{x_2}{k_2})$ truncated up to $o\big(\prod_{k,l} (-\rho_{k,l})^{m_{k,l}}\big)$ in the topological recursion formula \eqref{toporeaq} to obtain $\omega_n^{g}(\sheet{x_I}{k_I})$ up to the same order, recursively on $2g + n$.

In this example, we focused in the number of loops inside the enumerated maps. One could have chosen some other parameters, like $u$ (coupled to the number of vertices), or $u_k$ (coupled to the number of vertices of a specific color). One would find a similar recursive procedure to compute truncated versions of $\om_{1}^0$ and $\om_{2}^0$ generating $s$-colored maps with a loop configuration and a bounded number of vertices. Plugging these expressions in the topological recursion formula then gives rise to the generating functions of $s$-colored maps with a loop configuration of arbitrary topology and a number of vertices bounded by the order of approximation chosen.

In some other contexts, such as topological strings or Gromov-Witten theory, where one enumerates embeddings of surfaces into a target space, this procedure corresponds to fixing a bound on the degree of the embedding map or on the homology class of the embedded surface (see \cite{EOBKMP} for example where the same kind of induction procedure is performed). Indeed, in many applications for physics, one is interested in the coefficients of perturbative expansions, i.e. expansion in a small parameter which is often coupled to a geometric property, and the procedure we described is very efficient for such computations.

\subsection{Critical points and asymptotics of large maps}
\label{66}
When $u$ increases, Hypothesis~\ref{bus} may fail. In particular, if a zero of $R_{k,l}$ approaches $\Gamma_k\times\Gamma_l$ when the parameter of the model vary, we will reach at the limit a critical point which is characteristic of a loop model, i.e. cannot be reached in a model of usual random maps with bounded degree. Let us study qualitatively an example. First, let us focus on the master equation \eqref{jueqq} for $W_1^0$. We know that $W_1^0(\sheet{x}{k})$ is not analytic at $x = b_k$. When the model is offcritical, the last term in the left-hand side of \eqref{jueqq} is holomorphic in a neighborhood of $a_k$, while the right hand side is regular: the singular part near $x = a_k$ must satisfy
\beq
[W_1^0(\sheet{x}{k} + {\rm i}0)]_{\mathrm{sing}} + [W_1^0(\sheet{x}{k} - {\rm i}0)]_{\mathrm{sing}} = 0
\eeq
hence is of squareroot type, i.e. $W_1^0(\sheet{x}{k})\,\propto\,f(\sqrt{x - b_k})$ where $f$ is a holomorphic function in the vicinity of $0$. Coming back to the example considered in \S~\ref{comme}, assume that the parameters of the model are tuned so that we reach a critical point for which $s_{k,l,a}(\partial \Gamma_l)$ touches $\partial \Gamma_k$. To fix ideas, we assume that $s_{k,l,a}(b_k) = b_l$ for some triplets $k,l,a$. Then, the last term in \eqref{jueqq} is singular when $x = b_k$, and its singularity behaves like $[W_1^0(s_{k,l,a}(x))]_{\mathrm{sing}}$ when $x \rightarrow b_k$, and depends only on the local behavior of $s_{k,l,a}(x)$ when $x \rightarrow b_k$. Since $s_{k,l,a}(x)$ are roots of a polynomial, they are either regular or have an algebraic singularity at $b_k$.

When they are regular near $b_k$, we have $(s_{k,l,a}(x) - b_l) \propto (x - b_k)^{\mu_{k,l,a}}$ for some nonnegative integer $\mu_{k,l,a}$. Then, it is natural to make the ansatz of a power law singularity for $W_1^0$, namely we look for a local behavior:
\beq
\label{ssun2}[W_1^0(\sheet{x}{k})]_{\mathrm{sing}}\,\sim\,C_k\,(x - b_k)^{\nu_k}.
\eeq
Pluging \eqref{ssun2} into the master equation \eqref{jueqq} and identifying the leading singular part when $x \rightarrow b_k$, we find necessary conditions relating $\mu_{k,l,a}$ and $\nu_k$. For instance, let us consider the case of a single color ($s = 1$). If there is only $1$ element $a_0$ such that $s_{a}(\partial \Gamma) \cap \partial \Gamma \neq \emptyset$, we find that, if $\nu$ is not an half-integer, we must have:
\beq
\label{opro}\mu = 1,\qquad e^{2{\rm i}\pi \nu} + 1 + \rho\,m_{a_0}\,e^{{\rm }i\pi\nu} = 0.
\eeq
This gives the well-known parametrization of the critical exponent of the $O(-\rho)$ model:
\beq
\rho\,m_{a_0} = -2\cos\pi \nu.
\eeq
It corresponds to the case where $R(x_1,x_2)$ behaves locally near $x_1,x_2 \rightarrow b$ like $(x_1 + x_2 - 2b)^{m_{a_0}}$.

Thus, although it remains a difficult problem to solve \eqref{linsa} exactly, even at the critical point, it is always possible to perform case by case a local analysis on the singularities to deduce the values of the critical exponents $\nu_k$. By transfer theorems \cite{Flaj}, a behavior like \eqref{ssun2} implies that $T_{k,\ell}$, the number of planar $s$-colored maps of color $k$ with a loop configuration and a marked face of length $\ell$ is asymptotic to:
\beq
T_{k,\ell} \mathop{\sim}_{\ell \rightarrow \infty} \frac{C_k}{\Gamma(-\nu_k)}\,\frac{b_k^{\nu_k + \ell}}{\ell^{1 + \nu_k}}.
\eeq

Another interesting question is to find the asymptotics of the number of genus $g$ maps with $v$ vertices, i.e. of the coefficients of $F^{g}$ seen as a power series in $u$.  Such a singularity $u = u^*$ can only be reached when approaching a critical point as above. The spectral curves parametrized by $u$ becomes singular when $u \rightarrow u^*$, so the $W_k^{g}$ have a singularity as a function of $u$ at $u = u^*$, that we can describe thanks to Proposition~\ref{singp}. This in turns gives access to the asymptotic of maps with large number of vertices by transfer theorems \cite{Flaj}:
\begin{proposition}
Assume the existence of exponents $\alpha,\alpha' > 0$ such that:
\beq
[W_1^0(\sheet{x}{k})]_{\mathrm{sing}} \sim (u^* - u)^{\alpha}\,y^*(\sheet{x^*}{k}),\qquad \sheet{x^*}{k} = \frac{x - b_k}{b_k^* - b_k},\qquad |b_k^* - b_k|\,\propto\,(u_* - u)^{\alpha'},
\eeq
and:
\beq
W_2^0(\sheet{x_1}{k_1},\sheet{x_2}{k_2})\dd x_1\dd x_2 \sim (W_2^0)^*(\sheet{x_1^*}{k_1},\sheet{x_2^*}{k_2})\,\dd{}x^*_1\dd{}x^*_2.
\eeq
Then, for any $n,g$ such that $2g - 2 + n > 0$, we have:
\beq
W_n^g(\sheet{x_1}{k_1},\ldots,\sheet{x_n}{k_n}) \mathop{\sim}_{u \rightarrow u^*} (u^* - u)^{(\alpha + \alpha')(2 - 2g - n) - \alpha' n}\,(W_n^g)^*(\sheet{x_1^*}{k_1},\ldots,\sheet{x_n^*}{k_n}),
\eeq
where $(W_n^g)^*$ is computed by \eqref{leab} from $(\omega_n^g)^*$ obtained by applying the topological recursion formula \eqref{toporeaq} to the initial data $\omega_1^0(\sheet{x}{k}) = y^*(\sheet{x^*}{k})$ and:
\beq
(\omega_2^0)^*(\sheet{x_1^*}{k_1},\sheet{x_2^*}{k_2}) = \Big(W_2^0(\sheet{x_1^*}{k_1},\sheet{x_2^*}{k_2})\dd{}x_1^*\dd{}x_2^* + \frac{\delta_{k_1,k_2}}{(x^*_1 - x^*_2)^2}\Big)\dd{}x_1^*\dd{}x_2^*.
\eeq
Hence, the number of $s$-colored maps with a loop configuration, of genus $g$ with $v$ vertices, behaves for $g \geq 2$ as:
\beq
(F^{g})_v \mathop{\sim}_{v \rightarrow \infty} \frac{(F^g)^*}{\Gamma((\alpha + \alpha')(2g - 2))}\,v^{(\alpha + \alpha')(2g - 2) - 1}\,(u^*)^{(\alpha + \alpha')(2 - 2g) + v}.
\eeq
\end{proposition}
The blow-up $y(\sheet{x^*}{k})$, $F_g^*$ and the exponents $\alpha,\alpha'$ are universal, while $u^*$ depends on the model. This method has to be applied case by case, and in general the exponents $\alpha,\alpha'$ describing the approach of the critical point, are related to the exponent describing the behavior of the model at criticality (see for an illustration the discussion in \cite[Section 3]{BBG1}). For instance, consistency implies that, if $y(\sheet{x^*}{k}) \propto (x^*)^{\kappa}$ when $x^* \rightarrow \infty$, then:
\beq
\alpha'\kappa = \alpha.
\eeq

\subsection{Another combinatorial model: height model}
\label{SQADE}

In this paragraph, we apply the previous techniques to the study of the heights model introduced in \cite{Kostov:1989eg,Kostov:1991cg,KADE}, and show that it is solved by the topological recursion. Let $\mathfrak{G}$ be a finite graph with multiple edges, and $\mathbf{A} = (A_{v,v'})$ its adjacency matrix, i.e. $A_{v,v'}$ is the number of (unoriented) edges between two nodes $v,v'$ of $\mathfrak{G}$. We shall see in Lemma~\ref{uj} that it is meaningful to restrict oneself to $\mathfrak{G}$ being a Dynkin diagram of ADET type, or an extended Dynkin diagram of $\hat{\mathrm{A}}\hat{\mathrm{D}}\hat{\mathrm{E}}$ type (see Figure~\ref{ADEt}).

\subsubsection{Maps with a $\mathfrak{G}$-height configuration}

\begin{figure}[h]
\begin{center}
\begin{minipage}[c]{0.45\linewidth}
\includegraphics[width=1\textwidth]{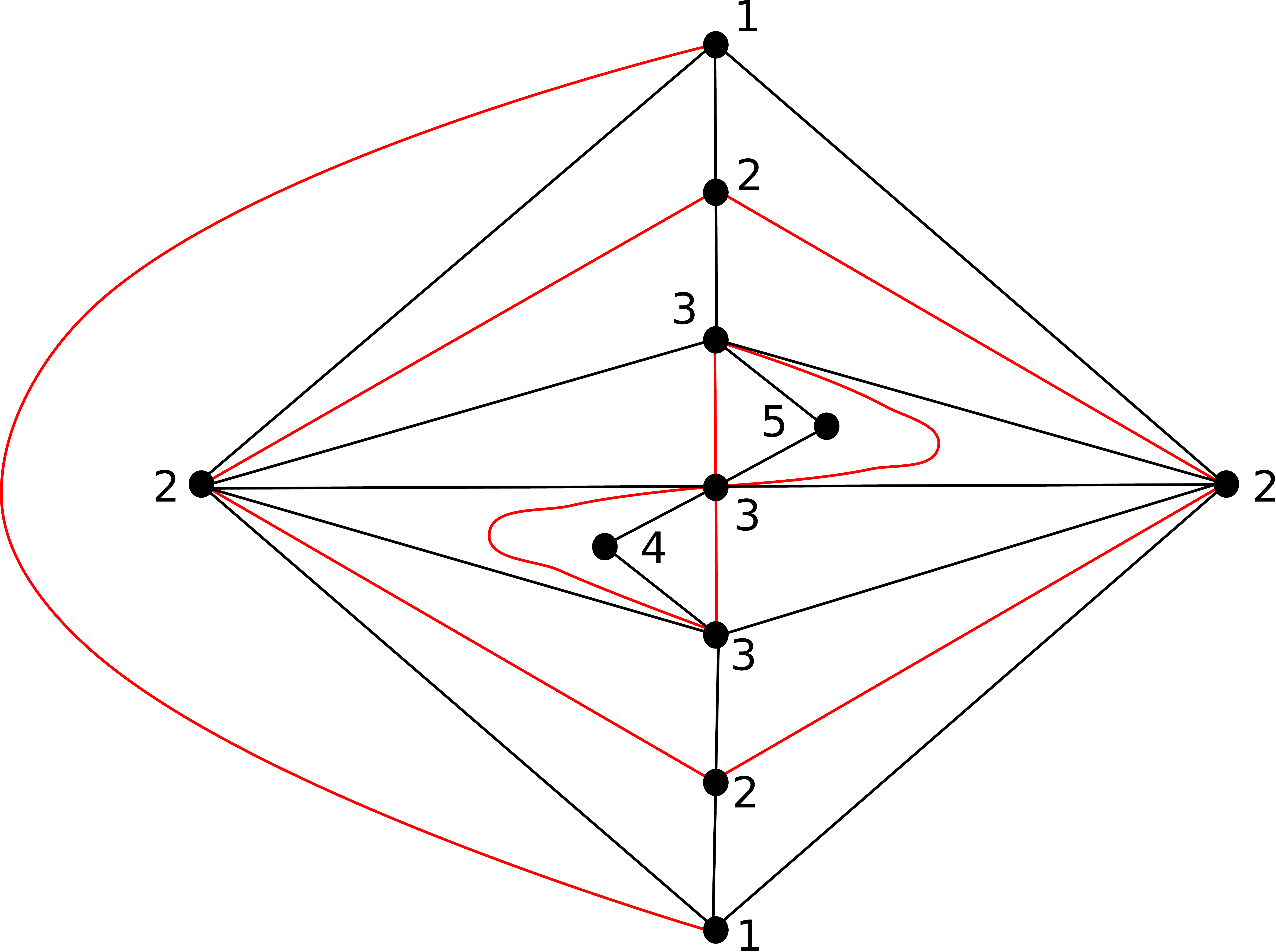}
\end{minipage} \hfill \begin{minipage}[c]{0.45\linewidth}
\includegraphics[width=1\textwidth]{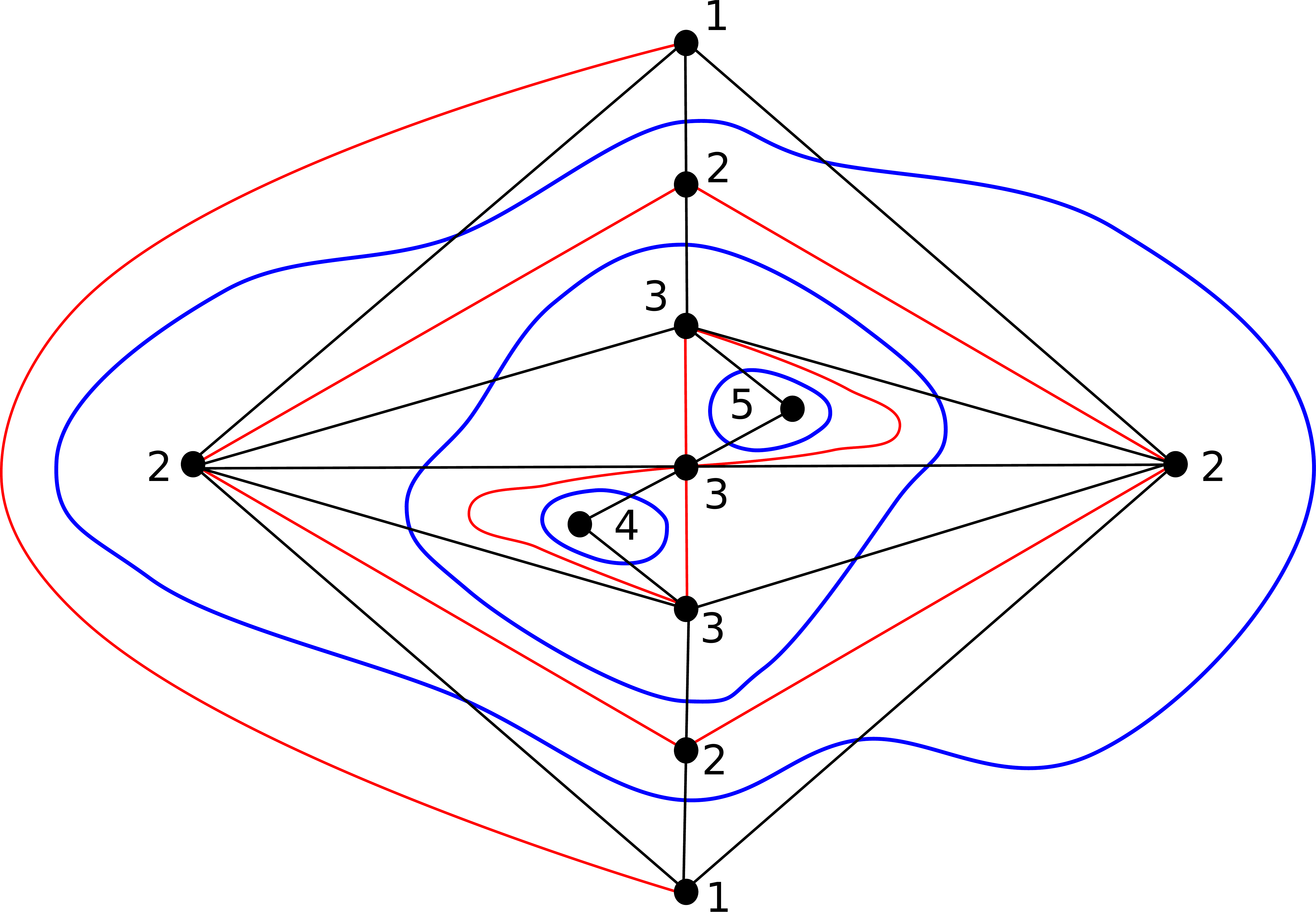}
\end{minipage}
\caption{\label{Dmap1} An example of map with a $\mathfrak{G} = D_5$-height configuration, and its loop representation.}
\end{center}
\end{figure}

If $\mathcal{M}$ is a map (in the usual sense) with faces of degree $4$ only (quadrangles), and if we denote $\mathcal{G}$ its underlying graph, a \emph{$\mathfrak{G}$-height configuration} is a map $\sigma\,:\,\mathcal{G} \rightarrow \mathfrak{G}$ (this means that it associates vertices to vertices, and edges to edges respecting the incidence relations) such that, the boundary of any face contains exactly two non-consecutive vertices with same height (see Fig.~\ref{Dmap1}). We associate to such a configuration the weight:
\beq
w(\mathcal{M},\sigma) = \prod_{v \in V(\mathcal{M})} uu_{\sigma(v)} \prod_{\substack{[v_1,v_2,v_3,v_4] \\ \mathrm{face}\,\,\mathrm{of}\,\,\mathcal{M}}} \Big(\frac{\delta_{\sigma(v_1),\sigma(v_3)}}{uu_{\sigma(v_1)}}+ \frac{\delta_{\sigma(v_2),\sigma(v_4)}}{uu_{\sigma(v_2)}}\Big).
\eeq
We collect in the set $\mathsf{M\mathfrak{G}H}^{g}_n(\mathbf{k};\mathbf{\ell})$ the connected maps of genus $g$, with $n$ marked faces consisting of vertices of the same height $\mathbf{k} = (k_1,\ldots,k_n)$ and of degree $\ell = (\ell_1,\ldots,\ell_n)$. We introduce the generating series:
\beq
F^{g} = \sum_{(\mathcal{M},\sigma) \in \mathsf{M\mathfrak{G}H}_0^g} \frac{w(\mathcal{M},\sigma)}{|\mathrm{Aut}\,\mathcal{M}|},
\eeq
and for any $n \geq 1$:
\beq
\label{genus}W_n^{g}(\sheet{x_1}{k_1},\ldots,\sheet{x_n}{k_n}) = \frac{\delta_{n,1}\delta_{g,0}\,uu_{k_1}}{x_1} + \sum_{\ell_1,\ldots,\ell_n \geq 1} \sum_{(\mathcal{M},\sigma) \in \mathsf{M}\mathfrak{G}\mathsf{H}_{g}^n(\mathbf{k},\mathbf{\ell})} \frac{w(\mathcal{M},\sigma)}{|\mathrm{Aut}\,(\mathcal{M},\sigma)|}\Big[\prod_{i = 1}^n \frac{1}{x_i^{\ell_i + 1}}\Big].
\eeq

\begin{figure}[h]
\begin{center}
\includegraphics[width=0.5\textwidth]{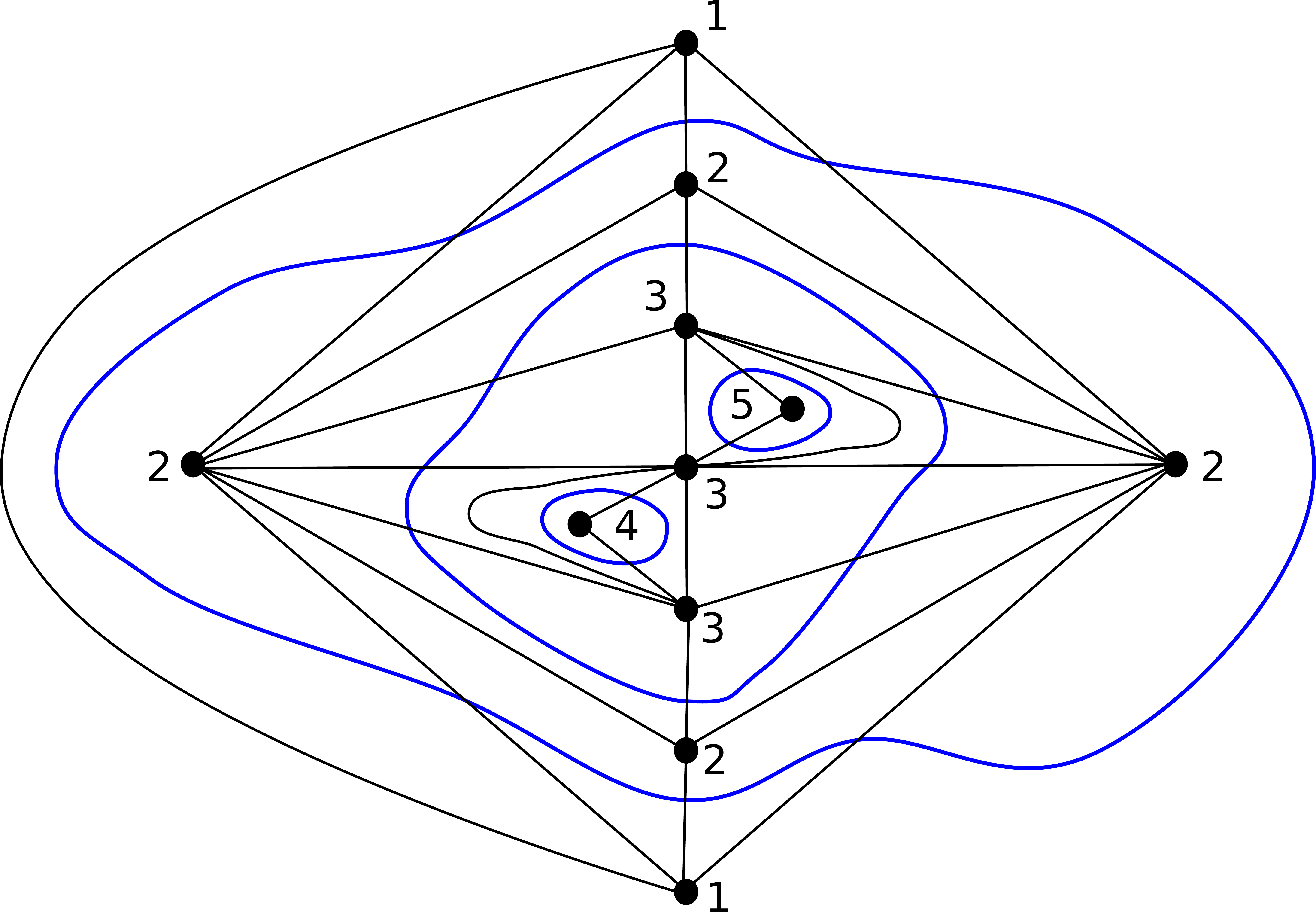}
\caption{\label{Dmap3} An example of map with a $\mathfrak{G} = D_5$-height configuration in loop representation.}
\end{center}
\end{figure}

Equivalently, we can split the degree $4$ faces into degree $3$ faces (triangles), by drawing a red diagonal between the two vertices whose heights are constrained to match (see Fig.~\ref{Dmap1} and \ref{Dmap3}). One can draw a path crossing the edges of the triangles which are not red, so that we obtain a colored map $\tilde{\mathcal{M}}$ with a loop configuration, where all faces are triangles and are crossed by a loop. One can go out of the fully-packed case by allowing faces of degree $j \geq 3$ whose boundaries consist of vertices of a given height $k$, with weight $t_{k,j}$ each. Hence, we retrieve a special case of the weight introduced in \S~\ref{612}, with interactions between colors prescribed by the adjacency matrix of $\mathfrak{G}$:
\beq
\label{fdijiji}\rho_{k,l} = -A_{k,l},\qquad R_{k,l}(x,y) = x + y,
\eeq
and where $g_{k,1,0} = g_{k,0,1} = \sqrt{t}$. 

\subsubsection{Matrix model representation}

In \cite{Kostov:1991cg}, a formal matrix model representation was proposed for the height model. It is based on the following measure:
\beq
\dd\varpi\,\propto\,\prod_{k = 1}^{s} \dd M_{k} \exp\big(-N\,\Tr \mathcal{V}_k(M_k)\big) \prod_{<k,l> \in \mathfrak{G}} \dd B_{k,l}\dd B_{k,l}^{\dagger} \exp\Big[- N\,\Tr\big(B_{k,l}B_{k,l}^{\dagger} M_k + B_{k,l}^{\dagger}B_{k,l} M_l\big)\Big]. 
\eeq
$M_k$ are hermitian matrices of size $N_k \times N_k$, and for each (unoriented) edge $<k,l>$ of $\mathfrak{G}$, $B_{k,l}$ are complex matrices of size $N \times N$, and:
\beq
N_k = Nu_k,\qquad V_k(x) = \frac{1}{uu_k}\Big(\frac{(x + t/2)^2}{2} - \sum_{j \geq 3} t_{k,j}\frac{(x + t/2)^j}{j}\Big).
\eeq
Integrating out the matrices $B_{k,l}$ and $B_{k,l}^{\dagger}$ who have Gaussian distribution, yields a measure:
\bea
\dd\varpi & \propto & \prod_{k = 1}^s \dd M_k \exp\big(-N\,\Tr \mathcal{V}_k(M_k)\big) \nonumber \\
& & \times \prod_{1 \leq k,l \leq s} \exp\Big[-N\,\frac{A_{k,l}}{2}\,\Tr \ln(M_k \otimes \mathbf{1}_N + \mathbf{1}_N\otimes M_l)\Big],
\eea
where we recognize \eqref{pouq}. The correlators of this formal matrix model give the generating series \eqref{genus}:
\beq
\Big\langle \prod_{i = 1}^n \Tr\,\frac{1}{x_i + t/2 - M_{k_i}} \Big\rangle_c = \sum_{g \geq 0} \Big(\frac{N}{u}\Big)^{2 - 2g - k}\,W_n^{g}(\sheet{x_1}{k_1},\ldots,\sheet{x_n}{k_n}).
\eeq
These ADE matrix models were first introduced in \cite{MMa}, and later appeared in $\mathcal{N} = 2$ supersymmetric gauge theories associated to ADE quivers \cite{DiVa} (see also \cite{Kle2} for the $A_2$ quiver matrix model, which is closely related to the $O(-2)$ model). The Schwinger-Dyson equations for such models have been previously written within the CFT formalism in \cite{Kle}.

\subsubsection{Strict convexity of the interactions}

\begin{figure}[h]
\includegraphics[width=\textwidth]{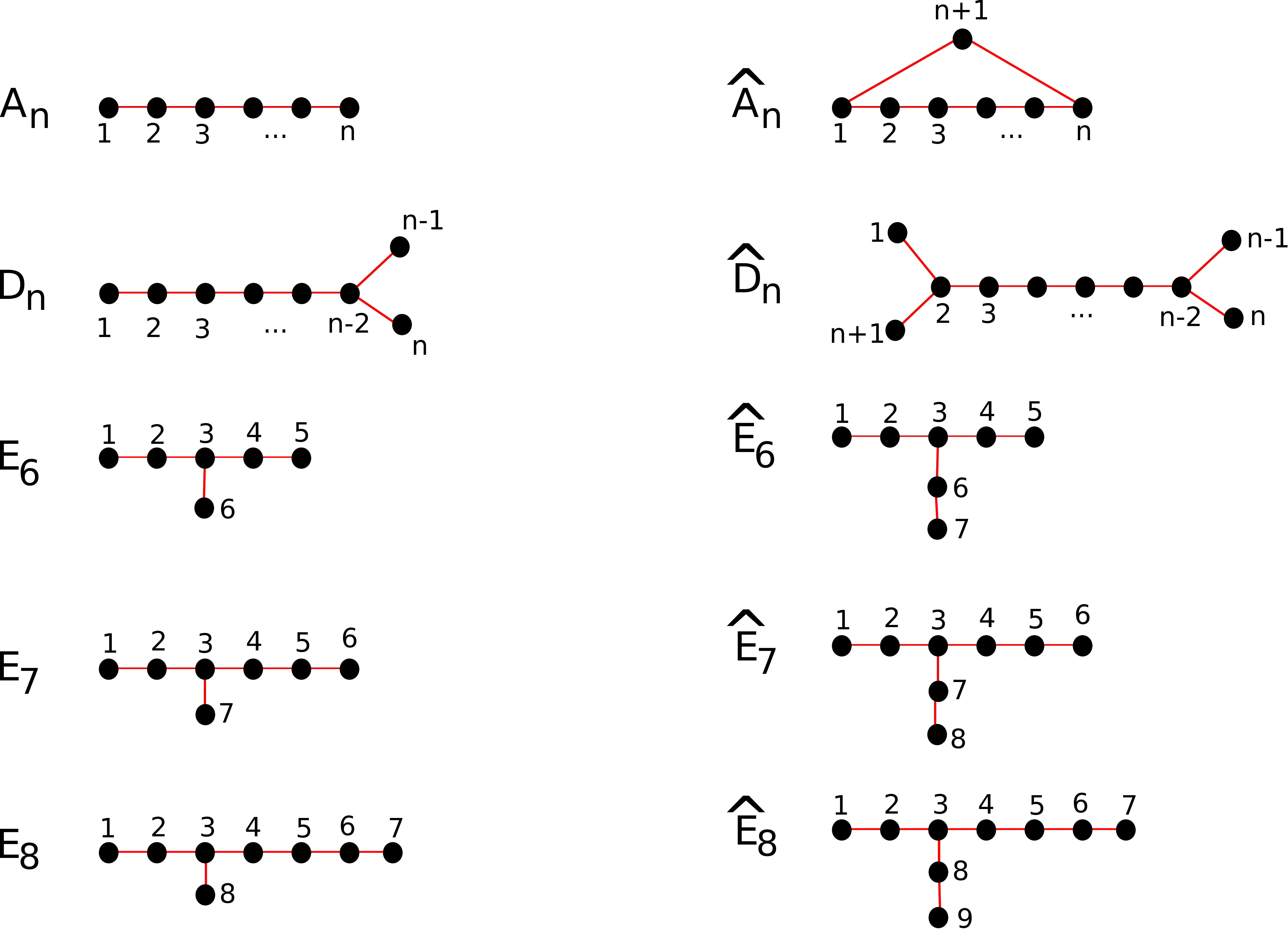}
\caption{\label{ADEt}On the left, Dynkin diagram with positive Cartan matrix. On the right, Dynkin diagram with nonnegative Cartan matrix, having one $0$ eigenvalue, and one has to add $T_{n}$, which is a cycle with $n$ vertices corresponding to the $A_{2n}/\mathbb{Z}_2$.}
\end{figure}

\begin{lemma}
\label{uj}The two-point interactions defined by \eqref{fdijiji} are strictly convex iff $\mathfrak{G}$ is a Dynkin diagram of ADET or an extended Dynkin diagram of $\hat{A}\hat{D}\hat{E}$ type (see Fig.~\ref{ADEt}).
\end{lemma}

\noindent\textbf{Proof.} We need to understand under which conditions, for any signed measures $(\nu_k)_k$ supported on $\mathbb{R}_+$, so that $\nu_k(\mathbb{R}_+) = 0$, we have
\beq
\label{sianq}\tilde{\mathcal{E}}(\nu) \equiv \iint_{\mathbb{R}^2}\Big( \sum_{k = 1}^s \dd\nu_k(x)\dd\nu_k(y)\ln |x - y| - \sum_{k,l = 1}^s A_{k,l}\,\dd\nu_{k}(x)\dd\nu_l(y) \ln|x + y|\Big) \leq 0,
\eeq
with equality iff $\nu_k = 0$ for any $k \in \ldbrack 1,s \rdbrack$. We use the representation:
\beq
\ln|x| = \lim_{\epsilon \rightarrow 0}\Big(\ln \epsilon - \mathrm{Re}\,\int_{0}^{\infty} \dd u\,e^{-\epsilon u}\,\frac{e^{{\rm i}u x} - 1}{u}\Big),
\eeq
and the fact that $\nu_k$ has total mass $0$ to rewrite:
\beq
\tilde{\mathcal{E}}(\nu) = -\frac{1}{2}\lim_{\epsilon \rightarrow 0} \int_{0}^{\infty} \dd u\,\frac{e^{-\epsilon u}}{u}\Big( \sum_{k,l = 1}^s (\mathbf{2} - \mathbf{A})_{k,l}\,\mathrm{Re}\,\hat{\nu}_k(u)\,\mathrm{Re}\,\hat{\nu}_l(u) + (\mathbf{2} + \mathbf{A})_{k,l}\,\mathrm{Im}\,\hat{\nu}_k(u)\,\mathrm{Im}\,\hat{\nu}_l(u)\Big),
\eeq
where $\hat{\nu}$ denotes the Fourier transform of the measure $\nu$.
Therefore, $\mathcal{E}(\nu) \leq 0$ iff $\mathbf{2} - \mathbf{A}$ and $(\mathbf{2} + \mathbf{A})$ are positive. Besides, due to the fact that a signed measure $\nu$ supported on $\mathbb{R}_+$ vanish iff $\mathrm{Im}\,\hat{\nu} \equiv 0$, we have $\tilde{\mathcal{E}}(\nu) = 0$ only for $(\nu_k)_k \equiv 0$ iff $\mathbf{2} + \mathbf{A}$ is positive definite. Notice that $\mathbf{C} = \mathbf{2} - \mathbf{A}$ is the Cartan matrix of the graph $\mathfrak{G}$. It is well-known that a finite graph with multiple edges has a positive definite Cartan matrix iff it is the Dynkin diagram of ADE type, and if we allow the Cartan matrix to be nonnegative, it can also be an extended Dynkin diagram of $\hat{A}\hat{D}\hat{E}$ type. This justifies a posteriori to restrict the study of height model based on such Dynkin diagrams $\mathfrak{G}$. In all those cases, the eigenvalues of $\mathbf{A}$ are of the form $2\cos(\pi m_{k}/h^{\vee})$, where $h^{\vee}$ is the Coxeter number, and $m_{k}$ are the Coxeter exponents (for extended Dynkin diagrams, one of the Coxeter exponent is $0$). So, we have a fortiori that $\mathbf{2} + \mathbf{A}$ is positive definite. Hence the result. \hfill $\Box$

\subsubsection{Topological recursion}

Since $A_{k,l}$ is nonnegative, the one-cut Lemma~\ref{lcut} can be applied whenever $u_k$, $t_{k,j}$ are nonnegative (and if it is not the case, we can still use the weaker version Lemma~\ref{lcut2}). Then, Hypotheses~\ref{ppp} is satisfied, so we can apply our Proposition~\ref{mik}: the generating series of maps in the height model for any $\mathfrak{G}$,
\beq
\omega_n^g(\sheet{x_1}{k_1},\ldots,\sheet{x_n}{k_n}) = W_n^{g}(\sheet{x_1}{k_1},\ldots,\sheet{x_n}{k_n})\dd x_1\cdots\dd x_n + \delta_{n,2}\delta_{g,0}\delta_{k_1,k_2}\,\frac{\dd x_1\dd x_2}{(x_1 - x_2)^2}
\eeq
satisfy the linear and quadratic loop equations in the sense of Definitions~\eqref{def10}-\eqref{def20}. Besides, if $\mathfrak{G}$ is a Dynkin diagram listed in Lemma~\ref{uj} (Fig.~\ref{ADEt}), Hypothesis~\ref{4331} is satisfied, so we can apply Corollary~\ref{aporo}: the generating series of maps in the height model are computed by the topological recursion
\bea
\label{toporeaq2}\omega_n^g(\sheet{z_0}{k_0},\sheet{z_I}{k_I}) & = & \sum_{\alpha \in \Gamma_{k_0}^{\mathrm{fix}}} \Res_{z \rightarrow \alpha} K_{k_0}(z_0,z) \\
& & \Big(\omega_{n + 1}^{g - 1}(\sheet{z}{k_0},\iota_{k_0}(\sheet{z}{k_0}),\sheet{z_I}{k_I}) 
+ \sum_{J \subseteq I,\,\,0 \leq h \leq g} \omega_{|J| + 1}^{h}(\sheet{z}{k_0},\sheet{z_J}{k_J})\omega_{n - |J|}^{g - h}(\iota_{k_0}(\sheet{z}{k_0}),\sheet{z_{I\setminus J}}{k_{I\setminus J}})\Big), \nonumber
\eea
where the recursion kernel is given by:
\beq
K_{k_0}(z_0,z) = \frac{-\frac{1}{2}\int_{\iota_{k_0}(z)}^{z} \omega_2^0(\sheet{z_0}{k_0},\sheet{\cdot}{k_0})}{\omega_1^0(\sheet{z}{k_0}) - \omega_1^0(\iota_{k_0}(\sheet{z}{k_0}))}.
\eeq

At present, an expression for $\omega_1^0(\sheet{x}{k})$ and $\omega_2^0(\sheet{x_1}{k_1},\sheet{x_2}{k_2})$ in full generality is not known, even at the critical points\footnote{It is known only for $\mathfrak{G} = A_{2}$, or $\mathfrak{G} = A_{s}$ with $s \geq 3$ but a symmetry assumption on the cuts, which both reduce to the solution of the $O(-\rho)$ model where loops live on triangles \cite{Kostovun,EKOn,EKOn2,BEOn}.}. Yet, we expect the problem to be solvable because (extended) Dynkin diagram are very special. We remind however that, if one is only interested in the generating series of maps with fixed number of level lines (in the loop representation, it corresponds to a fixed number of loops), the method described in \S~\ref{64} leads to explicit results. Let us mention that the analysis of singularities (see \S~\ref{66}) at the critical point in these models has been performed long ago \cite{Kostov:1989eg,Kostov:1991cg,KADE}, and the critical exponents are related to the spectrum of $\mathbf{A}$, in a way generalizing the relation~\ref{opro} valid for the universality class of the $O(-\rho)$ model where loops live on triangles.

\section{Chern-Simons invariants of torus knots}
\label{STknot}

We illustrate our method to give structural results on large $N$ expansion of Chern-Simons theory with gauge group $\mathrm{SU}(N)$ on a certain class of $3$-manifolds for which the partition function was known to be described by "repulsive particle systems".

\subsection{The model for torus knots}

\subsubsection{Definition}

For any knot $\mathsf{K}$ in a $3$-manifold $\mathsf{M}$, one can construct knot invariants $\mathsf{W}(\mathsf{G},\mathsf{R},q)$ indexed by a simply-laced group $\mathsf{G}$ and an irreducible representation $\mathsf{R}$, where $q$ is a variable. In quantum field theory, they can be defined as Wilson loops around $\mathsf{K}$ in Chern-Simons theory on $\mathsf{M}$ with gauge group $\mathsf{G}$ \cite{Witten89}. For $q$ equal to certain roots of unity, it was given a rigorous meaning in the work of Reshetikhin and Turaev \cite{TuraRe}. In particular, for $\mathsf{G} = \mathrm{SU}(2)$ and $\mathsf{R}$ its dimension $n$ irreducible representation, one retrieves the colored Jones polynomial $J_n(q)$ \cite{Jonesini}, and for $\mathsf{G} = U(N)$ and $\mathsf{R}$ the fundamental representation, one retrieves the HOMFLY-PT polynomial \cite{HOMFLY}.

Torus knots are knots which can be drawn on a torus $\mathbb{T} \subseteq \mathbb{S}_3$ without self-intersections. They are characterized by two coprime integers $(P,Q)$ describing the number of times the knot wraps around two independent non-contractible cycles in $\mathbb{T}$. They are in many regards the simplest knots among all. For instance, there exist closed formulas to compute all Wilson loops. It has been shown \cite{Law,Marinounp,Tierz,Beas,Kallen} that they can be rewritten as certain observables in a repulsive particle system. We shall restrict ourselves to the case of $\mathsf{G} = \mathrm{U}(N)$. Then:
\beq
\mathsf{W}(\mathrm{U}(N),\mathsf{R},q) = \langle s_{\mathsf{R}}(e^{T})\rangle,
\eeq
with respect to the measure on $\mathbb R^N$:
\bea
\label{fj}\dd\varpi(t_1,\ldots,t_N) & = & \frac{1}{\tilde{Z}_N^{(P,Q)}}\,\prod_{1 \leq i < j \leq N} \sinh\Big(\frac{t_i - t_j}{2P}\Big)\sinh\Big(\frac{t_i - t_j}{2Q}\Big)\,\prod_{i = 1}^N e^{-N\,\tilde{\mathcal{V}}(t_i)}\dd t_i, \\ \mathcal{V}(t) & = & \frac{t^2}{2uPQ}, \qquad u = N\ln q.
\eea
$s_{\mathsf{R}}$ denotes the character of the representation $\mathsf{R}$, which is here a Schur polynomial. We have set $T = \mathrm{diag}({t_1},\ldots,{t_N})$. We will focus on the case $q > 1$, hence $u > 0$.

\subsubsection{Relation between correlators and Wilson loops}

We define disconnected correlators in the model~\eqref{torus} as:
\beq
\overline{W}_n(x_1,\dots,x_n) = \Big\langle \prod_{i=1}^n \Tr \frac{1}{x_i- e^{T/PQ}}\Big\rangle,
\eeq
and we recall that they are related to the connected correlators by:
\beq
\overline{W}_{n}(x_1,\ldots,x_n) = \sum_{J_1 \dot{\cup} \cdots \dot{\cup} J_r = \ldbrack 1,n \rdbrack}  \prod_{j = 1}^r W_{|J_i|}\big((x_{j_i})_{j_i \in J_i}\big).
\eeq
They allow to compute any expectation values of traces of powers of $e^{T/PQ}$:
\beq
\langle \Tr(e^{k_1T/PQ}) \cdots \Tr(e^{k_n T/PQ}) \rangle = \oint_{\mathbb{R}_+^n} \prod_{i = 1}^n \frac{x_i^{k_i}\,\dd x_i}{2{\rm i}\pi}\,\overline{W}_n(x_1,\dots,x_n),
\eeq
and we now review how the expectation values of Schur polynomials can be deduced from them. Irreducible representations $\mathsf{R}$ of $\mathrm{U}(N)$ are in correspondence with Young tableaux with less than $N$ rows. If we denote $|\mathsf{R}|$ its number of boxes, it also determines an irreducible representation of the symmetric group $\mathfrak{S}_{|\mathsf{R}|}$, and by Schur-Weyl duality:
\beq
s_{\mathsf{R}}(e^{T}) = \frac{1}{|\mathsf{R}|!}\sum_{\mu\,\vdash\,|\mathsf{R}|} |C_{\mu}|\,\chi_{\mathsf{R}}(C_{\mu})\,p_{\mu}(e^{T}),
\eeq
where the sum runs over partitions $\mu = (\mu_1,\ldots,\mu_\ell)$ with $|\mathsf{R}|$ boxes, $C_{\mu}$ is the conjugacy class of the symmetric group $\mathfrak{S}_{|\mathsf{R}|}$ determined by $\mu$, $|C_{\mu}|$ the number of permutations in this class, $\chi_{\mathsf{R}}$ is the character of the symmetric group $\mathfrak{S}_{|\mathsf{R}|}$, and:
\beq
p_{\mu}(e^{T}) = \prod_{j = 1}^{\ell} p_{\mu_j}(e^{T}),\qquad p_j(e^{T}) = \Tr\,e^{jT},
\eeq
are the power-sums symmetric polynomials. Reminding the change of variable \ref{change}, we find:
\beq
\mathsf{W}(\mathrm{U}(N),\mathsf{R},q) = \frac{1}{|\mathsf{R}|!} \sum_{\mu\,\vdash\,|\mathsf{R}|} |C_{\mu}|\,\chi_{\mathsf{R}}(C_{\mu})\,\oint_{\mathbb{R}_+^{\ell(\mu)}} \prod_{j = 1}^{\ell(\mu)} \frac{x_j^{PQ\mu_j}\dd x_j}{2{\rm i}\pi}\,\,\overline{W}_{\ell(\mu)}(x_1,\ldots,x_{\ell(\mu)}).
\eeq

\subsubsection{Analytic continuation in $q$ and expansion of topological type}
\label{tue}

We justify in this paragraph the existence of an expansion of topological type (see Definition~\ref{top1}) for the correlators, where the variable of expansion $N$ is traded to $1/(\ln q)$.

Let us consider the measure \eqref{fj} with $u > 0$, i.e. $\ln q > 0$. At fixed $N$, one can perform the change of variable $\tilde{t_i} = (\ln q)^{-1/2}t_i$, and write:
\bea
\dd\tilde{\varpi}(t_1,\ldots,t_n) & = & \prod_{i = 1}^N e^{-\tilde{t}_i^2/2PQ}\dd\tilde{t}_i \cr
\label{usq} & & \times \prod_{1 \leq i < j \leq N} \Big(\sum_{k = 0}^{\infty} \frac{(\ln q)^{k}(\tilde{t}_i - \tilde{t}_j)^{2k}}{(2P)^{2k}(2k + 1)!}\Big)\Big(\sum_{l = 0}^{\infty} \frac{(\ln q)^l(\tilde{t}_i - \tilde{t}_j)^{2l}}{(2Q)^{2l}(2l + 1)!}\Big)(\tilde{t}_i - \tilde{t}_j)^2.
\eea
For our purposes, it is convenient to drop here the normalization factor, and define a new partition function as:
\beq
\tilde{Z}_N^{(P,Q)} = \int_{\mathbb{R}^N} \dd\tilde{\varpi}(t_1,\ldots,t_N).
\eeq
This integral is convergent for $\ln q > 0$, and from \eqref{usq}, it has an expansion in powers of $\ln q$ with positive coefficients. Hence, this series is absolutely convergent, and defines $\tilde{Z}_{N}^{(P,Q)}$ as an entire function of $\ln q$. In particular, it does not vanish for $\ln q$ small enough. Therefore, $F = \ln Z$ is an analytic function of $\ln q$ at least in a neighborhood of $\ln q = 0$. Now, the disconnected correlators can be defined as formal Laurent series in $x_1,\ldots,x_n$:
\beq
\label{Wnd}\overline{W}_n(x_1,\ldots,x_n) = \sum_{k_1,\ldots,k_n \geq 0} \frac{1}{x_1^{k_1 + 1}\cdots x_n^{k_n + 1}} \frac{1}{\tilde{Z}_N^{(P,Q)}}\int_{\mathbb{R}^N} \prod_{j = 1}^n\Big(\sum_{i_j = 1}^N e^{k_j\sqrt{\ln q}\,t_{i_j}/PQ}\Big)\dd\tilde{\varpi}(t_1,\ldots,t_N),
\eeq
whose coefficients can also be defined, from their series expansion in $\ln q$ (notice that, by parity, it is an expansion in $\ln q$ and not $\sqrt{\ln q}$), as analytic functions of $\ln q$ at least in a neighborhood of $0$. We can then deduce that the connected correlators $\overline{W}_n(x_1,\ldots,x_n)$, which can be expressed polynomially in terms of the disconnected correlators, are also formal Laurent series in $x_1,\ldots,x_n$ of analytic functions of $\ln q$ in a neighborhood of $0$. For any $\chi$, we can thus build formal Laurent series in $x_1,\ldots,x_n$ by collecting the coefficients of $(\ln q)^{\chi}$ in \eqref{Wnd}. It is clear from \eqref{Wnd} that the coefficient of $x_1^{-(k_1 + 1)}\cdots x_n^{-(k_n + 1)}$ grows atmost like $M_{n,\chi}^{(k_1 + \ldots + k_n)}$ for some $M_{n,\chi} > 0$ when $k_1,\ldots,k_n \rightarrow \infty$, so that those Laurent series actually define holomorphic functions at least in a neighborhood of $x_1,\ldots,x_n = \infty$. Since $W_n$ is initially holomorphic in $\mathbb{C}\setminus\mathbb{R}_+^{\times}$, we deduce that $W_n^g$ is also holomorphic at least in $\mathbb{C}\setminus\mathbb{R}_+^{\times}$. Their precise analytic structure will be determined in \S~\ref{anapa}.

The coefficients in the series representing $F \equiv W_{n = 0}$ and $W_n$ are finite sums of moments of Gaussian integrals, computed by Wick's theorem: they coincide with the generating series of connected maps with tubes and $n$ boundaries as explained in Section~\ref{S5}. In this case, since the potential is Gaussian, all faces are annular faces. This implies that, for a map of genus $g$ with $n$ marked faces, the counting of Euler characteristics gives $2 - 2g - n = v - e$, where $v$ is the number of vertices, and $e$ the number of edges. The weight of a map depends on $N = u/\ln q$ only through a factor $N^{v}$, and $\ln q$ appears also through a factor $(\ln q)^{e}$. Therefore, the coefficient of $(\ln q)^{\chi}$ in the series is a sum over maps with Euler characteristics $\chi$, as is well-known since t'Hooft \cite{tHooft}. This implies that the series defining $F$ and $W_n$ actually takes the form:
\beq
F = \sum_{g \geq 0} (\ln q)^{2g - 2}\,F^g,\qquad
W_n(x_1,\ldots,x_n) = \sum_{g \geq 0} (\ln q)^{2g - 2 + n}\,W_n^g(x_1,\ldots,x_n)
\eeq
where the coefficients $F^g$ is an analytic function of $u$ in a neighborhood of $0$, and $W_n^g$ is a formal Laurent series in $x_1,\ldots,x_n$, whose coefficients are analytic functions of $u$ in a neighborhood of $0$.

We recover by a direct method for torus knot complements some results known from the theory of LMO invariants for any $3$-manifold, namely that the $F^g$ can actually be defined for any knot complement using the theory of LMO invariants, and then shown to be analytic in a neighborhood of $u = 0$ \cite{GLMaa}. It is observed \cite{BEMknots} that the large $N$ expansion of Wilson loops in any representation are polynomial in $e^{u}$. As we shall see below, the stable $F^g$ and $W_n^g$ will be given by the topological recursion formula, from where their analytical structure in $u$ can be completely described, and one can prove the aforementioned observation, but this will not be addressed in this article.

\subsection{The spectral curve}

It is convenient to perform the change of variable:
\beq
\label{change}\lambda_i = e^{t_i/PQ}.
\eeq
We obtain the model with a measure on $\mathbb R_+^N$:
\bea
\label{torus}\dd\breve{\varpi}(\lambda_1,\ldots,\lambda_N) & = & \frac{1}{\breve{Z}_{N}^{(P,Q)}} 
\,\,\,\prod_{1 \leq i < j \leq N} (\lambda_i^{P} - \lambda_j^{P})(\lambda_i^{Q} - \lambda_j^{Q}) \prod_{i = 1}^N e^{-N\breve{\mathcal{V}}(\lambda_i)}\dd\lambda_i, \\
\breve{\mathcal{V}}(\lambda) & = & \frac{PQ(\ln \lambda)^2}{2u} + \frac{\ln \lambda}{N}\Big(1 + \frac{P + Q}{2}(N - 1)\Big)
\eea

\subsubsection{Properties}

\begin{lemma}
\label{L61} In the model \eqref{torus}, the interactions are strongly confining at $0$ and $\infty$, and strictly convex.
\end{lemma}
\textbf{Proof.} The two-point interaction is $R_0(x,y) = \sqrt{|x^{P} - y^{P}||x^{Q} - y^{Q}|}$, and we have:
\beq
\forall x,y > 0,\qquad \ln R_0(x,y) \leq \frac{P + Q}{2}\,(|\ln{x}| + |\ln y|).
\eeq
Since the potential grows like $(\ln x)^2$ when $x \rightarrow 0,\infty$, it implies that the interactions are strongly confining at $0$ and $\infty$ in the sense of Definition~\ref{sconf}. Besides, for any signed measure $\nu$ on $\mathbb{R}_{+}$ with total mass $0$, we have:
\beq
\frac{1}{2}\iint_{(\mathbb{R}_+^*)^2} \dd\nu(x)\big(\dd\nu(y)\big)^* \ln R_0(x,y) = - \int_{0}^{\infty} \frac{\dd s}{2s}\Big(|\widehat{\pi^*_{P}\nu}(s)|^2 + |\widehat{\pi^*_{Q}\nu}(s)|^2\Big) < 0,
\eeq
where $\pi_{\alpha}$ is the diffeomorphism of $\mathbb{R}_+^*$ defined by $\pi_{\alpha}(x) = x^{\alpha}$. Therefore, the two-point interaction is strictly convex in the sense of Definition~\ref{stconv}. \hfill $\Box$

Accordingly, the equilibrium measure $\breve{\mu}_{\mathrm{eq}}$ for the model \eqref{fj} on the real axis (or equivalently \eqref{torus} on the positive real axis) is unique (cf. Proposition~\ref{poeq}).

\begin{lemma}
\label{auea}For any $u > 0$, the support of the equilibrium measure of the model \eqref{torus} is a segment $[a,b]$, with $0 < a < b < \infty$.
\end{lemma}
\textbf{Proof.} Let $\tilde{\mu}_{\mathrm{eq}}$ denote the equilibrium measure of the model \eqref{fj}. The equilibrium measure of the model \eqref{torus} is just obtained by the change of variable \eqref{change}, i.e. is given by $\Phi_*\mu_{\mathrm{eq}}$ where $\Phi(t) = e^{t/PQ}$. From Theorem~\ref{poeq}, $\mu_{\mathrm{eq}}$ is characterized by the existence of a constant $C$ such that:
\beq
\label{saq22} \int \dd\mu_{\mathrm{eq}}(t')\Big[\ln \mathrm{sinh}\Big(\frac{|t - t'|}{2P}\Big) + \ln\mathrm{sinh}\Big(\frac{|t - t'|}{2Q}\Big)\Big] - \mathcal{V}(t) = C,
\eeq
with equality $\tilde{\mu}_{\mathrm{eq}}$ everywhere. We observe that, for any $\alpha > 0$, $\ln\mathrm{sh}\big(\frac{|t - t'|}{2\alpha}\big)$ is a concave function of $t$. Since $\tilde{\mu}_{\mathrm{eq}}$ is a nonnegative measure, this implies that the integral in the left-hand side is a concave function of $t$. Besides, $\tilde{V}(t) = t^2/2uPQ$ is strictly convex, hence the left-hand side is a concave function of $t$. This implies that, if the equality is realized for $t = t_1$ and $t = t_2$, it must be realized for $t \in [t_1,t_2]$. Hence, the support of $\breve{\mu}_{\mathrm{eq}}$ is connected. \hfill $\Box$

\subsubsection{The equilibrium measure}
\label{seqia}
\begin{figure}
$$\includegraphics[width=1\textwidth]{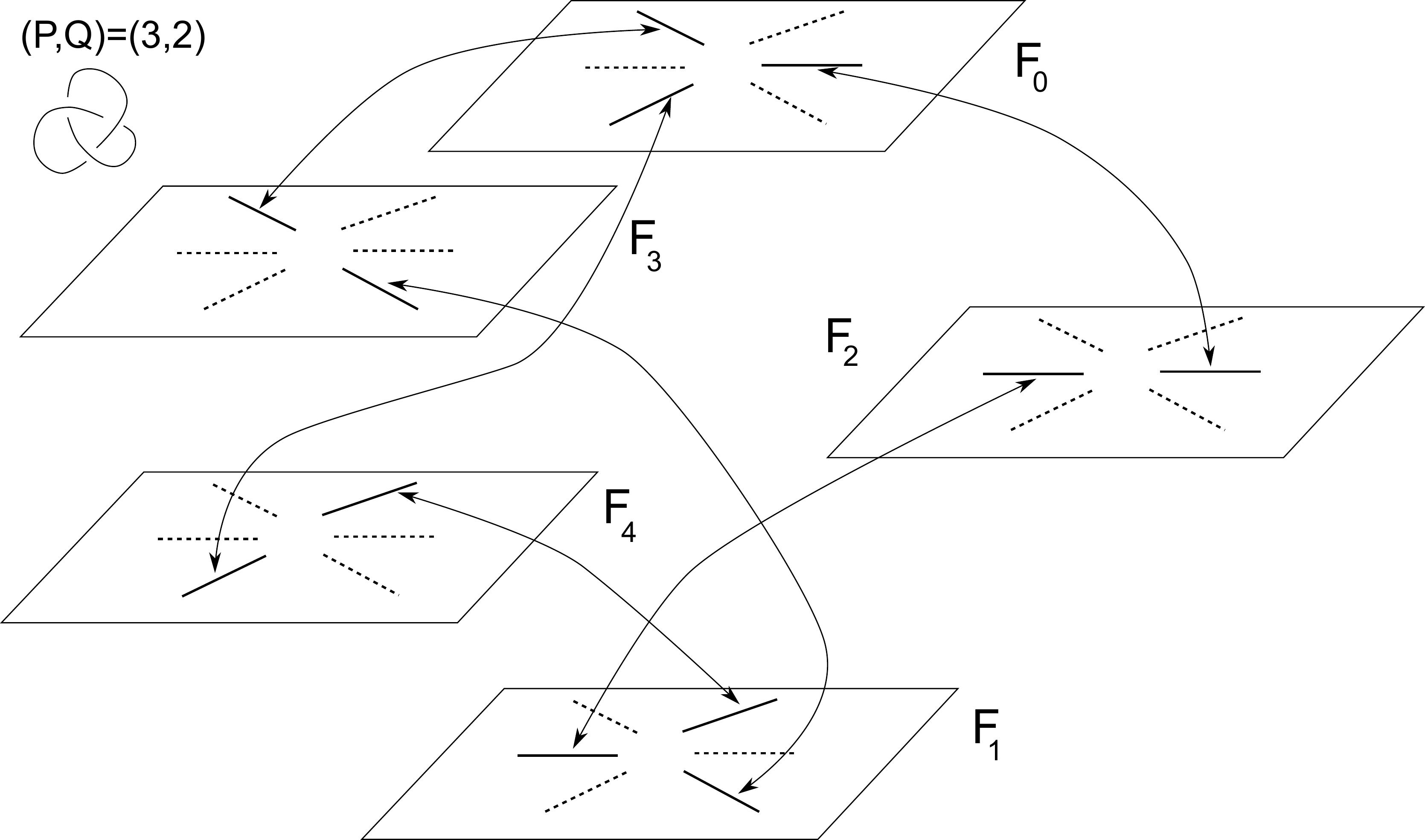}$$
\caption{\label{figknot32surf} For the $(3,2)$ knot (the trefoil knot), we have $P+Q=5$ functions $F_0,F_1,\dots,F_4$, with cuts indicated, and they transform among each other when crossing cuts. This defines a Riemann surface of genus $2$, as a degree 5 covering of the plane $\widehat{\mathbb C}$.
The 5 functions $F_0,\dots,F_4$ can be seen as 5 branches of a multivalued function on $\widehat{\mathbb{C}}$, which can be lifted to an analytic function on the Riemann surface.}
\end{figure}

Let us consider the Stieltjes transform:
\beq
\omega_{1}^0(x) = \Big(\int_a^b \frac{\dd\breve{\mu}_{\mathrm{eq}}(y)}{x - y}\Big)\dd x.
\eeq
It is characterized by \eqref{funcq} with a linear operator $\mathcal{O}$ defined by:
\beq
\mathcal{O} f(x) = \frac{1}{2{\rm i}\pi}\oint_{[a,b]}f(\xi)\left(-\sum_{j = 1}^{P - 1} \frac{1}{\xi - e^{2{\rm i}\pi j/P}x} - \sum_{j = 1}^{Q - 1} \frac{1}{\xi - e^{2{\rm i}\pi j/Q}x}\right),
\eeq
and for any $1$-form $f$ which is holomorphic in $\mathbb{C}\setminus[a,b]$ and is $O(\dd x)$ at $\infty$, we find:
\beq
\label{622O}\mathcal{O} f(x) = \sum_{j = 1}^{P - 1} f(e^{2{\rm i}\pi j/P}x) + \sum_{j = 1}^{Q - 1} f(e^{2{\rm i}\pi j/Q}).
\eeq
It consists of sums of rotations by angles $2\pi/Q$ and $2\pi/P$. In other words, we have the functional equation, for any $x \in ]a,b[$:
\beq
\label{rei}\omega_{1}^0(x + {\rm i}0) + \omega_{1}^0(x - {\rm i}0) + \sum_{j = 1}^{P - 1} \omega_{1}^0(e^{2{\rm i}\pi j/P}x) + \sum_{j = 1}^{Q - 1} \omega_{1}^0(e^{2{\rm i}\pi j/Q}x) = \Big(\frac{PQ}{u}\,\frac{\ln x}{x} + \frac{P + Q}{2x}\Big)\dd x.
\eeq
The solution in the one-cut regime (as required by Lemma~\ref{auea}) was found in \cite{BEMknots}:

\begin{proposition} \label{propknotspcurve}
We have:
\beq
\frac{1}{P}\,\sum_{j=0}^{P-1} \om_{1}^0(\ee{2\ii\pi\frac{j}{P}}\,x) = -\frac{1}{u}\,\frac{\dd x}{x} \ln\big[e^{-u}(-1)^{Q}\,z(x)\big],
\eeq
where $z(x)$ is an algebraic function:
\beq
\label{curve}x  = e^{\frac{u}{2}(1/P + 1/Q)}\,z^{-1/Q}\Big(\frac{1-e^{-u}z}{1-z}\Big)^{1/P}.
\eeq
The equilibrium measure is:
\beq
\label{622}\dd\breve{\mu}_{\rm eq}(x) = \frac{P}{2\ii\pi u}\,\frac{\dd x}{x}\,\ln\Big(\frac{z(x+\ii 0)}{z(x-\ii 0)}\Big).
\eeq
\end{proposition}
\noindent\textbf{Sketch of the proof of \cite{BEMknots}.} Let us define a function $Y(x)$ by setting:
\beq
\om_{1}^0(x) = \frac{PQ}{u(P+Q)}\,\big[\ln(e^{\frac{u}{2}(1/P + 1/Q)}x) - \ln(-Y(e^{\frac{u}{2}(1/P + 1/Q)}x)\big]\frac{{\mathrm d}x}{x}.
\eeq
and $[\breve{a},\breve{b}] = e^{\frac{u}{2}(1/P + 1/Q)}[a,b]$. Then, \eqref{rei} translates into:
\beq
\forall \xi \in ]\breve{a},\breve{b}[,\qquad Y(\xi +\ii 0)\,Y(\xi -\ii 0)\,\prod_{j=1}^{Q-1} Y(\ee{2\ii\pi j/Q}\,\xi)\,\,\prod_{j=1}^{P-1} Y(\ee{2\ii\pi j/P}\,\xi)
 = 1.
\eeq
The fact that $\om_{1}^0(x)$ is holomorphic on $\mathbb C\setminus [\breve{a},\breve{b}]$ and behaves like $\dd x/x$ when $x \rightarrow \infty$, implies that $Y(x)$ is holomorphic on $\mathbb C\setminus [\breve{a},\breve{b}]$, vanishes only at $x=0$, and behaves as:
\beq
\label{625}Y(\xi)\,\mathop{=}_{\xi \rightarrow 0}\,-\xi+O(\xi^2),
\qquad Y(\xi)\,\mathop{=}_{\xi \rightarrow \infty}\,-\xi\,e^{-u(1/P + 1/Q)} + O(1).
\eeq
Consider the $P+Q$ following functions:
\beq
F_k(\xi) = \prod_{j=0}^{P-1}\,Y(\ee{2\ii\pi k/Q}\,\ee{2\ii\pi j/P}\,\xi),
\qquad 0\leq k\leq Q-1.
\eeq
\beq
F_{Q+l}(\xi) = \prod_{j=0}^{Q-1}\,\frac{1}{Y(\ee{2\ii\pi j/Q}\,\ee{2\ii\pi l/P}\,\xi)},
\qquad 0\leq l\leq P-1.
\eeq
Notice that $F_k(\xi)$ has cuts along $\ee{-2\ii\pi k/Q}\,\ee{-2\ii\pi j/P}[\breve{a},\breve{b}]$ for $0\leq j\leq P-1$, and $F_{Q+l}(x)$ has cuts along $\ee{-2\ii\pi l/P}\,\ee{-2\ii\pi j/Q}[\breve{a},\breve{b}]$ for $0\leq j\leq Q-1$. In particular across the cut $\ee{-2\ii\pi l/P}\,\ee{-2\ii\pi k/Q}[\breve{a},\breve{b}]$, we have:
\beq
\forall \xi\in\,\ee{-2\ii\pi l/P}\,\ee{-2\ii\pi k/Q}]\breve{a},\breve{b}[,\qquad F_k(\xi-\ii 0) = F_{Q+l}(\xi+\ii 0).
\eeq
This implies that the functions $(F_k)_{0 \leq k \leq P + Q - 1}$ transform among themselves under cut crossings, see Fig.~\ref{figknot32surf} for the $(P,Q)=(3,2)$ case.

\begin{figure}
\begin{center}
\includegraphics[width=0.6\textwidth]{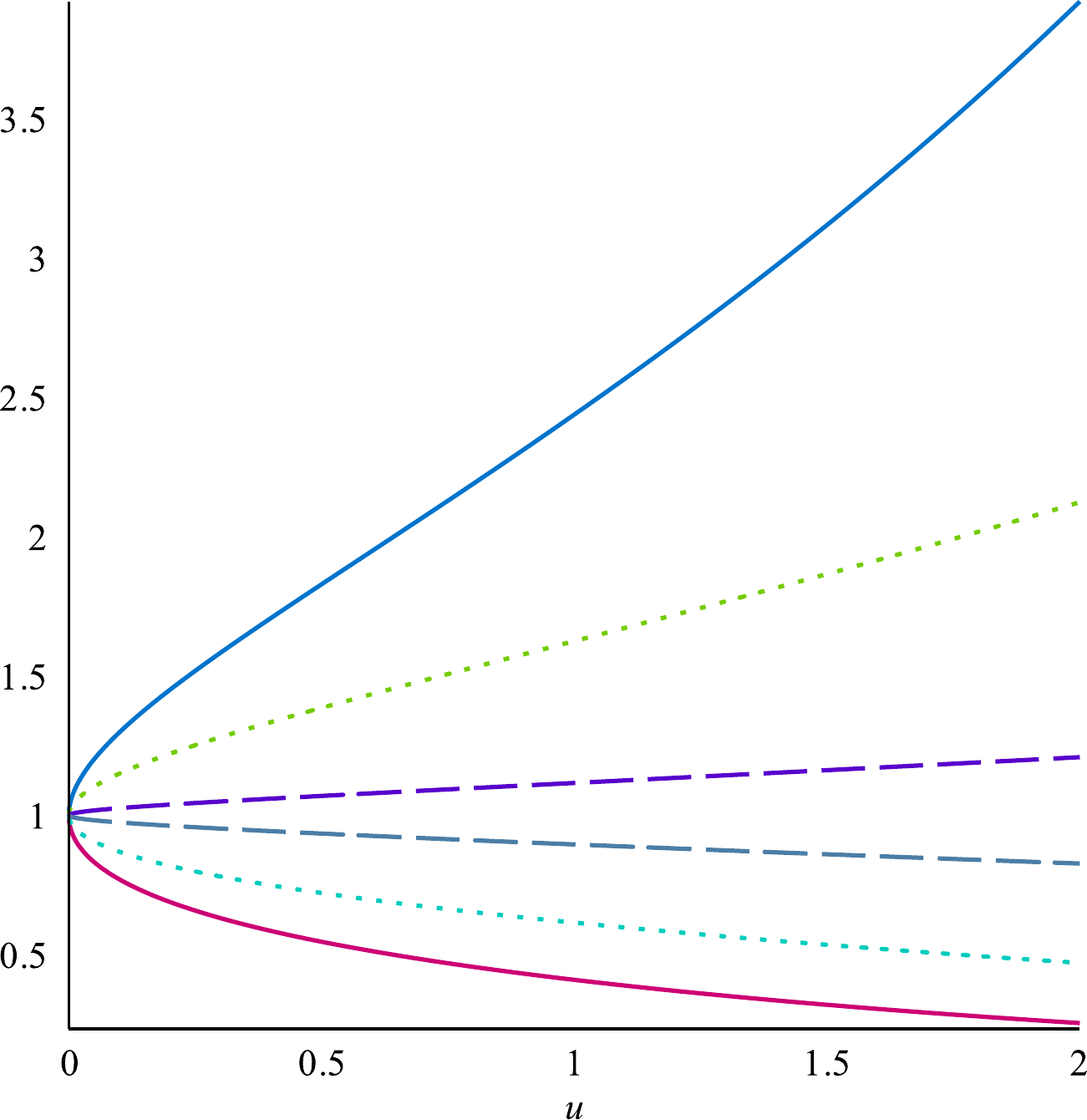}
\caption{\label{suppo1} Position of the support as a function of $u = N\ln q > 0$, $\breve{a}(u)$ (top lines) and $\breve{b}(u)$ (bottom lines) for $(P,Q) = (3,2)$ (solid line), $(P,Q) = (3,7)$ (dotted line) and $(P,Q) = (7,97)$ (dashed line).}
\end{center}
\end{figure}

Therefore, any polynomial symmetric function of the $F_k$'s is continuous across $[a,b]$ and all its images under rotations, i.e. must be holomorphic in $\mathbb{C}^{\times}$. Since $F_k$ behaves as integer powers of $\xi$ or $\xi^{-1}$ near zero and near $\infty$, it must be a polynomial in $\xi$ and $1/\xi$. This principle shows that 
\beq
\forall f\in \mathbb C\qquad
\Pi(f,\xi) = \prod_{k=0}^{P+Q-1} (f-F_k(\xi))
\eeq
is a polynomial in $\xi$ and $\xi^{-1}$, whose coefficients are polynomial in $f$. Moreover, we observe that
$\Pi(f,\ee{2\ii\pi\,\frac{1}{PQ}}\,\xi)=\Pi(f,\xi)$, and thus $\Pi(f,\xi)$ is actually a polynomial in $\xi^{PQ}$ and $\xi^{-PQ}$. The behaviors \eqref{625} at $\xi\to 0$ and at $\xi\to \infty$ imply:
\beq
\Pi(f,\xi) = f^{P+Q} + (-1)^{P+Q} - (-1)^{PQ}\,(-1)^P\,\xi^{-PQ}\,f^Q - (-1)^{PQ}\,(-1)^Q\,\ee{-u(P+Q)}\,\xi^{PQ}\,f^P  + \sum_{j=1}^{P+Q-1} \Pi_j\,f^j,
\eeq
for some coefficients $\Pi_j$. So far, we have thus shown that the function $F_0(x)$ satisfies an algebraic equation. It can be proved, see \cite[Section 4.2]{BEMknots} that, knowing that the cut locus of $\omega_{1}^0$ is a single segment (Lemma~\ref{auea}) determines uniquely all the coefficients $\Pi_j$. The solution can be written parametrically:
\beq
\label{alk}\xi^{PQ} = C_1\,\,z^{-P}\,\left(\frac{1-cz}{1-z}\right)^Q,\qquad F_0 = C_2\,z\,\frac{1- cz}{1-z}.
\eeq
and the conditions \eqref{625} at $\xi \rightarrow 0$ and $\xi \rightarrow \infty$ are fixing:
\beq
C_1 = e^{u(P + Q)},\qquad C_2 = -1,\qquad c = e^{-u}.
\eeq
If we remind the relation $\xi = e^{\frac{u}{2}(1/P + 1/Q)}x$, we obtain:
\beq
x = e^{\frac{u}{2}(1/P + 1/Q)}\,z^{-1/Q}\,\Big(\frac{1 - e^{-u}z)}{1 - z}\Big)^{1/P},
\eeq
as announced in \eqref{curve}. The position of the cut $[a,b]$ can be deduced as a function of $u$ (Fig.~\ref{suppo1}). What we have obtained is the symmetrization of $\omega_{1}^0$ (the Stieljes transform of $\mu_{\rm eq}$) under rotations by angle $2\pi/P$. However, since $\omega_{1}^0$ has a cut only along $[\breve{a},\breve{b}]$ and not along its rotated images, it is easy to recover $\breve{\mu}_{\rm eq}$ as the discontinuity along $\Gamma$.
\beq
\forall x \in ]a,b[,\qquad \dd\mu_{\rm eq}(x) = \frac{P}{2\ii\pi u}\,\ln{\left(\frac{z(x+\ii 0)}{z(x-\ii 0)}\right)}\,\frac{{\mathrm d}x}{x}.
\eeq

The branchpoints are the zeroes of $\dd x$, i.e. of $\dd x/x$. Their position in the $z$-variable satisfies the quadratic equation $-\frac{P}{z} + \frac{Q}{z - e^u} - \frac{Q}{z - 1} = 0$, whose solutions are
\beq
z_{\pm}(u) = \frac{(P + Q)e^u + (P - Q) \pm \sqrt{(e^u - 1)[(P + Q)^2e^{u} - (P - Q)^2]}}{P}.
\eeq
In particular, for any $u > 0$, we find $z_{\pm}(u)$ is real positive and such that $(1 - e^{-u}z_{\pm}(u))$ and $(1 - z_{\pm}(u))$ have same sign. Hence, we find that for any $u > 0$, the cut is a segment $[a(u),b(u)] = [x(z_{-}(u)),x(z_{+}(u))]$ on the positive real axis (see Fig.~\ref{suppo1}). Hence, \eqref{alk} gives a solution of \eqref{rei} (by construction), which leads to a measure (right-hand side of \eqref{622}) supported on a segment of the positive real axis. By unicity (deduced from Lemma~\ref{L61}), this measure must be the equilibrium measure sought for. \hfill $\Box$

\subsubsection{Fundamental $2$-form of the $2^{\mathrm{nd}}$ kind}

The method of the previous paragraph shows in general that, for any meromorphic $1$-form $f$ in $\mathbb{C}\setminus[a,b]$ which is solution of the master equation:
\beq
f(x + {\rm i}0) + f(x - {\rm i}0) + \sum_{j = 1}^{P - 1} f(e^{2{\rm i}\pi j/P}) + \sum_{j = 1}^{Q - 1} f(e^{2{\rm i}\pi j/Q}) = 0,
\eeq
the averages
\beq
\label{tyeu}\left\{\begin{array}{lll} \sum_{j = 0}^{P - 1} f(e^{2{\rm i}\pi(j/P + k/Q)}x) & &  0 \leq k \leq P - 1 \\ 
 -\sum_{j = 0}^{Q - 1} f(e^{2{\rm i}\pi(l/P + j/Q)}x) & & 0 \leq l \leq Q - 1
\end{array}\right.
\eeq
are actually branches of a unique meromorphic $1$-form on the curve $\mathcal{C}$ of equation \eqref{curve}, which is uniformized to $\widehat{\mathbb{C}}$ by the variable $z$. Therefore, there is a unique fundamental $2$-form of the $2^{\mathrm{nd}}$ kind on this curve:
\beq
\label{bob}B(z_0,z) = \frac{\dd z\,\dd z_0}{(z - z_0)^2},
\eeq
which defines the appropriate Cauchy kernel:
\beq
G(z_0,z) = -\int^{z} B(z_0,z) = \frac{\dd z_0}{z - z_0}.
\eeq
allowing to represent objects of type \eqref{tyeu}. As expected, we find that the $2$-form $\omega_2^0$ is closely related to $B(z_0,z)$:
\begin{proposition}
\beq
\sum_{j=0}^{P-1} \om_2^0(x_0,\ee{2\ii\pi j/P}x) = \frac{\dd z(x)\,\dd z(x_0)}{(z(x) - z(x_0))^2} - \,\frac{\dd (x^P)\,\dd(x_0^P)}{(x^P - x_0^P)^2}.
\eeq
\end{proposition}
\textbf{Proof.} We have from \eqref{pqu}, for any $x \in ]a,b[$:
\beq
\label{oi}\om_2^0(x_0,x+\ii 0)+\om_2^0(x_0,x-\ii 0)
+ \sum_{j=1}^{Q-1} \om_2^0(x_0,\ee{2\ii\pi j/Q}x)
+ \sum_{j=1}^{P-1} \om_2^0(x_0,\ee{2\ii\pi j/P}x)
= -\,\frac{\dd x\,\dd x_0}{(x-x_0)^2},
\eeq
and $\om_2^0(x_0,x)$ is holomorphic for $x\in \widehat{\mathbb C}\setminus [a,b]$. Let us first find a particular solution of the non-homogeneous equation. Let us introduce the following sequence of $1$-forms indexed by $m \in \ldbrack 1,PQ \rdbrack$:
\beq
f_m(x)  = \left\{\begin{array}{lll} \frac{x^{PQ}}{P + Q}\,\frac{\dd x}{x} & & \mathrm{if}\,\,m = PQ \\ \frac{x^{m}}{P}\,\frac{\dd x}{x} & & \mathrm{if}\,\,m\,\,\mathrm{is}\,\,\mathrm{a}\,\,\mathrm{multiple}\,\,\mathrm{of}\,\,P \\ \frac{x^{m}}{Q}\,\frac{\dd x}{x} & & \mathrm{if}\,\,m\,\,\mathrm{is}\,\,\mathrm{a}\,\,\mathrm{multiple}\,\,\mathrm{of}\,\,Q \\ 
\frac{1}{2{\rm i}\pi}\,\frac{1}{\frac{1}{e^{2{\rm i}\pi m/P} - 1} + \frac{1}{e^{2{\rm i}\pi m/Q} - 1}}\,x^m\ln x\,\frac{\dd x}{x} & & \mathrm{else} \end{array}\right. .
\eeq
They are constructed so that:
\beq
\forall x \in ]a,b[,\qquad f_m(x + {\rm i}0) + f_m(x - {\rm i}0) + \sum_{j = 1}^{P - 1} f_m(e^{2{\rm i}\pi j/P}x) + \sum_{j = 1}^{Q - 1} f_m(e^{2{\rm i}\pi j/Q}x) = x^m\,\frac{\dd x}{x}.
\eeq
Then, we can build:
\beq
G(x_0,x) = \sum_{k = 1}^{PQ} \frac{x_0^{PQ - k}\,f_k(x)}{x^{PQ} - x_0^{PQ}},
\eeq
which satisfies:
\beq
\forall x \in ]a,b[,\qquad G(x_0,x + {\rm i}0) + G(x_0,x - {\rm i}0) + \sum_{j = 1}^{P - 1} G(x_0,e^{2{\rm i}\pi j/P}x) + \sum_{j = 1}^{Q - 1} G(x_0,e^{2{\rm i}\pi j/Q}x) = \frac{\dd x}{x - x_0}.
\eeq
Therefore, $-\dd_{x_0}G(x_0,x)$ is a particular solution of \eqref{oi}, which is meromorphic on $\widehat{\mathbb{C}}\setminus\mathbb{R}_-$, with a double pole without residues at all $x^{PQ} = x_0^{PQ}$, and a logarithmic singularity on $\mathbb{R}_-$. 
We can decompose it:
\beq
\label{6412}\dd_{x_0}G(x_0,x) = -\,\frac{P^2+PQ+Q^2}{P^2 Q^2 (P+Q)}\,\frac{\dd(x_0^{PQ})\,\dd(x^{PQ})}{(x^{PQ} - x_0^{PQ})^2}+\frac{1}{P^2}\frac{\dd(x_0^P)\,\dd(x^P)}{(x^P - x_0^P)^2} + \frac{1}{Q^2}\frac{\dd(x_0^Q)\,\dd(x^Q)}{(x^Q - x_0^Q)^2} + \ln x\,R(x_0,x),
\eeq
where $R(x_0,x)$ is a rational function of $x_0$ and $x$. We observe however that, after averaging $G(x_0,x)$ over rotations of $x$ of angle $2\pi/P$ (or, over rotations of angle $2\pi/Q$), the logarithmic singularity disappears. The $2$-form:
\beq
\label{641}\breve{B}(x_0,x) = \omega_2^0(x,x_0) + \dd_{x_0} G(x_0,x)
\eeq
is a solution of \eqref{oi} with vanishing right-hand side. We now proceed like in \S~\ref{seqia} by defining:
\bea
H_{k}(x_0,x) & = & \sum_{j=0}^{P-1}  \breve{B}(x_0,\ee{2\ii\pi(k/Q+ j/P)}x),\qquad 0\leq k\leq Q - 1, \\
H_{Q+l}(x_0,x) & = & -\sum_{j=0}^{Q-1}  \breve{B}(x_0,\ee{2\ii\pi\,(j/Q+ l/P)}x),\qquad 0\leq l\leq P - 1.
\eea
For any $k \in \ldbrack 0, P + Q - 1\rdbrack$, $H_k(x_0,\cdot)$ is meromorphic on $\widehat{\mathbb{C}}$ (in particular has no logarithmic cut), with double poles without residues at $x^{PQ}=x_0^{PQ}$. For any $k \in \ldbrack 0,Q - 1 \rdbrack$, $H_k$ has cuts along $\ee{2\ii\pi(k/P+l/Q)}[a,b]$ for all $l \in \ldbrack 0,P - 1 \rdbrack$, whereas for any $l \in \ldbrack 0,P - 1 \rdbrack$, $H_{Q + l}(x_0,\cdot)$ has cuts along $\ee{2\ii\pi (k/P+l/Q)}[a,b]$ for all $k \in \ldbrack 0,Q - 1 \rdbrack$. And, more precisely:
\beq
\forall x\in\,\ee{-2\ii\pi(l/P+ k/Q)}]a,b[,\qquad H_k(x_0,x+\ii 0) = H_{Q+l}(x_0,x-\ii 0).
\eeq
This implies that $H_k(x_0,x)$ can be seen as the $k$-th branch of a meromorphic form $H(x_0,x)$ defined on the same Riemann surface as $\om_1^0(x)$. So, it must be a rational function of $z(x)$ and $z(x_0)$, with double poles at $x = e^{2{\rm i}\pi (j/P + k/Q)}$ for some $j,k$. Given its singularities, it takes the form:
\beq
\label{si}H(x,x_0) = A_1\,\frac{\dd z(x)\,\dd z(x_0)}{(z(x) - z(x_0))^2} + A_2\,\frac{\dd(x^{PQ})\,\dd(x_0^{PQ})}{(x^{PQ} - x_0^{PQ})^2}.
\eeq
Besides, we can deduce from \eqref{6412} and the observation that the logarithmic singularity in $G(x_0,x)$ disappears after average, that:
\beq
\label{jsq}H(x_0,x) = \sum_{j = 0}^{P - 1} \omega_2^0(x_0,e^{2{\rm i}\pi j/P}x) + \frac{\dd(x^P)\,\dd(x_0^P)}{(x^{P} - x_0^{P})^2} - \frac{1}{P+Q}\,\frac{\dd(x^{PQ})\,\dd(x_0^{PQ})}{(x^{PQ} - x_0^{PQ})^2}.
\eeq
In the first sheet, $H(x_0,x)$ assumes the values $H_0(x_0,x)$, and for a fixed $x_0$, we have $Q$ points such that $x^{PQ} = x_0^{PQ}$, which are actually those corresponding to $x^{Q} = x_0^Q$. The last term of \eqref{jsq} has a pole at $x = x_0$ in the first sheet, but is regular at the other aforementioned $(Q - 1)$ points. Therefore, matching with the poles of \eqref{si} yields $A_1 = 1$ and $A_2 = -1/(P+Q)$. Hence:
\beq
\label{sujj}\sum_{j = 0}^{P - 1} \omega_2^0(x_0,e^{2{\rm i}\pi j/P}x) = \frac{\dd z(x)\,\dd z(x_0)}{(z(x) - z(x_0))^2} - \frac{\dd (x^P)\,\dd(x_0^P)}{(x^P - x_0^P)^2}.
\eeq
Then, $\omega_2^0(x_0,x)$ can be recovered as the part of \eqref{sujj} which has a cut on $[a,b]$ and not on its rotations by $2\pi/Q$ angles. \hfill $\Box$

\subsection{Analytic structure of $W_n^g$}
\label{anapa} 
In the convergent model \eqref{torus}, since the potential is convex for $u>0$, $\Gamma$ is a single segment and $W_1^0(x)$ is discontinuous at any interior point of $\mathring{\Gamma}$. On the other hand, we could define $W_n^g(x_1,\ldots,x_n)$'s as holomorphic functions for $x_i$ in some neighborhood of $\infty$. We now claim:
\begin{lemma}
For any $(n,g)$, $W_n^g(x_1,\ldots,x_n)$ defines a holomorphic function in $(\mathbb{C}\setminus\Gamma)^n$.
\end{lemma}
\textbf{Proof.} The $W_n^g$ satisfy the Schwinger-Dyson equation order by order in $(\ln q)$, i.e. \eqref{jui} with $\rho = 1$ and $\mathcal{O}$ given by \eqref{odef} with:
\beq
R(x,y) = \frac{x^{P} - y^{P}}{x - y}\,\frac{x^{Q} - y^{Q}}{x - y}.
\eeq
We can rewrite them:
\bea
\label{kipao} W_n^g(x,x_I) & = & \big(2W_1^0(x) + \mathcal{O}W_1^0(x) - \mathcal{V}'(x)\big)^{-1} \nonumber \\
& & \times \Big(- W_{n + 1}^{g - 1}(x,x,x_I) - \sum_{\substack{J \subseteq I\,\,0 \leq h \leq g \\ (J,h) \neq (\emptyset,0),(I,g)}} W_{|J| + 1}^{h}(x,x_J)\,W_{n - |J|}^{g - h}(x,x_{I\setminus J}) \nonumber \\
& & - \mathcal{O}^{x,2}W_{n + 1}^{g - 1}(x,x,x_I) - \sum_{\substack{J \subseteq I\,\,0 \leq h \leq g \\ (J,h) \neq (\emptyset,0),(I,g)}} W_{|J| + 1}^{h}(x,x_J)\,\mathcal{O}^{x}W_{n - |J|}^{g - h}(x,x_{I\setminus J}) \nonumber \\
& & - W_1^0(x)\mathcal{O}^{x}W_n^g(x) - \sum_{i \in I} \frac{W_{n - 1}^{g}(x,x_{I\setminus\{i\}})}{(x - x_i)^2} - P_n^g(x;x_I)\Big),
\eea
where $P_n^g$ was defined in \eqref{3899}, and we recognize the prefactor
\beq
2W_1^0(x) + \mathcal{O}W_1^0(x) - \mathcal{V}'(x) = \Delta W_1^0(x).
\eeq
The equality \eqref{kipao} between holomorphic functions in $\mathbb{C}\setminus\mathbb{R}_+^{\times}$, extends to an equality valid in the maximal domain of analyticity of the functions at hand. We observe that $P_n^g$ here is a polynomial. Besides, if $f$ is holomorphic in $\mathbb{C}\setminus\mathbb{R}_+^{\times}$, $\mathcal{O}f$ is holomorphic in a neighborhood of $[\varepsilon,+\infty[$, since it can have singularities only on the rotations of $\mathbb{R}_+^{\times}$ by $2\pi/P$ and $2\pi/Q$ angles. So, the only singularity in the right-hand side of \eqref{kipao} comes from the singularities of $W_{m}^{h}$ for $2h - 2 + m < 2g - 2 + n$, or from the prefactor \mbox{$\big(2W_1^0(x) + \mathcal{O}W_1^0(x) - \mathcal{V}'(x)\big)^{-1}$} which has a singularity on $\Gamma \subset \mathbb{R}_+^{\times}$. Therefore, \eqref{kipao} implies by recursion of $2g - 2 + n$ that $W_n^g$ is holomorphic in $\mathbb{C}\setminus\Gamma$. \hfill $\Box$

\subsection{Result}

We have justified in \S~\ref{tue} and \ref{anapa} that the $W_n$ have an expansion of topological type. Therefore, we can apply Corollary~\ref{4333}, and find that the correlators are computed by the topological recursion. More precisely, let us define for any stable $n,g$:
\beq
\omega_n^g(x_1,\ldots,x_n) = W_n^{g}(x_1,\ldots,x_n)\dd x_1\cdots\dd x_n,
\eeq
and their averaged version $\check{\omega}_n^{g}(z_1,\ldots,z_n)$ which are meromorphic on $\mathcal{C}^n \simeq \widehat{\mathbb{C}}^{n}$, so that:
\beq
\check{\omega}_n^g(z_1,\ldots,z_n) = \sum_{0 \leq j_1,\ldots,j_n \leq P - 1} \omega_{n}^{g}(e^{2{\rm i}\pi j_1/P}x(z_1),\ldots,e^{2{\rm i}\pi j_n/P}x(z_n)),
\eeq
when $z_1,\ldots,z_n$ belong to the first sheet of $\mathcal{C}$. Then, Corollary~\ref{4333} tells us:
\begin{proposition}
The $\check{\omega}_n^g(z_1,\ldots,z_n)$ are determined by the topological recursion:
\bea
& & \check{\omega}_n^{g}(z_1,\ldots,z_n) \nonumber \\
& = & \Res_{z \rightarrow z_{\pm}(u)} \frac{\frac{1}{2}\Big(\frac{1}{z - z_0} - \frac{1}{\iota(z) - z_0}\Big)
}{\omega_1^0(z) - \omega_1^0(\iota(z))}\Big[\check{\omega}_n^{g - 1}(z,\iota(z),z_I) + \sum_{\substack{J \subseteq I,\,\,0 \leq h \leq g \\ (J,h) \neq (I,g),(\emptyset,0)}} \check{\omega}_{|J| + 1}^h(z,z_J)\check{\omega}_{n - |J|}^{g - h}(\iota(z),z_{I\setminus J})\Big]. \nonumber
\eea
\end{proposition}
Notice that we can use indifferently $\om_1^0$ or $\check{\omega}_1^0$ in the denominator. The description of $\omega_1^0$ (Proposition~\ref{propknotspcurve}) was actually obtained in \cite{BEMknots}, where it was conjectured (and checked for low $g$ and representation of small sizes) that the topological recursion with $B(z_0,z)$ given by \eqref{bob} would compute the topological expansion of the correlators, and hence the Wilson loops. Our present result fully justifies this prediction.

\section{Other examples}

We explain how to retrieve the 2-hermitian matrix model, and matrix models where eigenvalues live on a higher genus surface from the theory developed in Section~\ref{S1}. In this section, we rather revisit a few aspects of each problem at the light of our formalism rather than intend to present a detailed study.

\subsection{The 2-hermitian matrix model}
\label{2mm}
The 2-hermitian matrix model is defined by the measure:
\beq
\label{wei}\dd\varpi(M_1,M_2) = \dd M_1\,\dd M_2\,e^{-N\,\Tr[\mathcal{V}_1(M_1)+\mathcal{V}_2(M_2) + \alpha\,M_1 M_2]},
\eeq
where $M_1$ and $M_2$ are hermitian matrices, $\alpha$ is a coupling constant, and $\mathcal{V}_1$ and $\mathcal{V}_2$ are two polynomials. The Schwinger-Dyson equation of this model have been written down in \cite{Eynlp}. We introduce the correlators associated to the first matrix:
\beq
W_n(x_1,\ldots,x_n) = \Big\langle \prod_{i = 1}^n \Tr \frac{1}{x_i-M_1} \Big\rangle_c,
\eeq
and we assume that they have an expansion of topological type:
\beq
\label{texp}W_{n}(x_1,\ldots,x_n) = \sum_{g \geq 0} N^{2 - 2g - n}\,W_n^{g}(x_1,\ldots,x_n).
\eeq
If we consider \eqref{wei} as a convergent matrix model, we ask that $\mathcal{V}_1$ and $\mathcal{V}_2$ are chosen so that the weight \eqref{wei} is integrable over $H_N^2$, and \eqref{texp} would have to be justified under suitable assumptions. If we rather consider \eqref{wei} as a formal matrix model, one takes:
\beq
\mathcal{V}_j(x) = \frac{1}{u_j}\Big(\frac{x^2}{2} - \sum_{k \geq 3} \frac{t_{j,k}}{k}\,x^{k}\Big),\qquad j = 1,2,
\eeq
and $u_j$ and $t_{j,k}$ are considered as formal parameters. The combinatorial interpretation of the two hermitian matrix model is related to the enumeration of random maps whose faces carry an Ising variable, i.e. $+$ or $-$ \cite{KazIsing}.

A classical result \cite{Stau2,Eynlp} is that $Y(x) = V_1'(x) - W_1^{(0)}(x)$ satisfies an algebraic equation $E(x,Y(x)) = 0$, where:
\beq
E(x,y) = (\mathcal{V}'_1(x)-y)(\mathcal{V}'_2(y)-x)+1-\lim_{N \rightarrow \infty} \frac{1}{N}\Big\langle \Tr \frac{\mathcal{V}'_1(x)-\mathcal{V}'_1(M_1)}{x-M_1}\,\frac{\mathcal{V}'_2(y)-\mathcal{V}'_2(M_2)}{y-M_2}\Big\rangle
\eeq
is a polynomial of $x$ and $y$. This equation $E(x,y)=0$ defines a compact Riemann surface $\mathcal{C}$, endowed with a covering $x\,:\,\mathcal{C} \rightarrow \widehat{\mathbb{C}}$ of degree $d_2 = \mathrm{deg}\,\mathcal{V}_2$. In other words, $\mathcal{C}$ is realized as $d_2$ sheets $\mathcal{C}_k$ of $\mathbb{\mathbb{C}}$, glued together at the zeroes of $\dd x$ (the ramification points) along certain cuts $\gamma_j$ joining them. There is a distinguished sheet $\mathcal{C}_0$ for which $W_1^0(x(z)) \sim 1/x(z)$ when $x(z) \rightarrow \infty$ while $z \in \mathcal{C}_0$. We assume that the branchpoints are simple, so that we are in the framework of Section~\ref{S1} with $U = \mathcal{C}_0$, $V_j$ are neighborhoods of the cuts in $\mathcal{C}$, and the involution $\iota$ is the local exchange of sheets bordered by a cut. Let $V = \coprod_{j} V_j$. If $z \in V_j\cap\mathcal{C}_k$ and $\mathcal{C}_{k'}\cap\mathcal{C}_k \neq \gamma_j$, we also denote $\jmath_{k'}$ the map sending $z$ to $z' \in V \cap \mathcal{C}_{k'}$ so that $x(z) = x(z')$: it corresponds to sending $z$ to a distant sheet. The result of \cite{EORes,CEO06} can be rephrased as:
\begin{proposition}
\label{kiosae}
Introduce as usual:
\beq
\label{defsq2}\omega_n^{g}(z_1,\ldots,z_n) = W_n^{g}(x(z_1),\ldots,x(z_n))\dd x(z_1)\cdots\dd x(z_n) + \delta_{n,2}\delta_{g,0}\,\frac{\dd x(z_1)\,\dd x(z_2)}{(x(z_1) - x(z_2))^2}.
\eeq
and assume that $\omega_1^0$ is an off-critical $1$-form.  Then, $\omega_{\bullet}^{\bullet}$ satisfies solvable linear and quadratic loop equations. More precisely, they satisfy, for any $n,g$, any $z_I = (z_2,\ldots,z_n) \in U^{n - 1}$,
\beq
\label{sjqe} \forall z \in V_k,\qquad \omega_n^{g}(z,z_I) + \omega_n^g(\iota(z),z_I) + \sum_{k'} \omega_n^g(\jmath_{k'}(z),z_I) = \delta_{g,0}\Big(\delta_{n,1}\dd \mathcal{V}_1(z) + \delta_{n,2}\frac{\dd x(z)\,\dd x(z_2)}{(x(z) - x(z_2))^2}\Big),
\eeq
and $\omega_2^0$ is the unique fundamental $2$-form of the $2^{\mathrm{nd}}$ kind for the compact Riemann surface $\mathcal{C}$, whose period on the $\gamma_j$ vanish.
\end{proposition}
We observe that \eqref{sjqe} is very similar to the linear loop equation \eqref{sjq} satisfied by $\omega_1^0$ in the repulsive particle systems, where the operator $\mathcal{O}$ is replaced by the sum of evaluations at all sheets of $x\,:\,\mathcal{C} \rightarrow \widehat{\mathbb{C}}$. From Theorem~\ref{kiosae}, it was  shown in \cite{CEO06} that $\omega_n^g$ in the 2-hermitian matrix model is given by the topological recursion \eqref{topore}. In this article, we have seen that such a result can be unified with others in a more general theory.

\subsection{$1$-matrix model on elliptic curves}
\label{S455}
Let us revisit as a special case of repulsive particle systems, the case where the particles interact pairwise with the Coulomb interaction on the torus $\mathbb{T} = \mathbb{C}/\mathbb{L}$ where $\mathbb{L} =  \mathbb{Z} + \tau\mathbb{Z}$ and $\mathrm{Im}\,\tau > 0$. The model is defined by:
\bea
\label{kip}\dd\varpi(z_1,\ldots,z_n) & = & \prod_{i = 1}^N \dd \lambda_i\,e^{-\frac{N\beta \mathcal{V}(\lambda_i)}{2}} \prod_{1 \leq i < j \leq N} |\theta(\lambda_i - \lambda_j)|^{\beta}, \nonumber \\
Z_N & = & \int_{(\Gamma_0)^N} \dd\varpi(\lambda_1,\ldots,\lambda_N).
\eea
The theta function we consider is the first Jacobi theta function:
\beq
\theta(z) = \sum_{n \in \mathbb{Z} + 1/2} -{\rm i} e^{{\rm i}\pi\tau n^2 + 2{\rm i}\pi n (z + 1/2)},
\eeq
and it satisfies:
\beq
\theta(z + n + m\tau) = (-1)^n\,e^{-2{\rm i}\pi(z + m\tau/2)m}\,\theta(z).
\eeq
In particular, $\theta(\lambda_i - \lambda_j)$ has a simple zero when $\lambda_i \rightarrow \lambda_j$.
The conclusions of \S~\ref{S3} in the context of the topological expansion assuming $\beta = 2$, and leading to the topological recursion formula \eqref{topore}, can be applied to this case. In this paragraph, we will rather illustrate that it can be helpful to define the correlators slightly differently than \eqref{conera}, taking into account the underlying geometry, in order to write the Schwinger-Dyson equations in a simpler form than in \S~\ref{SDSQ}. We shall see the same trick at work in any genus in \S~\ref{SkI}, but we focus here on the genus $1$ case, since it allows to take a pedestrian route without too technical computations. This is a useful intermediate step in order to solve explicitly the master equation for $\omega_1^0$ and $\omega_2^0$, or to justify for convergent matrix models described by \eqref{kip} the existence of a topological expansion under suitable assumptions, following e.g. \cite{APS01,BG11}.

Let $M = \mathrm{diag}(\lambda_1,\ldots,\lambda_N)$. Here, it is natural to define the correlators by:
\beq
\label{cothe}\mathcal{W}_k(z_1,\ldots,z_k) = \Big\langle\prod_{j = 1}^{k} \Tr\,(\ln \theta)'(z_j - \Lambda)\Big\rangle_c.
\eeq
Notice that $\mathcal{W}_k(z_1,\ldots,z_k)$ are holomorphic on $\mathbb{C}\setminus\big(\Gamma_0 + \mathbb{L}\big)$. We can derive Schwinger-Dyson equations for the correlators, e.g. by performing the infinitesimal change of variable:
\beq
\lambda_i \rightarrow \lambda_i + \varepsilon\,(\ln\theta)'(z - \lambda),
\eeq
and express the invariance of the integral under change of variables. Assuming like in \S~\ref{SDSQ} that $\mathcal{V}$ is such that there is no boundary terms, we find:
\bea
\Big\langle \sum_{i = 1}^{N} -(\ln\theta)''(z - \lambda_i) + \frac{\beta}{2} \sum_{i \neq j} (\ln\theta)'(\lambda_i - \lambda_j)\big[(\ln\theta)'(z - \lambda_i) - (\ln\theta)'(z - \lambda_j)\big] & & \nonumber \\
- \sum_{i = 1}^{N} \frac{N\beta}{2}\,\mathcal{V}'(\lambda_i)\,(\ln\theta)'(z - \lambda_i)\Big\rangle & = & 0.
\eea
This equation can be simplified \cite{Bonelliandco} by a procedure analogous to partial fraction expansion for rational functions, based on the following relation:
\beq
\label{pfe}(\ln\theta)'(v - u)\big[(\ln\theta)'(u) - (\ln\theta)'(v)\big] = (\ln\theta)'(u)(\ln\theta)'(v) - \frac{1}{2}\Big(\frac{\theta''(u)}{\theta(u)} + \frac{\theta''(v)}{\theta(v)} + \frac{\theta''(v - u)}{\theta(v - u)}\Big) + \frac{1}{2}\frac{\theta'''(0)}{\theta'(0)}.
\eeq
Hence:
\bea
\Big\langle -\Big(1 - \frac{\beta}{2}\Big)\Tr (\ln \theta)''(z - M) + \frac{\beta}{2}\Tr (\ln\theta)'(z - M)\,\Tr(\ln\theta)'(z - M) & & \nonumber \\
- \frac{\beta N}{2}\,\Tr \frac{\theta''(z - M)}{\theta(z - M)} - \frac{\beta}{4}\,\Tr\Big(\frac{\theta''(M\otimes\mathbf{1} - \mathbf{1}\otimes M)}{\theta(M\otimes\mathbf{1} - \mathbf{1}\otimes M)} - \frac{\theta'''(0)}{\theta'(0)}\Big) & & \nonumber \\
- \frac{N\beta}{2}\,\Tr V'(M)(\ln \theta)'(z - M) \Big\rangle & = & 0.
\eea
Notice that both the measure and the observables we consider depend explicitly on the modulus $\tau$, since:
\beq
\theta''(z) = 4{\rm i}\pi\partial_{\tau}\theta(z).
\eeq
We observe that the measure \eqref{kip} depends explicitly on $\tau$. We denote $\partial_{\tau,\mathrm{mes}.}$ the derivative with respect to this dependence:
\beq
4{\rm i}\pi \partial_{\tau} \ln Z = 4{\rm i}\pi\partial_{\tau,\mathrm{mes}.}\ln Z = \Big\langle \frac{\beta}{2} \sum_{i \neq j} \frac{\theta''(\lambda_i - \lambda_j)}{\theta(\lambda_i - \lambda_j)}\Big\rangle.
\eeq
and recognize a part of the constant in the second line. Let us define:
\bea
\mathcal{V}\cdot W_1(z) & = & \Big\langle \Tr V'(M)\,(\ln\theta)'(z - M)\Big\rangle, \nonumber \\
Q_1(z) & = & \Big\langle \Tr \frac{\theta''(z - M)}{\theta(z - M)}\Big\rangle.
\eea
Notice that $Q_1(z)$ is holomorphic in the fundamental domain, and is such that:
\beq
Q_1(z + m + n\tau) - Q_1(z) = -4\pi^2n^2 - 4{\rm i}\pi n\,W_1(z).
\eeq
We thus have:
\bea
\Big(\frac{\beta}{2} - 1\Big)\mathcal{W}_1'(z) + \frac{\beta}{2}\big(\mathcal{W}_2(z,z) + \mathcal{W}_1^2(z)\big) - \frac{N\beta}{2} \oint_{\gamma} \frac{\dd \zeta}{2{\rm i}\pi} \mathcal{V}'(\zeta)\,(\ln \theta)'(z - \zeta)\,\mathcal{W}_1(\zeta) & & \nonumber \\
\label{SDY1}  + \frac{\beta N(N - 1)}{4}\,\frac{\theta'''(0)}{\theta'(0)} - 2{\rm i}\pi\,\partial_{\tau}\ln Z - \frac{N\beta}{2}\,Q_1(z) & = & 0. \nonumber
\eea
To derive higher Schwinger-Dyson equations,  we define the insertion operator $\partial/\partial \mathcal{V}(z_j)$ by perturbing the potential
\beq
\mathcal{V}(z) \rightarrow \mathcal{V}(z) - \frac{2\varepsilon}{\beta N}\,(\ln\theta)'(z_j - z),
\eeq
and differentiating with respect to $\varepsilon$, and then setting $\varepsilon = 0$. When we apply it on correlators, we find:
\beq
\frac{\partial}{\partial \mathcal{V}(z_k)} \mathcal{W}_{k - 1}(z_1,\ldots,z_k) = \mathcal{W}_k(z_1,\ldots,z_k).
\eeq
On the derivative of the potential:
\beq
\frac{\partial}{\partial \mathcal{V}(z_j)} \frac{\beta N \mathcal{V}'(z)}{2} = (\ln\theta)''(z_j - z).
\eeq
So, if we apply $\prod_{j \in I} \frac{\partial}{\partial \mathcal{V}(z_j)}$ on \eqref{SDY1}, we find:
\bea
\label{cum} \Big(\frac{\beta}{2} - 1\Big) + \partial_z W_k(z,z_I) + \frac{\beta}{2}\Big(W_{k + 1}(z,z,z_I) + \sum_{J \subseteq I} W_{|J| + 1}(z,z_J)W_{n - |J|}(z,z_{I\setminus J})\Big) & & \\
- \frac{N\beta}{2} \oint_{\gamma} \frac{\dd\zeta}{2{\rm i}\pi}\,(\ln \theta)'(z - \zeta)\,\mathcal{W}_k(z,z_I) - 2{\rm i}\pi\,\partial_{\tau,\mathrm{mes}.} W_{k - 1}(z_I) - \frac{N\beta}{2}\,Q_k(z;z_I) & & \nonumber \\
 - \sum_{j \in I} \Big\langle \Tr\,\partial_{z_j}\big[(\ln\theta)'(z_j - M)\,(\ln\theta)'(z - M)\big]\,\prod_{j' \neq j} \Tr(\ln \theta)'(z_{j'} - M)\Big\rangle_c & = & 0, \nonumber
\eea
where:
\beq
Q_k(z;z_I) = \Big\langle \Tr\,\frac{\theta''(z - M)}{\theta(z - M)}\,\prod_{j \in I} \Tr (\ln\theta)'(z_j - M)\Big\rangle_c
\eeq
is again holomorphic in the variable $z$ in the fundamental domain, and satisfies for $k \geq 2$:
\beq
Q_k(z + m + n\tau;z_I) - Q(z;z_I) = -4{\rm i}\pi n\,\mathcal{W}_{k}(z,z_I).
\eeq
We transform the last term of Eqn.~\ref{cum} thanks to the partial fraction expansion identity (Eqn.~\ref{pfe}):
\bea
\label{saju}& & -\partial_{z_j}\big[(\ln\theta)'(z_j - M)\,(\ln\theta)'(z - M)\big] \\
& = & \partial_{z_j}\big\{(\ln\theta)'(z - z_j)\big[(\ln\theta)'(z - M) - (\ln\theta)'(z_j - M)\big]\big\} - \frac{1}{2}\partial_{z_j}\Big(\frac{\theta''(z_j - M)}{\theta(z_j - M)}\Big) + f(z,z_j). \nonumber
\eea
Because only cumulants are involved in Eqn.~\ref{cum}, the term $f(z,z_j)$ will disappear from the computation. The correlators \eqref{cothe} depend on $\tau$ via the measure \eqref{kip}, and also via the logarithmic derivative of the theta function. We denote $\partial_{\tau,\mathrm{obs}.}$ the derivative with respect to this last dependance only. In other words, $\partial_{\tau} = \partial_{\tau,\mathrm{mes}.} + \partial_{\tau,\mathrm{obs}.}$. We observe that, if we differentiate $\mathcal{W}_{k - 1}(z_I)$ with respect to the $\tau$-dependence of the observable:
\bea
4{\rm i}\pi\partial_{\tau,\mathrm{obs}.} W_{k - 1}(z_I) & = & \sum_{j \in I} \Big\langle\Tr\Big(\frac{\theta'''(z_j - M)}{\theta(z_j - M)} - \frac{\theta'(z_j - M)\theta''(z_j - M)}{\theta^2(z_j - M)}\Big)\prod_{j' \neq j} \Tr(\ln \theta)'(z_{j'} - M)\Big\rangle_c \nonumber \\ 
& = & \sum_{j \in I} \Big\langle \Tr\partial_{z_j}\Big(\frac{\theta''(z_j - M)}{\theta(z_j - M)}\Big)\prod_{j' \neq j} \Tr (\ln\theta)'(z_{j'} - M)\Big\rangle_c,
\eea
we can retrieve the second term in \eqref{saju}. Thus, the loop equation at rank $k$ reads:
\bea
\Big(\frac{\beta}{2} - 1\Big)\partial_z\mathcal{W}_{k}(z) + \partial_z \mathcal{W}_k(z,z_I) + \frac{\beta}{2}\Big(\mathcal{W}_{k + 1}(z,z,z_I) + \sum_{J \subseteq I} \mathcal{W}_{|J| + 1}(z,z_J)\mathcal{W}_{n - |J|}(z,z_{I\setminus J})\Big) & & \\
- \frac{N\beta}{2} \oint_{\gamma} \frac{\dd\zeta}{2{\rm i}\pi}(\ln \theta)'(z - \zeta)\,\mathcal{W}_k(z,z_I) + \sum_{j \in J} \partial_{z_j}\big\{(\ln\theta)'(z - z_j)\big[\mathcal{W}_{k - 1}(z,z_{I\setminus\{j\}}) - \mathcal{W}_{k - 1}(z_I)\big]\big\} & & \nonumber \\
 -  2{\rm i}\pi\partial_{\tau} \mathcal{W}_{k - 1}(z_I) - \frac{N\beta}{2}\,Q_k(z;z_I) & = & 0. \nonumber
\eea
If one wishes to take a potential with logarithmic singularities:
\beq
\mathcal{V}(z) = \mathcal{V}_0(z) + \sum_{l = 1}^{L} \frac{2\alpha_l}{N\beta} \ln\theta(\xi_l - z),
\eeq
we may decompose again using \eqref{pfe}:
\bea
\oint_{\gamma} \frac{\dd\zeta}{2{\rm i}\pi}\,(\ln\theta)'(z - \zeta)\,\mathcal{V}'(\zeta)\,\mathcal{W}_1(\zeta) & \!\!\!\!\! = \!\!\!\!\!\!\! & \oint_{\gamma} \frac{\dd\zeta}{2{\rm i}\pi}\,(\ln\theta)'(z - \zeta)\,\mathcal{V}'_0(\zeta)\,\mathcal{W}_1(\zeta) \\
& &  + \sum_{l = 1}^{L} \alpha_l\Big\{(\ln\theta)'(z - \xi_l)\big[W_1(z) - W_1(\xi_l)\big] \nonumber \\
& & - 2{\rm i}\pi\partial_{\tau,\mathcal{V} - \mathcal{V}_0} \ln Z - 2{\rm i}\pi\partial_{\tau,\mathrm{mes}.} \ln Z - \frac{N}{2}\Big(\frac{\theta''(z - \xi_l)}{\theta(z - \xi_l)} - \frac{\theta'''(0)}{\theta'(0)}\Big)\Big\}, \nonumber 
\eea
where $\partial_{\tau,\mathcal{V} - \mathcal{V}_0}$ denotes the differentiation with respect to the $\tau$-dependence of $(\mathcal{V} - \mathcal{V}_0)$. This computation can be carried on to higher correlators:
\bea
\oint_{\gamma} \frac{\dd\zeta}{2{\rm i}\pi}\,(\ln \theta)'(z - \zeta)\,\mathcal{V}'(\zeta) \mathcal{W}_k(\zeta,z_I) & = & \oint_{\gamma} \frac{\dd\zeta}{2{\rm i}\pi}\,(\ln \theta)'(z - \zeta)\,\mathcal{V}'_0(\zeta) \mathcal{W}_k(\zeta,z_I) \\
& & + \sum_{l = 1}^{L} \alpha_l\Big\{(\ln\theta)'(z - \xi_l)\big[\mathcal{W}_k(z,z_I) - \mathcal{W}_k(\xi_l,z_I)\big] \nonumber \\
& & - 2{\rm i}\pi \partial_{\tau,\mathcal{V} - \mathcal{V}_0} \mathcal{W}_{k - 1}(z_I) - 2{\rm i}\pi\partial_{\tau,\mathrm{mes}.} \mathcal{W}_{k - 1}(z_I) \Big\}.\nonumber
\eea
The $\alpha_l$ are called \emph{momenta}. In particular, we observe that, when the momenta sum up to $1$, the term involving $\partial_{\tau,\mathrm{mes}.}$ disappear from the loop equations.

\subsection{Liouville theory on higher genus surface}
\label{SkI}
Consider a given Riemann surface $\Sigma_{\mathfrak g}$ of genus $\mathfrak g$ with $L$ marked points $z_1,\dots,z_L$, equipped with a symplectic basis of cycles $(\mathcal{A}_{\mathfrak{h}},\mathcal{B}_{\mathfrak{h}})_{1 \leq \mathfrak{h} \leq \mathfrak{g}}$. Given $n$ momenta $\alpha_1,\ldots,\alpha_n$, one wishes to compute the $n$-point functions in Liouville field theory:
\beq
\langle V_{\alpha_1}(z_1) \dots V_{\alpha_n}(z_n) \rangle_{\Sigma_{\mathfrak g}}.
\eeq
We introduce a variable $N$ such that:
\beq
N b = \,\sum_{l=1}^L \alpha_j + (1-\mathfrak{g})\,\left(b+\frac{1}{b}\right).
\eeq
It is known \cite{Bonelliandco,DotsenkoFateev,DotsenkoFateev2}, that this correlation function can be retrieved from the analytic continuation of the following integrals, first defined for a nonnegative integer $N$:
\bea
\label{inge} & & Z_{k_1,\dots,k_r}(p_1,\dots,p_{\mathfrak g}) \nonumber \\
&=& \int_{\gamma_1^{k_1}\times\dots\times \gamma_r^{k_r}} \prod_{i = 1}^N \Big[\Omega(\lambda_i)\,\ee{4\pi\, b \sum_{i=1}^N\sum_{j=1}^{\mathfrak g} p_j a_j(\lambda_i)}\,\prod_{l=1}^L E(\lambda_i,z_l)^{2b\alpha_j}\prod_{j=1}^{2\mathfrak g-2}\,\left(E(\lambda_i,\xi_j) \right)^{-1-b^2}   \Big] \nonumber \\
& & \times \prod_{1 \leq i<j \leq N} E(\lambda_i,\lambda_j)^{-2b^2},
\eea
where $\gamma_i$ are some paths on the Riemann surface, $E(\lambda,\lambda')$ is the prime form, and $\mathbf{a}(\lambda) = (a_1(\lambda),\dots,a_{\mathfrak g}(\lambda))$ is the Abel map associated to the $\mathcal{A}$-cycles, $\Omega$ is an arbitrary but fixed holomorphic 1-form on $\Sigma_{\mathfrak g}$, whose $2\mathfrak g-2$ zeroes are denoted $\xi_1,\dots, \xi_{2\mathfrak g-2}$. If $\mathbf{c}$ is an odd non-singular characteristics, the prime form is defined as \cite{MumTata}:
\beq
E(\lambda,\lambda') = \frac{\theta(\mathbf{a}(\lambda) - \mathbf{a}(\lambda') + \mathbf{c})}{\sqrt{\dd h_{\mathbf{c}}(\lambda)\,\dd h_{\mathbf{c}}(\lambda')}},\qquad \dd h_{\mathbf{c}}(\lambda) = \sum_{i = 1}^{\mathfrak{g}} \partial_{a_i}\theta(\mathbf{c})\,\dd a_i(\lambda).
\eeq
In the definition of the Abel map and of the prime form, the choice of the $(\mathcal{A},\mathcal{B})$ cycles is implicit. Notice that, since the prime form is a $(-1/2,-1/2)$ form, the integrand in \eqref{inge} is indeed a 1-form in each variable. This model is a repulsive particle system in the sense of Section~\ref{S3}, with Dyson index $\beta = -2b^2$. $Q = b + 1/b$ is called the background charge.

The case $\beta = 2$ correspond to $b^2 = -1$, i.e. Liouville theory without zero background charge. In this case, \eqref{inge} simplifies to:
\beq
\label{nge}Z_{k_1,\dots,k_r}(p_1,\dots,p_{\mathfrak g}) 
= \int_{\gamma_1^{k_1}\times\dots\times \gamma_r^{k_r}}  \prod_{1 \leq i < j \leq N} E(\lambda_i,\lambda_j)^{2}\,\,
 \prod_{i=1}^N \Big[\prod_{j=1}^n E(\lambda_i,z_j)^{2\ii \alpha_j}\,\ee{4\ii\pi\,\sum_{i=1}^N\sum_{j=1}^{\mathfrak g} p_j a_j(\lambda_i)}\Big].
\eeq
The large $N$ techniques make sense in the regime where the momenta $\alpha_j$ are large, i.e. we write:
\beq
\alpha_j = {\rm i}N\,\tilde{\alpha}_j,\qquad p_j = {\rm i}N\,\tilde{p}_j.
\eeq

Our purpose is to illustrate rather than study in detail, we shall ignore here the issues about the choice of contours, of convergence, of strict convexity of the interactions, of rigorous proof of existence of a large $N$ expansion of topological type, which should actually be nested problems. We rather want to show how the techniques of Section~\ref{S3} leading to a topological recursion apply, and focus on the description of the spectral curve, i.e. of the initial data $\omega_1^0$ \cite{Bonelliandco} and $\omega_2^0$. As in \S~\ref{S455}, it is convenient to define the correlators not by \eqref{conera}, but taking into account the geometry, by:
\beq
\mathcal{W}_{n}(x_1,\ldots,x_n) = \Big\langle \prod_{i = 1}^n \Tr\,\dd_{x_i}\ln \theta(\mathbf{a}(x_i) - \mathbf{a}(M) + \mathbf{c}) \Big\rangle_c.
\eeq
where averages are understood with respect to the measure \eqref{nge}. We assume an expansion of topological type:
\beq
\mathcal{W}_n(x_1,\ldots,x_n) = \sum_{g \geq 0} N^{2 - 2g - n}\,\mathcal{W}_{n}^{g}(x_1,\ldots,x_n),
\eeq
and we set:
\beq
\omega_n^{g}(x_1,\ldots,x_n) = \mathcal{W}_{n}^{g}(x_1,\ldots,x_n) + \delta_{n,2}\delta_{g,0}\,B(x_1,x_2),
\eeq
where:
\beq
\label{bre}B(x_1,x_2) = \dd_{x_1}\dd_{x_2} \ln \theta(\mathbf{a}(x_1) - \mathbf{a}(x_2) + \mathbf{c})
\eeq
is the fundamental $2$-form of the $2^{\mathrm{nd}}$ kind of $\Sigma_{\mathfrak{g}}$ associated to the basis of cycles $(\mathcal{A},\mathcal{B})$. $\omega_n^g$ satisfy solvable linear loop equations and quadratic loop equations in the sense of Definitions~\ref{def10}-\ref{def20}, and more precisely if we denote $\Gamma_0 \subseteq \Sigma_{\mathfrak{g}}$ the cut locus of $\omega_n^g$ and $x_I = (x_2,\ldots,x_n)$ a set of spectator variables, we have for any $x \in \mathring{\Gamma_0}$:
\beq
\label{relia}\omega_n^{g}(x_1 + {\rm i}0,x_I) + \omega_n^{g}(x_1 - {\rm i}0,x_I) = \delta_{g,0}\big(\delta_{n,1}\dd \mathcal{V}(x_1) + \delta_{n,2}\delta_{g,0}\,B(x_1,x_2)\big),
\eeq
where we introduced the potential:
\beq
\label{potL}\mathcal{V}(x) = -2\sum_{l = 1}^L \tilde{\alpha}_{j} \ln \theta(\mathbf{a}(x) - \mathbf{a}(z_j) + \mathbf{c}) - 4\pi \sum_{i = 1}^{\mathfrak{g}} \tilde{p}_{j}\,a_{j}(x).
\eeq

Let us assume that $\Gamma_0 = \bigcup_{j = 0}^r \gamma_j$ is a disjoint union of open arcs $\gamma_j$. We deduce that $\omega_n^g$ is a $n$-form on $\mathcal{C}^n$, where $\mathcal{C}$ is the \emph{Schottky double} of $\Sigma_{\mathfrak{g}}$, i.e. is a two-sheeted covering of $\Sigma_{\mathfrak{g}}$ obtained by gluing two copies $\Sigma_{\mathfrak{g}}^{(1)}$ and $\Sigma_{\mathfrak{g}}^{(2)}$ with opposite orientations along $\Gamma_0$. It has genus $2\mathfrak{g} + r - 1$. One can define a symplectic basis of cycles of $\mathcal{C}$, which consists of the two copies of the cycles $(\mathcal{A}_{\mathfrak{h}},\mathcal{B}_{\mathfrak{h}})_{1 \leq \mathfrak{h} \leq \mathfrak{g}}$, to which we add  $\mathcal{A}$-cycles surrounding $\gamma_j$ for $j \in \ldbrack 1,r \rdbrack$, and the $\mathcal{B}$-cycles going from $\gamma_j$ to $\gamma_{j + 1}$ in $\Sigma_{\mathfrak{g}}^{(1)}$ and coming back to its initial point in $\Sigma_{\mathfrak{g}}^{(2)}$. With this choice, $\omega_2^0$ is the unique fundamental $2$-form of the $2^{\mathrm{nd}}$ kind on $\mathcal{C}$ with zero periods along all $\mathcal{A}$-cycles, thus given by \eqref{bre} where the right-hand side is replaced by the theta function on $\mathcal{C}$ instead of $\Sigma_{\mathfrak{g}}$. Here, the involution $\iota$ is defined globally on $\mathcal{C}$ and correspond to the exchange of sheets. We can rewrite \eqref{relia} as the relation, for any $x \in \mathcal{C}$:
\beq
\label{ture}\omega_n^{g}(x_1,x_I) + \omega_n^{g}(\iota(x_1),x_I) = \delta_{g,0}\big(\delta_{n,1}\dd \mathcal{V}(x_1) + \delta_{n,2}\delta_{g,0}\,B(x_1,x_2)\big).
\eeq
$\omega_1^0$ can be constructed from \eqref{tildeaq}-\eqref{taq}, i.e. as the $1$-form in $\mathcal{C}$ having singularities in the second sheet prescribed by the right-hand side in \eqref{ture} and no singularities in the first sheet $\omega_1^0$ and $\omega_2^0$ are thus totally explicit in this case, once the cut locus $\Gamma_0$ is determined (see e.g. the discussion in \cite{Bonelliandco}). By the results of Section~\ref{S1}, we can deduce that topological recursion holds:
\beq
\omega_n^{g}(z_0,z_I) = \sum_{\alpha \in \Gamma^{\mathrm{fix}}} \Res_{z \rightarrow \alpha} K(z_0,z)\Big(\omega_{n + 1}^{g - 1}(z,\iota(z),z_I) + \sum_{\substack{J \subseteq I,\,\,0 \leq h \leq g \\ (J,h) \neq (\emptyset,0),(I,g)}} \omega_{|J| + 1}^{h}(z,z_J)\,\omega_{n - |J|}^{g - h}(\iota(z),z_{I\setminus J})\Big), \nonumber
\eeq
with recursion kernel:
\beq
K(z_0,z) = \frac{-\frac{1}{2}\int_{\iota(z)}^z \omega_2^0(z_0,\cdot)}{\omega_1^0(z) - \omega_1^0(\iota(z))},
\eeq
and this formula leads to an effective computation of the $\omega_n^g$, again provided the cut locus $\Gamma_0$ is known.

This generalizes the well-known situation of the 1-hermitian matrix model, for which we have $\Sigma_{\mathfrak{g} = 0} = \widehat{\mathbb{C}}$ and $\mathcal{C}$ is a hyperelliptic surface \cite{ACKM,ACM92,Ake96}. The description of the spectral curve as a Schottky double only relied on the expression of the 2-point interaction as a prime form. The results are also valid for general potentials $\mathcal{V}$.

\section{Conclusion}
\label{conclu}

We have proposed and studied the properties of a hierarchy of "abstract loop equations", which turns out to be solved by a topological recursion. The initial data is a $1$-point function $\omega_1^0$, and a 2-point function $\omega_2^0$. Actually, they only need to be defined in a neighborhood of ramification points for most of the properties of the topological recursion to hold. Saying that, we underline that we have not adressed here the issue of symplectic invariance, as we now comment.

From the physics point of view, $\omega_1^0 = y\dd x$ defines a spectral curve and encodes the geometry, while the data $\omega_2^0$ should define a way to quantize it. The stable $\omega_n^g$ are considered as quantum corrections determined by those two inputs. There are several (related) notions of quantization here. A first one is illustrated in the recent work of \cite{Dimofte}, i.e. replacing $x$ and $y$ by operators $\hat{x}$ and $\hat{y}$ so that $[\hat{y},\hat{x}] = g_s$, and the classical equation satisfied by $x,y$ becomes a $D$-module where wave functions sit. A second one is closer to the idea of quantum cohomology, and this picture is now well understood in the example of topological strings in toric Calabi-Yau $3$-folds $\mathfrak{X}$ after the recent work of two of the authors \cite{EOBKMP} proving the BKMP conjecture \cite{BKMP}. The perturbative partition function of these topological string theories are wave functions of a geometric quantization of the cohomology of the target space considered. Such a quantization is not unique and requires a choice of polarization to be performed. Following \cite{ABK,Wholo}, one can explicitly identify such a choice of quantization with a choice of $\om_2^0$. The symmetries exhibited by the perturbative wave function built in this way by the topological recursion then depend on this choice. In this case, $\omega_1^0$ is determined by the mirror curve of $\mathfrak{X}$. Besides, localization techniques can be used to compute generating series of Gromov-Witten invariants. The data of $\omega_2^0$ arise from the weights of the edges of the localization graphs, and it turns out in this case that it constructs a specific $\omega_2^0$ as a globally defined, fundamental $2$-form on the $2^{\mathrm{nd}}$ kind on the compactification of the mirror curve \cite{EOBKMP}. As explained in Section~\ref{S7}, we have described in this work hierarchies of loop equations where the weight of edges in the analog of those "localization graphs" is arbitrary. A "good quantization" should be covariant with respect to canonical transformations, i.e. transformations such that $(x,y) \rightarrow (x',y')$ such that $\dd x \wedge \dd y = \dd x' \wedge \dd y'$, also called symplectic transformations. This will certainly impose constraints on the possible choice of $\omega_2^0$ for a given geometry if we require that $F^g$ are invariant under those transformations. Actually, the only non trivial obstruction is the effect of $(x,y) \rightarrow (-y,x)$. Although the theory presented in Section~\ref{S1} makes sense \emph{for any $\omega_2^0$} (defined for example by formal series expansion at the ramification points, with arbitrary coefficients), it will not in general enjoy symplectic invariance\footnote{Let us give an example to argue that we do not expect in general symplectic invariance. Let us consider $x,y$ meromorphic functions on a compact curve $\overline{\Sigma}$, with $\omega_2^0$ a fundamental $2$-form of the $2^{\mathrm{nd}}$ kind globally defined on $\overline{\Sigma}^2$. \cite{EO2MM} tells us that $F^g$ are invariant under $(x,y) \rightarrow (-y,x)$. With $\omega_2^0$ being hold fixed, the topological recursion for the spectral curve $(\overline{\Sigma},x,y)$ involves residues at ramification points of $x\,:\,\overline{\Sigma} \rightarrow \widehat{\mathbb{C}}$, while that for the spectral curve $(\overline{\Sigma},x,y)$ involve residues at ramification points of $y\,:\,\overline{\Sigma} \rightarrow \widehat{\mathbb{C}}$. Let $\alpha$ be a ramification point for $x$, and replace now $\overline{\Sigma}$ by $\Sigma' = \overline{\Sigma}\setminus\{\alpha\}$. Since we are systematically forgetting a ramification point, it is likely that $F^g$ for $(\Sigma',-y,x)$ will be different from $F^g$ for $(\overline{\Sigma},-y,x)$, while $F^g$ for $(\Sigma',x,y)$ is obviously the same as $F^g$ for $(\overline{\Sigma},x,y)$. So, the $F^g$ with the open Riemann surface $\Sigma'$ as part of the initial data, and $\omega_2^0$ coming from $\overline{\Sigma}$, will not be symplectic invariants.}, so its application to quantum field theories may involve restricting oneself to the maximal subclass of $\omega_2^0$ for which symplectic invariance holds, which has not been yet clearly identified.  For example, for applying this procedure to the solving of integrable systems of topological type arising in the study of Frobenius manifold \cite{Dub} (and thus computing Gromov-Witten invariants in a larger setup), the choice of $\om_2^0$ is completely fixed by the one of $\om_1^0$ following the work of Givental \cite{G01} where both of these arguments correspond to a canonical transformation arising from a change of polarization in a geometrically quantized theory. This canonical transformation allows to go from a very specific point in the Frobenius manifold where the wave function is completely factorized as a product of KdV tau functions to an arbitrary point where the wave function is the generating function of Gromov-Witten invariants of a specific manifold. In general, we guess that symplectic invariance is only possible when there exists a globally defined underlying geometry.

Our present work extends the range of potential applications of the topological recursion, and includes the former applications to the large $N$ expansion of the 1-hermitian matrix model, the 2-hermitian matrix model, with possibly several cuts. In particular, it allows to treat "generalized matrix models" (also called "repulsive particle systems"), where the pairwise interaction of eigenvalues described by a squared Vandermonde only as an asymptotic behavior at short distances. We mention below some more examples (even with $\beta = 2$) of repulsive particle systems.
\begin{itemize}
\item[$\bullet$] Chern-Simons theory with general gauge group $\mathsf{G}$ on torus knot complement. The model is described by a measure on the Cartan subalgebra of the Lie algebra of $\mathsf{G}$:
\beq
\dd\varpi(t_1,\ldots,t_r) = \prod_{\alpha > 0} \mathrm{sinh}\Big(\frac{\alpha\cdot t}{2P}\Big)\mathrm{sinh}\Big(\frac{\alpha\cdot t}{2Q}\Big)\,\prod_{i = 1}^r e^{-\frac{Nt_i^2}{2PQu}}\dd t_i.
\eeq
\item[$\bullet$] Chern-Simons theory on Seifert manifolds $\mathsf{X}\big(\frac{P_1}{Q_1},\ldots,\frac{P_m}{Q_m}\big)$. It was shown in \cite{Law,MarinoCSM} that the contribution of the trivial flat connection  leads to a repulsive particle system. For $\mathsf{G} = \mathrm{U}(N)$, it is defined by the measure on $\mathbb{R}^N$:
\bea
 \dd\tilde{\varpi}(t_1,\ldots,t_N) & = & \frac{1}{\tilde{Z}_N^{(\mathbf{P},\mathbf{Q})}} \prod_{1 \leq i < j \leq N} \Big[\sinh^2\Big(\frac{t_i - t_j}{2}\Big) \prod_{l = 1}^m \frac{\sinh\big(\frac{t_i - t_j}{2P_l}\big)}{\sinh\big(\frac{t_i - t_j}{2}\big)}\Big]\,\prod_{i = 1}^N e^{-N\,\tilde{\mathcal{V}}(t)}\dd t_i, \nonumber \\
 \label{sefa} \tilde{\mathcal{V}}(t) & = & \Big(\sum_{i = 1}^m \frac{Q_i}{P_i}\Big)\frac{t^2}{2u}.
\eea
where $u = N\ln q$, while other contributions correspond to multispecies analogs of \eqref{sefa}.  The case $m = 1$ correspond to lens spaces $\mathsf{X}\big(\frac{P}{Q}\big)$, and coincides with the measure \eqref{fj} relevant for the $(P,Q)$ torus knot upon the change of variable $t_i \rightarrow t_i/Q$.
\item[$\bullet$] ABJM matrix model on $\mathbb{S}_3$ with $U(N_1)\times U(N_2)$ gauge group \cite{MarinoLec,MarPut}. The measure to study is:
\bea
\dd\varpi(t_1,\ldots,t_{N_1},s_1,\ldots,s_{N_2}) & = & \prod_{1 \leq i < j \leq N_1} \mathrm{sinh}^2\Big(\frac{t_i - t_j}{2}\Big)\prod_{1 \leq i < j \leq N_2} \mathrm{sinh}^2\Big(\frac{s_i - s_j}{2}\Big) \\
& & \times  \prod_{\substack{ 1 \leq i \leq N_1 \\ 1 \leq j \leq N_2}} \mathrm{cosh}^{-2}\Big(\frac{t_i - s_j}{2}\Big) \Big(\prod_{i = 1}^{N_1} e^{-\frac{t_i^2}{2g_s}}\dd t_i\Big)\Big(\prod_{i = 1}^{N_2} e^{\frac{s_i^2}{2g_s}}\dd s_i\Big). \nonumber
\eea
where the minus sign in the Gaussian potential for $s_i$'s has to be understood as an analytical continuation. $\omega_1^0$ has been computed (see e.g. \cite{MarinoLec}), and the same techniques of resolution would lead to an expression for $\omega_2^0$.
\end{itemize}
With Corollary~\ref{aporo}, we deduce that the topological expansion in those models is computed by the topological recursion.

Our formalism opens the way to a systematic study of the $(q,t)$ deformation of matrix models representations for various enumerative geometry problems \cite{Aga2,AgaSha}. In particular, although it has been found in \cite{BriSte} that the $\beta$ deformation alone \cite{CE06} of the usual topological recursion could not be used to extend the BKMP conjecture of \cite{BKMP} to compute refined topological strings amplitudes in toric Calabi-Yau $3$-folds, we expect that our formalism contain the appropriate deformation to handle it.

We let to a future work the study of the $\beta$ deformation of our construction, i.e. the cases where the short distance behavior of the two-point interaction is described by a Vandermonde to the power $\beta$. This appear for example to compute the conformal blocks of Liouville theory on positive genus surfaces (as discussed in \cite{Bonelliandco} and \S~\ref{SkI}), and Nekrasov partition functions \cite{Nekra,Sulko,BourgAGT}. The two models should be related by AGT conjecture \cite{AGTc}. We therefore hope that studying them both with the topological recursion would lead to some insight about the AGT conjecture. 

Repulsive particle systems correspond to the case of generalized matrix models which have an eigenvalue representation. We believe that many other matrix models where such diagonalization is not possible, should still be solvable by the topological recursion, as it has already been shown for the 2-hermitian matrix model \cite{EORes,CEO06} (see also \S~\ref{2mm}) and for the chain of matrices \cite{EPf}. The problem is reduced to that of putting the Schwinger-Dyson in the form of abstract loop equations. For instance, we hope that the present formalism will be applicable to all quiver matrix models.

We have shown in full generality in Section~\ref{S5} that maps endowed with self-avoiding loop configuration of all topologies are enumerated by the topological recursion, even in cases where $\omega_1^0$ and $\omega_2^0$ is not known in closed form. The same will be true for maps with a 6-vertex model, for which $\omega_1^0$ was found at the critical point \cite{PZJ6v}, and then in the general off-critical case \cite{Kos6v}. Since solving for $\omega_1^0$ is not more difficult than solving for $\omega_2^0$, we can now consider that the 6-vertex model is solved explicitly for maps of all topologies. There exists other statistical physics model on maps, like the Potts model on maps with controlled face degree \cite{KazPotts}, the asymmetric ABAB model \cite{KZinnABAB}, for which $\omega_1^0$ is known (and thus $\omega_2^0$ can be obtained by similarly techniques even if not found in the literature), and it would be interesting to know if combinatorics leads to loop equations for generating series of maps with certain boundary conditions, i.e. if the model can be solved by the topological recursion. We stress that, although the topological recursion \eqref{topore} for $\omega_n^g$ can be written as a sum over skeletons of genus $g$ surfaces with $n$ boundaries (see the diagrammatics in \cite[Section 3]{EORev}), it is not clear if each term counts a certain class of maps (with a statistical physics model or not). Hence, there is no known bijective interpretation of \eqref{topore}, although its geometric content seems clear.

Let us come to a few technical comments. The key idea in our approach was to define only local spectral curves, obtained by doubling an open Riemann surface across cuts. As explained earlier, this local approach is very powerful since it allows to effectively deal with supposedly higher dimensional version of the topological recursion, for example in the study of Gromov-Witten theories where the gluing of local spectral curves maps to gluing of target spaces. We have considered for simplicity only spectral curves having cuts ending at simple ramification points (see the definition of domains in \S~\ref{ss1}), but it seems possible to include ramification points of higher order, or branching cuts (for instance, tree-like cuts). The analog of the topological recursion in this case has been proposed in \cite{PratsFerrer3,Bouchetal}, and was then shown to arise naturally in a limit situation where branchpoints are simple but collide \cite{PratsFerrer3,BouEyn}.

Besides, we observe that cuts which are closed cycles do not give any contribution to the topological recursion, provided some analyticity assumptions. The form of the topological recursion suggest that $1/N$ corrections in generalized matrix models where eigenvalues live on a contour $\Gamma$ come only from fluctuations around the edges of the large $N$ support $\Gamma_0 \subseteq \Gamma$, while collective effects do not allow more than $O(N^{-\infty})$ contributions from the bulk. We are however not sure of the interpretation of this observation. For instance, in unitary matrix models (which are normal matrix models where eigenvalues live on the unit circle), would it mean that the partition function and the correlators cannot have an expansion in $1/N$, or that the first subleading term is actually a $O(N^{-\infty})$, unless a singularity in the potential allows for another behavior ?

When all quantities can be continued analytically on a compact Riemann surface $\mathcal{C}$ of genus $\mathfrak{g}$ (i.e. in the framework of the usual topological recursion of \cite{EORev}), the set $\mathfrak{B}$ of fundamental $2$-form of the $2^{\mathrm{nd}}$ kind is an affine space of dimension $\mathfrak{g}(\mathfrak{g} + 1)/2$, and specifying normalization on certain $\mathcal{A}$-cycles selects a unique $\omega_2^0 \in \mathfrak{B}$. Let us denote $\tau$ the matrix of periods of $\mathcal{C}$ with respect to a choice of symplectic basis $(\mathcal{A},\mathcal{B})$ of homology cycles on $\mathcal{C}$. One may consider the $\omega_n^g|_{(\epsilon,\kappa)}$ produced by the topological recursion from the initial data twisted by $3^{\mathrm{rd}}$ kind deformations:
\bea
\omega_1^0|_{\epsilon} & = & \omega_1^0 + \sum_{j = 1}^{\mathfrak{g}} {2{\rm i}\pi}\,\epsilon_j\,\dd a_j\\
\omega_2^0|_{\kappa} & = & \omega_2^0 + \sum_{1 \leq j,k\leq  \mathfrak{g}} {2{\rm i}\pi}\,\kappa_{j,k}\,\dd a_{j}\otimes\dd a_{k}
\eea
where $(\dd a_j)_{1 \leq j \leq \mathfrak{g}}$ is a basis of holomorphic $1$-forms dual to $\mathcal{A}$-cycles. The result is that stable $\omega_{n}^{g}|_{\epsilon,\kappa}$ are either modular but non holomorphic in $\tau$ and satisfy holomorphic anomaly equation, or holomorphic in $\tau$ but non-modular \cite{EMO}. One may wonder if similar "modular" and "holomorphic anomaly properties" could be formulated in the more general formalism presented here. "Holomorphic" here refers to the dependence in the moduli of the initial data, and actually it is not completely clear what should be the good moduli space(s) for the initial data $(\omega_1^0,\omega_2^0)$ in the framework of abstract loop equations.

In repulsive particle systems, the strict convexity assumption (Definition~\ref{stconv}) was essential to characterize completely the solution of loop equations: it implied the "solvability" of loop equations in the sense of Definition~\ref{def10}-$(iv)$. New large $N$ phenomena are expected when this assumption is not satisfied, and the problem is largely open since even the standard results of potential theory concerning the leading order cannot be applied. One may imagine a competition between several equilibrium measures, or/and that entropic effects become relevant at leading order. 

\newpage

\subsection*{Acknowledgments}

G.B. thanks A.~Guionnet, C.~Beasley, T.~Dimofte, P.~Zinn-Justin for discussions and references, and is supported by Fonds Europ\'een S16905 (UE7 - CONFRA) and Swiss NSF. N.O. thanks Ivan Kostov for discussions.

\appendix

\section{Some examples of strictly convex interactions}
\label{appA}
We first make obvious remarks. If $R_0(x,y)$ defines a strictly convex interaction on $I^2$, and $J \subseteq I$, then $R_0(x,y)$ defines a strictly convex interaction on $J^2$. The product of two strictly convex interaction is a strictly convex interaction. Any positive power of a strictly convex interaction is a strictly convex interaction. If $\varphi\,:\,I \rightarrow J$ is a diffeomorphism, $R_0(x,y)$ is a strictly convex interaction on $J^2$ iff then $R_0(\varphi(x),\varphi(y))$ is a strictly convex interaction on $I^2$.

\begin{lemma}
\label{sqkq}
If $L\,:\,\mathbb{R} \rightarrow \mathbb{R}$ be even function with negative Fourier transform, $R_{0}(x,y) = \exp(L(x - y))$ defines a strictly convex interaction on $\mathbb{R}$, and for any $\rho \in [-2,2]$, $R_0(x,y) = \exp\big(L(x- y) + \frac{\rho}{2}L(x + y)\big)$ defines a strictly convex interaction on $\mathbb{R}_+$. The same result holds if $\mathbb{R}$ is replaced by $\mathbb{R}/\mathbb{Z}$, and the Fourier transform by its discrete analog.
\end{lemma}
It is straightforward to generalize this result to $s$ species of particles for $s \geq 1$. For instance, $\rho$ becomes a matrix and a sufficient condition in the second case for having strictly convex interactions is that $-2 \leq \rho \leq 2$ as a matrix. We have seen an avatar of this fact in Lemma~\ref{uj}.

\vspace{0.2cm}

\noindent\textbf{Proof.} For any signed measure $\nu$ with total mass $0$, we may write:
\beq
\iint_{\mathbb{R}^2} \dd\nu(x)\dd\nu(y)\,L(x - y) = \int_{\mathbb{R}} \hat{L}(s)\,|\hat{\nu}(s)|^2
\eeq
This expression is nonpositive, and left-hand side is finite iff the right-hand side is finite. It vanishes iff $\hat{\nu} = 0$, i.e iff $\nu = 0$. Similarly, for $\nu$ a signed measure with total mass $0$ and supported on $\mathbb{R}_+$, and for any $\rho \in [-2,2]$, we may write:
\beq
\label{inda}\iint_{\mathbb{R}^2} \dd\nu(x)\dd\nu(y)\Big(L(x - y) + \frac{\rho}{2}L(x + y)\Big) =  \int_{\mathbb{R}} \Big[(2 + \rho)\big(\mathrm{Re}\,\hat{\nu}(s)\big)^2 + (2 - \rho)\big(\mathrm{Im}\,\hat{\nu}(s)\big)^2\Big]
\eeq
This expression is nonpositive, and left-hand side is finite iff the right-hand side is finite. When $\rho \in ]-2,2[$, it vanishes as before iff $\nu = 0$. When $\rho = -2$, it vanishes iff $\mathrm{Re}\,\hat{\nu} = 0$, which means that for any even bounded continuous function $\varphi$, we must have $\nu(\varphi) = 0$. Since $\nu$ is supported on $\mathbb{R}_+$, this implies $\nu = 0$. A similar proof works for $\rho = 2$, replacing $\mathrm{Re}$ by $\mathrm{Im}$ and "even" by "odd", showing that \eqref{inda} vanishes iff $\nu = 0$. \hfill $\Box$

We can apply Lemma \ref{inda} to the following function defined on $\mathbb{R}$:
\beq
L(x) = \ln |x| = \lim_{\eta \rightarrow 0}\Big(\ln\eta - \mathrm{Re}\,\int_{0}^{\infty} \frac{e^{-\eta s}}{s}(e^{isx} - 1)\dd s\Big),
\eeq
and to the following functions defined on $\mathbb{R}/\mathbb{Z}$:
\beq
L(x) = \ln |\sin\pi x| = -\sum_{m = 1}^{\infty} \frac{\cos(mx)}{m},\qquad L(x) = \ln|\vartheta_i(x|\tau)| = -\sum_{m = 0}^{\infty} b_{m,i}\,\cos(mx).
\eeq
where $q = e^{{\rm i}\pi\tau}$ and:
\beq
b_{m,1} = (-1)^mb_{m,2} = \frac{1}{m} + \frac{2q^{2m}}{m(1 - q^{2m})},\qquad \qquad b_{m,3} = (-1)^mb_{m,4} = \frac{2q^{2m}}{m(1 - q^{2m})}.
\eeq
Since we have:
\bea
\mathrm{sn}_{k}\Big(\frac{x}{2K(k)}\Big) & = & \frac{\vartheta_3(0)}{\vartheta_2(0)}\,\frac{\vartheta_1(x)}{\vartheta_4(x)}, \nonumber \\
\mathrm{sc}_{k}\Big(\frac{x}{2K(k)}\Big) & = & \frac{\vartheta_3(0)}{\vartheta_4(0)}\,\frac{\vartheta_1(x)}{\vartheta_2(x)}, \nonumber \\
\mathrm{sd}_k\Big(\frac{x}{2K(k)}\Big) & = & \frac{\vartheta_3^2(0)}{\vartheta_2(0)\vartheta_4(0)}\,\frac{\vartheta_1(x)}{\vartheta_3(x)},
\eea
we can also apply Lemma~\ref{inda} to the functions $L = \mathrm{sn}_{k}$, $\mathrm{sd}_{k}$ and $\mathrm{sc}_{k}$ defined on $\mathbb{R}/(2K(k)\mathbb{Z})$.

\section{Table of main notations and definitions}
\label{appB}

\begin{tabular}{lll}
\textit{Reference} & \textit{Notation} & \textit{Name} \vspace{0.2cm} \\
\hline & & \\
& $\widehat{\mathbb{C}}$ & Riemann sphere $\mathbb{C}\cup\{\infty\}$\\
Def.~\ref{dea1} & $\Ms(U)$  & space of meromorphic $1$-forms on $U$ \\
Def.~\ref{dea1} & $\Hs(U)$ & space of holomorphic $1$-forms on $U$ \\
Def.~\ref{dea1} & $\Ms'(\{p\})$ & space of germs of meromorphic $1$-forms near $p$ \\
Def.~\ref{dea1} & $\Ms'_{-}(\{p\})$ & negative Laurent polynomials at $p$ \\
Def.~\ref{dea2} & $\Hs_{\Gamma}^{\mathrm{inv}}(U)$ & space of holomorphic $1$-forms in $U$, continuable across $\Gamma$ \\
Def.~\ref{dea3} & $\Ls_{\Gamma}(U)$ & \\
Def.~\ref{Cau} & $G(z,z_0)$ & local Cauchy kernel \\
Def.~\ref{repa} & & space representable by residues \\
Def.~\ref{reapo} & & normalized space \\
\S~\ref{Htil} & $\tilde{\mathcal{H}}$ & image of $\mathcal{H}$ by the map \eqref{resp} \\
\S~\ref{Htil} & $\mathcal{H}_{G}$ & maximal space representable by residue for $G$ \\
\S~\ref{htil2} & $\mathcal{H}_n$ & symmetric $n$-forms in $n$ variables \vspace{0.2cm} \\ 
\hline & & \\
Def.~\ref{domainn} & $U$ & domain \\
\S~\ref{ss1} & $\Gamma = \bigcup_{j = 1}^r \gamma_j \subseteq \partial U$ & cuts \\
\S~\ref{ss1} & $U_{\Gamma}$ & open Riemann surface containing $U$ as physical sheet \\
\S~\ref{ss1} & $U_j$ & neighborhood of $\gamma_j$ in the physical sheet \\
\S~\ref{ss1} & $V_j = U_j \coprod_{\gamma_j} U_j'$ & annular neighborhood of $\gamma_j$ in $U_{\Gamma}$ \\
\S~\ref{ss1} & $V = \coprod_{j = 1}^r V_j$ & \\
\S~\ref{ss1}& $\iota$ & local involution across the cuts \\
Def.~\ref{afix} & $\Gamma^{\mathrm{fix}}$ & set of ramification points (fixed points of $\iota$) \\
Eq.~\ref{pola}  & $\mathcal{S}f(z) = f(z) + f(\iota(z))$ & analytic continuation of $2 \times$(principal value) \\
Eq.~\ref{pola} & $\Delta f(z) = f(z) - f(\iota(z))$ & analytic continuation of the discontinuity of the cut \vspace{0.2cm} \\
\hline & & \\
Def.~\ref{stable} & & stable $(n,g)$ \\
Def.~\ref{offc} & & off-critical $1$-form \\
Def.~\ref{def10} & & linear loop equations \\
Def.~\ref{def10} & & solvable linear loop equations \\ 
Def.~\ref{def20} & & quadratic loop equations \\
Eq.~\ref{topore} & & topological recursion formula \\
Eq.~\ref{reck} & & recursion kernel \\
\S~\ref{wdv} & & free energies \\
\S~\ref{spsa} & & spectral curve \\
\S~\ref{diq} & & WDVV-compatible variation \\
\S~\ref{diq} & $\delta_{\Omega}$ & infinitesimal deformation of the initial data \vspace{0.2cm} \\
\hline & & \\
Eq.~\ref{mes2} & & repulsive particle systems \\
Eq.~\ref{mes2} & $\mathcal{V}$ & potential \\
Eq.~\ref{mes2} & $N$ & number of particles \\
Eq.~\ref{OOdef} & $\mathcal{O}$ & a compact integral operator \\
Eq.~\ref{mes2} & $\beta$ & Dyson index (power of the Vandermonde determinant) \\
Eq.~\ref{mes2} & $R$ & two-point interactions (excluding Vandermonde) \\
\S~\ref{Popo} & $R_0$ & two-point interactions (including Vandermonde) \\
Eq.~\ref{mes2} & $\rho$ & power of the two-point interaction ($-$ loop fugacity) \vspace{0.2cm} \\
\hline & & 
\end{tabular}
\newpage
\begin{tabular}{lll}
\textit{Reference} & \textit{Notation} & \textit{Name} \vspace{0.2cm} \\
\hline & & \\ Def.~\ref{sconf} & & strongly confining interactions \\
Def.~\ref{stconv} & & strictly convex interactions \\
Prop.~\ref{saqq} & $\Gamma_0$ & range of integration \\
Prop.~\ref{saqq} & $\Gamma$ & support of the equilibrium measure \\
\S~\ref{Popo} & $\mathcal{P}_1(\Gamma)$ & set of probability measures on $\Gamma$ \\
Def.~\ref{defBerg} & $B(z_0,z)$ & fundamental $2$-form of the $2^{\mathrm{nd}}$ kind \\
Page~\pageref{1stk}& $h_{j}$ & $1^{\mathrm{st}}$ kind differentials \\
Def.~\ref{top1} & & expansion of topological type \vspace{0.2cm} \\
\hline & & \\
\S~\ref{61} & $s$ & number of colors \\
\S~\ref{scolorl} & $\mathsf{M}$ & set of maps \\
\S~\ref{611} & $s\mathsf{CMT}$ & set of $s$-colored maps with tubes \\
\S~\ref{maps2} & $s\mathsf{ML}$ & set of $s$-colored maps with a loop configuration \\
\S~\ref{61} & $u,u_k$ & weights per vertex \\
\S~\ref{61} & $t_{k,j}$ & weights per face \\
\S~\ref{61} & $g_{k,l;i,j}$ & weights per face carrying a loop \\
\S~\ref{61} & $-\rho,-\rho_{k,l}$ & loop fugacity (power of the two-point interaction) \\
\S~\ref{61} & $n$ & number of boundaries \\
\S~\ref{61} & $g$ & genus \vspace{0.2cm} \\
\hline & & 
\end{tabular}

\newpage

\newcommand{\etalchar}[1]{$^{#1}$}

\end{document}